\pdfoutput=1
\documentclass[12pt,preprint]{aastex}
\usepackage{natbib}
\usepackage{graphicx}
\usepackage{hyperref}

\newcommand{\teff}{\ensuremath{T_{\rm eff}}}
\newcommand{\mstar}{\ensuremath{M_\star}}
\newcommand{\msune}{$M_{\odot}$}
\newcommand{\ms}{\mbox{m\,s$^{-1}~$}}
\newcommand{\mse}{\mbox{m\,s$^{-1}$}}

\newcommand{\kms}{\mbox{km\,s$^{-1}~$}}
\newcommand{\mearth}{$M_\oplus$~}
\newcommand{\mearthe}{$M_\oplus$}
\newcommand{\mjup}{$M_{\rm J}~$}
\newcommand{\mjupe}{$M_{\rm J}$}
\newcommand{\msini}{\ensuremath{M_{\rm p} \sin i}}
\newcommand{\rphk}{\ensuremath{R'_{\mbox{\scriptsize HK}}}}
\newcommand{\lrphk}{\ensuremath{\log{\rphk}}}
\newcommand{\caii}{\ion{Ca}{2} H \& K}

\newcommand{\ninject}{5000~}
\newcommand{\recoverthresh}{25\%~}

\newcommand{\vsini}{\ensuremath{v \sin{i}}}

\begin{document}

\title{Limits on Planetary Companions from Doppler Surveys of Nearby Stars\altaffilmark{1}}
\author{
Andrew W.\ Howard, 
Benjamin J.\ Fulton 
}
\altaffiltext{1}{Based on observations obtained at the W.\,M.\,Keck Observatory, 
                      which is operated jointly by the University of California and the 
                      California Institute of Technology.  Keck time has been granted by  
                      NASA,  the University of California, and the University of Hawaii.} 
\affil{Institute for Astronomy, University of Hawaii, 2680 Woodlawn Drive, Honolulu, HI 96822, USA}

%
%
%
%
%




%
%
%
\begin{abstract}
Most of our knowledge of planets orbiting nearby stars comes from Doppler surveys.  
For spaced-based, high-contrast imaging missions, nearby stars with Doppler-discovered planets are attractive targets.  
The known orbits tell imaging missions where and when to observe, and the dynamically-determined masses provide important constraints for the interpretation of planetary spectra.  
Quantifying the set of planet masses and orbits that could have been detected will enable more efficient planet discovery and characterization.
We analyzed Doppler measurements from Lick and Keck Observatories collected by the California Planet Survey.  
We focused on stars that are likely targets for three space-based planet imaging mission concepts studied by NASA---WFIRST-AFTA, Exo-C, and Exo-S.  
The Doppler targets are primarily F8 and later main sequence stars, with observations spanning 1987--2014.
We identified 76 stars with Doppler measurements from the prospective mission target lists. 
We developed an automated planet search and a methodology to estimate the pipeline completeness using injection and recovery tests.
We applied this machinery to the Doppler data and computed planet detection limits for each star as a function of planet minimum mass and semi-major axis.  
For typical stars in the survey, we are sensitive to approximately Saturn-mass planets inside of 1 AU, Jupiter-mass planets inside of $\sim$3 AU, and our sensitivity declines out to $\sim$10 AU.  
For the best Doppler targets, we are sensitive to Neptune-mass planets in 3 AU orbits.  
Using an idealized model of Doppler survey completeness, we forecast the precision of future surveys of non-ideal Doppler targets that are likely targets of imaging missions. 
\end{abstract}




\section{Introduction}

NASA recently studied three mission concepts capable of  directly imaging extrasolar planets from space.   
The WFIRST-AFTA mission is envisioned as a wide-field, infrared imager on a 2.4-m telescope 
that could accomplish broad astrophysical goals \citep{Spergel2013}.
The baseline mission  includes a coronagraph in the instrument suite for exoplanet detection 
and characterization \citep{Traub2014,Goullioud2014}.  
In addition, two Science and Technology Definition Teams (STDTs) studied probe-scale 
(cost less than \$1B) 
mission concepts for the direct detection of extrasolar planets orbiting nearby stars.  
Exo-C is a mission concept based on a telescope with an internal coronagraph to 
generate the ultra-high contrast images needed for planet detection \citep{Stapelfeldt2014}. 
Exo-S would image extrasolar planets using a pair of spacecraft flying in formation --- 
an external occulter (starshade) and a telescope \citep{Seager2014AAS}. 

Teams studying these possible missions are considering the scientific return of each, 
including the discovery of new exoplanets orbiting nearby stars, 
spectroscopy of exoplanet atmospheres, and imaging of debris and protoplanetary disks.
Each mission needs a list of nearby stars to search for new planets and to characterize existing planetary systems.  
The characteristics of these stars (brightness, spectral type, distance, age, sky position, multiplicity, etc.) make 
some more favorable targets than others.  
Target selection will also be based on which stars have  planets and debris disks already known.  
Most of our knowledge of exoplanets orbiting nearby stars comes from Doppler surveys that started in the late 1980s and 
have improved in sensitivity since then.  These observational surveys have discovered hundreds of planets orbiting nearby stars, 
some of which are possible targets for Exo-C, Exo-S, and WFIRST-AFTA.   
Many of the Doppler target stars have clear non-detections of planets after hundreds of measurements spanning a decade or more.  
These non-detections can be quantified, as we do below, 
using injection-recovery tests to measure planet search completeness as a function of planet  
minimum mass ($\msini$) and orbital period (with assumptions about eccentricities).
Knowing that certain stars lack planets with particular masses and semi-major axes constitutes useful information for the imaging missions.  
The priority of some stars may be reduced  if the imaging search space in planet mass and semi-major axis has 
already been ruled out by  Doppler observations.
Conversely, other stars may be more attractive if Doppler surveys can rule out dynamically disruptive 
giant planets and the imaging missions are sensitive to smaller planets that are undetectable by the Doppler measurements.  

This paper is the product of a study 
titled `Radial Velocity Data Review in Support of Direct Imaging Mission Concept Reports'
that was carried out by the authors under contract from NASA Jet Propulsion Laboratory.  
We were charged with  
identifying the nearby stars that are likely targets of Exo-C, Exo-S, and WFIRST-AFTA and  
that have historical Doppler measurements at Lick Observatory and Keck Observatory taken 
by the California Planet Survey \citep{Howard10b}.  
For stars with Doppler measurements, we were to estimate the region of discovery space 
(planet masses and orbital semi-major axes) in which planets can be ruled out.  
These ``completeness'' estimates, and the lists of discovered 
planets orbiting those stars, provide a valuable input to the planning and eventual operation of 
space-based planet imaging missions.  


This report is organized as follows.  
The next two sections describe target lists for the Exo-S, Exo-C, and WFIRST-AFTA imaging studies (Sec.\ \ref{sec:targets_imaging}) and the 
California Planet Search using Doppler spectroscopy (Sec.\ \ref{sec:rv_surveys}). 
We review Keplerian orbits in Sec.\  \ref{sec:doppler}, describe 
our automated Doppler planet search in Sec.\ \ref{sec:rv_analysis}, 
and simulate the sensitivity of future Doppler searches in Sec.\ \ref{sec:recommendations}.  
Machine-readable data files included with this report are described in Appendix \ref{app:data} and can be downloaded from
\href{http://exoplanetarchive.ipac.caltech.edu/docs/contributed\_data.html}{http://exoplanetarchive.ipac.caltech.edu/docs/contributed\_data.html}.  
The remaining appendices provide exhaustive data that is too large for the main text. 
Appendix \ref{app:stellar_prop_nodata} includes a 
complete list of the Exo-C, Exo-S, and WFIRST-AFTA stars that are not part of our Lick/Keck Doppler survey and the likely reasons for their exclusion.  
Appendix \ref{app:completeness} contains plots summarizing the 
search results and completeness estimates for every star with Doppler measurements.


\section{Target Lists for the Exo-C, Exo-S, and WFIRST-AFTA Studies}
\label{sec:targets_imaging}

The Exo-S, Exo-C, and WFIRST-AFTA study teams  selected preliminary lists of nearby stars for their missions.  
These target lists contain real stars and allow for simulations of mission performance that depend on genuine stellar properties.
Target lists were provided to us by representatives of the study teams (M.\ Turnbull and K.\ Stapelfeldt)  
and are current as of July 31, 2014.
Stellar properties were taken from the ``Properties of Nearby Stars'' page  
on the NASA ExEP website\footnote{See \url{http://nexsci.caltech.edu/missions/EXEP/EXEPstarlist.html}}. 
This  catalog was assembled by M.\ Turnbull and includes catalog numbers from 
Hipparcos, Henry Draper (HD), and Gliese catalogs, as well as 
sky coordinates, distances, $V$-band brightnesses, $B-V$ colors, 
spectral types, and luminosities. 
These Exo-S, Exo-C, and WFIRST-AFTA target stars are listed in two tables below.   
Table \ref{tab:targets_with_data} includes stars for which we have Doppler measurements and provide completeness measurements. 
Stellar masses (to convert orbital periods into semi-major axes) are mostly derived from stellar synthesis fits using 
Spectroscopy Made Easy \citep{Valenti2005}.
Table \ref{tab:target_exclusion_reasons} gives a summary of the reasons that stars were not part of the Lick/Keck doppler planet search programs.
Table \ref{tab:targets_without_data} lists Exo-S, Exo-C, and WFIRST-AFTA targets without  Doppler measurements.
We adopt HD numbers as the primary stellar names throughout this report because they are the standard star names 
used internally for our Keck program.

\begin{figure}
         \begin{center}
              \includegraphics[width=0.5\textwidth]{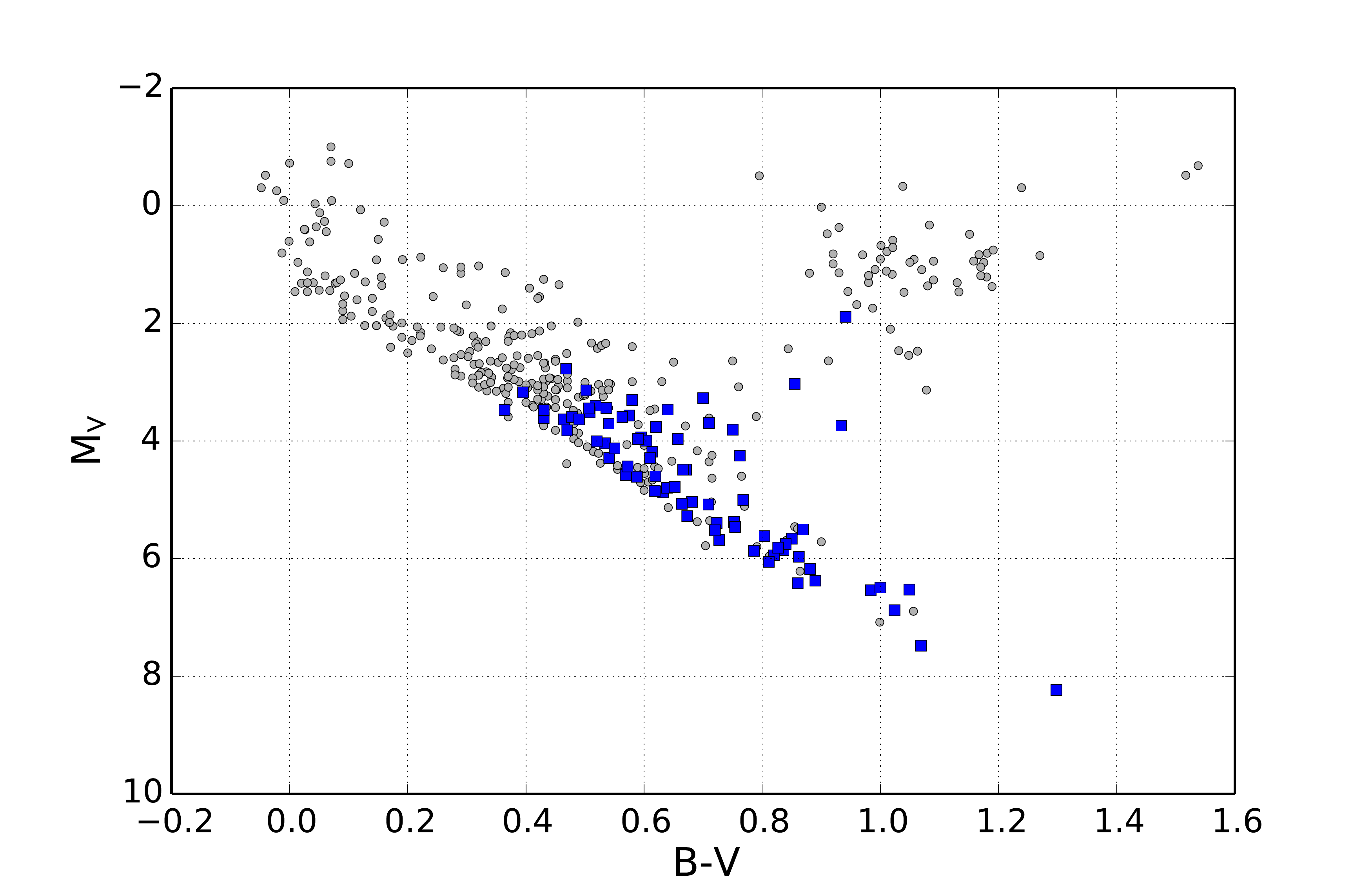}  
         \end{center}
         \caption{HR diagram of stars in the Exo-S, Exo-C, and WFIRST-AFTA target lists with (blue squares) and without (gray circles) 
         Doppler observations from Lick and Keck Observatories.  Doppler searches favor late-F through mid-M type dwarfs 
         for planet detectability.  Imaging searches for planets in reflected light typically prefer bright stars, 
         which are dominated by early type stars.  
         The region of overlap encompasses primarily F8--K0 dwarfs.  
         }
         \label{fig:hr_diag}
\end{figure}

\begin{deluxetable}{rrrlrrrrrll}
\tabletypesize{\footnotesize}
\tablecaption{Imaging Target Stars with Doppler Measurements
\label{tab:targets_with_data}}
\tablewidth{0pt}
\tablehead{
  \colhead{Hipp.}   & 
  \colhead{HD}   & 
  \colhead{Gliese}   & 
  \colhead{Target$^{\rm a}$} &
  \colhead{Dist.}  &
  \colhead{$V$}  &
  \colhead{$B$--$V$}  &
  \colhead{\teff}  &
  \colhead{\mstar}  &
  \colhead{Sp.\ T.}  & 
  \colhead{Notes} \\
  \colhead{No.}   & 
  \colhead{No.}   & 
  \colhead{No.}   & 
  \colhead{Lists} &
  \colhead{(pc)}  &
  \colhead{(mag)}  &
  \colhead{(mag)}  &
  \colhead{(K)}  &
  \colhead{(\msune)}  &
  \colhead{}  & 
  \colhead{} 
}
\startdata
    544 &     166 &         5 &      S &  13.671 &  6.060 &  0.752 &  5577 &  0.987 &       K0V &  \\ 
   3093 &    3651 &        27 &      S &  11.060 &  5.880 &  0.850 &  5221 &  0.920 &       K0V &  \\ 
   3821 &    4614 &       34A &     S, C, A &   5.944 &  3.452 &  0.569 &  5941 &  0.986 &       G3V &  \\ 
  3765 &    4628 &        33 &     S, C &   7.455 &  5.740 &  0.890 &  4944 &  0.728 &       K1V &  \\ 
   7513 &    9826 &        61 &      C, A &  13.492 &  4.090 &  0.536 &  6213 &  1.310 &       F8V &  \\ 
   7981 &   10476 &        68 &     S, C, A &   7.532 &  5.240 &  0.836 &  5181 &  0.863 &       K1V &  \\ 
   8102 &   10700 &        71 &     S, C, A &   3.650 &  3.490 &  0.727 &  5283 &  0.762 &     G8V &  \\ 
   8362 &   10780 &        75 &      S &  10.067 &  5.630 &  0.804 &  5327 &  0.879 &       K0V &  \\ 
  12114 &   16160 &      105A &      S &   7.180 &  5.819 &  0.984 &  4866 &  0.750 &       K3V &  \\ 
  13402 &   17925 &       117 &      S &  10.352 &  6.046 &  0.862 &  5236 &  0.800 &       K1V &  \\ 
  14632 &   19373 &       124 &     S, C, A &  10.541 &  4.050 &  0.595 &  6032 &  1.204 &       G0V &  \\ 
  14954 &   19994 &   128A &      A &  22.580 &  5.070 &  0.580 &  6188 &  1.300 &     F8.5V &   \\ 
  15457 &   20630 &       137 &     S, C, A &   9.140 &  4.842 &  0.681 &  5742 &  0.900 &       G5V &  \\ 
  16537 &   22049 &       144 &      C, A &   3.213 &  3.714 &  0.881 &  5146 &  0.815 &     K2V &  young \\ 
  16852 &   22484 &       147 &     S, C, A &  13.963 &  4.290 &  0.575 &  6038 &  1.206 &       F8V &  \\ 
  17378 &   23249 &       150 &     S, C &   9.041 &  3.518 &  0.934 &  5095 &  1.163 &      K0IV & subgiant \\ 
  18859 &   25457 &   159 &      A &  18.830 &  5.380 &  0.520 &  6308 &  1.179 &       F7V &   young \\
  19849 &   26965 &      166A &     S, C, A &   4.984 &  4.430 &  0.820 &  5151 &  0.775 &     K1V &  \\ 
  22263 &   30495 &       177 &      S &  13.277 &  5.486 &  0.632 &  5759 &  1.000 &       G3V &  \\ 
  22449 &   30652 &       178 &     S, C, A &   8.068 &  3.167 &  0.464 &  6424 &  1.236 &       F6V &  \\ 
  23311 &   32147 &       183 &      S &   8.708 &  6.225 &  1.049 &  4827 &  0.821 &     K3V &  \\ 
  23835 &   32923 &      188A &     S, C, A &  15.434 &  4.910 &  0.657 &  5694 &  1.001 &       G4V &  \\ 
  24813 &   34411 &       197 &     S, C, A &  12.631 &  4.691 &  0.614 &  5911 &  1.105 &       G0V &  \\ 
  26779 &   37394 &       211 &      S &  12.277 &  6.198 &  0.840 &  5351 &  0.919 &       K1V & young \\ 
  29650 &   43042 &   3390 &      A &  20.810 &  5.200 &  0.430 &  6418 &  1.000 &  F5.5IV-V &  early type \\  
  32480 &   48682 &   245 &      A &  16.720 &  5.240 &  0.550 &  6064 &  1.177 &       G0V &  \\
  39780 &   67228 &   \nodata &      A &  23.290 &  5.300 &  0.640 &  5862 &  1.220 &      G2IV &  \\
  40693 &   69830 &       302 &      S &  12.494 &  5.943 &  0.754 &  5361 &  0.871 &       K0V &  \\ 
  40843 &   69897 &   303 &      A &  18.270 &  5.130 &  0.470 &  6294 &  1.095 &       F6V &  early type \\
  42438 &   72905 &       311 &      S &  14.355 &  5.630 &  0.618 &  5920 &  1.000 &    G1.5Vb & young \\ 
  43587 &   75732 &      324A &      S &  12.341 &  5.960 &  0.869 &  5235 &  0.966 &       G8V &  \\ 
  47592 &   84117 &       364 &     S, C, A &  15.013 &  4.924 &  0.534 &  6152 &  1.159 &       G0V &  \\ 
  48113 &   84737 &   368 &      A &  18.370 &  5.080 &  0.620 &  5960 &  1.164 &    G0V &  \\
  49081 &   86728 &       376 &      S, A &  15.047 &  5.375 &  0.671 &  5700 &  1.095 &       G1V &  \\ 
  51459 &   90839 &       395 &     S, C, A &  12.780 &  4.820 &  0.541 &  6126 &  1.200 &       F8V &  \\ 
  53721 &   95128 &   407 &      A &  14.060 &  5.030 &  0.610 &  5882 &  1.083 &       G1V & \\ 
  56452 &  100623 &      432A &      S &   9.559 &  5.959 &  0.811 &  5189 &  0.747 &       K0V &  \\ 
  56997 &  101501 &       434 &     S, C, A &   9.612 &  5.307 &  0.723 &  5488 &  0.913 &    G8V &  \\ 
  57443 &  102365 &      442A &     S, C, A &   9.220 &  4.890 &  0.664 &  5630 &  0.863 &     G5V &  \\ 
  57757 &  102870 &       449 &     S, C, A &  10.929 &  3.590 &  0.518 &  6161 &  1.100 &       F9V &  \\ 
  58576 &  104304 &       454 &      S &  12.763 &  5.533 &  0.768 &  5565 &  1.018 &      G8V &   \\ 
  61317 &  109358 &       475 &     S, C, A &   8.440 &  4.241 &  0.588 &  5930 &  0.929 &       G0V &  \\ 
  64394 &  114710 &       502 &     S, C, A &   9.129 &  4.237 &  0.572 &  6075 &  1.150 &       G0V &  \\ 
  64408 &  114613 &   501.2 &      A &  20.670 &  4.850 &  0.700 &  5782 &  1.266 &       G3V &  \\
  64792 &  115383 &   504 &      A &  17.560 &  5.190 &  0.590 &  6234 &  2.310 &      G0IV & young \\
  64924 &  115617 &       506 &     S, C, A &   8.555 &  4.740 &  0.709 &  5571 &  0.954 &       G5V &    \\ 
  65721 &  117176 &   512.1 &      A &  17.990 &  4.970 &  0.710 &  5545 &  1.109 &       G5V & \\ 
  67275 &  120136 &      527A &     S, C, A &  15.618 &  4.479 &  0.508 &  6387 &  1.341 &       F7V &   \\ 
  71284 &  128167 &       557 &     S, C, A &  15.833 &  4.470 &  0.364 &  6566 &  1.500 &   F3V &  early type \\ 
  72659 &  131156 &   566A &     S, C, A &   6.776 &  4.675 &  0.720 &  5380 &  0.920 &     G7V &  young \\ 
  73184 &  131977 &    570A &     S, C &   5.861 &  5.720 &  1.024 &   4744 & 0.760 &          K4V & \\ 
  73996 &  134083 &   578 &      A &  19.550 &  4.930 &  0.430 &  6435 &  1.210 &       F5V &  early type \\
  75181 &  136352 &      582  &      S &  14.810 &  5.650 &  0.639 &  5672 &  0.848 &     G4V &   \\ 
  77257 &  141004 &       598 &     S, C, A &  12.124 &  4.413 &  0.604 &  5936 &  1.036 &    G0Vvar &   \\ 
  77760 &  142373 &       602 &     S, C, A &  15.893 &  4.599 &  0.563 &  5861 &  1.100 &       F9V &   \\ 
  78072 &  142860 &       603 &     S, C, A &  11.254 &  3.850 &  0.478 &  6262 &  1.300 &       F6V &   \\ 
  79672 &  146233 &       616 &      S &  13.900 &  5.496 &  0.652 &  5791 &  1.038 &       G5V &   \\ 
  81300 &  149661 &       631 &      S &   9.751 &  5.760 &  0.827 &  5277 &  0.883 &       K1V & young   \\ 
  84862 &  157214 &       672 &      S, A &  14.327 &  5.383 &  0.619 &  5697 &  0.871 &       G0V &   \\ 
  86974 &  161797 &      695A &     S, C, A &   8.310 &  3.405 &  0.750 &  5641 &  1.142 &      G5IV & subgiant  \\ 
  89962 &  168723 &       711 &      C &  18.543 &  3.232 &  0.941 &  4975 &  1.721 &  K0III-IV & subgiant \\ 
  91438 &  172051 &       722 &      S &  13.084 &  5.860 &  0.673 &  5564 &  0.855 &       G5V &   \\ 
  92043 &  173667 &     725.2 &      C, A &  19.209 &  4.189 &  0.468 &  6423 &  1.100 &       F6V &   \\ 
  95447 &  182572 &       759 &      S, A &  15.177 &  5.157 &  0.762 &  5656 &  1.141 &   G8IVvar &   \\ 
  96100 &  185144 &       764 &     S, C, A &   5.754 &  4.668 &  0.786 &  5246 &  0.801 &     K0V &   \\ 
  96441 &  185395 &      765A &      C, A &  18.335 &  4.490 &  0.395 &  6594 &  1.380 &       F4V & early type   \\ 
  97675 &  187691 &  768.1A &      A &  19.190 &  5.120 &  0.540 &  6139 &  1.370 &       F8V &  \\
  98036 &  188512 &      771A &     S, C &  13.699 &  3.711 &  0.855 &  5163 &  1.257 &   G8IVvar &    \\ 
  99461 &  191408 &      783A &     S, C, A &   6.015 &  5.317 &  0.860 &  4922 &  0.686 &     K2.5V &   \\ 
 104214 &  201091 &      820A &     S, C, A &   3.496 &  5.200 &  1.069 &  4655 &  0.662 &     K5V &   \\ 
 104217 &  201092 &      820B &      S &   3.496 &  5.950 &  1.298 &  4145 &  0.548 &     K7V &   \\ 
 109422 &  210302 &   849.1 &      A &  18.280 &  4.940 &  0.490 &  6339 &  1.299 &       F6V &  early type \\
 112447 &  215648 &      872A &     S, C, A &  16.297 &  4.200 &  0.502 &  6204 &  1.300 &       F7V &   \\ 
 113357 &  217014 &       882 &      S &  15.608 &  5.452 &  0.666 &  5787 &  1.064 &       G5V &   \\ 
 114622 &  219134 &       892 &     S, C &   6.543 &  5.570 &  1.000 &  4835 &  0.782 &     K3V &   \\ 
 116771 &  222368 &       904 &     S, C, A &  13.714 &  4.130 &  0.507 &  6204 &  1.170 &       F7V &    \\ 
 \enddata
\tablenotetext{a}{Target list code: S = starshade study mission target,  C = coronagraph study mission target, A = WFIRST-AFTA study mission target.}
\end{deluxetable}

\begin{deluxetable}{cccccccccc}
\tabletypesize{\footnotesize}
\tablecaption{Summary of Reasons for Excluding Stars from Lick/Keck Programs$^{\mathrm{a}}$
\label{tab:target_exclusion_reasons}}
\tablewidth{0pt}
\tablehead{
  \colhead{Mission}   & 
  \colhead{}   & 
  \colhead{Total Stars}   & 
  \colhead{}   & 
  \colhead{Have RVs$^{\mathrm{b}}$}   & 
  \colhead{}   & 
  \multicolumn{4}{c}{No RVs}  \\
  \cline{1-1} \cline{3-3}\cline{5-5} \cline{7-10} \\
  \colhead{}   & 
  \colhead{}   & 
  \colhead{}   & 
  \colhead{}   & 
  \colhead{}   & 
  \colhead{}   & 
  \colhead{Hot$^{\mathrm{c}}$}   & 
  \colhead{Southern$^{\mathrm{d}}$} &
  \colhead{Evolved$^{\mathrm{e}}$}  &
  \colhead{Binary$^{\mathrm{f}}$}
}
\startdata
Exo-S (S)   && 127 &&		 57 &&		 19 &		 24 &		 3 &		 22 \\
Exo-C (C) && 249 &&		 40 &&		 112 &		 43 &		 39 &		 33 \\
WFIRST-AFTA (A)        && 263 &&		 51 &&		 125 &		 51 &		 4 &		 38 \\
Total (S$+$C$+$A)       && 376 &&		 76 &&		 148 &		 71 &		 40 &		 51 
\enddata
\tablenotetext{a}{ Stars were not included in the Lick and Keck RV planet searches for a variety of reasons.  
Here we attempt to reconstruct the reasons that stars were not added to those programs, focussing on four non-exclusive and incomplete categories.}
\tablenotetext{b}{ All stars with RVs are listed in Table \ref{tab:targets_with_data}.}
\tablenotetext{c}{ The number of stars with $B-V < 0.44$, corresponding to spectral type F5 V.}
\tablenotetext{d}{ The number of stars with declination $< -30^\circ$.}
\tablenotetext{e}{ The number of stars not listed as dwarf spectral types in Simbad.}
\tablenotetext{f}{ A rough estimate of the number of stellar multiples.   
We include targets noted as 'spectroscopic binary' or with multiple, distinct spectral types listed in Simbad, 
or a listing in the Washington Double Star Catalog.}
\end{deluxetable}

Figure \ref{fig:hr_diag} shows the Exo-S, Exo-C, and WFIRST-AFTA targets in a Hertzsprung-Russell diagram.  
Stars with Doppler measurements from Lick or Keck Observatories are highlighted in blue.  
These stars are nearly all main sequence stars with spectral type F8 and later  ($B-V > 0.4$) 
with a handful of K-type subgiants and giants.  
Table \ref{tab:target_exclusion_reasons} summarizes the number of stars in each of the three imaging missions 
for which we have Lick and/or Keck Doppler measurements.  
The table also attempts to reconstruct the reasons that stars in the imaging mission target lists were not observed 
by the Lick/Keck searches.


\section{Doppler Targets and Measurements}
\label{sec:rv_surveys}

The historic Lick Planet Search \citep{Fischer2014} with the 
the Hamilton spectrograph \citep{Vogt1987} began in 1987 with a spectrum of $\tau$ Ceti.  
The original target list 
included 120 stars from the 
Bright Star Catalog \citep{Hoffleit1982} and the Gliese--Jahreiss catalog \citep{Gliese1969,Gliese1979}.  
In 1997 many fainter stars were moved to the new Keck Planet Search and 200 additional stars were added to the Lick Planet Search.  
An additional 67 metal-rich stars were added in 2001, bringing the total in the Lick Planet Search to 367 stars \citep{Fischer2014}.  
The Lick target list was dominated by hotter and brighter stars (typically $B - V$ = 0.4--0.7).  
Earlier type stars were generally excluded because of increased Doppler noise ($> 5$ \mse).  
Observations continued through 2011.

Doppler velocities were measured for the Lick and Keck Planet Searches with the aid of an iodine cell.  
These glass cells containing gaseous I$_2$ act a transmission filter,
imprinting thousands of narrow iodine absorption lines on the
stellar spectra in the wavelength region 5000--6200 \AA. 
The dense set of molecular absorption lines provide a robust wavelength fiducial 
against which Doppler shifts are measured, 
and place strong constraints on the shape of the spectrometer instrumental profile at 
the time of each observation \citep{Marcy92,Valenti95}.  
Relative radial velocities were measured using a forward-modeling technique that simultaneously solves 
for the spatially-varying instrumental profile, wavelength solution, and Doppler shift of each spectrum \citep{Butler96a}.  

The Lick Planet Search was world-leading in the discovery and characterization of exoplanets.  
Major discoveries include the confirmation of 51 Peg b \citep{Marcy1997}, 
the discovery of the first planet in an eccentric orbit \citep{Marcy96}, 
the first multi-planet system \citep{Butler1999}, 
the first sub-Saturn-mass planets \citep{Marcy2000}, 
the discovery of 70 of the first 100 extrasolar planets \citep{Marcy2000b}, and 
the first planet search completeness limits for a large sample of stars \citep{Cumming1999}.

The Keck Planet Search  has been in operation since 1996 July using the HIRES echelle spectrometer on the Keck I
telescope \citep{Vogt94}.  Selection of the target stars is described in \citet{Wright2004} and \citet{Marcy05b}. 
By 2004, the Keck target list included $\sim$1000 stars with spectral types F5--M5.  
These stars generally lie close to the main sequence and are chromospherically quiet. 
Most have $B-V > 0.55$, declination~$>$~$-35^\circ$, and have no stellar companion within 2$''$ that would introduce a second, 
complicating spectrum into the Doppler analysis.  
Young, magnetically active stars are excluded from the primary planet search because of increased Doppler noise. 
The Keck Planet Search was expanded over the years with bright stars added from the original Lick program as it winded down, 
as well as large samples of 
subgiants \citep{Johnson07},  metal-rich main sequence stars \citep{Fischer2005,Robinson2007},  
young stars \citep{Hillenbrand2014}, and other stellar populations.

The Keck Planet Search was also highly successful at discovering and characterizing exoplanets.  
Highlights include discoveries of the 
the first Neptune-size planets \citep{Maness2007} and 
numerous super-Earths \citep{Howard09,Howard11,Howard2011-97658,Howard2014-gl15a}, 
demonstration of the planet-metallicity correlation \citep{Fischer2005},  
statistical studies of planet occurrence \citep{Marcy05a,Cumming08,Howard10a}, 
and recently measurements of transiting planets discovered by NASA's Kepler Mission \citep{Marcy2014}.  

Combining RVs and high-contrast imaging to discover planets, brown dwarfs, and other orbiting companions 
has a strong heritage at Keck Observatory.   
The TRENDS program has used the long RV time series to detect accelerations from wide companions.  
Followup imaging with Keck-NIRC2 and other instruments has 
revealed several brown dwarfs and more massive companions 
\citep{Crepp2012-7672,Crepp2012-trends1,Crepp2013-trends3,Crepp2013-trends2,Crepp2014-trends5}.  
Similarly, statistical analyses of the long-period accelerations in the Keck RV time series have been used to estimate the 
fraction of stars with wide-orbit planets  \citep{Knutson2014-trends,Montet2014-trends4}

For this Doppler planet search completeness study, 
we analyzed RVs from four instruments with independent RV zero points at Lick and Keck.
Instrument codes p and l refer to data from the Hamilton spectrograph on the Shane Telescope at 
Lick Observatory before and after a major detector upgrade \citep{Fischer2014}. 
Similarly, codes k and j refer to data taken with  HIRES  on the Keck I telescope before 
and after an upgrade in 2004 \citep{Vogt94}. 
The detector and/or optics in the Hamilton spectrograph were upgraded several times during its 25 years of operation. 
The  commissioning of the Lick Planet Search in 1987 used a TI 800$\times$800 pixel CCD that was 
upgraded to a 2048$\times$2048 pixel detector in 1990 (dewar 16). 
During this era, nightly velocity corrections were made at the level of 10 m s$^{-1}$ to account for 
uncontrolled changes in the instrumental setup \citep{Fischer2014}.
In November 1994 the Schmidt camera optics were replaced and a field flattener and new CCD were installed.  
We treat data prior to November of 1994 as a single instrument with instrument code ``p''. 
After the 1994 camera and optics upgrade the CCD was changed another two times in February of 1998, and 2001. 
We treat these CCDs as a single instrument in our analysis with telescope code ``l''. 
RV zero-point offsets between the three dewars on the telescope from 1998 to 2011 have been 
solved for and removed as best as possible by \citep{Fischer2014}. They found negligible average offsets ($<0.5$ \ms) between all cameras except that RVs 
collected using the final CCD (dewar 8) were on average 13.1 \ms higher than the rest. \citet{Fischer2014} do not mention the variance on the mean offset measured
for each instrument so we can not say how robustly these offsets have been removed for any given star.
Residual uncorrected offsets can lead to spurious planet candidate detections in our automated planet detection pipeline 
(see Figure \ref{fig:completeness_10700}) and these cases are all mentioned explicitly in Section \ref{sec:known_signals}.
Dewar 8 was in operation until 2011 when the iodine cell failed and the Lick Planet Search ceased. 
Lick RVs were taken from \citet{Fischer2014} with no additional processing 
(with the exception of binning measurements in 2 hour intervals, as needed).  
The CCD on HIRES was upgraded and a new field flattener was installed on 19 August 2004, 
introducing an RV zero-point uncertainty.  
Data taken on or before the HIRES upgrade are assigned the instrument code ``k'', 
and post-upgrade data use the code ``j''.
Keck RVs reported here run through 26 August 2014.  

Table \ref{tab:RV_RMS} summarizes the observational histories of each star, 
including the time baseline, UT date of the first RV measurement, and the number and scatter (RMS) of the RVs for each instrument code.
When multiple measurements were gathered in short succession for the same star, we binned the data in 2 hr intervals.

\begin{deluxetable}{rrrrrrrrrrrrr}
\tabletypesize{\footnotesize}
\tablecaption{Properties of Doppler Measurements
\label{tab:RV_RMS}}
\tablewidth{0pt}
\tablehead{
\colhead{} &
\colhead{} &
\colhead{} &
\colhead{} &
\multicolumn{4}{c}{Number of RVs} &
\colhead{} &
\multicolumn{4}{c}{RMS of RVs (\mse)\,$^{\rm a}$}  \\
\cline{5-8} \cline{10-13} \\
\colhead{Hipp.} &
\colhead{HD} &
\colhead{$t_{\rm span}$} &
\colhead{Start Date} &
\multicolumn{2}{c}{\underline{Lick Dewars}}  &
\multicolumn{2}{c}{\underline{Keck Dewars}} & 
\colhead{} &
\multicolumn{2}{c}{\underline{Lick Dewars}}  &
\multicolumn{2}{c}{\underline{Keck Dewars}} \\
  \colhead{no.}   & 
  \colhead{no.}   & 
  \colhead{(yr)}  & 
  \colhead{}  & 
  \colhead{p}  & 
  \colhead{l}  & 
  \colhead{k}  & 
  \colhead{j} & 
  \colhead{}  & 
  \colhead{p}  & 
  \colhead{l}  & 
  \colhead{k}  & 
  \colhead{j} 
}
\startdata
    544 &     166 &     27.1 &  1987-06-13 &      22 &      22 & \nodata &      17 &       &          18.8 &          20.7 &       \nodata &          17.5 \\
   3093 &    3651 &     26.3 &  1987-09-09 &      16 &     139 &      29 &      54 &       &          11.1 &           7.2 &           3.7 &           3.3 \\
   3821 &    4614 &     18.9 &  1995-09-09 & \nodata &      21 & \nodata &      48 &       &       \nodata &          10.4 &       \nodata &          10.2 \\
   3765 &    4628 &     26.9 &  1987-09-08 &      15 &      37 & \nodata &      89 &       &          10.4 &           6.1 &       \nodata &           2.7 \\
   7513 &    9826 &     26.9 &  1987-09-08 &      17 &     312 & \nodata &       8 &       &          30.7 &          13.0 &       \nodata &           2.9 \\
   7981 &   10476 &     26.8 &  1987-09-09 &      15 &      49 &      65 &      69 &       &           8.9 &           8.1 &           5.1 &           2.7 \\
   8102 &   10700 &     27.0 &  1987-09-08 &      95 &     535 &      87 &     191 &       &           8.9 &           6.0 &           3.9 &           2.5 \\
   8362 &   10780 &     15.6 &  1998-12-05 & \nodata &      16 & \nodata &      16 &       &       \nodata &           6.2 &       \nodata &           4.1 \\
  12114 &   16160 &     25.9 &  1987-09-10 &      14 &      48 & \nodata &      61 &       &          12.4 &           5.8 &       \nodata &           2.5 \\
  13402 &   17925 &     10.9 &  1995-02-20 & \nodata &      15 & \nodata & \nodata &       &       \nodata &          32.1 &       \nodata &       \nodata \\
  14632 &   19373 &     27.0 &  1987-09-08 &      71 &     158 &       8 &      74 &       &           9.0 &           8.8 &           4.7 &           2.9 \\
  15457 &   20630 &     24.1 &  1987-09-08 &      25 &      19 & \nodata & \nodata &       &          21.9 &          25.6 &       \nodata &       \nodata \\
  16537 &   22049 &     27.0 &  1987-09-08 &      49 &     127 & \nodata &      61 &       &          17.8 &           9.6 &       \nodata &           8.3 \\
  16852 &   22484 &     18.8 &  1995-02-18 & \nodata &      34 & \nodata &       5 &       &       \nodata &          10.0 &       \nodata &           5.3 \\
  17378 &   23249 &     16.8 &  1997-01-14 & \nodata & \nodata &      20 &      20 &       &       \nodata &       \nodata &           4.1 &           3.5 \\
  19849 &   26965 &     19.5 &  1995-02-21 & \nodata &      77 &       7 &      92 &       &       \nodata &           7.8 &           2.0 &           3.3 \\
  22263 &   30495 &     11.0 &  2000-10-19 & \nodata &      42 & \nodata & \nodata &       &       \nodata &          16.4 &       \nodata &       \nodata \\
  22449 &   30652 &     11.1 &  2000-09-09 & \nodata &      48 & \nodata & \nodata &       &       \nodata &          31.0 &       \nodata &       \nodata \\
  23311 &   32147 &     26.3 &  1987-09-08 &       7 &      23 & \nodata &      96 &       &           7.8 &           3.6 &       \nodata &           2.8 \\
  23835 &   32923 &     14.7 &  1998-12-05 & \nodata &      24 & \nodata &      55 &       &       \nodata &           8.5 &       \nodata &           5.7 \\
  24813 &   34411 &     26.0 &  1987-09-10 &      17 &     259 &       8 &      60 &       &           8.6 &           6.7 &           8.6 &           2.9 \\
  26779 &   37394 &     16.8 &  1995-11-11 & \nodata &       8 & \nodata &      12 &       &       \nodata &          10.7 &       \nodata &          15.4 \\
  40693 &   69830 &     13.3 &  2000-11-12 & \nodata &      31 & \nodata &     106 &       &       \nodata &           7.4 &       \nodata &           3.2 \\
  42438 &   72905 &     11.0 &  2002-10-25 & \nodata & \nodata &       4 &      12 &       &       \nodata &       \nodata &          45.0 &          40.9 \\
  43587 &   75732 &     25.3 &  1989-02-21 &      14 &     274 &      24 &     193 &       &          19.7 &           6.8 &           4.4 &           3.0 \\
  47592 &   84117 &      9.2 &  2004-11-29 & \nodata & \nodata & \nodata &      44 &       &       \nodata &       \nodata &       \nodata &           4.1 \\
  49081 &   86728 &     26.5 &  1987-06-13 &      15 &      59 & \nodata &      44 &       &           9.2 &           8.0 &       \nodata &           3.2 \\
  51459 &   90839 &     17.0 &  1987-12-20 &      16 &      12 & \nodata & \nodata &       &          18.5 &           9.3 &       \nodata &       \nodata \\
  56452 &  100623 &     17.1 &  1996-12-01 & \nodata & \nodata &      16 &       3 &       &       \nodata &       \nodata &          19.0 &           6.1 \\
  56997 &  101501 &     24.8 &  1987-06-12 &      20 &      69 & \nodata &       4 &       &          17.8 &          10.6 &       \nodata &          15.9 \\
  57443 &  102365 &      6.6 &  2007-05-26 & \nodata & \nodata & \nodata &      16 &       &       \nodata &       \nodata &       \nodata &           2.5 \\
  57757 &  102870 &     21.4 &  1987-06-13 &      28 &      70 & \nodata & \nodata &       &          20.2 &          12.6 &       \nodata &       \nodata \\
  58576 &  104304 &     13.2 &  2001-05-11 & \nodata &      23 & \nodata &      36 &       &       \nodata &           8.6 &       \nodata &           2.5 \\
  61317 &  109358 &     14.4 &  2000-02-09 & \nodata & \nodata &      10 &      69 &       &       \nodata &       \nodata &           3.9 &           3.2 \\
  64394 &  114710 &     21.7 &  1987-06-13 &      35 &      66 & \nodata & \nodata &       &          25.8 &          11.7 &       \nodata &       \nodata \\
  64924 &  115617 &     23.2 &  1991-04-28 &      14 &      95 & \nodata &     103 &       &          10.8 &           9.1 &       \nodata &           3.0 \\
  67275 &  120136 &     23.7 &  1987-06-12 &      24 &      99 & \nodata & \nodata &       &          92.7 &          24.9 &       \nodata &       \nodata \\
  71284 &  128167 &     20.2 &  1988-03-04 &      18 &      36 & \nodata & \nodata &       &          53.4 &          51.1 &       \nodata &       \nodata \\
  72659 &  131156 &     19.4 &  1995-02-21 & \nodata &      15 & \nodata &       5 &       &       \nodata &          56.8 &       \nodata &          41.9 \\
  73184 &  131977 &     14.8 &  1993-08-01 &       5 &      77 & \nodata & \nodata &       &          18.7 &           8.2 &       \nodata &       \nodata \\
  75181 &  136352 &      7.2 &  2007-05-26 & \nodata & \nodata & \nodata &      23 &       &       \nodata &       \nodata &       \nodata &           4.4 \\
  77257 &  141004 &     26.9 &  1987-09-09 &      16 &     149 &       8 &      81 &       &          16.1 &          10.2 &           2.7 &           6.2 \\
  77760 &  142373 &     20.8 &  1987-09-10 &      34 &      61 & \nodata & \nodata &       &          16.8 &           8.8 &       \nodata &       \nodata \\
  78072 &  142860 &     21.1 &  1987-06-11 &      27 &      30 & \nodata & \nodata &       &          56.7 &          19.2 &       \nodata &       \nodata \\
  79672 &  146233 &     18.0 &  1996-07-19 & \nodata & \nodata &      27 &      79 &       &       \nodata &       \nodata &           3.7 &           6.2 \\
  81300 &  149661 &     19.0 &  1991-04-28 &      14 &      67 &      31 &       3 &       &          14.7 &          10.4 &           7.6 &           9.6 \\
  84862 &  157214 &     26.9 &  1987-09-08 &      25 &      71 &       8 &      42 &       &          10.5 &           7.8 &           1.6 &           3.4 \\
  86974 &  161797 &     26.9 &  1987-09-10 &      29 &     132 &      13 &      40 &       &          13.1 &           8.5 &           3.7 &           4.4 \\
  89962 &  168723 &     24.1 &  1987-09-10 &      17 &      81 &      47 &      27 &       &          10.6 &           7.8 &          14.5 &           5.9 \\
  91438 &  172051 &     18.0 &  1996-07-19 & \nodata & \nodata &      32 &      36 &       &       \nodata &       \nodata &           4.0 &           3.0 \\
  92043 &  173667 &     24.4 &  1987-06-11 &      31 &      56 & \nodata & \nodata &       &          89.3 &          76.0 &       \nodata &       \nodata \\
  95447 &  182572 &     17.8 &  1996-10-10 & \nodata & \nodata &      49 &      33 &       &       \nodata &       \nodata &           4.4 &           3.6 \\
  96100 &  185144 &     23.2 &  1991-04-30 &       5 &      18 &      21 &     229 &       &           4.5 &           4.1 &           2.4 &           2.1 \\
  96441 &  185395 &     10.2 &  2001-07-21 & \nodata &     223 & \nodata & \nodata &       &       \nodata &          57.1 &       \nodata &       \nodata \\
  98036 &  188512 &     27.2 &  1987-06-13 &      28 &     155 &       7 &      20 &       &           8.0 &           7.4 &           3.9 &           3.9 \\
  99461 &  191408 &     10.0 &  2004-08-20 & \nodata & \nodata & \nodata &      45 &       &       \nodata &       \nodata &       \nodata &           2.3 \\
 104214 &  201091 &     27.1 &  1987-06-11 &      44 &      24 & \nodata &      77 &       &          13.3 &          10.9 &       \nodata &           5.4 \\
 104217 &  201092 &     27.1 &  1987-06-12 &      30 &      26 & \nodata &      74 &       &          17.0 &          11.7 &       \nodata &           7.0 \\
 112447 &  215648 &     23.3 &  1988-06-19 &      23 &      77 & \nodata & \nodata &       &          17.3 &          11.9 &       \nodata &       \nodata \\
 113357 &  217014 &     18.8 &  1995-10-12 & \nodata &     202 & \nodata &      41 &       &       \nodata &           6.2 &       \nodata &           2.4 \\
 114622 &  219134 &     21.9 &  1992-10-11 &      11 &      35 &       1 &     113 &       &          11.8 &           6.4 &       \nodata &           7.8 \\
 116771 &  222368 &     24.1 &  1987-09-08 &      30 &      23 & \nodata & \nodata &       &          52.7 &         134.5 &       \nodata &       \nodata \\
  14954 &   19994 &     24.1 &  1987-09-09 &      16 &      82 & \nodata & \nodata &       &          33.2 &          13.4 &       \nodata &       \nodata \\
  18859 &   25457 &     11.1 &  2002-08-29 & \nodata & \nodata &       7 &      14 &       &       \nodata &       \nodata &          27.5 &          16.9 \\
  29650 &   43042 &     10.9 &  2000-11-19 & \nodata &      29 & \nodata & \nodata &       &       \nodata &          13.7 &       \nodata &       \nodata \\
  32480 &   48682 &     27.0 &  1987-09-10 &      16 &      74 & \nodata &      48 &       &          13.0 &          10.2 &       \nodata &           4.4 \\
  39780 &   67228 &     12.8 &  1998-12-26 & \nodata &      39 &      16 &      21 &       &       \nodata &          11.8 &          16.5 &           8.9 \\
  40843 &   69897 &     21.4 &  1987-09-10 &      23 &      46 & \nodata & \nodata &       &          22.2 &           9.6 &       \nodata &       \nodata \\
  48113 &   84737 &     26.5 &  1987-06-13 &      22 &      66 & \nodata &      44 &       &          14.2 &           7.8 &       \nodata &           3.4 \\
  53721 &   95128 &     27.1 &  1987-06-13 &      19 &     215 & \nodata &      48 &       &          15.5 &           8.2 &       \nodata &           2.9 \\
  64408 &  114613 &      5.7 &  2007-05-26 & \nodata & \nodata & \nodata &      24 &       &       \nodata &       \nodata &       \nodata &           5.9 \\
  64792 &  115383 &     21.7 &  1987-06-12 &      21 &      16 & \nodata & \nodata &       &          28.8 &          23.6 &       \nodata &       \nodata \\
  65721 &  117176 &     26.6 &  1988-02-03 &      21 &      98 & \nodata &      54 &       &           9.5 &           9.1 &       \nodata &           3.5 \\
  73996 &  134083 &     21.0 &  1987-06-11 &      22 &      23 & \nodata & \nodata &       &         218.2 &         204.8 &       \nodata &       \nodata \\
  97675 &  187691 &     11.1 &  2000-09-09 & \nodata &      15 & \nodata & \nodata &       &       \nodata &          15.2 &       \nodata &       \nodata \\
 109422 &  210302 &     15.1 &  1999-06-12 & \nodata & \nodata &      11 &      30 &       &       \nodata &       \nodata &          22.7 &           9.3 \\
\enddata
\tablenotetext{a}{The RMS of the RVs is computed as the standard deviation of the velocities 
after subtracting the signals discussed in Sec.\ \ref{sec:known_signals}.}
\end{deluxetable}

This study is limited to Lick and Keck RVs gathered by the California Planet Search and its predecessor the 
California/Carnegie Planet Search.  
For a complete planet census and search completeness estimate, we urge other groups to undertake 
analyses for the historic Doppler measurements from the 
CFHT planet search \citep{Walker1995}, 
the Anglo-Australian Planet Search \citep{Tinney2001}, 
the ELODIE, CORALIE, and HARPS planet searches \citep{Mayor2011}, 
the McDonald planet search \citep{Wittenmyer2006}, 
the Lick giant star planet search \citep{Quirrenbach2011}, and others.


\section{Doppler Planet Detection}
\label{sec:doppler}

A planetary orbit is defined by an orbital period $P$ (or semi-major axis $a$), 
eccentricity $e$, inclination to the line of sight $i$, 
argument of periastron $\omega$, and longitude of the ascending node $\Omega$ \citep[see e.g.,][]{Murray1999}.
Imaging measurements of planetary (and stellar companions) consist of 
measuring the projected separation and position angle of the secondary relative to the primary source of light.  
Given a sufficient number, precision, and timing of such two-dimensional measurements, 
all five Keplerian orbital elements and the primary-secondary mass ratio can be inferred.  

Doppler planet searches measure the line-of-sight velocity (radial velocity, RV) of the bright primary 
star and over time fit the measurements with a Keplerian model.  
Doppler measurements are sensitive to $P$, $e$, and $\omega$, but not $i$ or $\Omega$.  
They also are sensitive to the Doppler amplitude $K$ and an orbital reference time 
(typically a time of periastron passage, $t_{\rm p}$, or a time of transit, $t_{\rm c}$).
For a planetary system with $n$ planets, the 
stellar reflex velocity is the sum of Keplerian contributions from each planet, 
\begin{equation}
\label{eq:model}
v(t) = \sum_{j=1}^n \Big(K_j(\cos(\omega_j+f_j(t))+e_j\cos\omega_j)\Big) +
\gamma + \dot{\gamma} \cdot (t-t_0),
\end{equation}
where 
the term in big parentheses refers to the Keplerian signal from planet $j$,
$f_j(t)$ is the true anomaly of planet
$j$ at time $t$,
$\gamma$ is the time-independent velocity offset (often degenerate with  instrument-specific velocity zero points), 
$\dot{\gamma}$ describes a linear velocity term 
(``slope,'' ``trend,'' or constant acceleration, often from a massive, secondary companion whose orbit is only partially sampled), 
and $t_0$ is a conveniently chosen epoch of the observations.  
The assumption in Eq.\ \ref{eq:model} that the total Doppler signal is the superposition of each planet's contribution implies 
that the planets are not dynamically interacting with one another.  This is a good assumption because 
detectable planet-planet interactions require planets to be near or in mean-motion resonances, which have 
only been detected for a handful of Doppler systems \citep[e.g.,][]{Rivera2010,Tan2013}, and those systems can be {\em discovered}
using their Keplerian approximations (Eq.\ \ref{eq:model}).

\begin{figure}
         \begin{center}
              \includegraphics[width=0.45\textwidth]{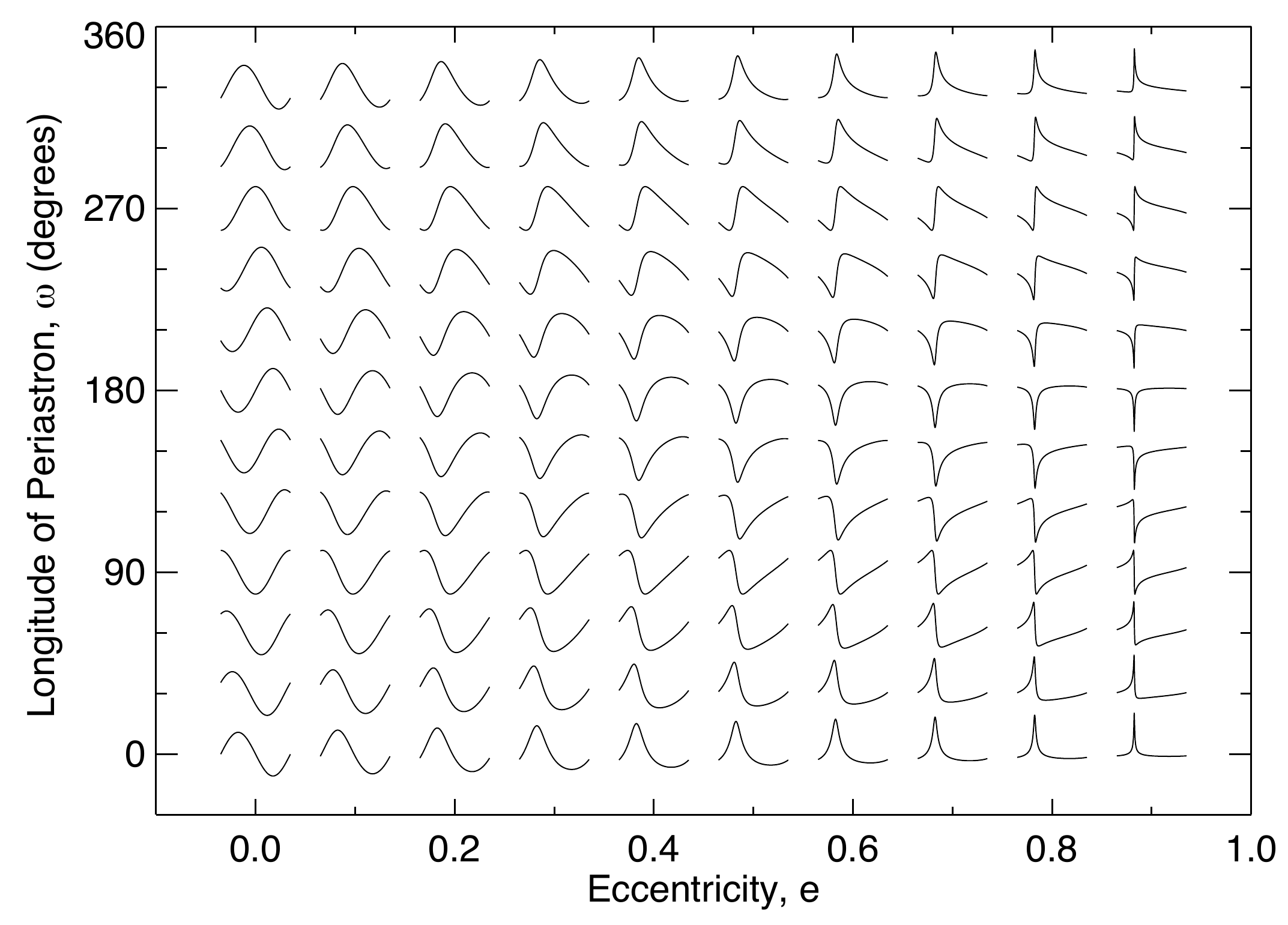}  
         \end{center}
         \caption{Keplerian function shape versus orbital eccentricity, $e$ (0-0.9 in 0.1 steps), 
         and longitude of periastron, $\omega$ (0--$330 ^{\circ}$ in $30 ^{\circ}$ steps).  
         This function represents the Doppler shift of a planet host star. 
         It is sinusoidal for $e = 0$, but becomes closer to a $\delta$-function as $e \rightarrow 1$.
         Orbits with $e \lesssim 0.4$ are reasonably well fit by single sinusoids, but more eccentric 
         orbits are increasingly less likely to be detected by orbit fitting techniques that assume circular orbits.  
         Each curve represents a Keplerian function with unit amplitude ($K$), unit orbital period ($P$), and 
         a time of periastron passage ($t_p$ = $t_c$ + $P/4$ is shown, where $t_c$ is a time of transit), 
         evaluated on the grid of $e$ and $\omega$.
         }
         \label{fig:kepzoo}
\end{figure}

The true anomaly in Eq.\ \ref{eq:model} is defined implicitly in terms of 
the other three Keplerian parameters $P_j, t_{{\rm p},j},$ and $e_j$ through
the relations  
\begin{equation}
\label{edef}
\tan\frac{f_j(t)}{2} = \sqrt{\frac{1+e_j}{1-e_j}} \tan \frac{E_j(t)}{2},
\end{equation}
\begin{equation}
\label{Kepler}
E_j(t)-e_j \sin E_j(t) = \frac{2\pi(t-t_{{\rm p},j})}{P_j} = M_j(t).
\end{equation}
Here $E_j$ is called the eccentric anomaly of planet $j$, $M_j$ is 
the mean anomaly, and Eq.~\ref{Kepler} is Kepler's Equation.
The Keplerian function $v$($t$) is sinusoidal for a single planet and $e=0$, 
but the function becomes significantly cusp-like as $e \rightarrow 1$ and the planet's orbital speed 
varies dramatically over one orbit (Figure\ \ref{fig:kepzoo}).
The RV semi-amplitude of the star, $K$, can be expressed in units of ${\rm m\,s^{-1}}$ with the planet 
mass in units of Jupiter masses (\mjupe),
\begin{equation}
K =  \frac{28.4\,\mathrm{m\,s^{-1}}}{\sqrt{1\!-\!e^2}} \, \frac{M_p \sin{i}}{M_{\rm J}} 
\! \left( \frac{M_\star\!+\!M_p}{M_\odot}\right)^{-2/3} \left(\frac{P}{\mathrm{yr}} \right)^{-1/3}, 
\label{eq:K}
\end{equation}
using Kepler's Third Law, $K \propto a^{-1/2}$.
Thus,  Doppler signals from giant planets in few AU orbits are of order $\sim$10\,\mse, 
while Neptune-mass planets and super-Earths produce 10--100$\times$ smaller signals.  
The timescales to discover a planet using Doppler velocities are daunting for $a$ larger than a couple AU; 
recall that Jupiter has $a = 5.2$ AU and $P = 11.9$ yr and Saturn has $a = 9.6$ AU and $P = 29.5$ yr.  

Also note that $K \propto \msini$, the planet's ``minimum mass,'' where $\sin i$ is in the range 0--1 
and accounts for the unknown orbital inclination $i$.  
For example, $i = 45 ^{\circ}$ corresponds to $\sin i = 1/\sqrt{2}$ and the true mass $M_{\rm p}$ being 41\% larger than \msini.
Large $\sin i$ corrections are rare because orbital orientations are randomly distributed on the celestial sphere. 
(The distribution of $\cos i$ is uniform.)  
The statistical probability that the orbit inclination
is within a range $i_1 < i < i_2$ is ${P}_{\rm incl} = | \cos (i_2) - \cos (i_1) |$.  
This means that 87\% of orbits have $30^{\circ}  < i < 90 ^{\circ}$, and $M_{\rm p}$ within a factor of two of \msini.  
Randomly finding $i < 30^{\circ}$ is equivalent to randomly selecting a location in the Northern Hemisphere with latitude $>$ 60$^{\circ}$, 
i.e.\ nearly in the Arctic Circle (latitude = 66.5$^{\circ}$).


\section{Radial Velocity Analysis}
\label{sec:rv_analysis}



\subsection{Automated Planet Search}
\label{sec:automated_planet_search}

We search for planets in the RV data using an iterative multi-planet detection algorithm 
based on the two-dimensional Keplerian Lomb-Scargle (2DKLS) periodogram \citep{OToole09}. 
We create the periodogram by fitting a Keplerian RV model to the dataset at many starting points on a 
two-dimensional grid of $P$ and $e$. 
The primary advantages of the 2DKLS periodogram over the Lomb-Scargle periodogram \citep{Lomb76,Scargle82}
are that it is more sensitive to eccentric planets, and both measurement errors and 
zero point offsets between instruments ($\gamma_i$) can be incorporated directly into the periodogram.  



We fit  models to the data using the Levenberg-Marquardt (L-M) $\chi^2$-minimization in the RV fitting package 
\citep[RVLIN,][]{WrightHoward09}. 
Each  RV model is a sum of single planet models ($P$, $t_{\rm p}$, $e$, $\omega$, and $K$), 
but with $\gamma_i$ (and $\dot{\gamma}$, if needed) shared by all planets (Eq.\ \ref{eq:model}). 
We define a grid of search periods following the prescription of \citet{Horne86} for optimal frequency sampling. 
At each period we start an L-M fit at five eccentricities between 0.05 and 0.7. All parameters are free to vary in each fit, 
but $P$ and $e$ are constrained to intervals that allow them to vary only half the distance to adjacent search 
$P$ and $e$ ranges in parameter space.  All parameters for any previously detected planets are simultaneously 
re-fit so that slightly incorrect fits (sometimes caused by the presence of other planets) do not lead to the false detection 
of additional planets. The periodogram power is 
\begin{equation}
Z(P,e) = \frac{\chi^2 - \chi^2_B}{\chi^2_B},
\label{eq:Z}
\end{equation}
where $\chi^2$ is the sum of the squared, error-normalized residuals to the current $N$-planet Keplerian fit, and $\chi^2_B$ 
is for the best $N$-1-planet fit. In the first iteration of the planet search 
(comparing 1-planet models to a 0-planet fit), $\chi^2_B$ is simply the sum of the squared, error-normalized 
residuals to the mean or a linear fit.

We start the iterative planet search by fitting for any known planets in the system using the 
orbital parameters cataloged in the Exoplanet Orbit Database \citep{Wright2011} and 
well-established planet candidates (Howard et al. 2014 in prep.) as initial guesses for RVLIN. 
We then look for a significant linear trend by fitting a line to the RVs and checking if 
the total change in RV due to the fitted line is greater than 10 times the median of the individual 
measurement errors. If a significant trend is detected we allow $\dot{\gamma}$ to vary at each 
point in the 2DKLS periodogram, or if no trend is detected we fix $\dot{\gamma} = 0$.
For systems with known planets, we start by creating a periodogram to test if a $N$+1 planet model 
is a better fit than the model of $N$ known planets; otherwise we start by searching for a single planet 
and compare to the null hypothesis (the $N=0$ planet model).
Note that for cases where the orbital period is between half and the full baseline of the dataset we can suffer a modest reduction in sensitivity.
In the case of a partially sampled orbit the data may show an apparent linear trend that is partially absorbed into $\dot{\gamma}$.
This effect dictates the smooth rolloff in sensitivity to long-period, low-mass planets as discussed in Section \ref{sec:idealized_completeness}.

\begin{figure}
         \begin{center}
              \includegraphics[width=0.49\textwidth]{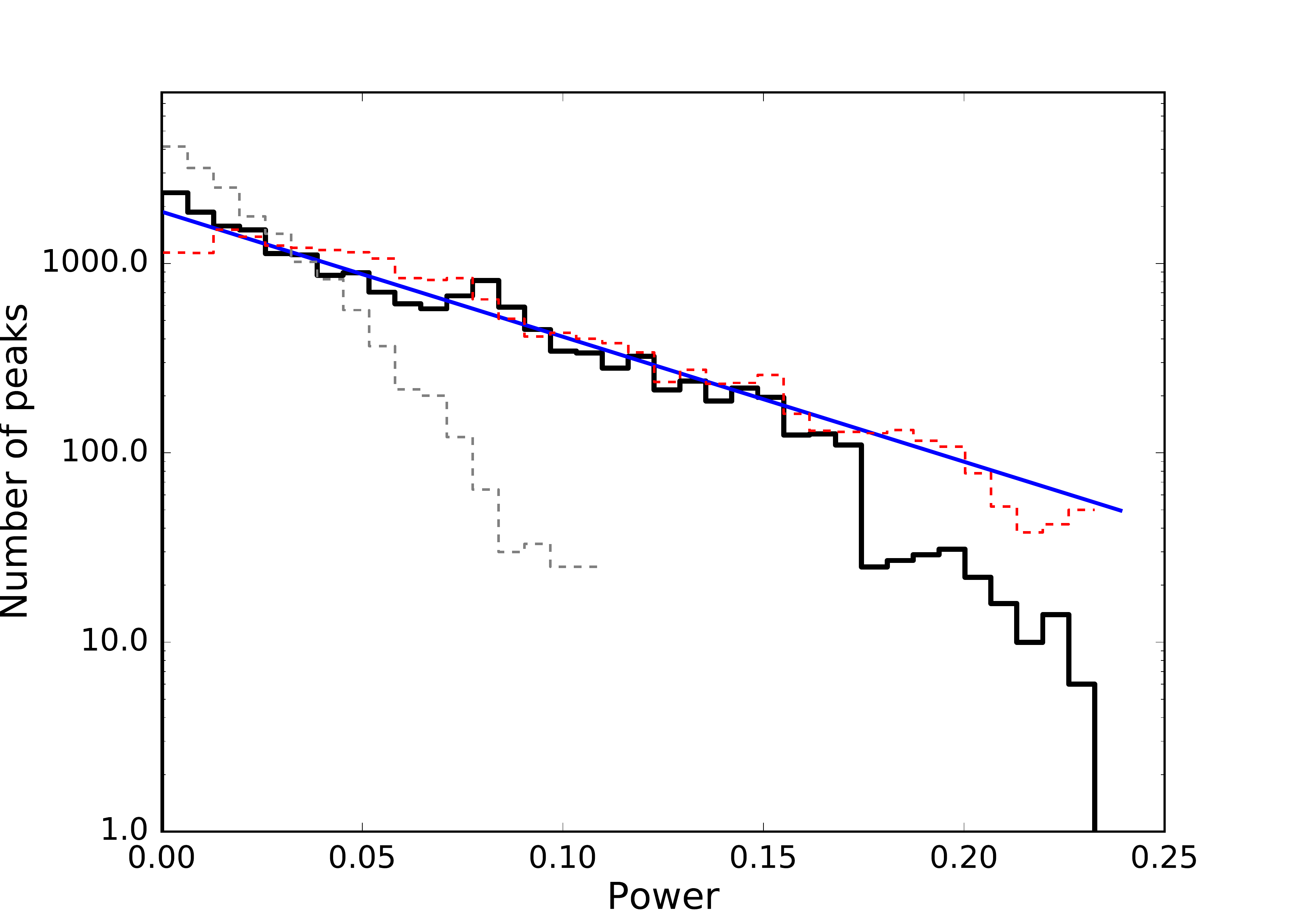} 
         \end{center}
         \caption{Distribution of 2KDLS periodogram peak heights for the HD 10476.  
         (The periodogram is showing the top panel of Fig.\ \ref{fig:search_example}.)  
         We fit a linear profile to this distribution (blue line) and define the FAP = 1\% threshold 
         as the periodogram power where the linear extrapolation intersects N = 0.01.  
         As shown by the red line in the top panel of Fig.\ \ref{fig:search_example}, this FAP threshold is at a power of $Z = 0.85$, 
         well separated from the forrest of peaks that have a maximum value of $Z$ = 0.25. The grey dashed line shows the periodogram
         value distribution for Gaussian noise, and the red dashed line shows the distribution for 1/f (``pink") noise simulation as described 
         in \S \ref{sec:automated_planet_search}. It is interesting to note that our best fit model and the histogram of 
         periodogram peaks for the real dataset
         more closely match the pure ``pink" noise dataset then the ``white" noise dataset. The histogram produced from the periodogram of
         ``white" noise has much many more peaks at low power (note log y-axis) and drops off much faster with no periodogram peaks above
         a power of ~0.1 in contrast to the observations. This indicates that the RVs for this star (and likely most stars) are 
         dominated by non-Gaussian noise that arises from a combination of instrumental and astrophysical processes.
         }
         \label{fig:peakhist}
\end{figure}

\begin{figure}
         \begin{center}
              \includegraphics[width=0.5\textwidth]{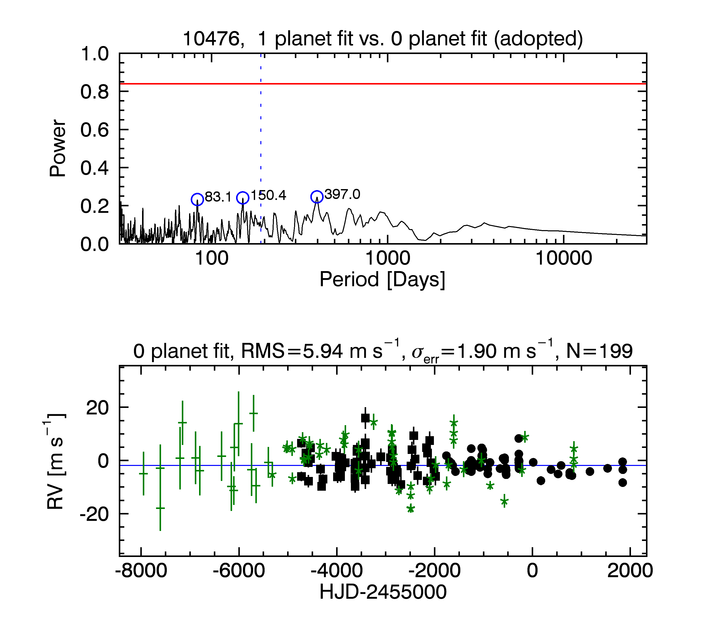} 
         \end{center}
         \caption{Two panel plot showing the results from an automated  search for periodic signals in Lick and Keck RVs 
         for the star HD 10476.   
         Appendix \ref{app:completeness} includes plots like this one for all stars without detected signals 
         or Fig.\ \ref{fig:search_example2} for stars with significant signals.  
         \emph{Top panel:} 2DKLS periodogram of the RVs from Lick and Keck. 
         Periodogram power is calculated using Eq.\ \ref{eq:Z} by comparing $\chi^2$ for prospective single planet fits 
         having the range of periods on the horizontal axis to $\chi^2$ of the best-fit model without any planets 
         (but with floating, telescope-specific RV zero points).  
         The three most significant periodogram periods are labeled and circled in blue. 
         In this case none of the peaks exceeds the 1\% empirical FAP (solid red line) required for a solution to be adopted.  
         The dashed vertical line marks the one year alias with the most significant peak (397 days).  
         No planets were found significant in this fit, as indicated by ``0 planet fit (adopted)'' in the title of the top plot.
         \emph{Bottom panel:} RVs as a function of time with  $\gamma$ offsets between datasets minimized are plotted (the best-fit ``0 planet fit''). 
	The plot symbols are: green crosses (Lick ``p''), green stars (Lick ``l''), 
	black squares (Keck ``k''), black circles (Keck ``j'').  
         Annotations above the bottom panel indicate the 
         RMS of the RVs, the median estimated Doppler uncertainty ($\sigma_{\rm err}$), 
         and the number of RVs plotted ($N_{\rm obs}$).
         }
         \label{fig:search_example}
\end{figure}

\begin{figure}
         \begin{center}
              \includegraphics[width=0.5\textwidth]{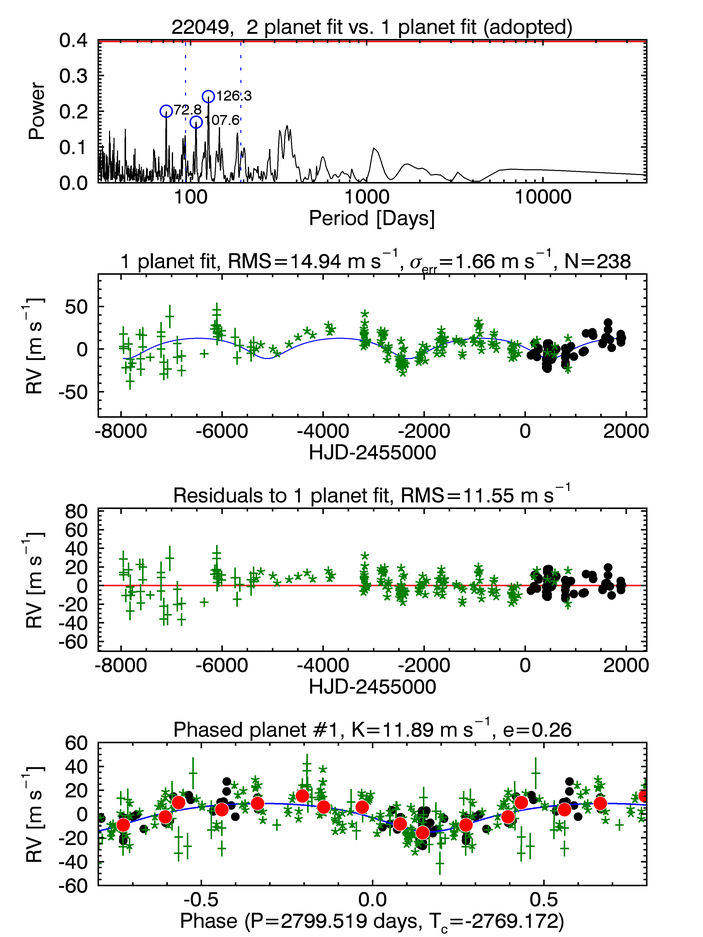} 
         \end{center}
         \caption{Four panel plot showing the results from an automated search for periodic signals in Lick and Keck RVs 
         for the star HD 22049 ($\epsilon$ Eridani; see Sec.\ \ref{sec:known_signals} for a discussion).   
         Appendix \ref{app:completeness} includes plots of this type for all stars with detected planets.
	\emph{Top panel:} 2DKLS periodogram of the RVs from Lick and Keck, as in Fig.\ \ref{fig:search_example}. 
         Periodogram power is calculated using Eq.\ \ref{eq:Z} by comparing $\chi^2$ for prospective two-planet fits 
         having the range of periods on the horizontal axis to $\chi^2$ of the best-fit single planet model.  
         The three most significant periodogram periods are labeled and circled in blue. 
         In this case none of the peaks exceeds the 1\% empirical FAP (solid red line) required for a solution to be adopted.  
         The dashed vertical lines mark the one year aliases with the most significant peak (126.3 days).  
         The single planet fit is adopted, as indicated by ``1 planet fit (adopted)'' in the title of the top plot.
	\emph{Second panel:} RVs as a function of time with the best-fit single planet model overplotted (blue line).  
	Plot symbols are the same as in Fig.\ \ref{fig:search_example}.
         Annotations above the bottom panel indicate the 
         RMS of the RVs, the median estimated Doppler uncertainty ($\sigma_{\rm err}$), 
         and the number of RVs plotted ($N_{\rm obs}$).
	\emph{Third panel:} RV time series showing residuals to the best-fit single planet model in the second panel.  
	The RMS of these residuals is indicated in the panel title.  
	\emph{Bottom panel:} RVs phased at the period of the adopted single planet model (2799 days).  
	The panel title lists $K$ and $e$ for this planet.
         }
         \label{fig:search_example2}
\end{figure}

Since RV measurement errors are often underestimated due to the presence of systematic 
and/or poorly-understood astrophysical noise, periodogram power can not be directly converted into a 
significance estimate using traditional $\chi^2$ statistics. 
Instead we derive an empirical false alarm probability (FAP) 
by fitting a histogram of the periodogram amplitudes higher than the median power value to a linear function 
in $\log{n}$ vs.\ $Z$ (Figure \ref{fig:peakhist}). 
This provides an estimate of the number of peaks that should fall above a given value, 
and when multiplied by the number of independent test periods, the approximate probability that we would find a peak of a 
given value within the particular periodogram. We use this to define a threshold of FAP\,=\,1\% 
above which spikes in the periodogram are marked as a planet candidates and we 
continue searching for the next potential planet in the system. If no significant signals are detected, we stop the search.   

In order to compare an observed distribution of periodogram peak heights to some idealized cases we ran 2DKLS periodograms
for simulated white and 1/f (``pink") noise datasets (see Figure \ref{fig:peakhist}).This was done by taking the real dataset for
HD 10476 used to produce the plots in Figure \ref{fig:search_example} and replacing the velocities with Gaussian noise using
sigmas equal to the measurement errors for each individual measurement. We then ran our 2DKLS periodogram on this dataset
to measure the distribution of periodogram values in the case of pure white noise (grey dashed line in Figure \ref{fig:peakhist}).
We also generated 1/f (``pink") noise using the
Voss-McCartney algorithm (http://www.firstpr.com.au/dsp/pink-noise/) and scaled it to match the standard deviation of the real
post-upgrade Keck RVs for the same star. We added this ``pink" noise to the white noise dataset and ran the periodogogram again
to measure the distribution of periodogram values in the case of pure white+pink noise (red dashed line in Figure \ref{fig:peakhist}).

If a periodogram peak is found to have an empirical FAP less than the 1\% threshold some sanity checks are applied 
before it is considered a viable candidate. If greater than 20 peaks are found above the threshold we do not consider 
any of them viable unless the highest peak has an empirical FAP $\leq$ 10$^{-4}$. This is caused by a lack of phase
coverage in cases with a low number of observations ($N_{\rm obs} \lesssim20$) and is not common among the generally
well-observed stars studied in this work.
If the period corresponding to the highest peak is within 10\% of the periods of any previously found planets we mask this 
period and continue searching for additional periods until none are above the FAP threshold 
or a viable candidate is identified at a different period. 
Finally, if the L-M fit seeded at the period corresponding to the highest peak outputs a period that is $\geq$50\% away 
from the input period we consider the period very poorly constrained and not a viable candidate.

Note that our empirical approach of estimating a FAP threshold differs from the bootstrapping strategy of scrambling RVs  
and searching for periodic signals in many realizations of the data, each with the RVs scrambled differently \citep[e.g.,][]{Howard10b}.  
One advantage of our FAP calculation here is that time-correlated RV noise is reflected in the distribution of periodogram peak heights, 
while the scrambling technique does not preserve that information.

See Figure \ref{fig:search_example} for an example of the automated planet search algorithm on the star HD~10476.
The example shows data from all four instrument codes and the search did not detect any credible planetary signals.
Figure \ref{fig:search_example2} provides a second example, in this case a search for a second planet orbiting 
HD~22049 ($\epsilon$ Eridani), which is known to host a single giant planet \citep{Hatzes2000}.  
A complete set of automated planet search plots for all stars in Table \ref{tab:targets_with_data} is included 
in Appendix \ref{app:completeness}.

\subsection{Known Signals in Doppler Data}
\label{sec:known_signals}

The automated planet searches for each star describe the completeness of our Doppler data 
in a search for planets in addition to the already known planets.  
That is, the completeness limits are for an $N+1$ planet model, compared to an $N$ planet model.  
We adopt the Exoplanet Orbit Database at exoplanets.org 
as the definitive source of ``known planets'' \citep{Wright2011}.  
In addition, we list below the other ``known signals'' (wide stellar binaries, activity correlations, etc.) 
that were seeded into our automated planet searches and that our completeness estimates are with respect to.  
We also list stars with interesting histories relevant to our RV planet searches.  Not all stars with Doppler observations are 
listed below, only those with known signals/planets and a few for which we make comments.

\textit{HD 3651 (HIP 3093; program:\ S)}---\citet{Fischer2003} discovered an eccentric ($e$ = 0.63) 
sub-Saturn mass planet with $P$ = 62 day orbiting  54 Piscium.   
A T dwarf in a wide orbit was later discovered, 
making this the first known planetary system with an imaged brown dwarf 
\citep{Mugrauer2006,Luhman2007,Liu2007}.  
See Fig.\ \ref{fig:completeness_3651}.

\textit{HD 4614 (HIP 3821; programs:\ S, C, A)}---$\eta$ Cassiopeiae shows a significant linear trend with no detectable curvature, 
presumably due to its K7V stellar companion, HD 4614B ($\eta$ Cassiopeiae B).   
See Fig.\ \ref{fig:completeness_4614}.

\textit{HD 9826 (HIP 7513; programs:\ C, A)}---$\upsilon$ Andromedae hosts 
three known planets with masses of 0.67, 1.9, and 4.1 \mjup with 
orbital periods of 4.6, 241, and 1278 days \citep{Butler1997,Butler1999,Wright2009}.  
We detect an additional periodicity at $\sim$4000 days and interpret this as the signature of a 
stellar magnetic activity cycle.  See Fig.\ \ref{fig:completeness_9826}.

\textit{HD 10700 (HIP 8102; programs:\ S, C, A)}---$\tau$ Ceti is an extremely quiet RV standard star. 
Five low-mass planets were reported to orbit $\tau$ Ceti \citep{Tuomi2013}. 
However, we detect no periodic signals. 
The automated pipeline does detect a long-period signal but it appears to be caused by a poorly constrained offset between datasets.  
$\tau$ Ceti has long been an attractive planet search target because of its proximity (3.7 pc) and brightness ($V = 3.5$).
Frank \citet{Drake1961} started modern SETI by searching for radio signals from $\tau$ Ceti  and $\epsilon$ Eridani. 
See Fig.\ \ref{fig:completeness_10700}.

\textit{HD 16160 (HIP 12114; program:\ S)}---The RV time-series for this star shows significant curvature, 
likely from a late M-type companion \citep{Golimowski1995b,Golimowski1995a,Tanner2010}. 
The RV curvature makes it difficult to detect small, long-period planets ($P \gtrsim t_{\rm span}$) in this system.
See Fig.\ \ref{fig:completeness_16160}.

\textit{HD 17925 (HIP 13402; program:\ S)}---The automated pipeline detects a marginally significant linear trend in the 
RV time series.
See Fig.\ \ref{fig:completeness_17925}.

\textit{HD 19994 (HIP 14954; program:\ A)}---The star 94 Ceti hosts a giant planet in a 1.4 AU orbit \citet{Mayor2004}.
See Fig.\ \ref{fig:completeness_19994}.

\textit{HD 22049 (HIP 16537; programs:\ C, A)}---$\epsilon$ Eridani is a 
young K2 dwarf that exhibits an RV period of $\sim$2500 days, 
interpretable as a Jovian-mass planet \citep{Hatzes2000}.  
This star is one of the best prospects for directly imaging planets \citep{Marengo2009} 
due to the star's proximity (3.2 pc) and youth, 
and the detection of a debris disk \citep{Greaves1998,Moran2004,Backman2009}.
The planet has remained controversial because of high stellar magnetic activity \citep{Metcalfe2013,Jeffers2014}.  
However, our measurements of the activity-sensitive \caii\ lines \citep{Isaacson2010} have a period of $\sim$3 yr and 
do not correlate with RVs spanning 6 yr in our Keck spectra.  
While such a correlation would have cast doubt on the planet, the absence of one strengths the case for a giant planet 
orbiting this star.
See Fig.\ \ref{fig:completeness_22049}.

\textit{HD 25457 (HIP 18859; program:\ A)}---This young, 
F6 dwarf is a T Tauri star and a member of the AB Dor moving group \citep{LopezSantiago2006}.  
Ground-based direct imaging searches for planets having yielded non-detections to date, with planet mass limits of 
$\sim$7 \mjup at 1 AU \citep{Maire2014}.
Our pipeline formally identifies a planet candidate based on 21 Keck RVs. 
Given the high jitter and small number of observations, we deem this candidate not credible with the current data.  
See Fig.\ \ref{fig:completeness_25457}.

\textit{HD 30652 (HIP 22449; programs:\ S, C, A)}---$\pi^3$ Orionis hosts no known planets. 
We detect a marginally significant periodic signal with a period of 338 days. 
However, the poor observing history and the proximity of the period to one year, 
we conclude that this signal is not caused by a real planetary companion. 
The nature of the signal is most likely non-astrophysical in nature.  
\citet{Wittenmyer2006} did not detect companions in their McDonald Observatory Doppler search.  
See Fig.\ \ref{fig:completeness_30652}.

\textit{HD 34411 (HIP 24813; programs:\ S, C, A)}---The automated pipeline picks up a 
long-period, high-eccentricity signal that is likely 
due to poorly constrained offsets between instruments for $\lambda$ Aurigae.  See Fig.\ \ref{fig:completeness_34411}.

\textit{HD 37394 (HIP 26779; program:\ S)}---This young star is in the Pleiades association \citep{LopezSantiago2006} 
and shows a small, marginally significant linear trend.  
See Fig.\ \ref{fig:completeness_37394}.

\textit{HD 48682 (HIP 32480; program:\ A)}---56 Aurigae has a formally adopted signal that appears to be due to uncorrected zero-point offsets in the Lick RVs and not due a planet.  
See Fig.\ \ref{fig:completeness_48682}.

\textit{HD 69830 (HIP 40693; program:\ S)}---This star hosts three Neptune-mass planets \citep{Lovis2006}.  
However, the automated pipeline can only pick out the planets at 8.7 and 197 days. 
While we do not detect the 31.6 day period in the Lick or Keck data, 
we have not performed an analysis to show that our non-detection is dispositive.  
See Fig.\ \ref{fig:completeness_69830}.

\textit{HD 75732 (HIP 43587; program:\ S)}---55 Cancri hosts five known planets \citep{Fischer2009} with at least one that transits \citep{Winn2011}.  
The orbital periods are $P$ = 0.74, 14.6, 44, 260, and 5200 days. 
The middle three planets (b, c, and f) are approximately Saturn-mass while the innermost, transiting planet (e) 
is a super-Earth and the outermost planet (d) is a super-Jupiter.  See Fig.\ \ref{fig:completeness_75732}.

\textit{HD 84737 (HIP 48113; program:\ A)}---This star has a 
formally adopted signal that appears to be due to uncorrected zero-point offsets in the Lick RVs 
and not due a planet.  
See Fig.\ \ref{fig:completeness_84737}.

\textit{HD 95128 (HIP 53721; program:\ A)}---47 Ursae Majoris has hosts two well-known giant planets with semi-major axes of 2 and 3.6 AU \citep{Fischer02}, 
and possibly a third planet at $\sim$11 AU \citep{Gregory10}.  Our automated search prefers a model with three planets, 
although the outer most planet has a poorly constrained orbit.
See Fig.\ \ref{fig:completeness_95128}.

\textit{HD 100623 (HIP 56452; program:\ S)}---This star shows a significant linear trend with no detectable curvature. 
The Keck RVs (code = j) are very sparse and only 19 observations were collected in total for this star.  
See Fig.\ \ref{fig:completeness_100623}.

\textit{HD 104304 (HIP 58576; program:\ S)}---This system shows a strong 
long-term linear trend with curvature, 
likely due to a detected low-mass, stellar companion \citep{Tanner2010,Schnupp2010}.  
Our automated search prefers a model with a linear velocity trend (constant acceleration) 
in addition to the orbit segment from the companion (three bodies total), 
with considerable model degeneracy between the slope and the mass of the companion causing the RV curvature.
See Fig.\ \ref{fig:completeness_104304}.

\textit{HD 115617 (HIP 64924; programs:\ S, C, A)}---\citet{Vogt2010} reported three small planets orbiting 61 Virginis with $P$ = 4.2, 38, and 124 days.  
We see evidence for only the first two planets in our Keck RVs.   
The inner planet b is a small super-Earth with a mass of 5.3 M$_{\oplus}$ and 
planet c is approximately Neptune-mass (19 M$_{\oplus}$).  See Fig.\ \ref{fig:completeness_115617}.

\textit{HD 117176 (HIP 65721; program:\ A)}---70 Virginis hosts a  $\sim$7 \mjup planet in a 0.5 AU orbit \citep{Marcy96}.  
See Fig.\ \ref{fig:completeness_117176}.

\textit{HD 120136 (HIP 67275; programs:\ S, C, A)}---$\tau$ Bootis b was one of the first exoplanets discovered \citep{Butler1997}. 
It is an extreme hot Jupiter with a mass 6 \mjup and an orbital period of 3.3 days. 
We also detect a significant linear trend in the RV data and a periodicity at $\sim$5000 days. 
However, due to the possibilities of offsets within the Lick data, it is difficult to trust a long-period signal 
in the Lick data alone \citep{Fischer2014}. 
The trend is likely real, but the 5000 day periodicity could be due to instrumental effects or a  
stellar magnetic activity cycle.  See Fig.\ \ref{fig:completeness_120136}.

\textit{HD 131156 (HIP 72659; program:\ S, C, A)}---$\xi$ Bootis A is a young (200 Myr), late G star in a binary stellar system
with the late K star, $\xi$ Bootis B \citep{Mamajek2008} . 
The significant linear trend with no detectable curvature in the RV time series of $\xi$ Bootis A is likely due to the stellar companion. 
See Fig.\ \ref{fig:completeness_131156}.

\textit{HD 131977 (HIP 73184; programs:\ S, C)}---This star has a significant linear RV trend, likely due to a binary companion (Gl 570B).  
See Fig.\ \ref{fig:completeness_131977}.

\textit{HD 161797 (HIP 86974; programs:\ S, C, A)}---The RV time-series for $\mu$ Herculis shows a strong linear trend and significant curvature (or a possibly closed orbit).  
$\mu$ Herculis is a hierarchical triple stellar system.  
The data can be well-fit by a Keplerian orbit with a period of 9800 days, 
but this is very poorly constrained and the period is likely to be much longer.  
See Fig.\ \ref{fig:completeness_161797}.

\textit{HD 168723 (HIP 89962; program:\ C)}---The Lick data for $\eta$ Serpentis show a marginal linear trend while the long-baseline 
pre-upgrade Keck data does not show the trend.  
See Fig.\ \ref{fig:completeness_168723}.

\textit{HD 185395 (HIP 96441; program:\ C, A)}---This early-type star (F4 V) has high jitter 
and a claimed, controversial planet with $P \approx 150$~days \citep{Desort2009}.  
We see evidence in our Lick data for RV variation at this period 
and other periods related by the yearly alias, however these signals are not statistically significant.
See Fig.\ \ref{fig:completeness_185395}.

\textit{HD 191408 (HIP 99461; programs:\ S, C, A)}---This star shows a slight linear trend with no detectable curvature, presumably due to its 
common proper motion companion LHS 487 (K4V).  
See Fig.\ \ref{fig:completeness_191408}.

\textit{HD 201091 (HIP 104214; programs:\ S, C, A)}---61 Cygni A shows a strong linear trend caused by 61 Cygni B.  
Under the direction of Peter van de Kamp, \citet{Strand1943,Strand1957} suggested that 61 Cygni A was orbited by one or two giant planets.  
These claims were later rejected by high precision RVs \citep{Walker1995,Cumming08}.
See Fig.\ \ref{fig:completeness_201091}.

\textit{HD 201092 (HIP 104217; program:\ S)}---61 Cygni B shows a strong linear trend caused by 61 Cygni A.  
See Fig.\ \ref{fig:completeness_201092}.

\textit{HD 217014 (HIP 113357; program:\ S)}---51 Pegasi is the host to the first known exoplanet; 
a hot Jupiter with a mass of 0.5 \mjup and 
an orbital period of 4.2 days \citep{Mayor1995}.  
We also detect a long-period signal but this is likely caused by instrumental offsets within the Lick dataset.  
See Fig.\ \ref{fig:completeness_217014}.

\textit{HD 219134 (HIP 114622; programs:\ S, C)}---This star has candidate planets including a giant planet in a 3 AU orbit that we will continue to examine as more RVs are gathered.

\subsection{Injection-Recovery Completeness}
\label{sec:injection}
To determine our sensitivity to planets as a function of $\msini$ and $P$ 
we injected synthetic planetary RV signals into the real RV data for each star, 
preserving the actual times of observation, the previously measured velocities, and the measurement uncertainties.
We then used the iterative automated planet search algorithm (Sec.\ \ref{sec:automated_planet_search}) 
to attempt to recover the injected signals. We injected planets on circular orbits uniformly distributed 
in log($K$) and log($P$) centered around the sensitivity threshold line determined from the 
method of \citet{Howard10a}. 
We did not study sensitivity as a function of eccentricity because $>$80\% of known Doppler-discovered 
planets have $e < 0.4$ \citep{Wright2011}, for which a circular orbit approximation is adequate (Fig.\ \ref{fig:kepzoo}). 

Note that the completeness plots are a function of minimum mass, $\msini$.  While $\msini$ gives a good estimate for $M$ usually, it is a significant underestimate for face-on orbits.  These cases are rare though for the same reason that the fractional area of the EarthÕs surface near the poles is small.  For example, 87\% of random inclinations are between 30 and 90 degrees (where the true mass is within a factor of two of the minimum mass).  And there is only a 0.5 chance of $i < 5.7^\circ$, where $\msini$ underestimates $M$ by a factor of ten or more \citep{Fischer2014}.

We injected \ninject synthetic planets into the RV time-series for each star in the sample. 
The synthetic planets are injected in addition to any known planets in the system and the 
search algorithm starts by searching for one additional planet while simultaneously fitting for all known 
planets in the system.  An injected planet is considered recovered if the highest peak in the periodogram 
is above the 1\% FAP detection threshold, the period of that peak is within \recoverthresh of the injected period, 
and the phase of the recovered orbit is within $\pi/6$ of the injected phase.  
In some cases multiple planets are recovered when only a single planet is injected. 
We still consider these cases good recoveries if any of the detected periods are within \recoverthresh 
of the injected period.

\begin{figure}
         \begin{center}
              \includegraphics[width=0.50\textwidth]{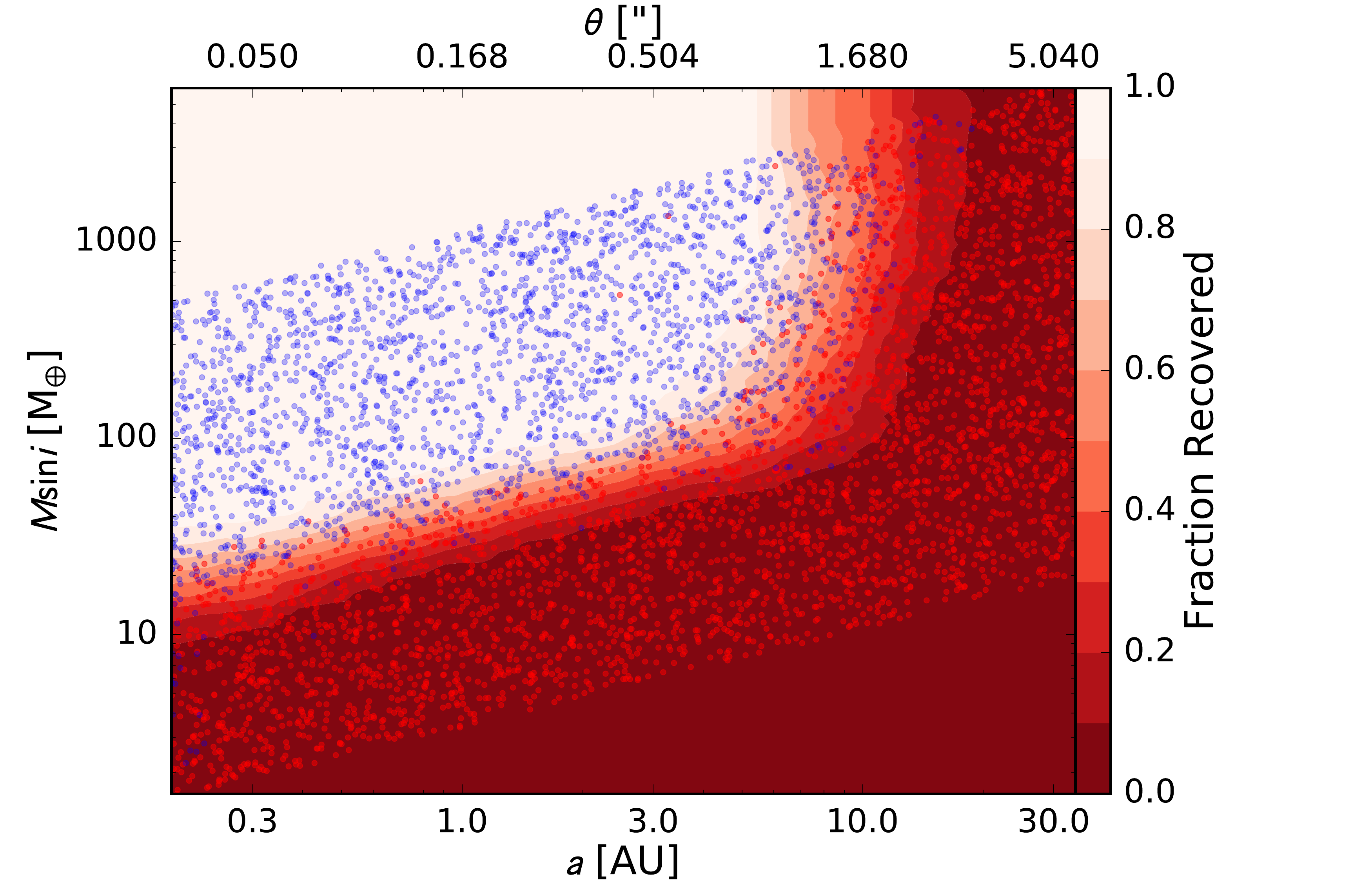} 
         \end{center}
         \caption{Example completeness plot for HD~4614 showing the results of our injection/recovery tests. 
         Blue dots indicate the \msini\ and $a$ combinations of injected planets that were successfully recovered by our automated planet search, 
         while red dots indicate injections that were not recovered. The contours are computed from a two-dimensional 
         moving average over a 100$\times$100 grid with window widths of 0.5 dex in \msini\ and $a$. 
         Completeness in regions of the grid outside the limits of the injection-recovery tests are extrapolated 
         vertically along columns from the nearest point in the grid that contains at least 20 injections within the averaging window.  
         Appendix \ref{app:completeness} includes a completeness plot for every star with Doppler measurements.
         }
         \label{fig:completeness_example}
\end{figure}

Completeness contours are derived from the injection-recovery tests by computing a two-dimensional moving average 
of the recovery rate over a 100$\times$100 grid in semi-major axis ($a$) and $\msini$ 
with an averaging window of width 0.5 dex.  
Figure \ref{fig:completeness_example} shows an example completeness plot.  
The full set of completeness contours for all stars in the sample are in Appendix \ref{app:completeness}.

In some cases anomalies are visible in the shape of the completeness contours derived from the injection-recovery tests. 
A feature of enhanced sensitivity is sometimes seen at long periods. 
This is caused by  stellar magnetic activity masquerading as the injected planet. 
Although we require that the period of the recovered planet is within 25\% of the period of the injected planet and the 
recovered phase is within $\pi/6$ of the injected phase, some of the injected planet signals will happen to look very 
similar to the magnetic activity signal. 
Planets that are well below the local completeness threshold will still be ``recovered'' when the algorithm finds the 
magnetic activity cycle as the most significant periodic signal and finds that it has the correct period and phase to 
match the injected signal. This situation can occur in $\lesssim10\%$ of cases and can corrupt the 16\% and 84\% completeness contours. 
In other cases narrow spikes of decreased sensitivity are seen well above the local completeness threshold. 
This occurs in systems with known planets.  For example, see Fig.\ \ref{fig:completeness_75732}.
When a planet is injected at nearly the same period as a known planet, it can be absorbed by the fit for the known planets. Fortunately for the estimate, it is unlikely that two planets with very similar orbital periods could remain dynamically stable.  
Kepler-36bc has the smallest period ratio detected to date with $P_c/P_b \simeq 7/6$ \citep{Carter2012}. 

The one-dimensional contours provided in machine-readable tables as a supplement to this report were derived by scanning the 
2D completeness grid column-by-column starting at low then progressively higher \msini\ values until recovery rates 
of 16\%, 50\%, and 84\% are first reached. We report \msini\ as a function of $P$, $a$, and projected on-sky separation 
($\theta$) for each of the three 1D completeness lines which corresponds to the minimum \msini\ that is detectable in 
the given fraction of injections.  
In separate machine-readable files, 
we include the two-dimensional completeness for each star using the grid of semi-major axes and \msini\ values in the completeness plots.  
See Appendix \ref{app:data}.

Appendix \ref{app:completeness} includes completeness estimates for every star with RVs.

\subsection{Summary Results}
\label{sec:summary_results}

\begin{figure}
         \begin{center}
              \includegraphics[width=0.49\textwidth]{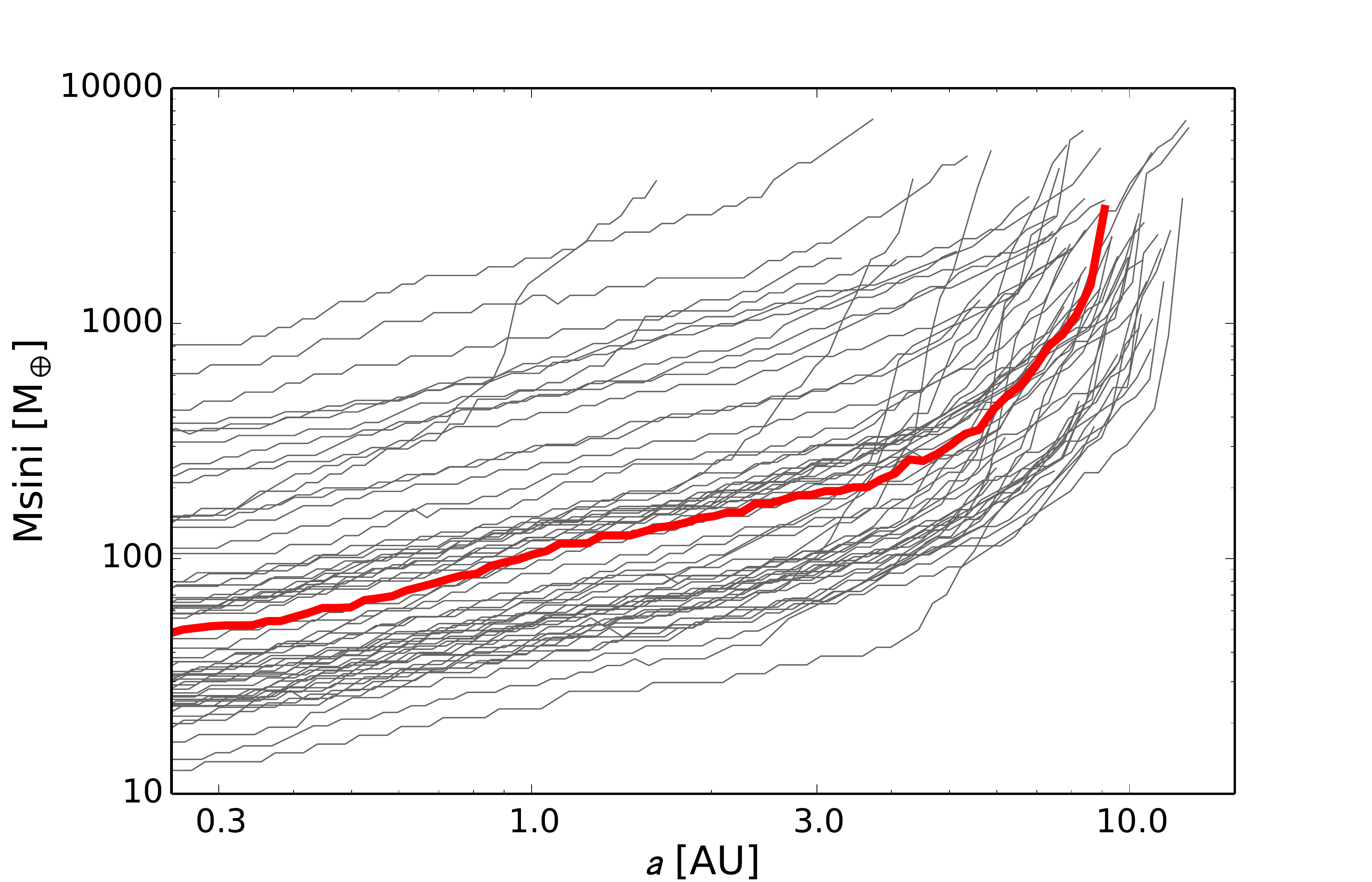} 
              \includegraphics[width=0.49\textwidth]{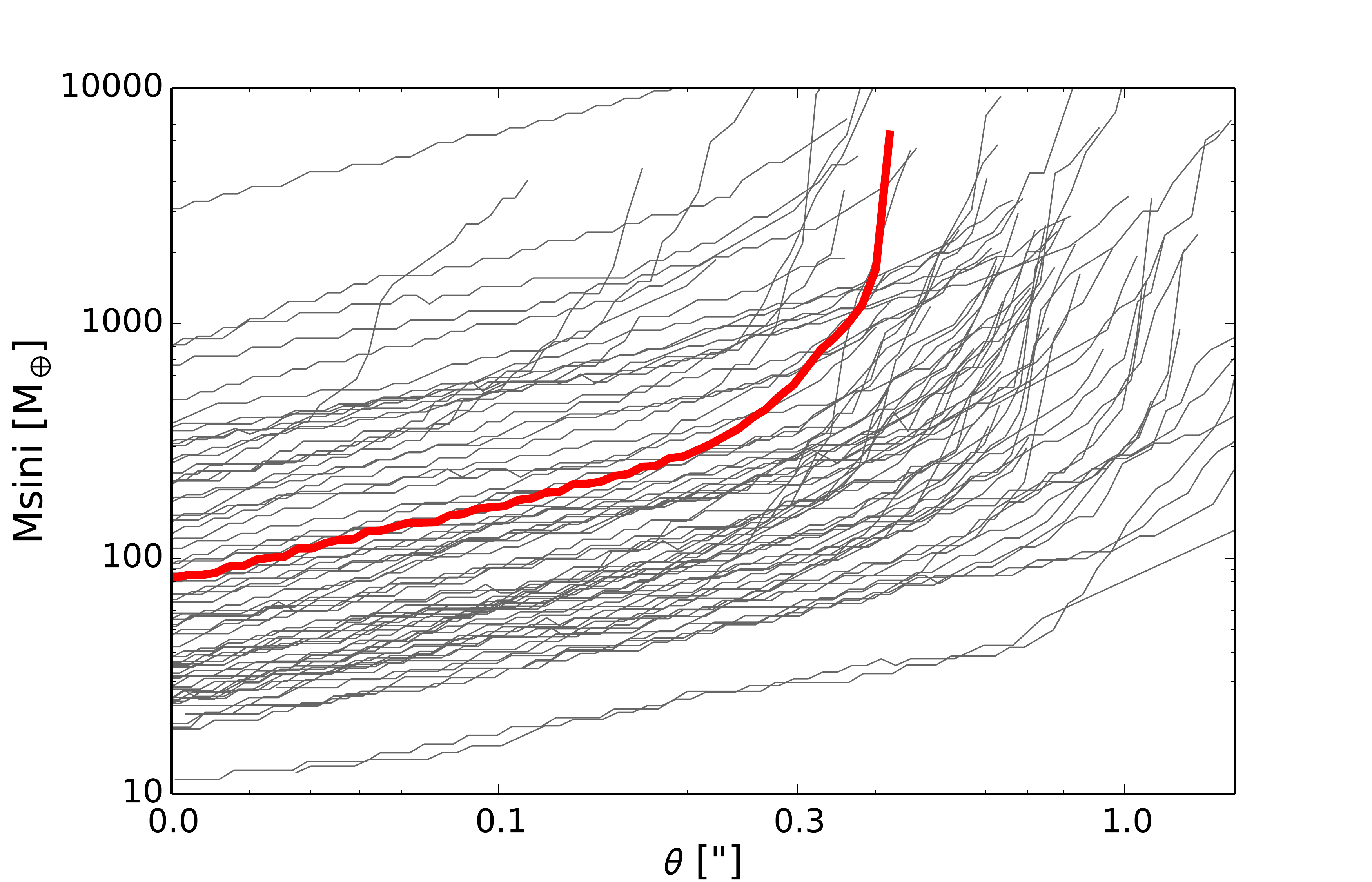} 
         \end{center}
         \caption{Contours showing 50\% completeness   
         for every star with Doppler measurements (gray lines) and the median of those  contours (red line). 
         The panels show the same completeness curves as a function of semi-major axis (left) and projected separation (right).
         }
         \label{fig:completeness_all_stars}
\end{figure}

\begin{figure}
         \begin{center}
              \includegraphics[width=0.49\textwidth]{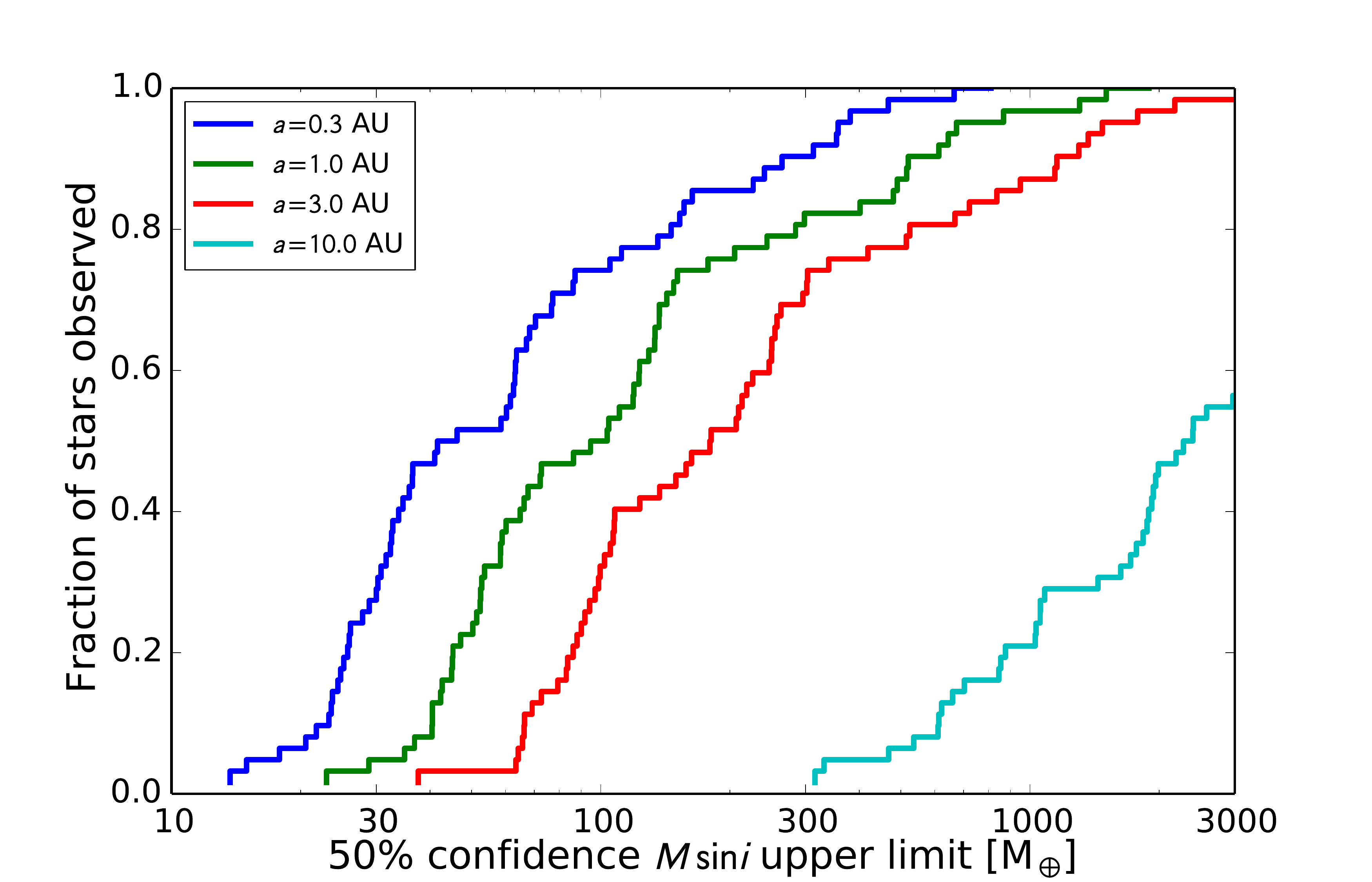} 
              \includegraphics[width=0.49\textwidth]{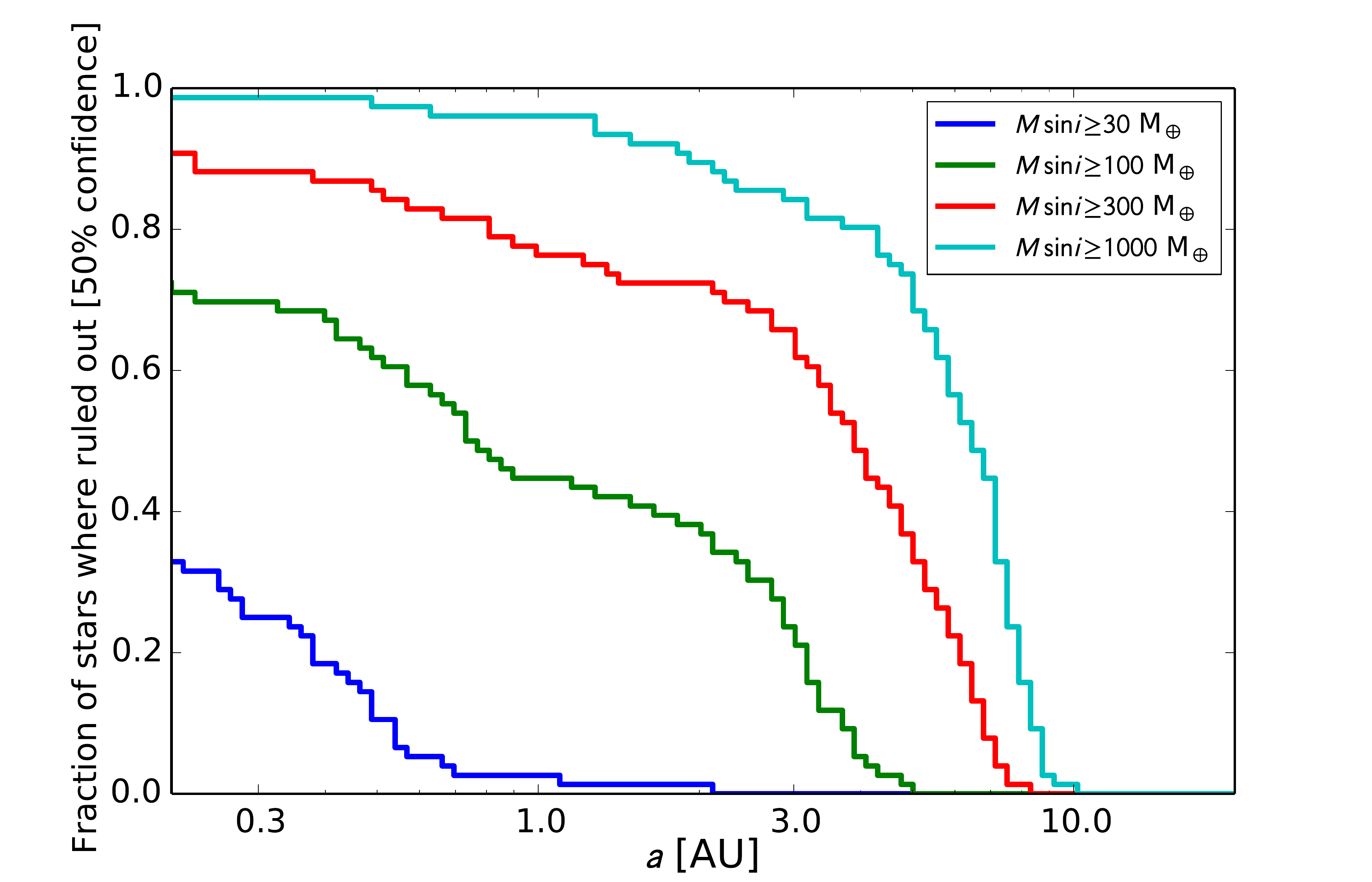} 
         \end{center}
         \caption{Fraction of stars observed by Lick/Keck  Planet Searches whose RV observations can 
         exclude planets of particular masses  and semi-major axes.  
         The left panel depicts the mass upper limits (\msini\ at 50\% confidence) at four semi-major axes while the 
         right panel shows the fractions of stars with ruled out companions for four \msini\ values, as a function of semi-major axis.}
         \label{fig:summary_results}
\end{figure}

Figures  \ref{fig:completeness_all_stars} and \ref{fig:summary_results} summarize the  results of our completeness 
estimates.  
Figure \ref{fig:completeness_all_stars} shows the 50\% detection thresholds for every star in the sample, 
with the median completeness contour shown in red.  
Note that the completeness contours span more than order of magnitude in \msini, while having similar shapes.  
This variation is mostly due to the differing jitter values for stars observed in the Lick and Keck surveys.  
Figure \ref{fig:summary_results} summarizes the full set of contours, slicing in both \msini\ and semi-major axis.

\subsection{Idealized Completeness}
\label{sec:idealized_completeness}

An idealized Doppler planet search can be characterized by the number of observations ($N_{\rm obs}$), their time span ($t_{\rm span}$), 
and the precision of a single measurement  ($\sigma_{\rm RV}$).  
Precision stems from astrophysical noise sources (``jitter'') that produce apparent Doppler shifts and measurements uncertainties 
stemming from the signal-to-noise ratio and the Doppler information content (number of lines, $\vsini$, etc.) of each spectrum.  
In practice, $\sigma_{\rm RV}$ for a particular star can be empirically estimated as the 
standard deviation of the RVs measured over a long time span (assuming no orbital companions are present). 
That is, single measurement errors are approximately the measurement scatter.
To improve sensitivity, several RVs of the same star are often gathered 
on a given night to average over stellar noise and to improve signal-to-noise.  
In such cases, $\sigma_{\rm RV}$ represents the scatter in the nightly averages (or the appropriate binning timescale).

In the limit of large $N_{\rm obs}$, uniform orbital phase coverage (random times of observation), and an orbital period 
$P < t_{\rm span}$,  50\% completeness can be characterized by a Doppler amplitude $K_{50}$ that is related to the survey parameters by 
\begin{equation}
K_{50} = \alpha \frac{\sigma_{\rm RV}}{\sqrt{N_{\rm obs}}},
\label{eq:K_50}
\end{equation}
where $\alpha$ is a dimensionless parameter that 
represents the signal-to-noise ratio of a detectable plant, accounting for $\sqrt{N_{\rm obs}}$ measurements.
Naively, we expect $\alpha$ to be greater than one and of order a few.  
We estimated $\alpha$ using three techniques described below.  

First, we estimated $\alpha$ using the completeness plots in Appendix \ref{app:completeness} 
using $N_{\rm obs}$ and $t_{\rm span}$ from our measurements and taking  $\sigma_{\rm RV}$ to be the RMS of the RVs after subtracting 
all significant signals.  
Using this technique we found $\alpha$ in the range $\sim$5--20.  This range accounts for the ways that 
real observations violate our assumptions of the idealized survey: 
telescope/instrument combinations have heterogenous Doppler precision 
(choosing the ``typical'' $\sigma_{\rm RV}$ difficult to estimate),  
RV zero point offsets between datasets must be fitted for, and 
non-random observing cadences.  
Nevertheless, this factor of four range in $\alpha$ suggests that we might simulate completeness with similar mass precision 
simply by estimating survey and stellar parameters.  

Our second method to estimate $\alpha$ was calculating completeness for an ideal observing campaign.
Note in Eq.\ \ref{eq:K_50} that $K_{50}$ and $\sigma_{\rm RV}$ have the same units and scale linearly with one another.  
Searching for a $K = 10$ \ms signal with $\sigma_{\rm RV} = 1$ \ms precision is equivalent to searching for 
a $K = 100$ \ms signal with $\sigma_{\rm RV} = 10$ \mse.  
Making the problem dimensionless, we define a survey sensitivity,
\begin{equation}
\kappa_{50} = \frac{K_{50}}{\sigma_{\rm RV}} =  \frac{\alpha}{\sqrt{N_{\rm obs}}},
\label{eq:kappa}
\end{equation}
that is valid for dimensionless orbital periods $\tau = P/t_{\rm span} \lesssim 1$.

\begin{figure}
         \begin{center}
              \includegraphics[width=0.50\textwidth]{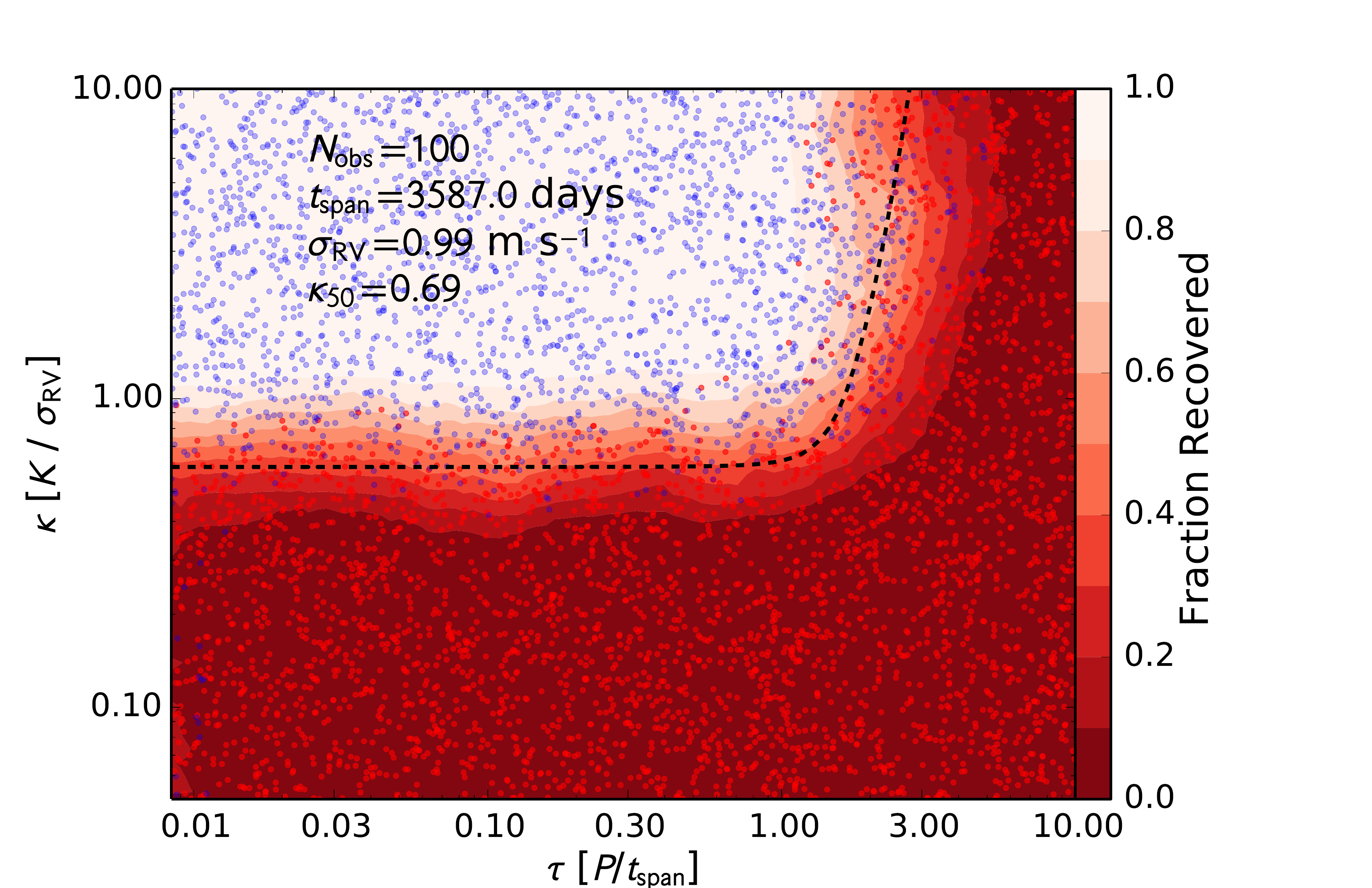} 
              \includegraphics[width=0.50\textwidth]{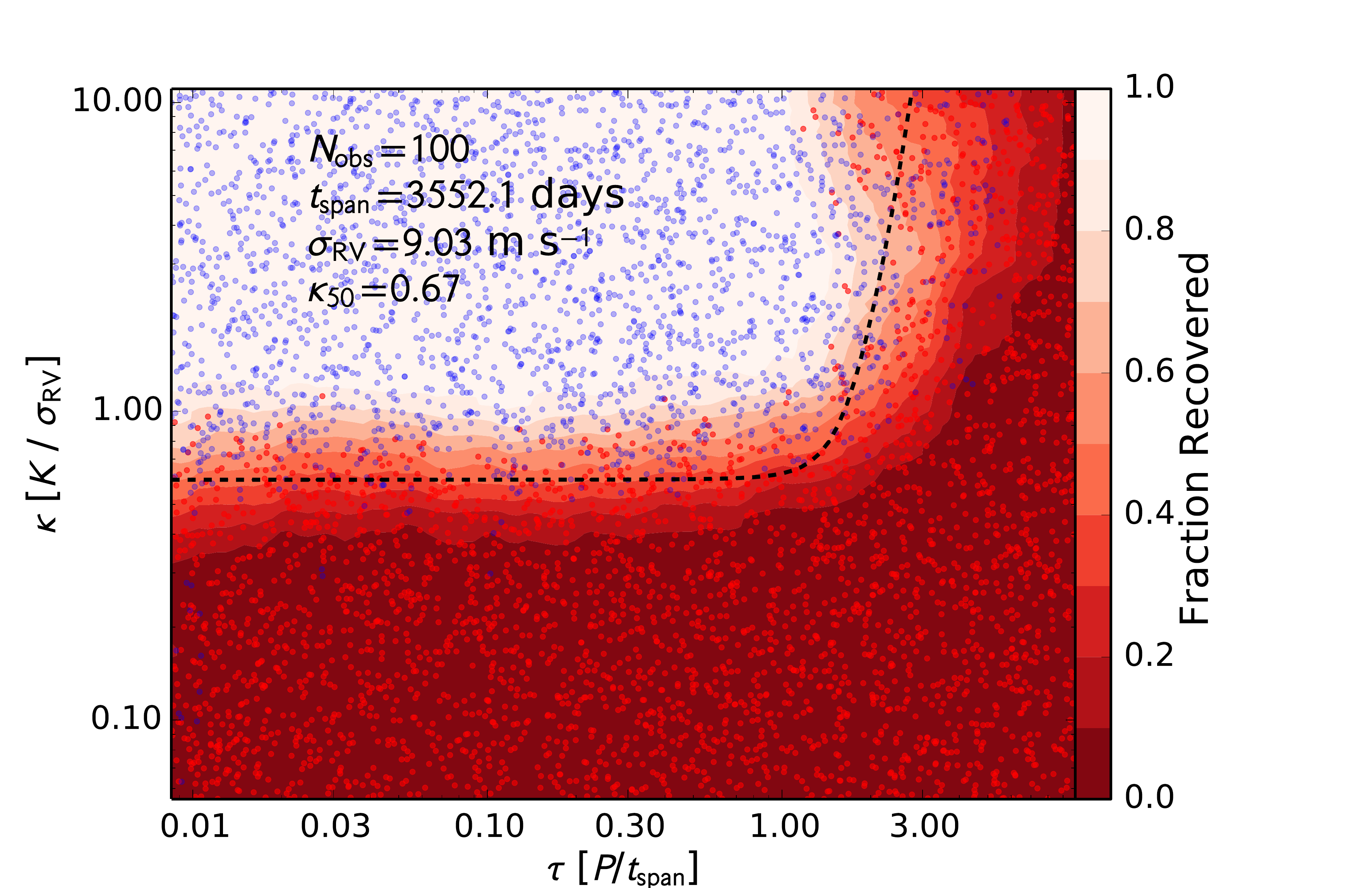} 
              \includegraphics[width=0.50\textwidth]{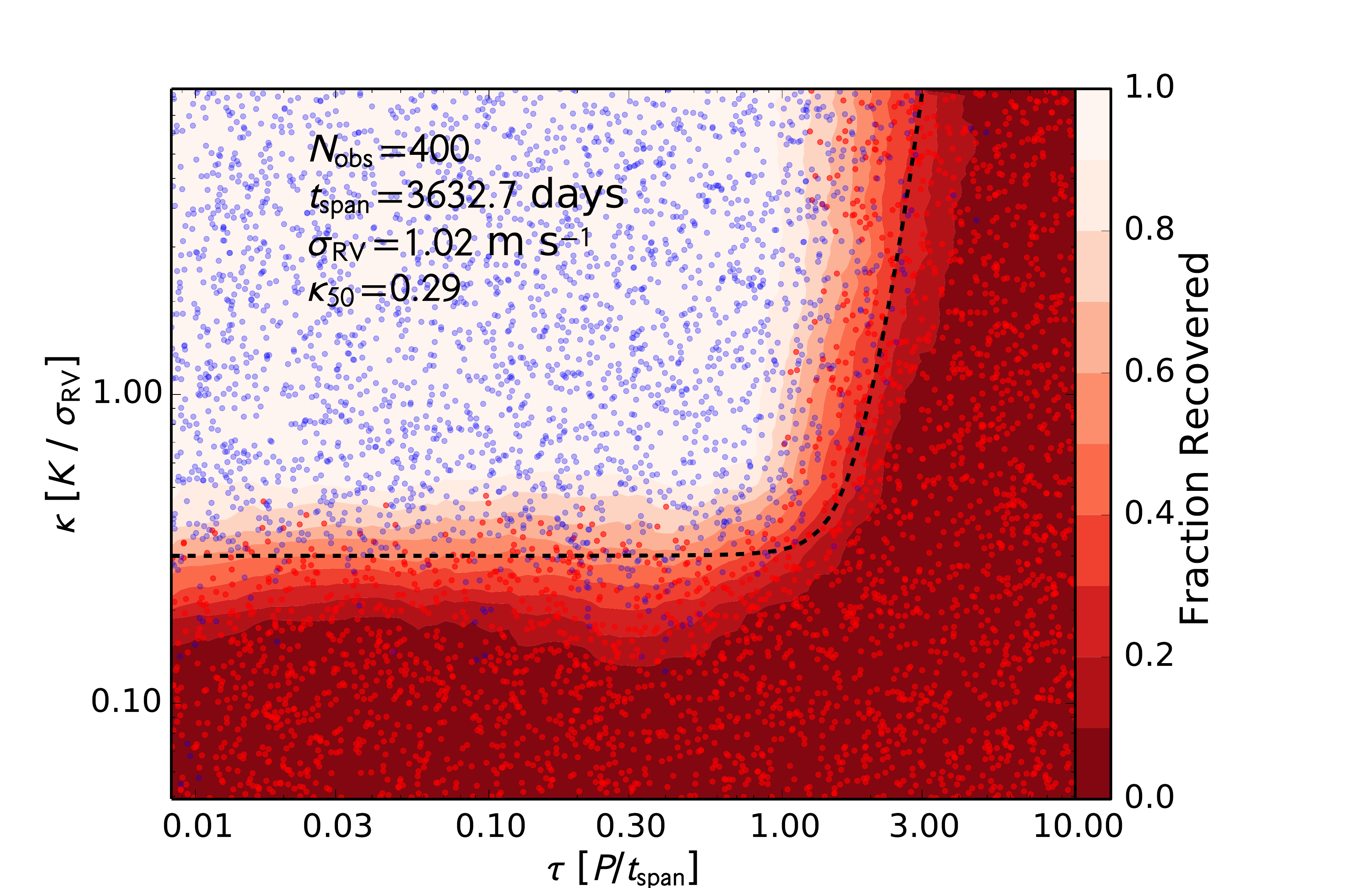} 
         \end{center}
         \caption{Planet search completeness in dimensionless Doppler amplitude $\kappa= K/\sigma_{\rm RV}$ 
         versus dimensionless orbital period $\tau= P/t_{\rm span}$ 
         for three idealized observational surveys.  
         The survey parameters are:
         $N_{\rm obs}=100$, $\sigma_{\rm RV} = 1$ \mse, $t_{\rm span} = 10$ yr (top); 
         $N_{\rm obs}=100$, $\sigma_{\rm RV} = 10$ \mse, $t_{\rm span} = 10$ yr  (middle);  
         $N_{\rm obs}=400$, $\sigma_{\rm RV} = 1$ \mse, $t_{\rm span} = 10$ yr  (bottom).  
         The three surveys give consistent $\alpha = \kappa_{50} \cdot \sqrt{N_{\rm obs}} \approx 6$ for $\tau < 1$, 
         showing that idealized completeness is scale invariant (see caveats in Sec.\ \ref{sec:idealized_completeness}).
         Dashed black lines show  $\kappa_{50}(\tau)$ estimated from Eq.~\ref{eq:kappa_tau}.
         }
         \label{fig:kappa_recovery}
\end{figure}

We tested and calibrated this idealized completeness model with injection-recovery simulations.  
Figure\ \ref{fig:kappa_recovery} shows the simulation results and their parameters, 
with the plots demonstrating that $\kappa \propto K/(\sigma_{\rm RV}\cdot\sqrt{N_{\rm obs}})$.  
$\kappa_{50}(\tau)$  is flat in these three simulations for short period orbits.  
Based on these simulations, we estimate $\alpha \approx 6$.
For $\tau \gtrsim 1.5$, the $\kappa_{50}$ contour rises rapidly; 
we estimate a slope  of $\sim$10 and $\kappa_{50}(\tau) = \alpha \cdot 10^{\tau-1.5}/\sqrt{N_{\rm obs}}$.  
We can combine these two functions smoothly using quadrature addition, 
\begin{equation}
\kappa_{50}(\tau) = \frac{\alpha}{\sqrt{N_{\rm obs}}} \cdot \sqrt{1+(10^{\tau-1.5})^2},
\label{eq:kappa_tau}
\end{equation}
giving
\begin{equation}
K_{50}(\tau) = \frac{\sigma_{\rm RV} \, \alpha}{\sqrt{N_{\rm obs}}} \cdot \sqrt{1+(10^{\tau-1.5})^2}.
\label{eq:K_tau}
\end{equation}

Eq.\ \ref{eq:K_tau} nicely encapsulates the 50\% percentile search completeness as a function of a three observational parameters: 
$\sigma_{\rm RV}$, $N_{\rm obs}$, and $t_{\rm span}$.
However, the simplistic form is only a fit to an idealized observational campaign 
and is likely  accurate to a factor of a few (perhaps good enough for planning observations).  
Real life observing is more complicated.  One should consider the following caveats 
when applying this idealized completeness model:

\begin{enumerate}
\item Measurement errors are rarely Gaussian distributed.  
Velocity errors from stellar jitter and instrument systematics are often time correlated.  
Periodic astrophysical noise sources (i.e., magnetic activity cycles) 
can masquerade as planets and/or the hinder detectability of small planets.
Thus, $\sigma_{\rm RV}$ doesn't fully encapsulate the Doppler noise sources.
\item Times of observation are non-random, depending on daily, monthly, and yearly cycles, not to mention scientific interests and 
the vicissitudes of telescope time allocation committees.  Our assumption of uniform phase coverage (from random times of observation 
and large $N_{\rm obs}$) is rarely achieved for real surveys.  Non-randomly timed clusters of observation times inject 
period aliases into the periodograms, which can suppress or enhance the detectability at particular periods \citep{Dawson2010}.  
Furthermore, combining RVs from multiple telescopes can reduce sensitivity 
to long-period signals since such signals can be absorbed into the fitted zero-point offsets.
\item Completeness estimates using quantified detection criteria often fail to account for 
human evaluation of data.   Many observers are cautious about announcing planet discoveries, having seen 
instrumental systematics and correlated astrophysical noise exceed their prior expectations.  
\end{enumerate}

\begin{figure}[t]
         \begin{center}
              \includegraphics[width=0.45\textwidth]{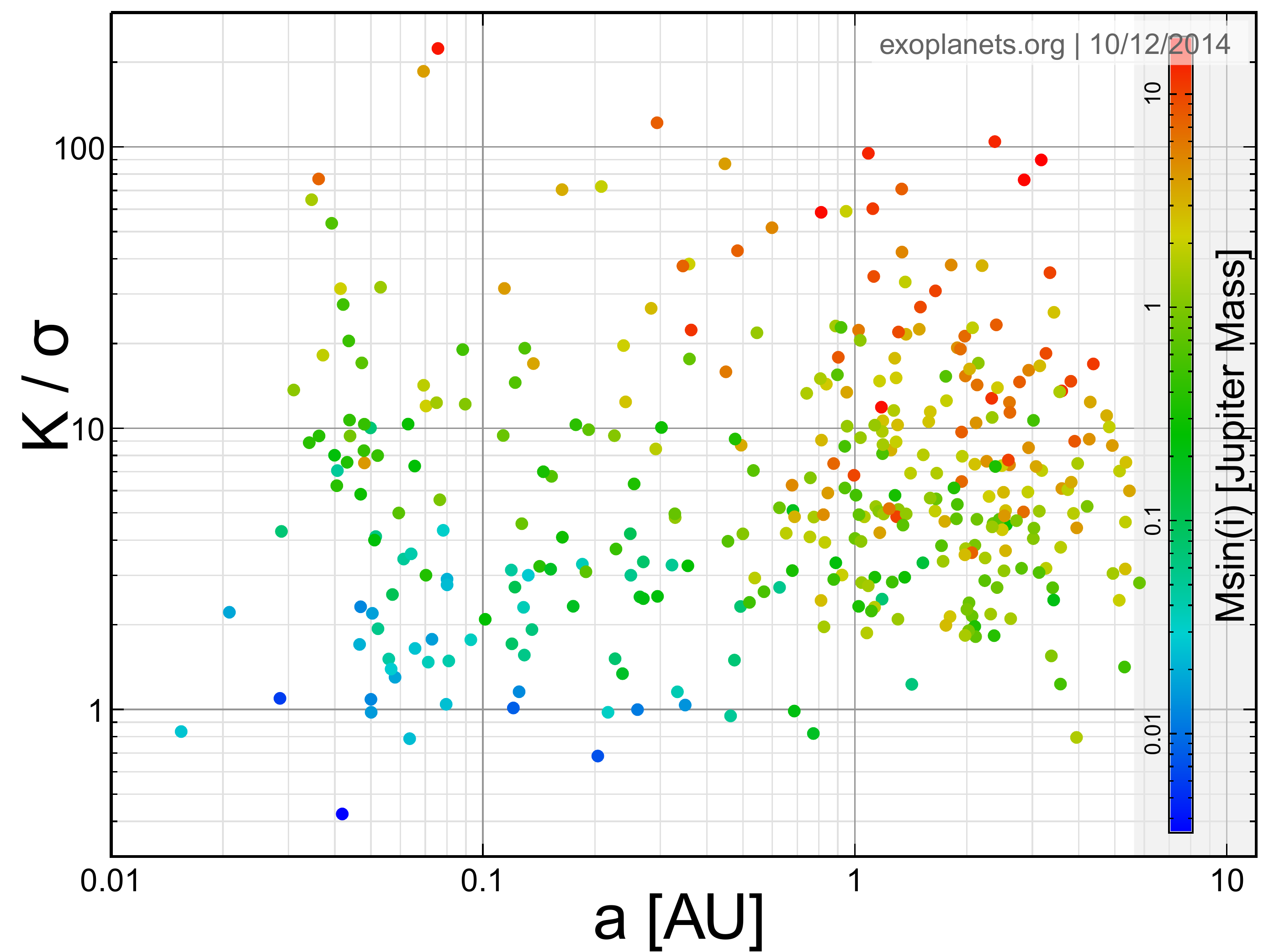} 
              \includegraphics[width=0.45\textwidth]{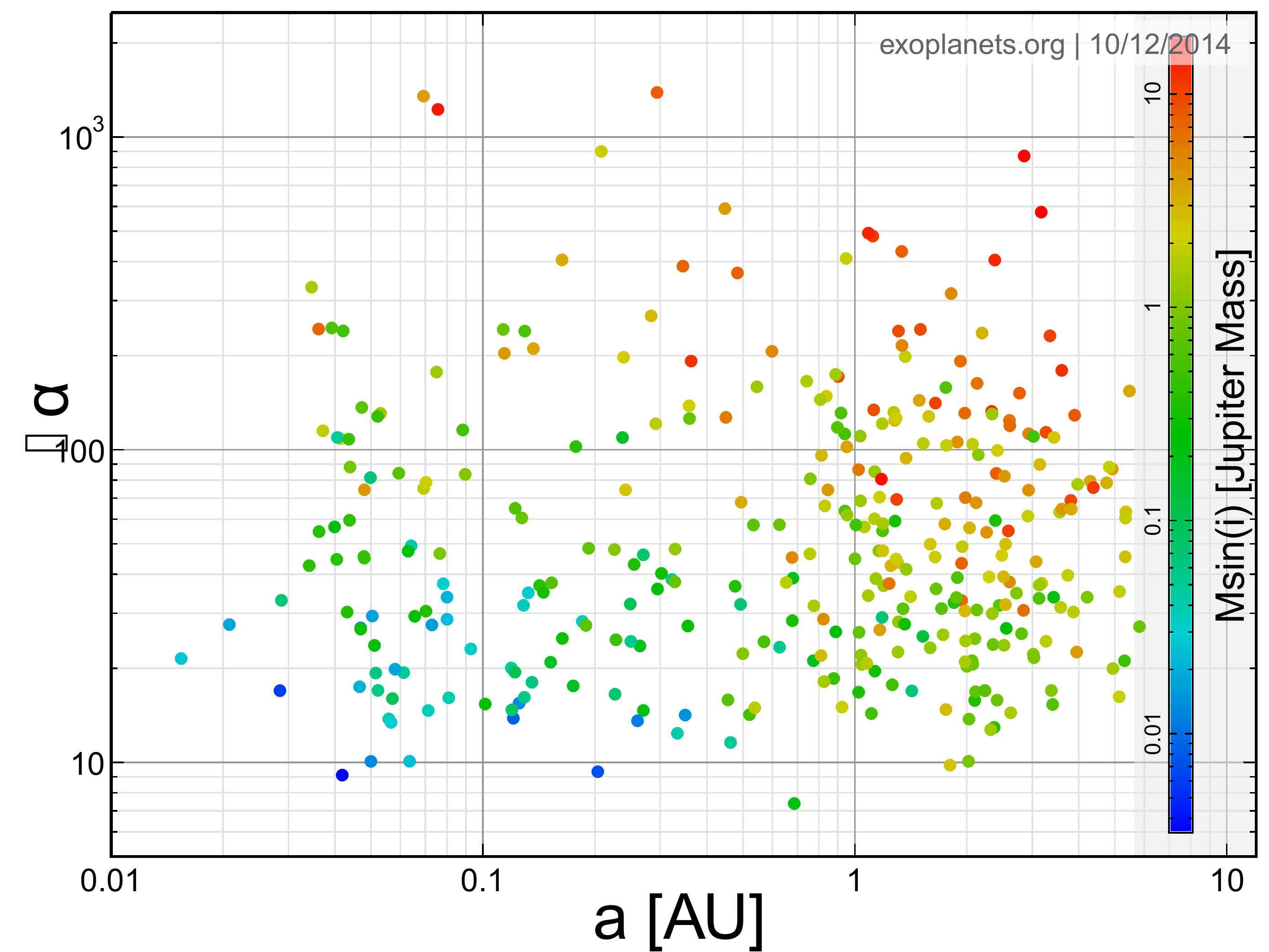} 
         \end{center}
         \caption{
         Distributions of  $\kappa = K / \sigma_{\rm RV}$ (left) and $\alpha = \sqrt{N_{\rm obs}} \, K  / \sigma_{\rm RV}$ (right) 
         vs.\ semi-major axis $a$ for real planets within 100 pc with Doppler-measured masses on exoplanets.org \citep{Wright2011}.  
         The plot includes discoveries from HIRES, HARPS, and other facilities.  
         Point color indicates \msini. 
         }
         \label{fig:kappa_exoplanets_org}
\end{figure}

Our third method of estimating  $\alpha$ involves examining the parameters of known planets on exoplanets.org \citep{Wright2011}.
Figure\ \ref{fig:kappa_exoplanets_org} shows estimates of $\kappa$ and $\alpha$ 
for planets with Doppler-measured masses with 100 pc.  
Only a handful of planets having been discovered with $\kappa < 1$, i.e.\ a signal smaller than the noise.  Such planets 
include HD 85512b \citep{Pepe2011} and HD 156668b \citep{Howard11}.
Our estimate of $\alpha \gtrsim 6$ for successful detection based on injection-recovery simulations appears to be 
slightly optimistic compared to the distribution of discovered planets that have a discovery threshold of $\alpha \gtrsim 10$ 
more commonly.  Note that Fig.\ \ref{fig:kappa_exoplanets_org} plots $\kappa$ and $\alpha$ that are defined slightly differently than in 
Eq.\ \ref{eq:kappa} and \ref{eq:K_tau}.  The figure shows the Doppler amplitude $K$, not the 50\%  completeness value $K_{50}$.  
And $N_{\rm obs}$ is the number of measurements in the ``orbit reference'' paper (ORBREF on exoplanets.org), 
not the number of measurements when the planet was first detectable. 
Still, the discovery of some exoplanets with $\alpha \approx 10$ suggests that our estimate of $\alpha \approx 6$ is not wildly off. 


\subsection{Prospective Completeness for Unobserved Stars}
\label{sec:completeness_nodata}

For stars without Doppler observations, it is often helpful to estimate the sensitivity of a prospective Doppler campaign.  
With the above caveats in mind, 
we can use the idealized completeness formalism of Sec.\ \ref{sec:idealized_completeness} to estimate completeness.  
The campaign is characterized by $N_{\rm obs}$ and $t_{\rm span}$, while the 
the RV observations are characterized by $\sigma_{\rm RV}$.  The general procedure to estimate the 50\% completeness in mass 
as a function of semi-major axis, $M_p \sin{i}_{50}(a)$, is: 
\begin{enumerate}
\item Choose $N_{\rm obs}$ and $t_{\rm span}$ for the survey.  Choose a star with a particular $M_\star$ and estimate $\sigma_{\rm RV}$ from the expected jitter
and measurement uncertainties.  
\item Compute $K_{50}(P)$ from Eq.\ \ref{eq:K_tau} using $\alpha = 6$ (idealized simulations) or $\alpha = 10$ 
(the threshold of detection for historical discoveries).
\item Convert $K_{50}(P)$ into a 50\% mass contour $M_p \sin{i}_{50}$($P$) using the inverted form of Eq.~\ref{eq:K} for circular orbits, 
\begin{equation}
\frac{M_p \sin{i}_{50}(P)}{M_{\rm J}}  = \frac{K_{50}(P)}{28.4\,\mathrm{m\,s^{-1}}}  \left( \frac{M_\star\!+\!M_p}{M_\odot}\right)^{2/3} \left(\frac{P}{\mathrm{yr}} \right)^{1/3}.
\end{equation}
For planetary mass companions ($M_\star \gg M_p$), the first term in parentheses simplifies to $(M_\star / M_p)^{2/3}$.
\item Convert $P$ to $a$ using Kepler's Third Law, $a = (G M_\star P^2 / 4\pi^2)^{1/3}$, again assuming $M_\star \gg M_p$.
\end{enumerate}

Computing idealized completeness for all unobserved stars (Table \ref{tab:targets_without_data}) requires good estimates of 
$M_\star$ and $\sigma_{\rm RV}$ for every star.  $M_\star$ could be reasonably estimated from photometry and other sources, 
but the expected RV scatter, $\sigma_{\rm RV}$, is difficult to estimate accurately without a detailed study of each star.  
In lieu of detailed completeness estimates for every star in Table \ref{tab:targets_without_data}, 
we provide the above recipe for estimating completeness with due caution that the 
$M_p \sin{i}_{50}(P)$ limits scale linearly with $\sigma_{\rm RV}$.  
Underestimated astrophysical and/or measurement errors will give false hope for planet detectability.  

Stars were not included in our historical Lick/Keck Doppler search for several reasons enumerated in 
Table \ref{tab:targets_without_data}.  
These include: 
the star is too far South for Keck/Lick; 
the star has evolved into a subgiant or giant;
the star is earlier spectral type  than $\sim$F8V;
the star is young and chromospherically active;
and the star is a spectroscopic binary.  
We discuss the expected Doppler precision $\sigma_{\rm RV}$ for each class below.

\textbf{Southern Hemisphere}---Many excellent main sequence G and K dwarfs are simply too far South to be easily observed at 
Keck and/or Lick Observatory (the cutoff is $-$30$^\circ$ to $-$40$^\circ$ declination).  Observations  of these stars 
could yield completeness curves comparable to our best cases in Fig.\ \ref{fig:completeness_all_stars}.  
Other teams are likely observing many of these stars \citep[e.g.,][]{Mayor2011}.

\textbf{Early Spectral Type}---Spectral types earlier than $\sim$F8V have a decreasing density of spectral lines and 
an increasing average \vsini, both of which degrade Doppler precision to $\gtrsim$5 \ms  for the sample identified 
for Lick and transferred to Keck.  
\citet{Galland2005} developed a Doppler planet search for A--F stars.  They measured the RV scatter for a sample 
of stars observed with ELODIE and HARPS finding an RV scatter of $\sigma_{\rm RV} \approx 0.16 \times \vsini^{1.54}$ 
(for ELODIE observations with S/N = 200).  
This relationship provided an accurate precision forecast at the factor of two level. 
The \vsini\ scaling to the $\sim$1.5 power is consistent with a study by \citet{Bouchy2001} 
on the fundamental photon noise limit to radial velocity measurements, in the case of early F type main sequence stars.
In the examples below, we adopt the \citet{Galland2005} prescription for $\sigma_{\rm RV}$.
Note however that this relation fails for late A and early F type stars ($B-V$ between 0.2 and 0.4) that are often highly RV variable. 
This $B-V$ range corresponds to the intersection of the instability strip and the main sequence, 
where pulsators including $\delta$ Scuti and $\gamma$ Doradus stars are found.  
The most massive stars in the Exo-C/Exo-S/WFIRST-AFTA target lists are B stars ($B-V < 0$; Fig.\ \ref{fig:hr_diag}).
Doppler measurements have achieved $\sigma_{\rm RV}$ = 0.8 to 2.0 \kms\ for such targets, 
depending on \vsini\ (J.\ Johnson, personal communication).

\textbf{Evolved Stars}---Surface gravity is reduced as stars evolve off the main sequence.  
The resulting surface oscillations produce RV variations that can be used to measure precise stellar properties (asteroseismology), and
that serve as a noise source for Doppler planet searches.  \citet{Kjeldsen1995} estimated the Doppler oscillation amplitude, 
finding a dependence on the light-to-mass ratio, $v_{\rm osc} = 0.234(L_\star/M_\star)$ \mse.  
We adopt this estimate of $v_{\rm osc}$ for $\sigma_{\rm RV}$ for giant stars.
As a check, this formula predicts $v_{\rm osc} = 6$ \ms for Pollux and \citet{Reffert2006} observed an RV scatter of 9 \mse, 
which is the level of agreement we expect for estimates like these.

\textbf{Young Stars}---Line shape distortions due to rotationally modulated stellar surface features (e.g., spots and plage) 
make young stars difficult targets for planets searches  \citep{Crockett2012}.  
\citet{Hillenbrand2014} measured the RV scatter for a set of young stars observed by HIRES, 
finding $\sim$100 \ms scatter for young stars ($\sim$30 Myr) with the activity measure \lrphk\ = $-$4.0, 
and RV scatter decreasing approximately linearly with  \lrphk\  to $\sim$3 \ms for old, quiet stars with \lrphk\ = $-$5.0.

\textbf{Spectroscopic Binaries}---Spectroscopic binaries are routinely excluded from Doppler surveys 
because their spectra cannot be modeled by standard cross-correlation or forward-modeling techniques.  
However, techniques developed by \citet{Konacki2009} show promise for 
detecting $\sim$\mjup planets for close-in orbits and several \mjup planets for $\sim$AU orbits.  
We decline to provide specific advice for the expected Doppler precision and completeness for these stars 
since such estimates would be highly uncertain.

\begin{figure}
         \begin{center}
              \includegraphics[width=0.49\textwidth]{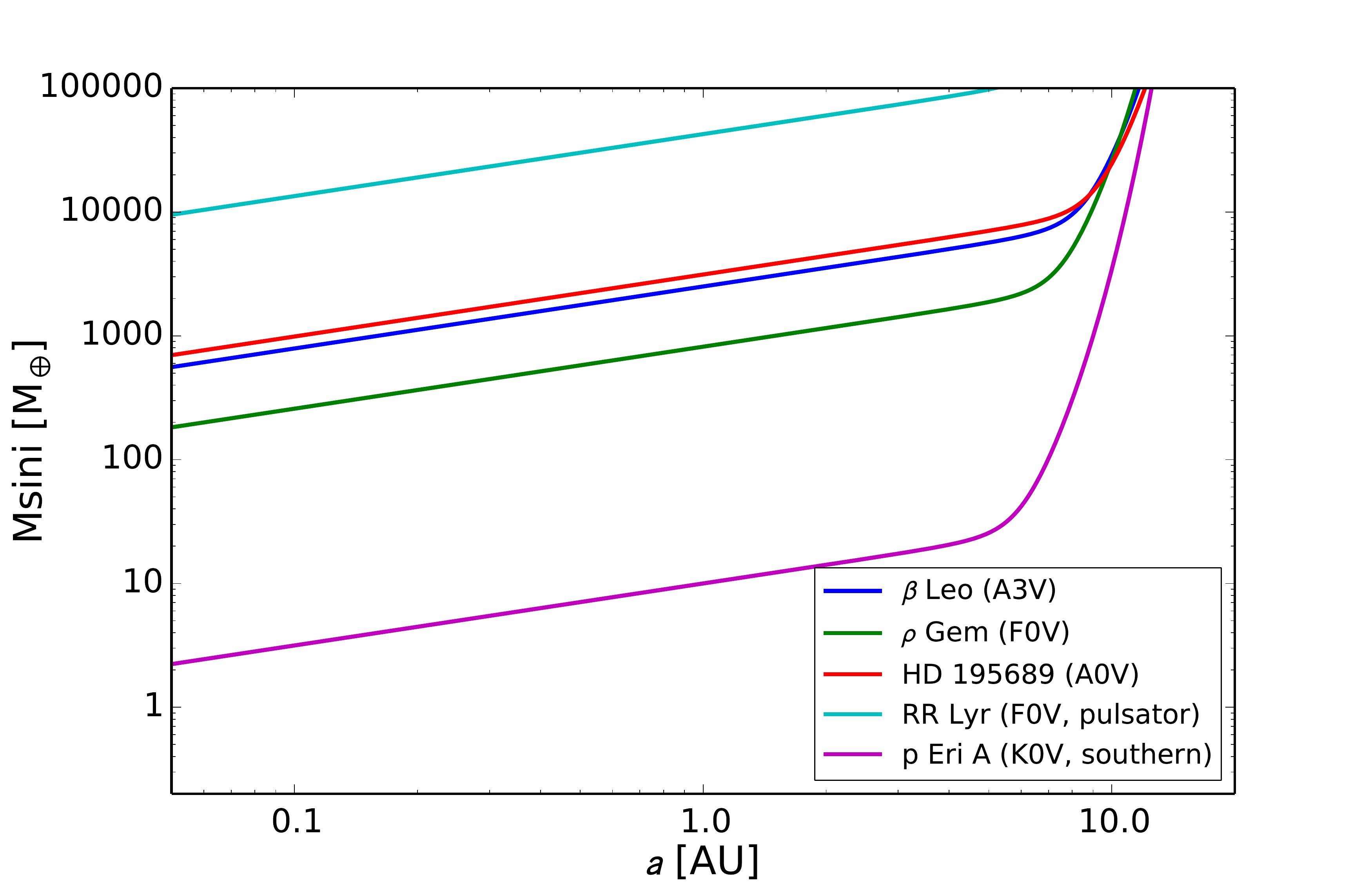} 
              \includegraphics[width=0.49\textwidth]{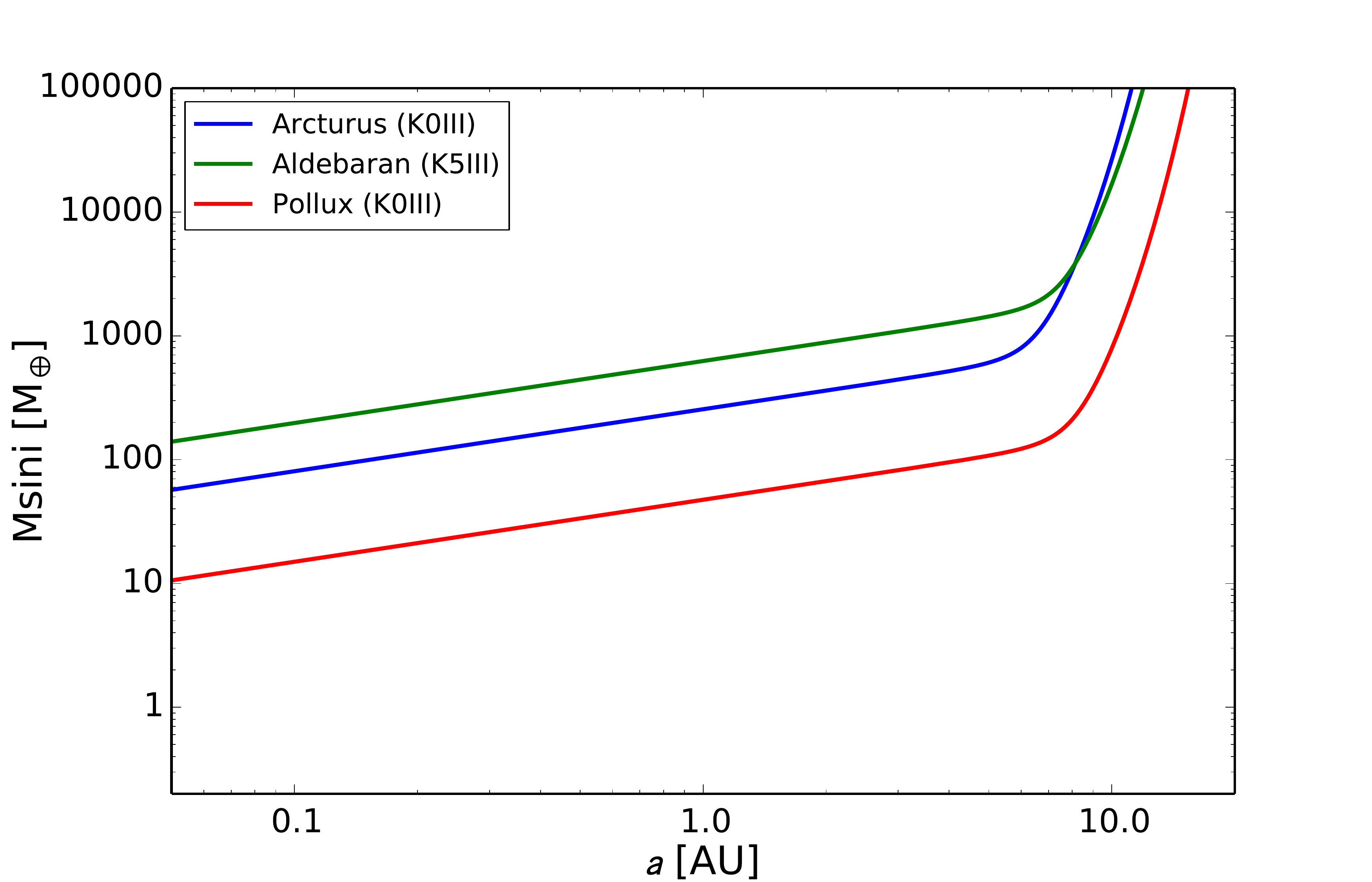} 
         \end{center}
         \caption{Simulated planet sensitivity thresholds, $M_p \sin{i}_{50}(P)$, for a representative sample of 
         early-type and Southern Sun-like stars that were not observed by the Lick and Keck planet searches (left) 
         and evolved stars (right).  
         See Sec.\ \ref{sec:completeness_nodata} for details on the selected stars and the method to compute the 
         sensitivity thresholds.  We adopt $\alpha = 6$  for these plots.  
         These predictions depend critically on the estimated RV scatter, $\sigma_{\rm RV}$, 
         which is uncertain at the factor of two or more level for most stars.  
         The excellent sensitivity for p Eri A ($\sim$10~\mearth at 1 AU) is driven by our adopted  $\sigma_{\rm RV} = 1$~\mse, 
         a factor of 2--2.5 better than we routinely achieve with HIRES.
         See also the caveats to the idealized completeness method in Sec.\ \ref{sec:idealized_completeness}.  
         }
         \label{fig:completeness_sample}
\end{figure}

Using the above estimates for $\sigma_{\rm RV}$, we computed prospective completeness curves for eight stars that 
are representative of the imaging targets that lack Doppler measurements.  
These idealized completeness curves are based on a hypothetic survey with $n_{\rm obs} = 100$ and $t_{\rm span} = 10$ yr 
and $\sigma_{\rm RV}$ dependent on stellar characteristics.  
See Figure \ref{fig:completeness_sample}, which includes four early type stars 
($\beta$ Leo, A3V; $\rho$ Gem, F0V; HD 195689, A0V; RR Lyr, F0V, pulsator), 
a quiet dwarf in the Southern Hemisphere (p~Eri~A, K0V), and  
and the three brightest giant stars in Table \ref{tab:targets_without_data} (Arcturus, K0III; Aldebaran, K5III; Pollux, K0III).
These simulations suggest that giant planets orbiting giant stars in $\sim$AU orbits can be detected by dedicated Doppler 
campaigns, as indeed they have been \citep[e.g.,][]{Quirrenbach2011,Trifonov2014}.
Estimating completeness for all stars in Table \ref{tab:targets_without_data} can be obtained by applying 
the formulas for $\sigma_{\rm RV}$ (above) or measuring it directly (preferred), 
estimating $M_\star$ for each star, and carefully identifying pulsating 
and young stars that will be particular unpredictable.



\section{Recommendations for Future Doppler Surveys}
\label{sec:recommendations}

\subsection{Limits to Doppler Precision}

The Doppler uncertainty of a single measurement is often expressed as the quadrature sum of three terms:\
stellar jitter, 
Poisson uncertainty stemming from the signal-to-noise ratio of the spectrum and the information content 
in spectrum \citep{Bottom2013}, 
and instrumental uncertainty.  

Next-generation Doppler instruments are pushing for $\leq$0.5 \ms precision \citep[e.g.,][]{Pepe2014} 
with the ultimate goal of reaching $<$0.1 \ms precision \citep{Pepe2008,Pasquini2010}.  
These are engineering challenges that will only bear fruit if the stars themselves are quiet enough for planet detection 
at those levels.  

Stellar jitter represents a major challenge for Doppler planet detection.  
All stars except for old G and K dwarfs have jitter 
that precludes the detection of small planets in few AU orbits (Fig.\ \ref{fig:completeness_sample}).  
There are many sources of jitter.  
Acoustic oscillations including p-modes cause the stellar surface to oscillate with a characteristic 
timescale of a few minutes and a characteristic amplitude of $\sim$1 \ms for Solar-type stars.  
The convective overturning of granules on the stellar surface (granulation) changes the flux balance 
between sinking cool regions and rising hot cells, producing a jitter of a similar scale.  
These effects can be mitigated by observational strategies that average over the relevant timescales 
with multiple RV measurements taken per night \citep{Dumusque2011b,Dumusque2011a}.
Jitter from rotationally modulated stellar surface features (spots, plage, faculae) 
cause apparent Doppler shifts due to distorted line profiles at the $\sim$1 \ms level even 
for quiet stars.  These signals can be mitigated by measuring stellar activity proxies 
and de-correlating the RVs \citep{Dumusque2011c}.  
In some sense, these short-term jitter signals are less troublesome for searches for long-period planets.  
Short-timescale jitter sets an error floor for single measurements, but multiple measurements on the 
appropriate timescales can average over these noise sources. 

\begin{figure*}
         \begin{center}
              \includegraphics[width=0.65\textwidth]{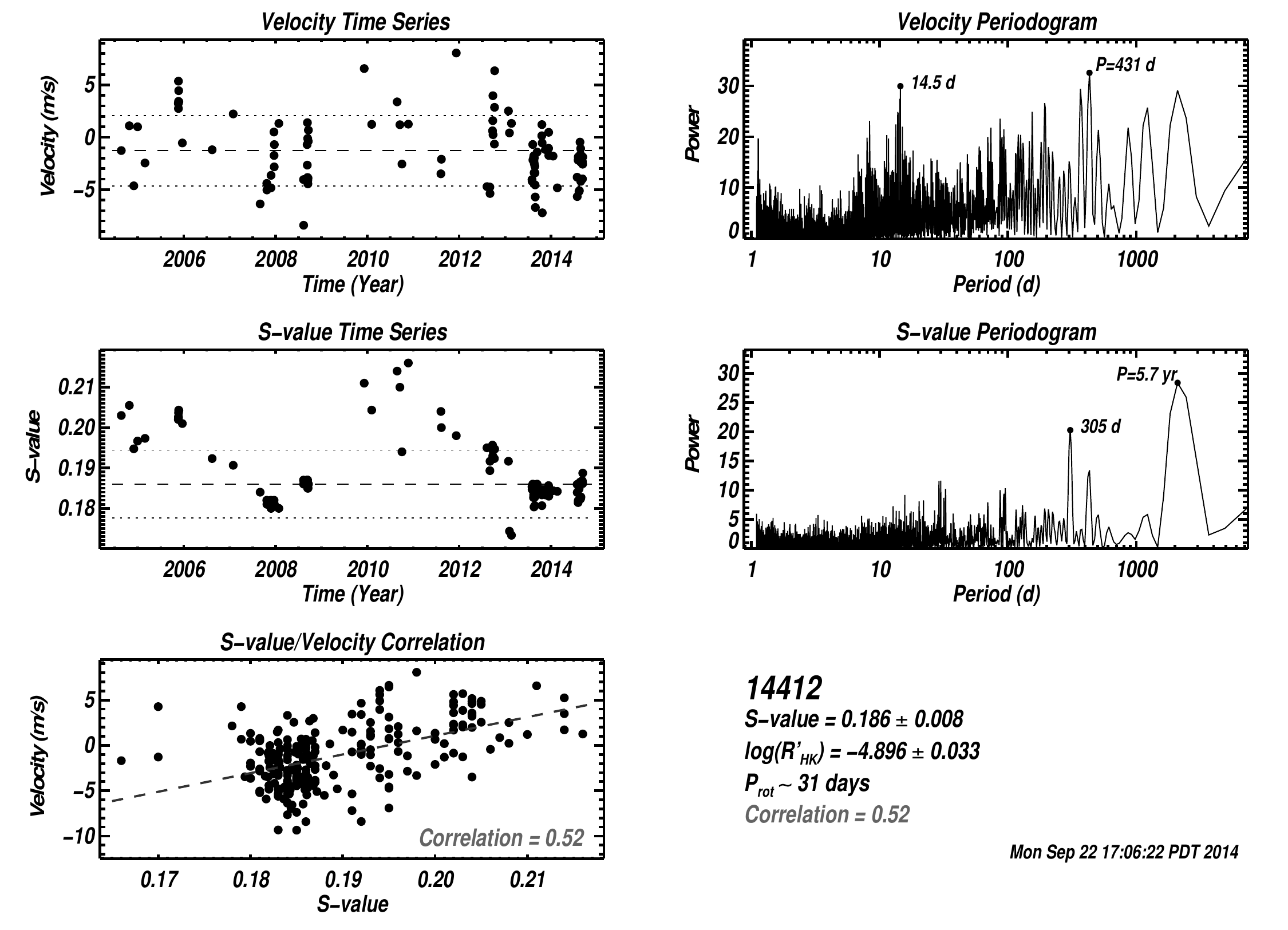}\\ \vspace*{0.45in} 
              \includegraphics[width=0.65\textwidth]{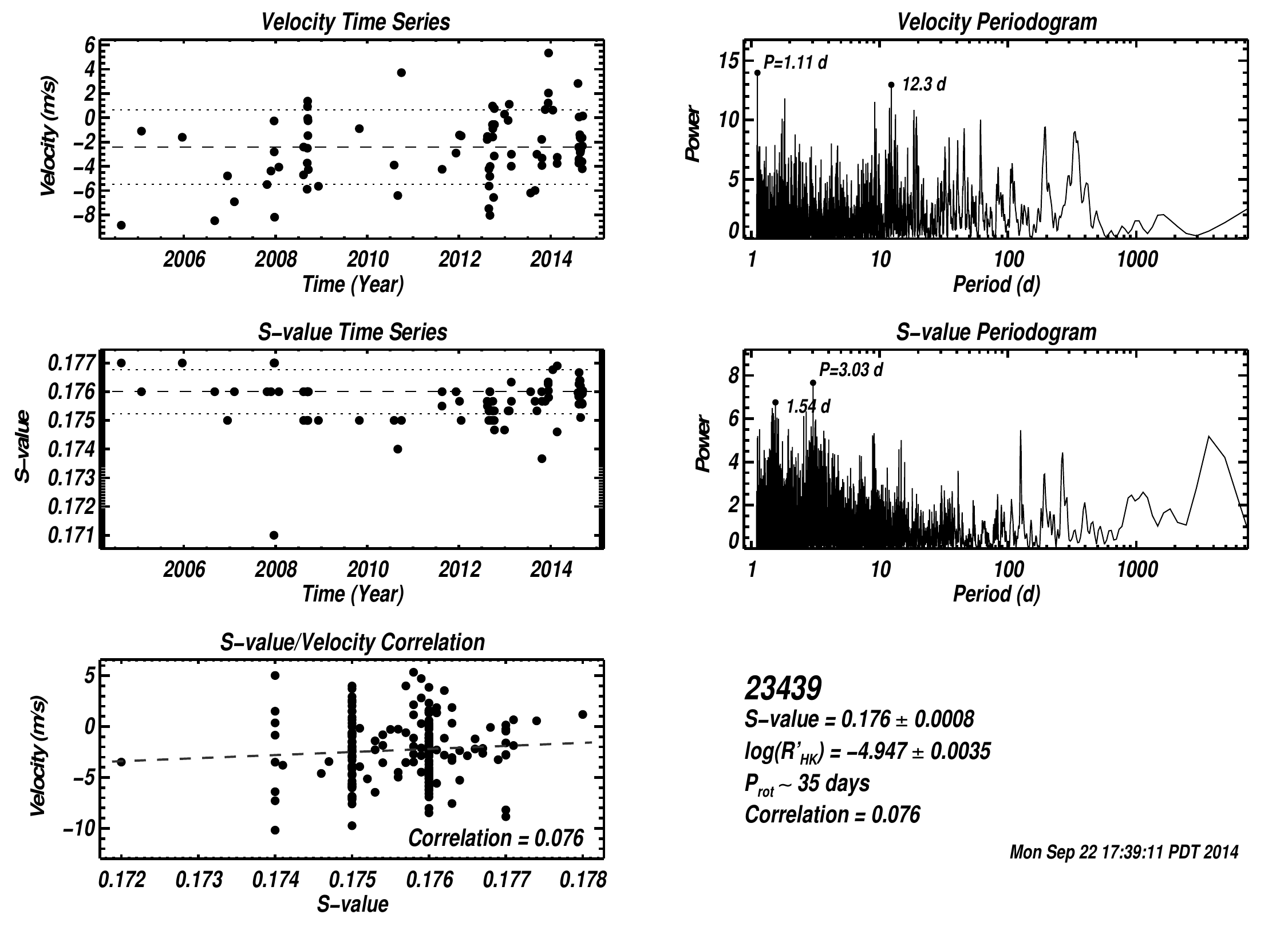}\vspace*{-0.15in} 
         \end{center}
         \caption{Automated reports on the correlations between RVs and $S_{\rm HK}$ activity measurements  for two stars observed by 
         the California Planet Survey using Keck-HIRES.   Each six-panel plot (HD 14412 -- top; HD 23439 -- bottom) 
         shows the RV time series (top left) and its periodogram (top right), the $S_{\rm HK}$ time series (middle left) 
         and its periodogram (middle right),  a linear correlation between them (lower left), and annotations (lower right).  
         Both stars are considered ``inactive,'' yet HD 14412 displays a clear RV--$S_{\rm HK}$ correlation resulting in $\sim$3 \ms 
         jitter with a timescale of 5.7 yr (presumably the stellar rotation period).  
         HD 23429 is less active (\lrphk = $-$4.95 vs.\ $-$4.90) and has no detectable RV--$S_{\rm HK}$ correlation.
         }
         \label{fig:svals}
\end{figure*}

Long-term magnetic  cycles analogous to our Sun's activity cycle 
pose perhaps the greatest challenge in the search for long-period planets 
\citep{Isaacson2010,Lovis2011}.  
Coherent, nearly sinusoidal Doppler shifts of a few \ms amplitude with 5--20 yr periods are common even for 
``inactive'' G and K dwarfs and have led to a small number of apparent false planet claims.  
For example, one of the best Jupiter analog exoplanets is HD~154345~b \citep{Wright08}, 
whose existence has been questioned because of a strong correlation between the RVs 
and the $S_{\rm HK}$ measure of stellar activity.  
Current Doppler planet searches routinely monitor activity diagnostics, including the \caii\ lines 
(from which the  $S_{\rm HK}$ values are derived) and the width and bisectors of the spectrum's cross correlation function.  
Figure \ref{fig:svals} shows two example stars form our Keck-HIRES planet search.  Both are ``inactive stars.''  
The top star (HD 14412) has a $K \approx 3$ \ms signal with $P \approx 6$ yr, the signature of an apparent 
Saturn-mass planet. This signal is mirrored in the $S_{\rm HK}$ time series, which has the same period and phase as 
the RVs.  The bottom star (HD 23439) is even less active and has no detectable variation in $S_{\rm HK}$, 
and a reduced overall jitter.

Doppler planet searches are pushing the sensitivity limits by modeling and subtracting jitter, de-correlating it, 
selecting spectral lines and regions less sensitive to jitter, and selecting low-jitter stars for searches.  
Some groups are developing techniques to suppress stellar activity including modeling magnetic activity 
with time-series methods \citep{Haywood2014,Grunblatt2015}, identification and removal of photospheric signals \citep{Hebrard2014}, 
and Bayesian methods to identify noise by its temporal characteristics.  
For imaging search to take advantage of Doppler searches with $K \lesssim 0.5$ \ms sensitivity, 
these developing techniques must be incorporated.  

\subsection{Ten Year Forecast}

\begin{figure*}
         \begin{center}
              \includegraphics[width=0.49\textwidth]{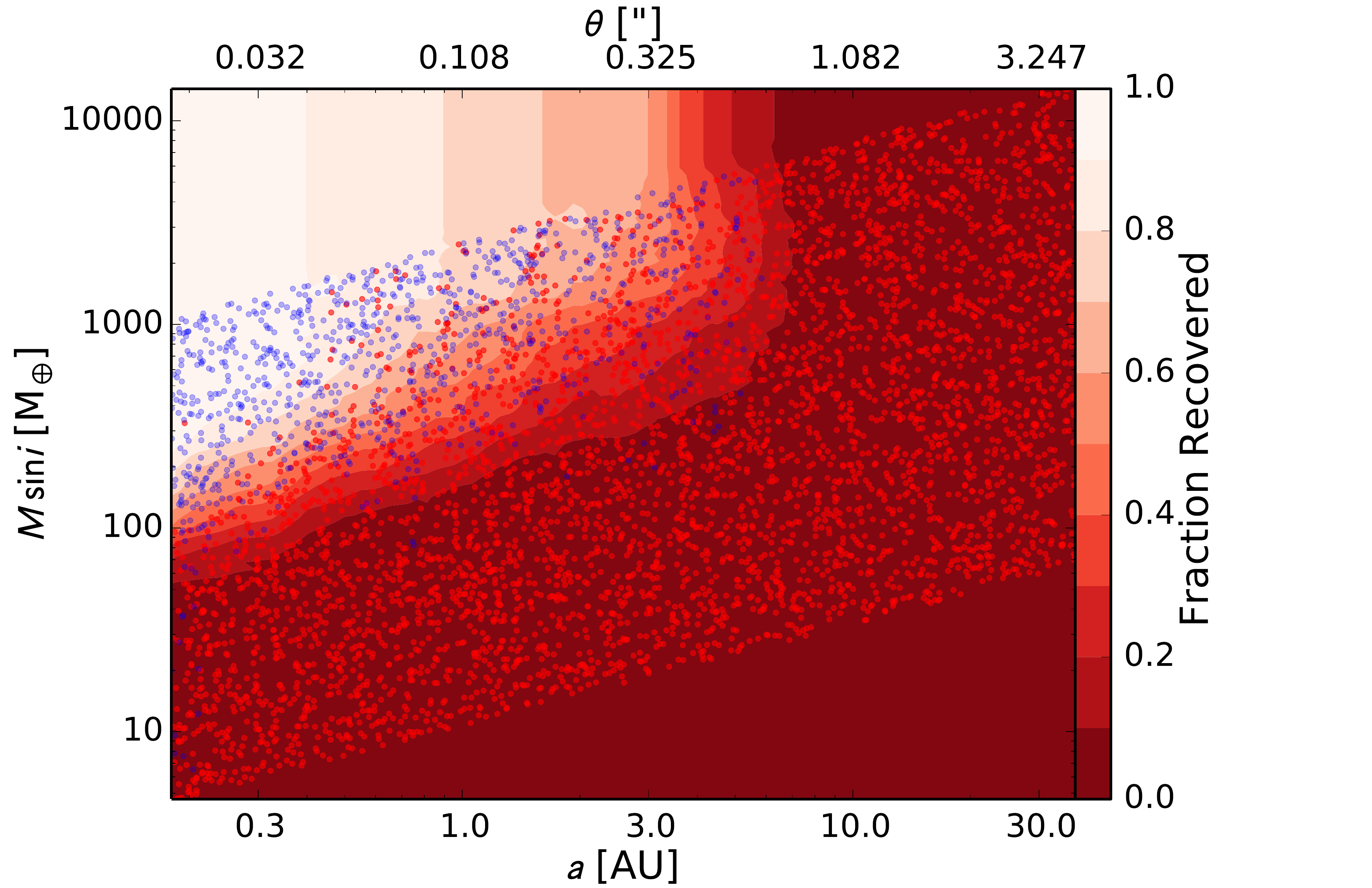} 
              \includegraphics[width=0.49\textwidth]{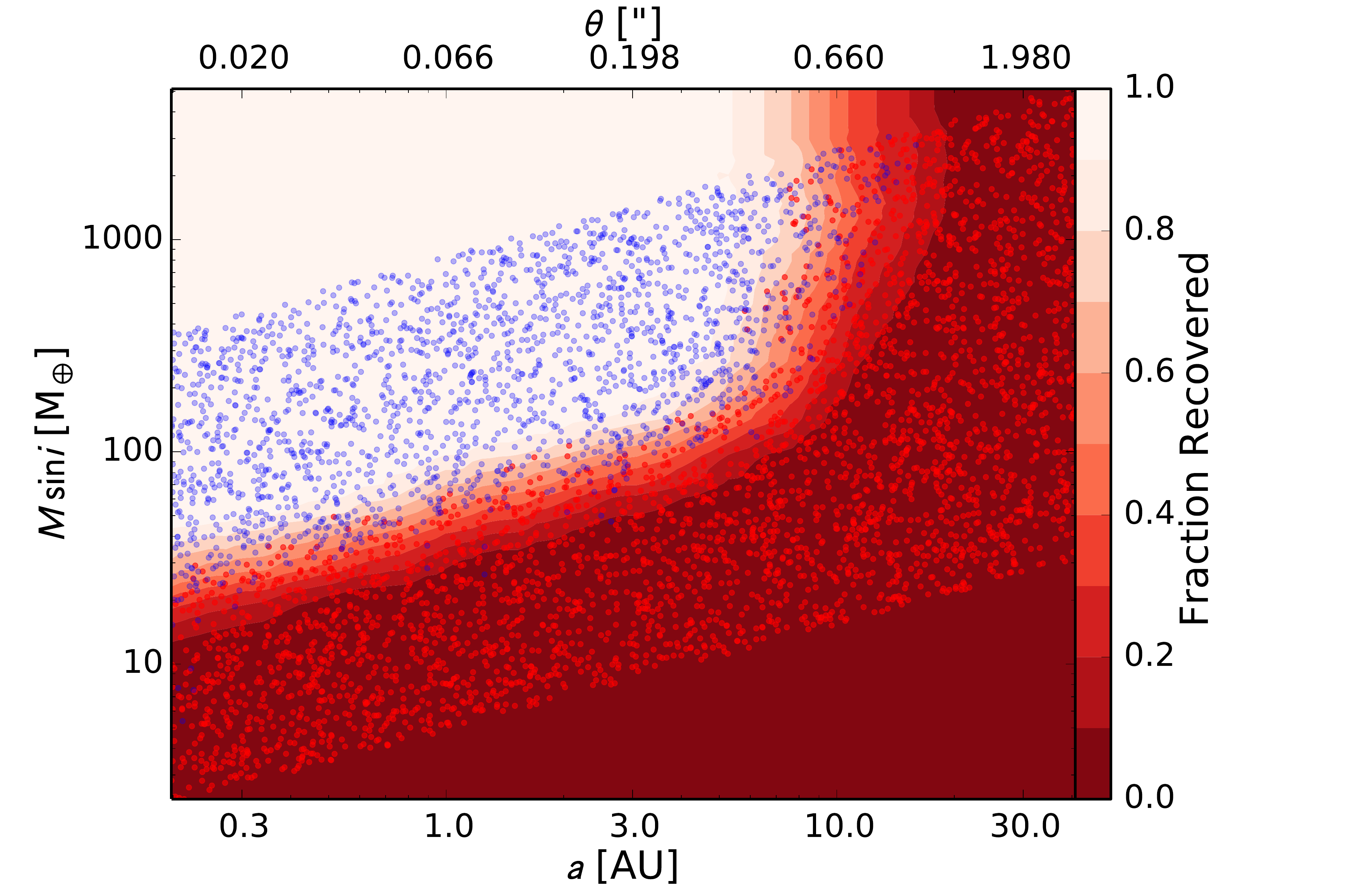} \\ \vspace*{0.25in}  
              \includegraphics[width=0.49\textwidth]{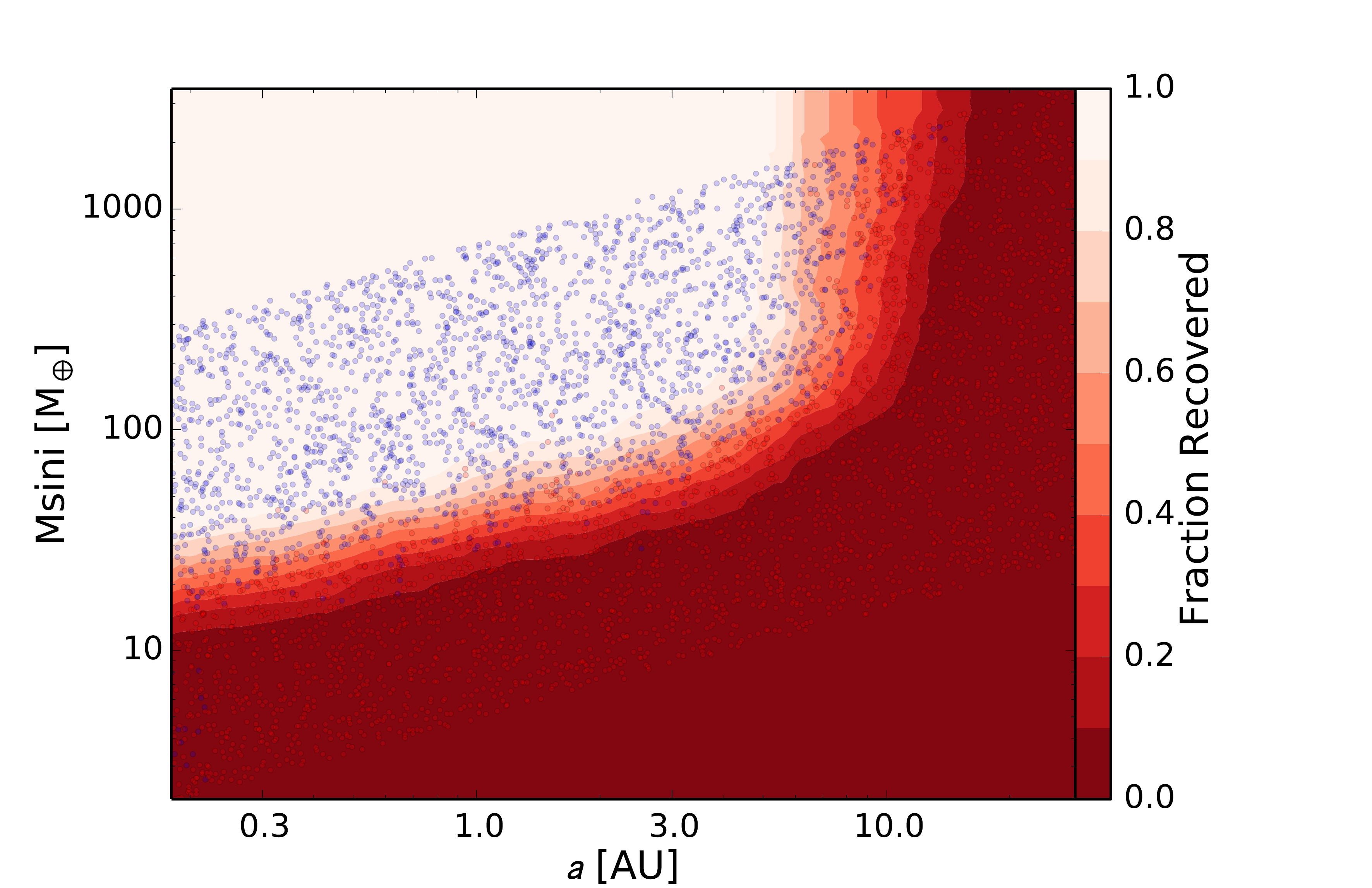}   
              \includegraphics[width=0.49\textwidth]{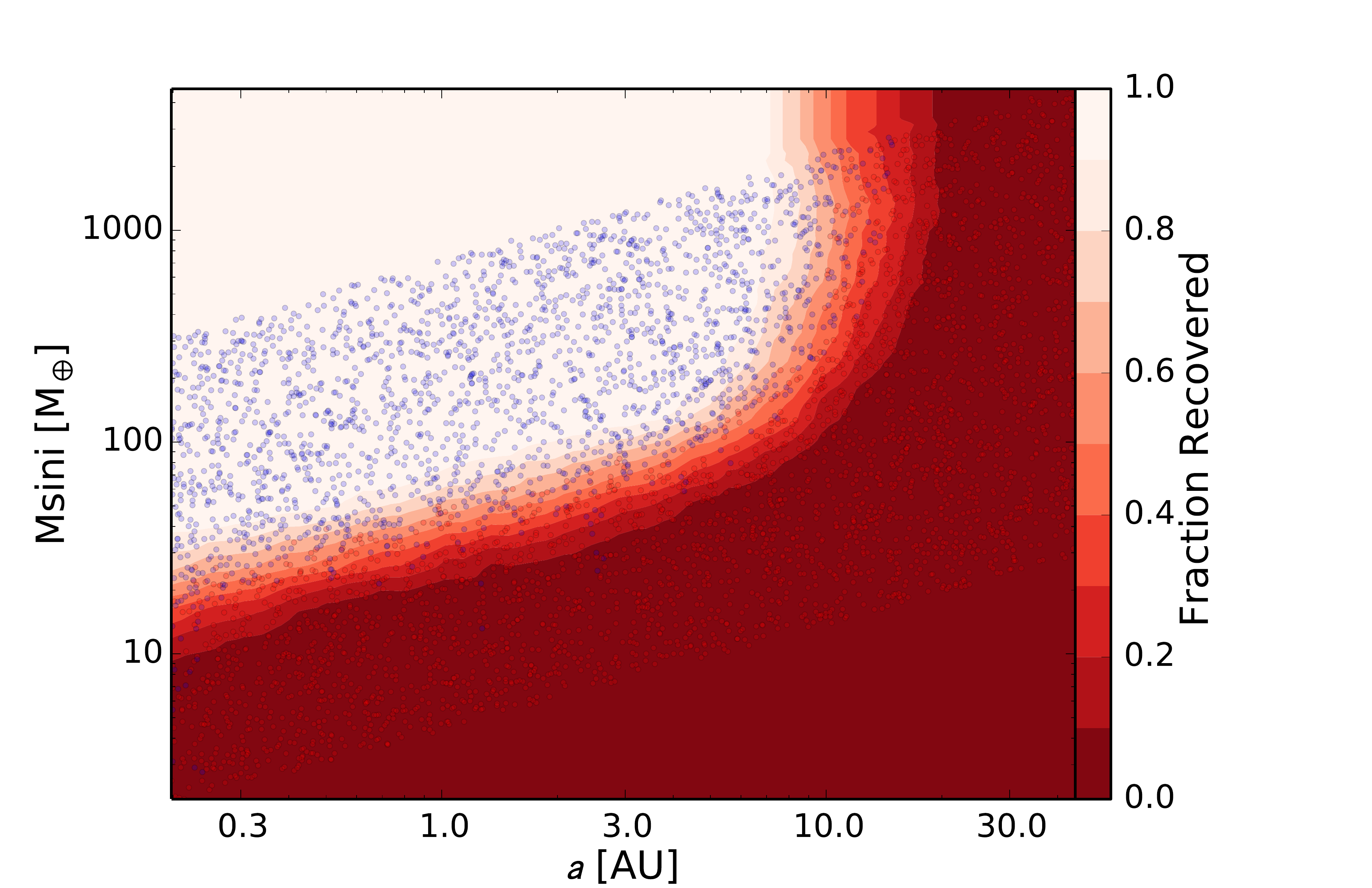} \\ \vspace*{0.25in}  
              \includegraphics[width=0.49\textwidth]{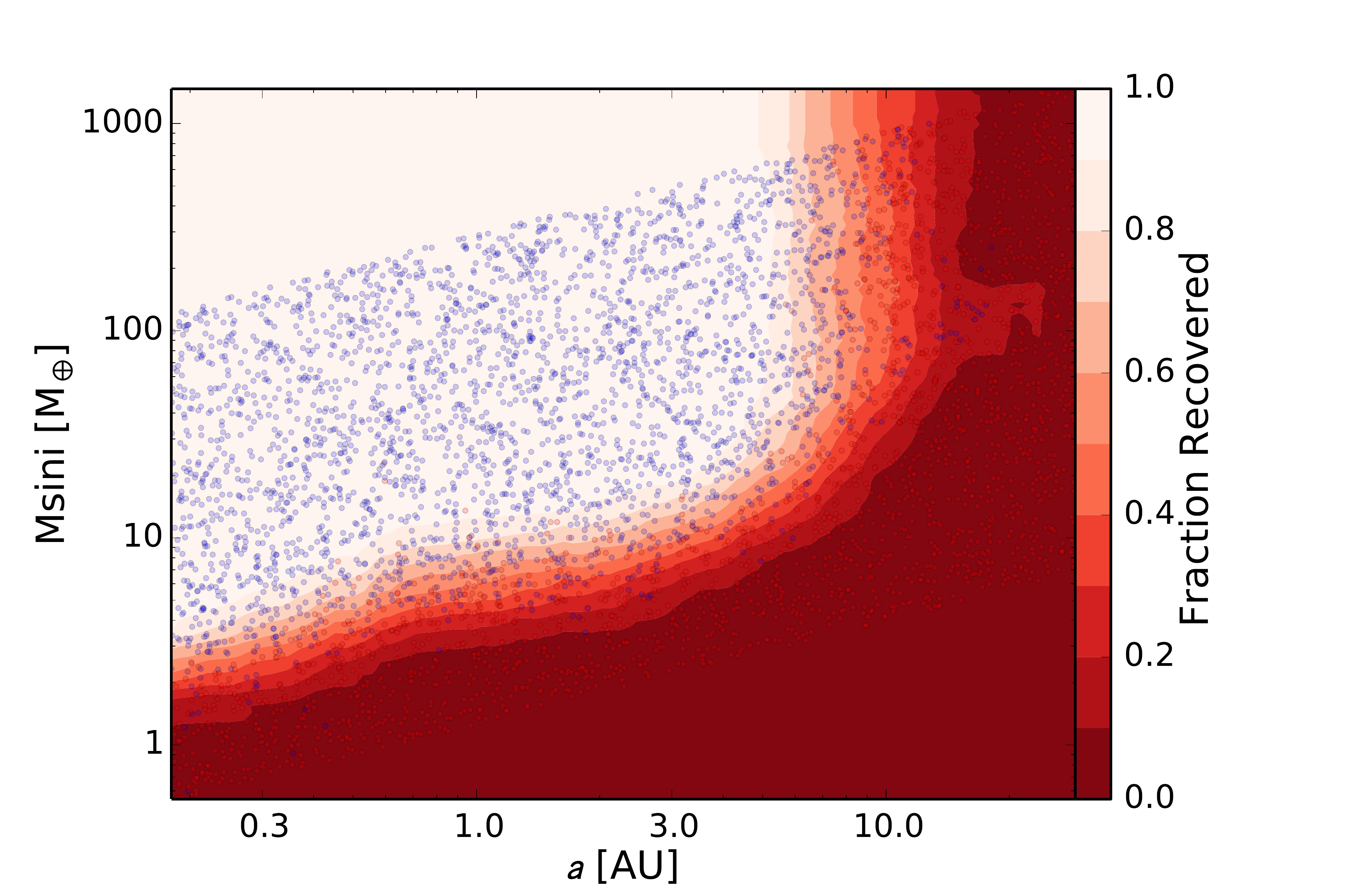}  
              \includegraphics[width=0.49\textwidth]{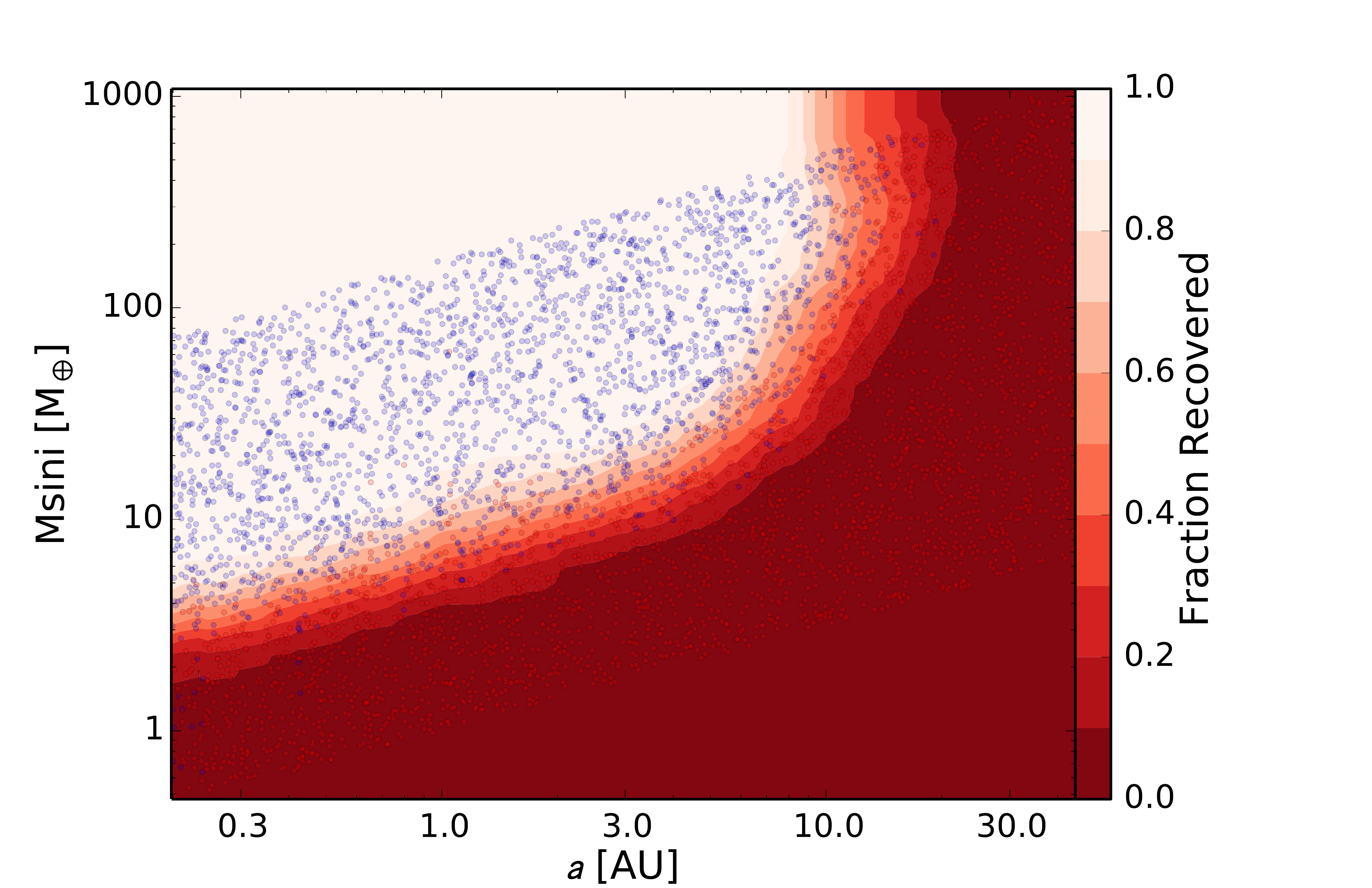}  
         \end{center}
         \caption{Completeness for two stars, HD 102365 (left) and HD 182572 (right).  
         The top row shows the completeness with the observations to date (through 2014).  
         The middle row shows the expected completeness of the same stars after adding ten years of measurements  
         with 3 RVs per year having the same precision as the recent Keck-HIRES observations. 
         The bottom row shows the expected completeness after adding 10 years of 10 RVs per year with $\sigma_{\rm RV}$ = 0.5 \mse.  
         Such a high-precision planet search would be sensitive to $\lesssim$10~\mearth super-Earth planets in 3 AU orbits.
         }
         \label{fig:completeness_future}
\end{figure*}

We estimated the improvement in completeness that could be produced by continuing our Doppler observations for the next ten years.  
This estimate only assumes current techniques to identify low-jitter stars and to obtain RVs at 0.5 \ms precision.  
Future jitter-suppressing techniques could improve planet sensitivity further.
Figure \ref{fig:completeness_future} shows our current completeness estimates (top row) for two stars observed at Keck (but not Lick).  
HD 102365 (left column in Figure \ref{fig:completeness_future}) is one of the most poorly observed stars in our sample with 
only $N_{\rm obs}$ = 16 measurements spanning the last 6.6 yr with an RMS of 2.5 \mse.  
(It has been poorly observed because it is relatively far South, $\delta \simeq -40^\circ$.)
HD 182572 (right column in Fig. \ref{fig:completeness_future}) has $N_{\rm obs}$ = 82 measurements spanning 17.8 yr with an RMS of 4.0 \ms (3.6 \ms for HIRES instrument code j).  

The Fig.\ \ref{fig:completeness_future} plots also show the results of two simulations.  
In the middle row, we simulated adding $N_{\rm obs}$ = 30 measurements over 10 yr to the existing RVs with 
$\sigma_{\rm RV}$ of the new measurements equal to the RMS of the most recent HIRES RVs (code = j).  
This simulation represents the improvement in completeness from continuing observations with the current precision 
and a cadence of $N_{\rm obs}$ = 3 RVs per year.  

The bottom row of Fig.\ \ref{fig:completeness_future} shows the gain in completeness resulting from improved Doppler precision.  
Here we adopted $\sigma_{\rm RV}$ = 0.5 \ms with $N_{\rm obs}$ = 100 over 10 yr.  
This level of precision is possible, but on the cutting edge of current Doppler techniques.  
It requires the most Doppler quiet stars---old, 
choromospherically inactive late-G and early-K dwarfs---and observational strategies that average over stellar noise.
To date, the lowest noise RV planet detections have been with HARPS.  
For example, HD 88512 b was detected with RVs whose per-night measurement uncertainties are $< 0.3$~\ms and 
the data set has an RMS of 0.75~\ms after subtracting the single planet model with $K = 0.77$~\ms 
\citep{Pepe2011}.  
Achieving this level of precision would enable 50\% completeness to $\sim$10~\mearth planets (typical for super-Earths) at 3 AU, 
a factor of $\sim$6 improvement in mass sensitivity compared adding $N_{\rm obs}$ = 30 RVs 
over 10 yr with the current HIRES precision.

\acknowledgments{We thank Karl Stapelfeldt and Sara Seager for 
instigating us and the JPL Program Office to make this study happen. 
Thanks to Steve Unwin for cheerfully and skillfully shepherding our work on this study and to   
Margaret Turnbull for providing target lists and associated stellar properties and for helpful conversations.  
We thank the many observers who contributed to the measurements reported here, 
including Geoff Marcy, Debra Fischer, Jason Wright, John Johnson, R.\ Paul Butler, and Steve Vogt.  
We gratefully acknowledge the efforts and dedication of the staffs of Lick Observatory and Keck Observatory  
and the time assignment committees of the 
University of Hawaii, the University of California, NASA, the California Institute of Technology, and Yale University
for their generous allocations of observing time over the past two decades that enabled these measurements.  
We thank Geoff Marcy for helpful comments.  
We acknowledge NASA-JPL award  \#1505707.
This work made use of the SIMBAD database (operated at CDS, Strasbourg, France), 
NASA's Astrophysics Data System Bibliographic Services, the NASA Star and Exoplanet Database (NStED), 
the Exoplanet Orbit Database and the Exoplanet Data Explorer at exoplanets.org.
Finally, we extend special thanks to those of Hawai`ian ancestry 
on whose sacred mountain of Maunakea we are privileged to be guests.  
Without their generous hospitality, the Keck observations presented herein
would not have been possible.}

Facilities:\ \facility{Lick (Hamilton Spectrograph)}, \facility{Keck (HIRES)}





\clearpage

\appendix
\renewcommand\thefigure{\thesection.\arabic{figure}}    
\renewcommand\thetable{\thesection.\arabic{table}}    
\setcounter{figure}{0} 
\setcounter{table}{0}


\section{Machine-readable Data}
\label{app:data}

The set of RV time series for every star with measurements is supplied as a supplement to this report.\footnote{All machine-readable data products can be downloaded from\\ \href{http://exoplanetarchive.ipac.caltech.edu/docs/contributed\_data.html}{http://exoplanetarchive.ipac.caltech.edu/docs/contributed\_data.html}}
Each star has a separate ASCII file for RV data that is named {\it starname}\_rv.csv (e.g., 10700\_rv.csv).  
The file header lists the star name and instrument codes.  
Each column is briefly labeled in the first row. 
The four data columns are Heliocentric Julian Date (HJD) less 2,440,000, relative radial velocity (\mse), uncertainty in the radial velocity (\mse), 
and instrument code.  Statistical uncertainties in determining the Doppler shifts are accounted for, but astrophysical jitter is not. 

We  summarize the completeness contours for each star (Appendix \ref{app:completeness}) with two sets of machine-readable files
using a format similar to the RV files.  
The first set of ASCII files are named {\it starname}\_contours.csv (e.g., 10700\_contours.csv).  
The header lists the star name as well as the stellar mass and distance used to convert between orbital period, 
semi-major axis, and projected sky separation.
Each column is briefly labeled in the first row. 
The six data columns are orbital period (days), semi-major axis (AU), sky-projected angular distance (arcsec), 
and  the $M\sin{i}$ (\mearthe) values that corresponds to completeness levels of 16\%, 50\%, and 84\%, respectively. 

A second set of ASCII files provide complete, two-dimensional sampling of the completeness measurements.  
In files named {\it starname}\_2D\_recovery.txt (e.g., 10700\_2D\_recovery.txt), 
we list the completeness values (the red shading) at every cell in the completeness plots. 
These files have headers listing the star masses, distances and column names.  
The columns are sky-projected angular distance (arcsec), 
and  the $M\sin{i}$ (\mearthe), and fractional completeness (0--1).


\section{Properties of Stars without Doppler Measurements}
\label{app:stellar_prop_nodata}

Table \ref{tab:targets_without_data} contains the complete list of  244 stars that are on the Exo-S and Exo-C target lists  
and were not observed at Lick or Keck Observatories.  
The Notes column lists the reason(s) that the stars were {\em likely} not included in our Keck/Lick Planet Searches. 
This table is the complement to Table \ref{tab:targets_with_data}, which includes stars with observations.  
Because of the quantity of this data, we relegated Table \ref{tab:targets_without_data}  to this appendix.  

\begin{deluxetable}{rrrlrrrrll}
\tabletypesize{\footnotesize}
\tablecaption{Imaging Target Stars Without Doppler Measurements
\label{tab:targets_without_data}}
\tablewidth{0pt}
\tablehead{
  \colhead{Hipp.}   & 
  \colhead{HD}   & 
  \colhead{Gliese}   & 
  \colhead{Target$^{\rm a}$} &
  \colhead{Dist.}  &
  \colhead{$V$}  &
  \colhead{$B$--$V$}  &
  \colhead{\teff}  &
  \colhead{Sp.\ T.}  & 
  \colhead{Notes$^{\rm b}$}  \\
  \colhead{No.}   & 
  \colhead{No.}   & 
  \colhead{No.}   & 
  \colhead{List} &
  \colhead{(pc)}  &
  \colhead{(mag)}  &
  \colhead{(mag)}  &
  \colhead{(K)}  &
  \colhead{}  & 
  \colhead{} 
}
\startdata
    171 &   224930 & 914A &      S &  12.170 &  5.800 &  0.690 &   5502 &             G3V & B \\
    677 &      358 & \nodata &      C, A &  29.740 &  2.060 & $-$0.048 &  13098 &             B9p &H  \\ 
    746 &      432 &   8  &      C, A &  16.780 &  2.260 &  0.365 &   6915 &        F2III-IV & H\\
    910 &      693 &  10  &      C, A &  18.750 &  4.890 &  0.487 &   6255 &             F5V &H  \\
    950 &      739 &  3013 &      A &  21.280 &  5.240 &  0.430 &   6548 & F5V & H, S \\ 
   1599 &     1581 &  17  &     S, C, A &   8.590 &  4.230 &  0.576 &   5948 &             G0V & S \\
   2021 &     2151 &  19  &     S, C, A &   7.460 &  2.820 &  0.618 &   5873 &            G1IV & S \\
   2072 &     2262 &  20  &      C, A &  23.810 &  3.930 &  0.175 &   7922 &            A6VN & S \\
   2081 &     2261 & \nodata &      C &  25.970 &  2.400 &  1.083 &   4436 &          K0IIIB & E \\ 
   2762 &     3196 &    23A &      A &  21.250 &  5.200 &  0.570 &   6067 & F8V & B \\
   2941 &     3443 &  25A &      S &  15.400 &  5.570 &  0.715 &   5480 &           K1V+G & B \\
   3419 &     4128 &  31  &      C &  29.530 &  2.020 &  1.038 &   4944 &           K0III & E \\
   3505 &     4247 &     31.3 &      A &  26.740 &  5.220 &  0.320 &   6903 & F3V & H \\
   3810 &     4676 &     34.1 &      A &  23.450 &  5.070 &  0.500 &   6254 & F8V & H\\
   3909 &     4813 &   37 &      S, A &  15.753 &  5.164 &  0.514 &   6250 &             F7V &  H, $N_{\rm obs}=2$ \\
   4151 &     5015 &  41  &      C, A &  18.740 &  4.800 &  0.540 &   6196 &             F8V & B \\
   5336 &     6582 &  53A &     S, C, A &   7.550 &  5.170 &  0.704 &   5526 &            G5Vp & B \\
   5799 &     7439 &    54.2A &      A &  23.380 &  5.140 &  0.450 &   6465 & F5V & H \\
   5862 &     7570 &  55  &     S, C, A &  15.110 &  4.960 &  0.571 &   6116 &             G0V & S \\ 
   5896 &     7788 &    55.3A &      A &  20.950 &  4.250 &  0.450 &   6505 & F5V & H, S \\
   6537 &     8512 & \nodata &      C &  34.900 &  3.800 &  1.070 &    \nodata &           K0III & E \\
   6686 &     8538 & \nodata &      C &  30.500 &  2.700 &  0.160 &    \nodata &         A5Vv & B, H \\
   6706 &     8723 & \nodata &      A &  25.210 &  5.350 &  0.370 &   6690 & F2V & H \\
   6813 &     8799 &   \nodata  &      A &  28.620 &  4.830 &  0.420 &   6555 & F4V & H \\  
   7751 &    10360 & 66AB &      S &   7.820 &  5.680 &  0.864 &   5016 &            K0/4 & S \\
   7918 &    10307 &  67  &     S, C, A &  12.740 &  4.960 &  0.618 &   5874 &             G2V & B \\
   8209 &    10830 & \nodata &      A &  28.110 &  5.290 &  0.400 &   6740 & F2V & H \\
   8497 &    11171 & \nodata &      C, A &  23.190 &  4.650 &  0.333 &   7087 &             F0V & H \\
   8796 &    11443 & 78.1  &      C, A &  19.420 &  3.420 &  0.488 &   6273 &            F6IV & B \\
   8903 &    11636 &  80  &      C, A &  17.990 &  2.630 &  0.156 &   8300 &          A5V & H \\
   9007 &    11937 &  81A &      C &  17.850 &  3.690 &  0.844 &   5182 &      G8III & S, E \\
   9236 &    12311 &  83  &      C, A &  22.010 &  2.860 &  0.290 &   7201 &             F0V & S, H \\
   9884 &    12929 & 84.3  &      C &  20.180 &  2.010 &  1.151 &   4504 &           K2III & E \\
  10138 &    13445 &  86  &      S &  10.780 &  6.120 &  0.812 &   5151 &             K1V & S \\
  10306 &    13555 & \nodata &      A &  28.870 &  5.230 &  0.440 & 6496 & F5V & H \\
  10644 &    13974 &  92  &     S, C, A &  10.780 &  4.860 &  0.607 &   5667 &             G0V & B \\
  10670 &    14055 & \nodata &      C &  34.400 &  4.000 &  0.020 &    \nodata &           A1Vnn & H \\
  11072 &    14802 &     97 &      A &  21.950 &  5.190 &  0.610 &   5854 & G1V & B \\
  11783 &    15798 & \nodata &      A &  26.690 &  4.740 &  0.450 & 6502 & F5V & H \\
  12390 &    16620 &   105.4A &      A &  26.960 &  4.830 &  0.430 &   6516 & F4V & H \\
  12623 &    16739 &    105.6 &      A &  24.190 &  4.910 &  0.580 &   5973 & F9V & H \\
  12706 &    16970 & 106.1A &      C, A &  24.410 &  3.470 &  0.093 &   8673 &            A2Va & H \\
  12777 &    16895 & 107A &     S, C, A &  11.127 &  4.098 &  0.489 &   6344 &             F7V & H, $N_{\rm obs}=2$ \\
  12828 &    17094 &\nodata &      C, A &  25.770 &  4.270 &  0.311 &   7225 &        F1III-IV & H \\ 
  12843 &    17206 & 111  &     S, C, A &  14.220 &  4.470 &  0.481 &   6378 &           F5/6V & H \\
  14146 &    18978 & 121  &      C, A &  27.170 &  4.080 &  0.163 &   8045 &             A4V & H \\
  14576 &    19356 & \nodata &      C, A &  27.570 &  2.110 & $-$0.010 &   9634 &             B8V & H \\
  14668 &    19476 & \nodata &      C &  34.600 &  4.000 &  0.980 &    \nodata &           K0III & E \\
  14879 &    20010 & 127A &     S, C, A &  14.240 &  3.800 &  0.543 &   6258 &             F8V & B \\
  15330 &    20766 & 136  &      S &  12.010 &  5.530 &  0.641 &   5699 &           G3/5V & S \\
  15371 &    20807 & 138  &     S, C, A &  12.030 &  5.240 &  0.600 &   5845 &             G2V & S \\
  15510 &    20794 & 139  &     S, C, A &   6.040 &  4.260 &  0.711 &   5401 &           G8.0V & S \\   
  16245 &    22001 & 143.2A &      C, A &  21.680 &  4.700 &  0.410 &   6629 &          F3IV/V & S, H \\
  17440 &      \nodata & \nodata &      C &  29.860 &  3.840 &  1.133 &   4514 &    K0IV       & B, S \\
  17651 &    23754 & 155  &      C, A &  17.630 &  4.220 &  0.434 &   6631 &           F3/5V & H \\
  18907 &    25490 & \nodata &      C &  35.900 &  3.900 &  0.030 &    \nodata &             A1V & H \\
  19205 &    25867 & \nodata &      A &  27.600 &  5.210 &  0.340 &   6979 & F1V & H \\
  19747 &    26967 & \nodata &      C &  35.300 &  4.000 &  1.090 &    \nodata &           K1III & E \\
  19893 &    27290 & 167.1  &      C, A &  20.460 &  4.250 &  0.312 &   7060 &             F0V & S, H \\
  19921 &    27442 & 167.3  &      C &  18.240 &  4.440 &  1.078 &   4846 &         K1/2III & S, E \\
  19990 &    27045 & \nodata &      A &  28.940 &  4.930 &  0.260 &   7384 & A3M & H \\
  21421 &    29139 & 171.1A &      C &  20.430 &  0.870 &  1.538 &   3889 &           K5III & E \\
  21547 &    29391 & \nodata &      A &  29.430 &  5.220 &  0.280 &   7257 & F0V & H \\
  21770 &    29875 & 174.1A &      C, A &  20.170 &  4.440 &  0.342 &   6991 &             F1V & H \\
  21861 &    29992 &    176.1 &      A &  28.660 &  5.040 &  0.390 &   6666 & F3IV & H, S \\
  23482 &    32743 &    187 &      A &  26.070 &  5.370 &  0.420 &   6624 & F5V & H, S \\
  23693 &    33262 & 189  &     S, C, A &  11.650 &  4.710 &  0.526 &   6246 &           F6/7V & S, H \\
  23783 &    32537 &   187.2A &      A &  26.290 &  4.980 &  0.320 &   7018 & F2V & H \\
  23875 &    33111 & \nodata &      C, A &  27.400 &  2.760 &  0.150 &   8377 &           A3IVn & H \\
  23941 &    33256 &    189.2 &      A &  25.460 &  5.110 &  0.430 &   6411 & F5.5V & H \\
  24608 &    34029 & 194A &      C &  13.120 &  0.080 &  0.795 &   5356 &         M1 &  B \\
  25110 &    33564 &    196 &      A &  20.890 &  5.080 &  0.480 &   6394 & F7V & H \\
  25278 &    35296 & 202  &     S, C, A &  14.390 &  5.000 &  0.523 &   6202 &           F8V &  Y, B \\
  27072 &    38393 & 216A &     S, C, A &   8.927 &  3.590 &  0.481 &   6372 &             F7V &  Y, $N_{\rm obs}=8$ \\
  27288 &    38678 & 217.1  &      C, A &  21.610 &  3.550 &  0.104 &   8337 &          A2 & H \\
  27321 &    39060 & 219  &      C, A &  19.440 &  3.850 &  0.171 &   8052 &             A5V & S, H \\
  27628 &    39425 &\nodata &      C &  26.730 &  3.100 &  1.175 &   4545 &         K1.5III & E \\ 
  27654 &    39364 & \nodata &      C &  34.900 &  3.900 &  0.980 &    \nodata &        G8III/IV & E \\
  27890 &    40409 &224.1  &      C &  26.250 &  4.640 &  1.048 &   4661 &        K1III/IV & S, E \\
  27913 &    39587 & 222AB &     S, C, A &   8.663 &  4.395 &  0.594 &   5882 &             G0V & B \\ 
  28103 &    40136 & 225  &     S, C, A &  14.880 &  3.710 &  0.337 &   7069 &             F1V & H \\
  28360 &    40183 & \nodata &      C, A &  24.870 &  1.890 &  0.071 &   9024 &             A2V & H \\
  29271 &    43834 & 231  &     S, C, A &  10.200 &  5.080 &  0.714 &   5587 &             G6V & S \\
  29800 &    43386 &   9207 &      A &  19.250 &  5.040 &  0.430 &   6602 & F5V & H \\ 
  31592 &    47205 & 239.1  &      C &  19.750 &  3.950 &  1.063 &   4799 &       K1III & E, B \\
  32349 &    48915 & 244A &     S, C, A &   2.630 & $-$1.440 &  0.009 &   9580 &           A1.0V & H \\
  32362 &    48737 & 242  &      C, A &  18.000 &  3.320 &  0.443 &   6455 &            F5IV & H \\
  32607 &      \nodata & \nodata &      C, A &  29.600 &  3.230 &  0.222 &   7536 &    A7IV         & S, H \\
  32765 &    50223 &    249.1 &      A &  25.260 &  5.140 &  0.450 &   6482 & F5.5V & H, S \\
  33202 &    50635 & \nodata &      C, A &  25.630 &  4.730 &  0.321 &   7064 &            F0Vp & H \\
  33302 &    51199 & \nodata &      A &  29.580 &  4.660 &  0.370 &   6790 & F3V & H \\
  34065 &    53705 & 264.1A &      S &  16.520 &  5.560 &  0.624 &   5827 &             G3V & B, S \\
  34834 &    55892 &  268.1  &      C, A &  21.430 &  4.490 &  0.324 &   6907 &            F0IV & H \\
  35350 &    56537 & \nodata &      C &  30.900 &  3.600 &  0.110 &    \nodata &          A3V & H \\
  35550 &    56986 & 271A &      C, A &  18.540 &  3.500 &  0.374 &   6906 &         F0IV & H \\
  36046 &    58207 & \nodata &      C &  36.900 &  4.000 &  1.020 &    \nodata &       G9III & E \\
  36366 &    58946 & 274A &      C, A &  18.050 &  4.160 &  0.320 &   7035 &          F0V & H \\
  36439 &    58855 & \nodata &      A &  20.240 &  5.350 &  0.450 & 6457 & F6V & H \\
  36795 &    60532 & 279  &      C, A &  25.300 &  4.440 &  0.521 &   6262 &             F6V & H \\ 
  36850 &    60179 & 278A &     S, C, A&  15.600 &  1.580 &  0.034 &   8932 &            A2Vm & H \\
  37279 &    61421 & 280A &     S, C, A &   3.510 &  0.400 &  0.432 &   6543 &          F5IV-V & H \\
  37606 &    62644 &    284 &      A &  24.670 &  5.040 &  0.760 &   5343 & G8IV-V & S \\
  37826 &    62509 & 286  &      C &  10.360 &  1.160 &  0.991 &   4850 &        K0IIIvar & E \\
  37853 &    63077 & 288A &      S, A &  15.210 &  5.360 &  0.589 &   6002 &             G0V &  B \\
  38382 &    64096 & 291A &      S, A &  16.500 &  5.160 &  0.600 &   5826 &             G2V & B \\
  38423 &    64379 &   292A &      A &  17.940 &  5.010 &  0.430 &   6554 & F5V  & H, S, B \\
  38908 &    65907 & 294A &      S &  16.200 &  5.580 &  0.573 &   5949 &             G0V & S \\
  39757 &    67523 & \nodata &      C, A &  19.480 &  2.790 &  0.456 &   6449 &        F2 & H \\
  39903 &    68456 & 297.1  &      C, A &  19.980 &  4.740 &  0.437 &   6467 &             F5V & S, H \\
  40167 &    68255 & \nodata &      C, A &  25.080 &  5.240 &  0.531 &   5741 &             G0V & B \\
  40702 &    71243 & 305  &      C, A &  19.560 &  4.050 &  0.404 &   6625 &             F5V & S, H \\
  40706 &    70060 & 1109  &      C, A &  28.630 &  4.440 &  0.222 &   7790 &      A4 & H \\
  41312 &    71878 & \nodata &      C &  33.000 &  3.900 &  1.130 &    \nodata &        K2IIIvar &  S, E \\
  42430 &    73752 &   314A &      A &  19.400 &  5.050 &  0.710 &   5499 & G5IV & E  \\
  42913 &    74956 & 321.3A &      C, A &  24.700 &  1.930 &  0.043 &   9021 &             A0V & S, H \\
  44127 &    76644 & 331A &     S, C, A &  14.510 &  3.100 &  0.207 &   7769 &            A7IV & H \\
  44143 &    77370 &    333.1 &      A &  26.420 &  5.170 &  0.420 &   6690 & F4V & H, S \\
  44248 &    76943 & 332A &     S, C, A &  16.070 &  3.970 &  0.443 &   6538 &             F5V & H \\
  44901 &    78209 & \nodata &      C, A &  28.820 &  4.440 &  0.288 &   7231 &              Am & H \\
  45038 &    78154 & 335A &      C, A &  20.380 &  4.800 &  0.489 &   6180 &          F7IV-V & B \\
  45238 &    80007 & \nodata &      C &  34.700 &  1.700 &  0.070 &    \nodata &            A2IV & S, H \\
  45333 &    79028 &    337.1 &      A &  19.570 &  5.180 &  0.590 &   5871 & G0V & B \\
  46509 &    81997 & 348A &      C, A &  17.330 &  4.590 &  0.411 &   6488 &             F5V & H \\
  46651 &    82434 & 351A &      C, A &  18.810 &  3.600 &  0.371 &   6837 &            F2IV & H \\
  46733 &    81937 & \nodata &      C, A &  23.820 &  3.640 &  0.360 &   6875 &            F0IV & H \\
  46853 &    82328 & 354A &     S, C, A &  13.480 &  3.160 &  0.469 &   6334 &            F6IV & H \\
  47080 &    82885 & 356A &     S, C, A &  11.370 &  5.390 &  0.770 &   5370 &          G8IV-V & B \\
  48319 &    84999 & \nodata &      C &  35.600 &  3.800 &  0.290 &    \nodata &            F0IV & H \\
  48833 &    86146 & \nodata &      A &  28.130 &  5.110 &  0.470 & 6393 &F6V & H \\
  49593 &    87696 & 378.3  &      C, A &  28.240 &  4.490 &  0.190 &   7839 &             A7V & H \\
  49669 &    87901 &\nodata&      C, A &  24.310 &  1.410 & $-$0.041 &  11962 &             B7V & H \\ 
  49809 &    88215 & \nodata &      A &  27.730 &  5.300 &  0.370 &   6776 & F3V & H \\
  50191 &    88955 & \nodata &      C &  31.100 &  3.900 &  0.050 &    \nodata &             A2V & H \\
  50564 &    89449 & 388.1  &      C, A &  21.370 &  4.780 &  0.452 &   6476 &            F6IV &  E \\
  50954 &    90589 & 391  &     S, C, A &  16.220 &  3.980 &  0.369 &   6885 &        F2/3IV/V & S, H \\
  51502 &    90089 &    392.1 &      A &  21.480 &  5.250 &  0.370 &   6762 & F4V & H \\
  51523 &    91324 &    397.2 &      A &  21.810 &  4.890 &  0.500 &   6287 & F9V & S \\
  51814 &    91480 & \nodata &      A &  26.520 &  5.160 &  0.330 &   6972 & F2V & H \\
  51986 &    92139 & \nodata &      C, A &  26.840 &  3.830 &  0.299 &   7274 &        A3 & H \\
  52727 &    93497 & \nodata &      C &  35.900 &  2.800 &  0.900 &    \nodata &        G5III  & B \\
  53229 &    94264 &\nodata&      C &  29.090 &  3.790 &  1.040 &   4833 &        K0III-IV & E \\ 
  53253 &    94510 & 404.1  &      C &  29.130 &  3.780 &  0.945 &   5014 &           K1III & S \\
  53910 &    95418 & \nodata &      C, A &  24.450 &  2.350 &  0.026 &   9342 &             A1V & H \\
  54182 &    96097 &\nodata &      C, A &  28.990 &  4.620 &  0.332 &   7010 &     F2III-IVvar & H \\ 
  54872 &    97603 & 419  &      C, A &  17.910 &  2.560 &  0.128 &   8037 &             A4V & H \\
  55642 &    99028 & 426.1A &      C, A &  23.670 &  4.000 &  0.423 &   6600 &          F2IV & B, H \\
  55705 &    99211 & \nodata &      C, A &  25.240 &  4.070 &  0.216 &   7805 &             A9V & H \\
  55779 &    99453 &   3663 &      A &  27.220 &  5.180 &  0.500 &   6361 & F7V & H, S \\ 
  57632 &   102647 & 448  &     S, C, A &  11.000 &  2.140 &  0.090 &   8378 &          A3Vvar & H \\
  58001 &   103287 & \nodata &      C, A &  25.500 &  2.390 &  0.045 &   9272 &           A0V & B, H \\
  58803 &   104731 &   3701  &      A &  25.320 &  5.150 &  0.420 &   6638 & F5V & H, S \\ 
  59072 &   105211 & 455.2  &      C, A &  19.760 &  4.140 &  0.353 &   6950 &            F0IV & S, H \\
  59199 &   105452 & 455.3  &     S, C, A &  14.940 &  4.020 &  0.334 &   7081 &          F0IV/V & H \\
  59774 &   106591 & 459  &      C, A &  24.690 &  3.280 &  0.077 &   8613 &          A3Vvar & H \\
  60965 &  108767A &\nodata &      C, A &  26.630 &  2.930 & $-$0.013 &  10207 &           B9.5V & H \\ 
  61084 &   108903 & 470  &      C &  27.150 &  1.650 &  1.517 &   3385 &           M4III & S \\
  61174 &   109085 & 471.2  &      C, A &  18.280 &  4.300 &  0.388 &   6784 &             F2V & H \\
  61941 &   110379 & 482A &     S, C, A &  11.680 &  3.440 &  0.362 &   5674 &             F1V & H \\
  62956 &   112185 & \nodata &      C, A &  25.310 &  1.760 & $-$0.022 &   9020 &             A0p & H \\
  63076 &   112429 & \nodata &      A &  29.290 &  5.230 &  0.290 &   7129 & F0V & H \\
  63125 &   112413 & \nodata &      C &  35.200 &  2.800 &  0.120 &    \nodata &        A0spe & H \\
  63503 &   113139 & \nodata &      A &  25.440 &  4.930 &  0.370 &   6829 & F2V & H \\
  63608 &   113226 & \nodata &      C &  33.600 &  3.000 &  0.930 &    \nodata &        G8IIIvar & E \\
  63613 &   112985 &\nodata&      C &  27.870 &  3.600 &  1.189 &   4390 &           K2III & S, E\\ 
  64241 &      \nodata & \nodata &      C, A &  17.830 &  4.320 &  0.455 &   6399 &    F5V          & H \\
  64583 &   114837 & 503  &      C, A &  18.200 &  4.910 &  0.469 &   6390 &             F5V & S, H \\
  65109 &   115892 & 508.1  &      C, A &  18.020 &  2.720 &  0.068 &  10207 &             A2V & H \\
  65378 &   116656 & \nodata &      C, A &  26.310 &  2.220 &  0.051 &   9330 &             A2V & H \\
  65477 &   116842 & \nodata &      C, A &  25.060 &  3.980 &  0.169 &   7955 &           A5V & B, H \\
  66249 &   118098 &\nodata &      C, A &  22.710 &  3.380 &  0.114 &   8633 &          A0/1IV & H \\ 
  67153 &   119756 &  525.1  &      C, A &  19.400 &  4.230 &  0.375 &   6781 &             F3V & H \\
  67301 &   120315 & \nodata &      C &  31.900 &  1.800 &  0.100 &    \nodata &          B3V  & B, H \\
  67927 &   121370 & 534  &     S, C, A &  11.400 &  2.680 &  0.580 &   6116 &            G0IV & B \\
  68895 &   123123 & \nodata &      C &  31.000 &  3.400 &  1.090 &    \nodata &           K2III & E \\
  68933 &   123139 & 539  &      C &  18.030 &  2.060 &  1.011 &   4823 &          K0IIIB & E \\
  69673 &   124897 & 541  &      C &  11.260 & -0.050 &  1.239 &   4336 &          K2IIIp & E \\
  69701 &   124850 &\nodata &      C, A &  22.240 &  4.070 &  0.511 &   6234 &           F6III & H \\ 
  69713 &   125161 & \nodata &      A &  29.070 &  4.750 &  0.240 &   7700 & A7IV & H \\
  69732 &   125162 & \nodata &      C &  30.400 &  4.200 &  0.090 &    \nodata &            A0sh & H \\
  70497 &   126660 & 549A &     S, C, A &  14.530 &  4.040 &  0.497 &   6192 &             F7V & B \\
  71075 &   127762 & \nodata &      C, A &  26.610 &  3.040 &  0.191 &   8047 &        A7IIIvar & H \\
  71681 &   128621 & 559B &     S, C, A &   1.340 &  1.350 &  0.900 &   5178 &             K0V & S \\
  71683 &   128620 & 559A &     S, C, A &   1.340 & -0.010 &  0.710 &   5801 &           G2.0V & S \\
  71908 &   128898 & 560A &     S, C, A &  16.570 &  3.160 &  0.256 &   7631 &      \nodata  & S, H \\  
  71957 &   129502 &\nodata &      C, A &  18.270 &  3.860 &  0.385 &   6751 &             F2V & H \\ 
  72603 &   130819 &  563.4 &      A &  22.980 &  5.150 &  0.400 &   6745 & F4V & H \\
  72622 &   130841 &  564.1  &      C, A &  23.240 &  2.750 &  0.147 &   8128 &         A3III/V & H \\
  72848 &   131511 &  567 &      S &  11.510 &  5.996 &  0.841 &   5335 &             K2V & B? \\
  73165 &   132052 & \nodata &      C, A &  26.900 &  4.460 &  0.318 &   7079 &             F2V & H \\
  73695 &   133640 & 575A &     S, C, A &  12.510 &  4.830 &  0.647 &   5533 &         G2V+G2V & B \\
  74395 &   134505 & \nodata &      C &  36.000 &  3.600 &  0.920 &    \nodata &           G8III & S \\
  74605 &   136064 &    580.2 &      A &  25.340 &  5.150 &  0.540 &   6152 & F8V & H \\
  74824 &   135379 & \nodata &      C &  30.600 &  4.100 &  0.090 &    \nodata &             A3V & S, H \\
  74975 &   136202 &  \nodata &      A &  25.380 &  5.040 &  0.540 &   6119 & F8IV & H  \\ 
  75458 &   137759 & \nodata &      C &  31.000 &  3.500 &  1.170 &    \nodata &           K2III & E \\
  75695 &   137909 & \nodata &      C &  34.300 &  3.700 &  0.320 &    \nodata &             F0p & H \\
  76267 &   139006 & \nodata &      C, A &  23.010 &  2.210 &  0.025 &   9584 &             A0V & H \\
  76829 &   139664 & 594  &      C, A &  17.440 &  4.630 &  0.413 &   6649 &           F3/5V & H \\
  77070 &   140573 & 596.2  &      C &  22.680 &  2.610 &  1.167 &   4548 &           K2III & E \\
  77622 &   141795 & \nodata &      C, A &  21.600 &  3.710 &  0.147 &   8257 & \nodata   & H \\  
  77952 &   141891 & 601A &     S, C, A &  12.380 &  2.810 &  0.315 &   7109 &        F0III/IV & S, H \\
  78527 &   144284 & 609.1  &      C, A &  21.030 &  3.990 &  0.528 &   4642 &          F8IV-V & B \\
  79822 &   148048 &   3950A &      A &  29.730 &  4.950 &  0.360 &   6868 & F2V & H \\ 
  79882 &   146791 & \nodata &      C &  32.600 &  3.400 &  0.970 &    \nodata &           G8III & E \\
  80179 &   147449 & \nodata &      A &  27.270 &  4.820 &  0.340 &   6981 & F0V & H \\
  80331 &   148387 &  624.1A &      C &  28.230 &  2.730 &  0.910 &   4941 &           G8III & B \\
  80337 &   147513 &   620.1A &      A &  12.780 &  5.370 &  0.630 &   5930 & G5V & S \\
  80686 &   147584 & 624  &     S, C, A &  12.120 &  4.900 &  0.555 &   6107 &             G0V & S \\
  81693 &   150680 & 635A &     S, C, A &  10.720 &  2.810 &  0.650 &   5820 &            F9IV & B \\
  81833 &   150997 & \nodata &      C &  33.300 &  3.600 &  0.920 &    \nodata &        G8III-IV &E \\
  82020 &   151613 & \nodata &      A &  26.730 &  4.840 &  0.380 &   6805 & F2V & H \\
  82396 &   151680 & 639.1  &      C &  19.540 &  2.260 &  1.181 &   4703 &          K2IIIB & E \\
  82587 &   152598 & \nodata &      A &  29.190 &  5.340 &  0.310 &   7136 & F0V & H \\
  82860 &   153597 & 648  &      C, A &  15.260 &  4.880 &  0.481 &   6146 &          F6Vvar & B \\
  83000 &   153210 &\nodata&      C &  28.040 &  3.180 &  1.158 &   4564 &        K2IIIvar & E \\ 
  83431 &   153580 & \nodata &      A &  27.220 &  5.270 &  0.470 & 6501 & F5V & H, S \\
  84012 &   155125 &  656.1A &      C, A &  27.090 &  2.430 &  0.059 &   8788 &          A2.5VA & H \\
  84143 &   155203 & 657  &      C, A &  22.530 &  3.310 &  0.423 &   6519 &             F2V & H \\
  84379 &   156164 & \nodata &      C, A &  23.040 &  3.120 &  0.080 &   8879 &         A3IVv & B, H \\
  84405 &   155885 & 663A &     S, C, A &   5.950 &  4.330 &  0.855 &   5119 &           K1.5V & B \\
  84709 &   \nodata &  667AB &      S &   6.836 &  6.255 &  0.999 &   4810 &           M1.5V &   S \\
  84720 &   \nodata &  666AB &     S, C &   8.800 &  5.520 &  0.791 &   5052 &    M0V          &   S \\ 
  84893 &   156897 & 670A &      C, A &  17.360 &  4.390 &  0.366 &   6704 &           F2/3V & H \\
  85340 &   157792 & 673.1  &      C, A &  25.500 &  4.150 &  0.283 &   7440 &       A3 & H \\ 
  85667 &   158614 & 678A &      S &  16.340 &  5.310 &  0.715 &   5538 &             G6V & B \\
  86032 &   159561 & 681  &     S, C &  14.900 &  2.080 &  0.155 &   8225 &           A5III & H \\
  86036 &   160269 &   684A &      A &  14.190 &  5.230 &  0.600 &   5898 & G0V & B \\
  86201 &   160922 &\nodata &      C &  23.160 &  4.770 &  0.430 &   6560 &             F5V & H \\ 
  86486 &   160032 & 686.2  &      C &  21.450 &  4.750 &  0.403 &   6678 &             F2V & H \\
  86614 &   162003 & 694.1A &      C &  22.840 &  4.550 &  0.433 &   6435 &          F5IV-V & H \\
  86736 &   160915 & 692  &      C &  17.650 &  4.860 &  0.469 &   6465 &           F6/7V & H \\
  86742 &   161096 & \nodata &      C &  25.090 &  2.750 &  1.191 &   4571 &           K2III & E \\
  86796 &   160691 &    691 &      A &  15.510 &  5.120 &  0.690 &   5784 & G3IV-V & S \\
  87108 &   161868 & \nodata &      C &  31.500 &  3.800 &  0.040 &    \nodata &             A0V & H \\
  87585 &   163588 & \nodata &      C &  34.500 &  3.900 &  1.180 &    \nodata &           K2III & E \\
  88175 &   164259 & 699.2  &      C, A &  23.550 &  4.620 &  0.367 &   6771 &            F2IV & H \\
  88601 &   165341 & 702A &     S, C, A &   5.100 &  4.030 &  0.860 &   5019 &             K0V & E \\
  88635 &   165135 &\nodata &      C &  29.700 &  2.950 &  1.021 &   4914 &           K0III & E \\ 
  88745 &   165908 & 704A &      S, A &  15.640 &  5.070 &  0.504 &   5925 &             F7V & B \\
  88771 &   165777 & \nodata &      C, A &  26.630 &  3.700 &  0.140 &   8400 &           A4IVs & H \\
  89348 &   168151 &    708.1 &      A &  22.920 &  4.990 &  0.430 &   6404 & F5V & H \\
  89937 &   170153 & 713AB  &      C, A &   8.060 &  3.560 &  0.489 &   6122 &          F7Vvar & B \\
  90139 &   169414 & \nodata &      C &  36.500 &  4.000 &  1.170 &    \nodata &           K2III & E \\
  90496 &   169916 & 713.1  &      C &  23.970 &  2.810 &  1.057 &   4809 &          K1IIIB & E \\
  91262 &   172167 & 721  &     S, C, A &   7.680 &  0.030 & $-$0.001 &   9519 &          A0Vvar & H \\
  92024 &   172555 & \nodata &      A &  28.550 &  4.780 &  0.200 &   7846 & A7V & H, S \\
  92161 &   173880 & \nodata &      C, A &  28.890 &  4.340 &  0.127 &   8334 &           A5III & H \\
  93017 &   176051 & 738A &      S, A &  14.870 &  5.280 &  0.555 &   6064 &             G0V & B \\
  93506 &   176687 & \nodata &      C &  27.040 &  2.600 &  0.062 &   8799 &            A3IV & H \\
  93747 &   177724 & \nodata &      C, A &  25.460 &  2.990 &  0.014 &   9190 &            A0Vn & H \\
  93825 &   177474 & 743.1A &      C, A &  17.300 &  4.230 &  0.523 &   6202 &          F8/G0V & B \\
  94083 &   180777 &    748.1 &      A &  27.300 &  5.110 &  0.310 &   7129 & F0Vs & H \\
  94376 &      \nodata & \nodata &      C &  29.870 &  3.050 &  1.001 &   4966 &    G9III        & E \\
  95501 &   182640 & 760  &     S, C, A &  15.530 &  3.360 &  0.319 &   7074 &            F2IV & H \\
  97295 &   187013 &   767.1A &      A &  21.230 &  5.000 &  0.470 &   6401 & F5.5IV-V & H, B \\
  97649 &   187642 & 768  &     S, C, A &   5.120 &  0.760 &  0.221 &   7800 &          A7IV-V & H \\
  97650 &   187532 &             \nodata &      A &  27.870 &  5.380 &  0.350 &   6812 & F5V & H, B \\
  98066 &   188376 &    770.1 &      A &  25.840 &  4.700 &  0.750 &   5425 & G5IV & B, $N_{\rm obs}$=13 \\
  98495 &   188228 & \nodata &      C &  32.200 &  4.000 &  0.030 &    \nodata &             A0V & S, H \\
  99240 &   190248 & 780  &     S, C, A &   6.110 &  3.530 &  0.765 &   5590 &          G8.0IV & S \\
 101612 &   195627 &             \nodata &      A &  27.790 &  4.750 &  0.290 &   7201 & F0V & H, S \\
 101983 &   196378 &    794.2 &      A &  24.660 &  5.110 &  0.510 &   6040 & G0V & S \\
 102333 &   197157 & \nodata &      C, A &  24.170 &  4.500 &  0.278 &   7448 &        A7III/IV & S, H \\
 102422 &   198149 & 807  &     S, C &  14.270 &  3.410 &  0.912 &   4940 &            K0IV & E \\
 102431 &   198084 &\nodata&      C, A &  27.290 &  4.520 &  0.535 &   6138 &          F8IV-V & B \\ 
 102485 &   197692 & 805  &     S, C, A &  14.680 &  4.130 &  0.426 &   6633 &             F5V & H \\
 102488 &   197989 & 806.1A &      C &  22.290 &  2.450 &  1.021 &   4799 &           K0III & E \\
 104858 &   202275 & 822A &      C, A &  18.490 &  4.470 &  0.529 &   6238 &         F5V & H \\
 104887 &   202444 & 822.1A &      C, A &  20.340 &  3.740 &  0.393 &   6621 &            F1IV & H \\
 105199 &   203280 & 826  &     S, C, A &  15.040 &  2.430 &  0.243 &   7773 &          A7IV-V & H \\
 105858 &   203608 & 827  &     S, C, A &   9.260 &  4.220 &  0.469 &   6205 &             F7V & S \\
 107089 &   205478 & 835.1  &      C &  21.200 &  3.730 &  1.017 &   4769 &           K0III & S \\
 107310 &   206826 & 836.6A &      C, A &  22.240 &  4.690 &  0.454 &   6309 &             F6V & B \\
 107556 &      \nodata & \nodata &     S, C, A &  11.870 &  2.850 &  0.305 &   7301 &    A5mF2   & H \\
 107649 &   207129 & 838  &      S &  15.990 &  5.570 &  0.601 &   5889 &             G0V & Y \\
 108036 &   207958 &    838.5 &      A &  26.610 &  5.080 &  0.380 &   6799 & F2V & H \\
 108870 &   209100 & 845  &     S, C, A &   3.620 &  4.690 &  1.056 &   4683 &             K4V &  Y, S \\
 108917 &      \nodata & \nodata &      C, A &  29.590 &  4.400 &  0.341 &   6964 &    Am           & H \\
 109176 &   210027 & 848  &     S, C, A &  11.730 &  3.770 &  0.435 &   6442 &             F5V & H \\
 109268 &   209952 & \nodata &      C &  31.000 &  1.700 &  0.070 &    \nodata &            B7IV & H \\
 109427 &   210418 &\nodata &      C, A &  28.300 &  3.520 &  0.086 &   8569 &             A2V & H \\ 
 109857 &   211336 &\nodata&      C, A &  26.200 &  4.170 &  0.278 &   7283 &            F0IV & H \\ 
 110109 &   211415 & 853A &      S, A &  13.790 &  5.360 &  0.614 &   5837 &             G3V & S \\
 110618 &   211998 &   855.1A &      A &  28.700 &  5.280 &  0.630 &   5486 & G9V& S \\
 110649 &   212330 &    857 &      A &  20.560 &  5.310 &  0.670 &   5739 & G2IV-V & S \\
 110960 &      \nodata & \nodata &      C &  28.170 &  3.650 &  0.406 &   6619 &    F3III-IV     & H, E \\
 111169 &   213558 & \nodata &      C &  31.500 &  3.800 &  0.030 &    \nodata &             A1V & H \\
 111449 &   213845 &    863.2 &      A &  22.680 &  5.210 &  0.450 &   6597 & F5V & H \\
 112724 &   216228 & \nodata &      C &  35.300 &  3.700 &  1.050 &    \nodata &           K0III & E \\
 112748 &   216131 & \nodata &      C &  32.500 &  3.700 &  0.930 &    \nodata &           M2III & E \\
 112935 &   216385 &  9801A  &      A &  27.280 &  5.160 &  0.470 &   6250 & F6V & H  \\ 
 113368 &   216956 & 881A &     S, C, A &   7.700 &  1.230 &  0.140 &   8399 &             A3V & H \\
 113638 &   217364 & \nodata &      C &  33.400 &  4.300 &  0.960 &    \nodata &           G8III & S \\
 113860 &   217792 &    886.2 &      A &  29.400 &  5.120 &  0.280 &   7143 & F1V & H, S \\
 114570 &   219080 & 891.1  &      C, A &  24.590 &  4.520 &  0.302 &   7176 &             F0V & H \\
 114996 &   219571 & \nodata &      C, A &  23.060 &  3.990 &  0.410 &   6618 &          F3IV/V & S, H \\
 115126 &   219834 & \nodata &      A &  21.050 &  5.200 &  0.790 &   5461 & G8.5IV & \\ 
 116584 &   222107 &\nodata&      C &  26.410 &  3.850 &  0.987 &   4636 &        G8III-IV & B \\ 
 116727 &   222404 & 903  &      C &  14.100 &  3.210 &  1.031 &   4761 &            K1IV & E \\
 118268 &   224617 & \nodata &      C &  32.000 &  4.100 &  0.420 &    \nodata &            F4IV & H \\
\enddata
\tablenotetext{a}{Target list code: S = Starshade study mission target,  C = Coronagraph study mission target, A = WFIRST-AFTA study mission target.}
\tablenotetext{b}{Stars were not observed by Lick and Keck Doppler programs because of 
these (non-exhaustive) list of reasons: star too hot (H), typically earlier than F8V; star too far south (S); 
star is evolved (E) into a giant or subgiant; star too young (Y), with chromospheric activity substantially increasing jitter; 
or binary (B) or higher stellar multiple noted in the literature and/or our RVs.  
These reasons for a lack of observations were inferred in 2014 based on stellar properties and not based on 
a target down-selections when observing lists were created.
}
\end{deluxetable}



\section{Completeness of Individual Stars}
\label{app:completeness}

In this appendix show the results of automated searches for planets in our Doppler data.  
This search methodology is described fully in Sec. \ref{sec:automated_planet_search}.
We also show the completeness limits for each star computed by injection-recovery tests (Sec.\ \ref{sec:injection}).
The figures below show these results in graphical form for the 76 stars for which we have Doppler data.  
Each figure caption lists the HD number, Hipparcos number, and the codes for the imaging program target lists 
(S = starshade, C = coronagraph, A = WFIRST-AFTA).

\begin{figure}
\begin{centering}
\includegraphics[width=0.45\textwidth]{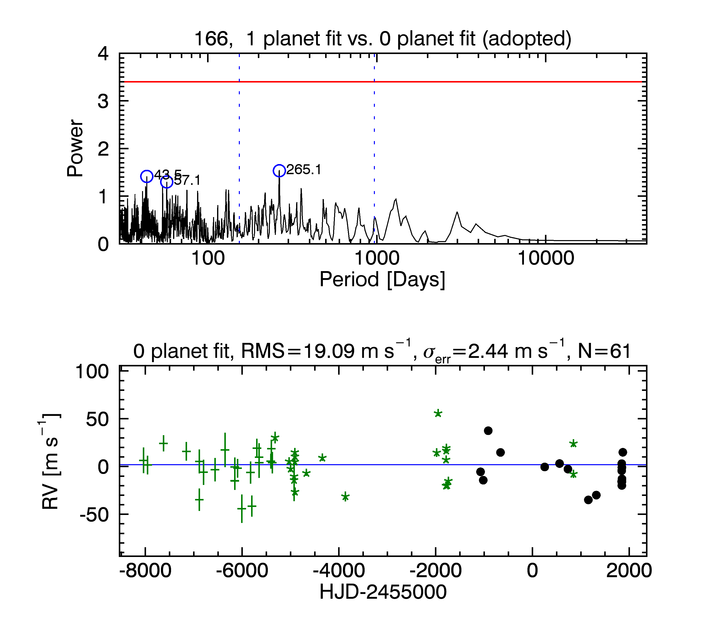}
\includegraphics[width=0.50\textwidth]{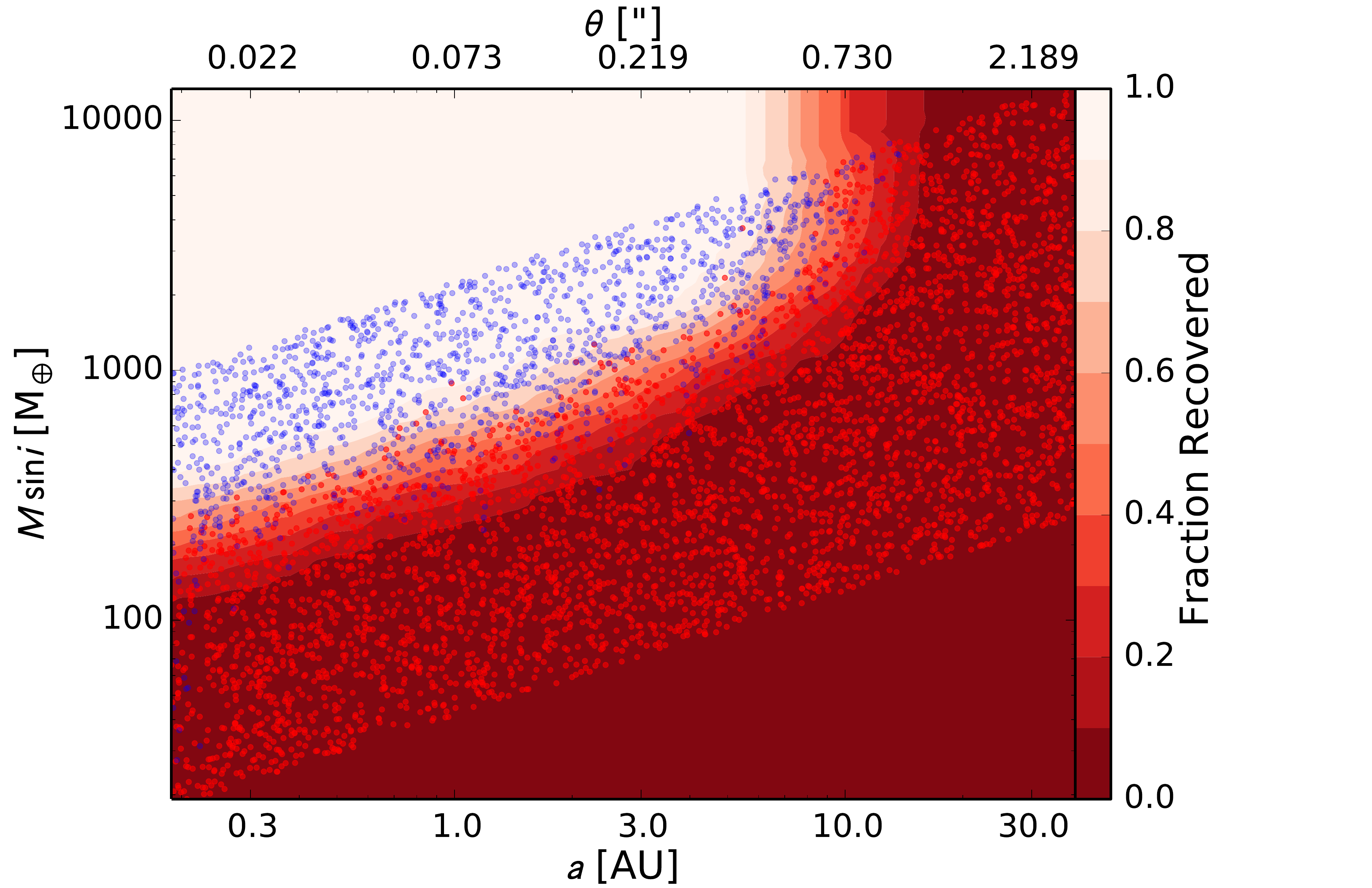}
\end{centering}
\caption{Results from an automated search for planets orbiting the star 
HD~166 (HIP~544; program = S) 
based on RVs from Lick and/or Keck Observatory.
The set of plots on the left (analogous to Figures \ref{fig:search_example} and \ref{fig:search_example2}) 
show the planet search results 
and the plot on the right shows the completeness limits (analogous to Fig.\ \ref{fig:completeness_example}). 
See the captions of those figures for detailed descriptions.  
}
\label{fig:completeness_166}
\end{figure}

\begin{figure}
\begin{centering}
\includegraphics[width=0.45\textwidth]{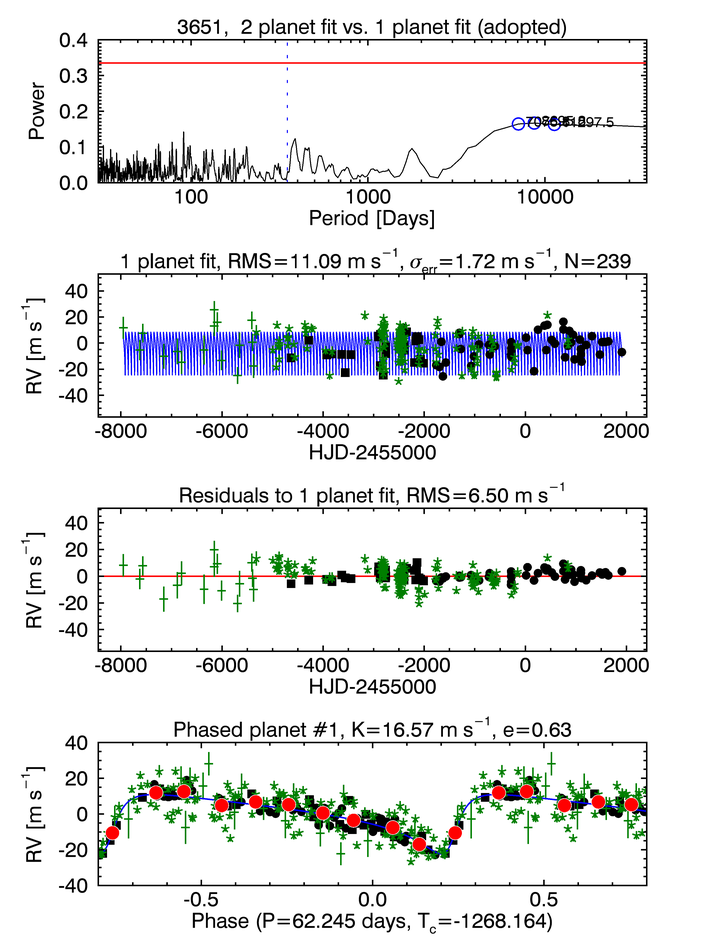}
\includegraphics[width=0.50\textwidth]{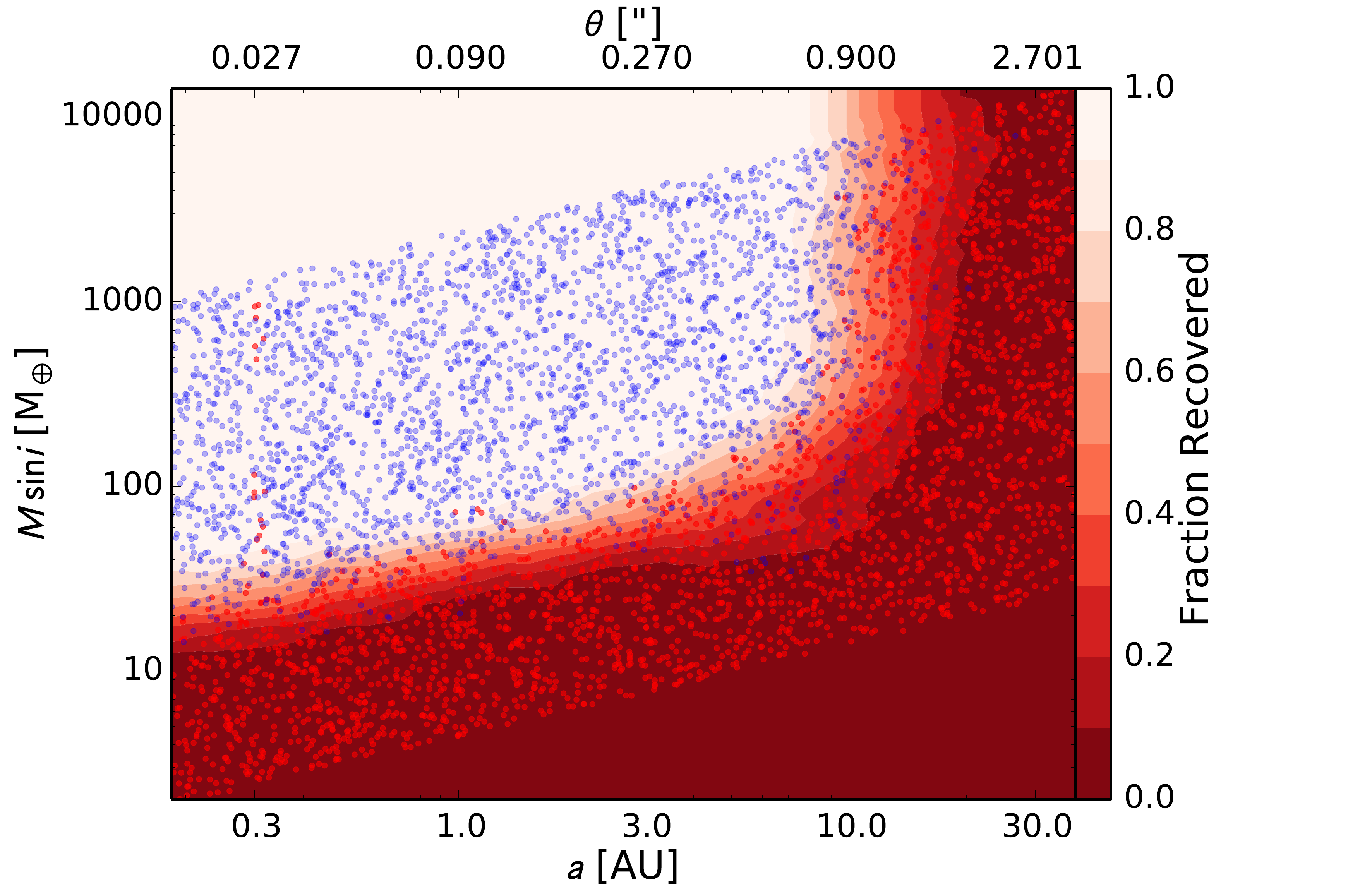}
\end{centering}
\caption{Results from an automated search for planets orbiting the star 
HD~3651 (HIP~3093; program = S) 
based on RVs from Lick and/or Keck Observatory.
The set of plots on the left (analogous to Figures \ref{fig:search_example} and \ref{fig:search_example2}) 
show the planet search results 
and the plot on the right shows the completeness limits (analogous to Fig.\ \ref{fig:completeness_example}). 
See the captions of those figures for detailed descriptions.  
This star has one known planet.
}
\label{fig:completeness_3651}
\end{figure}
\clearpage

\begin{figure}
\begin{centering}
\includegraphics[width=0.45\textwidth]{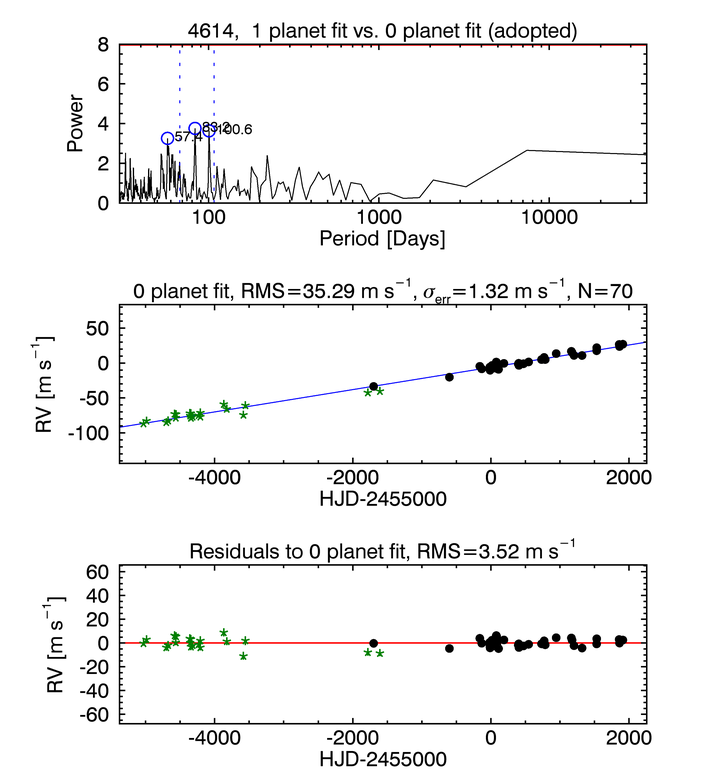}
\includegraphics[width=0.50\textwidth]{4614-recovery.pdf}
\end{centering}
\caption{Results from an automated search for planets orbiting the star 
HD~4614 (HIP~3821; programs = S, C, A) 
based on RVs from Lick and/or Keck Observatory.
The set of plots on the left (analogous to Figures \ref{fig:search_example} and \ref{fig:search_example2}) 
show the planet search results 
and the plot on the right shows the completeness limits (analogous to Fig.\ \ref{fig:completeness_example}). 
See the captions of those figures for detailed descriptions.  
This star shows a significant linear trend with no detectable curvature, presumably due to its known stellar companion.
}
\label{fig:completeness_4614}
\end{figure}

\begin{figure}
\begin{centering}
\includegraphics[width=0.45\textwidth]{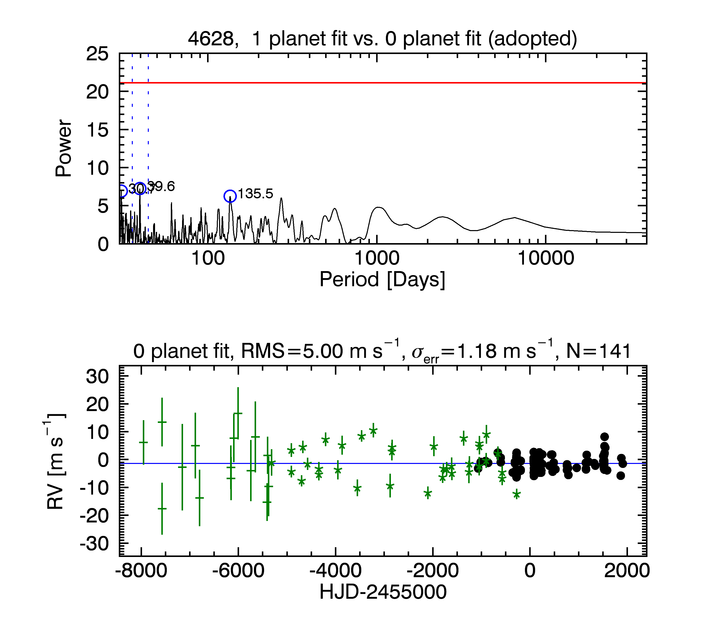}
\includegraphics[width=0.50\textwidth]{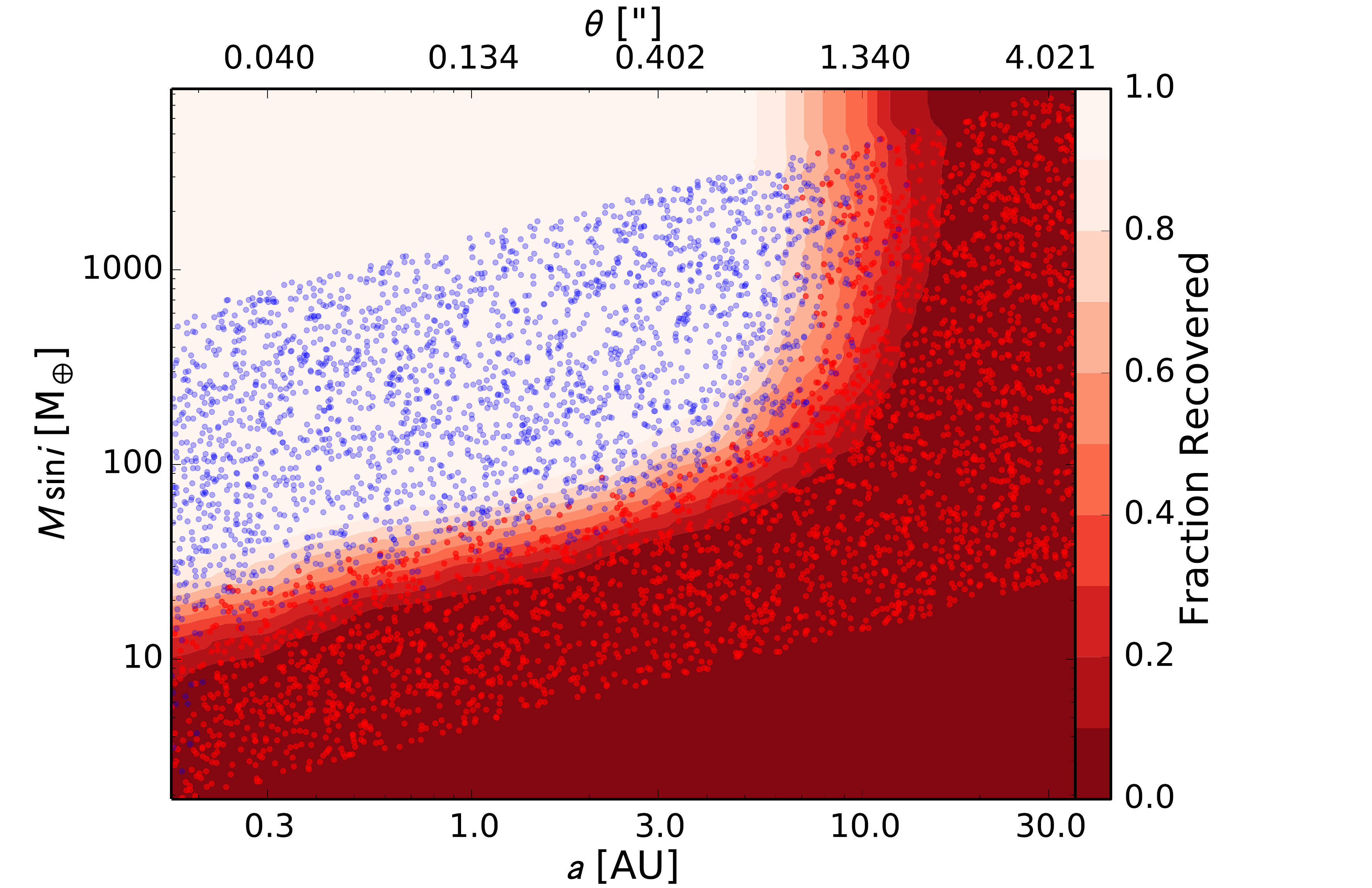}
\end{centering}
\caption{Results from an automated search for planets orbiting the star 
HD~4628 (HIP~3765; programs = S, C) 
based on RVs from Lick and/or Keck Observatory.
The set of plots on the left (analogous to Figures \ref{fig:search_example} and \ref{fig:search_example2}) 
show the planet search results 
and the plot on the right shows the completeness limits (analogous to Fig.\ \ref{fig:completeness_example}). 
See the captions of those figures for detailed descriptions.  
}
\label{fig:completeness_4628}
\end{figure}
\clearpage

\begin{figure}
\begin{centering}
\includegraphics[width=0.45\textwidth]{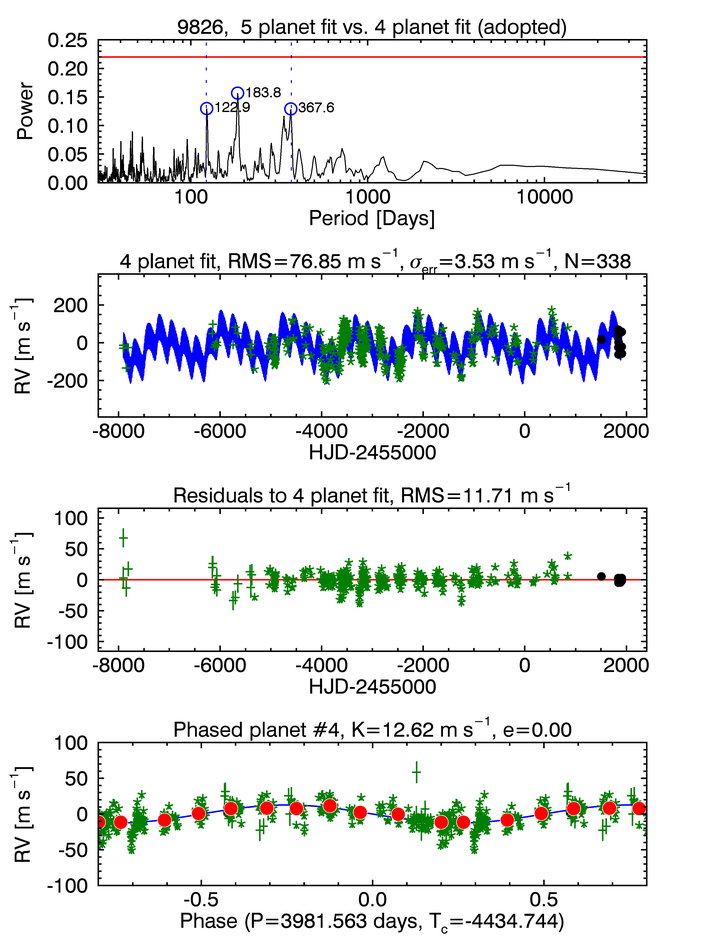}
\includegraphics[width=0.50\textwidth]{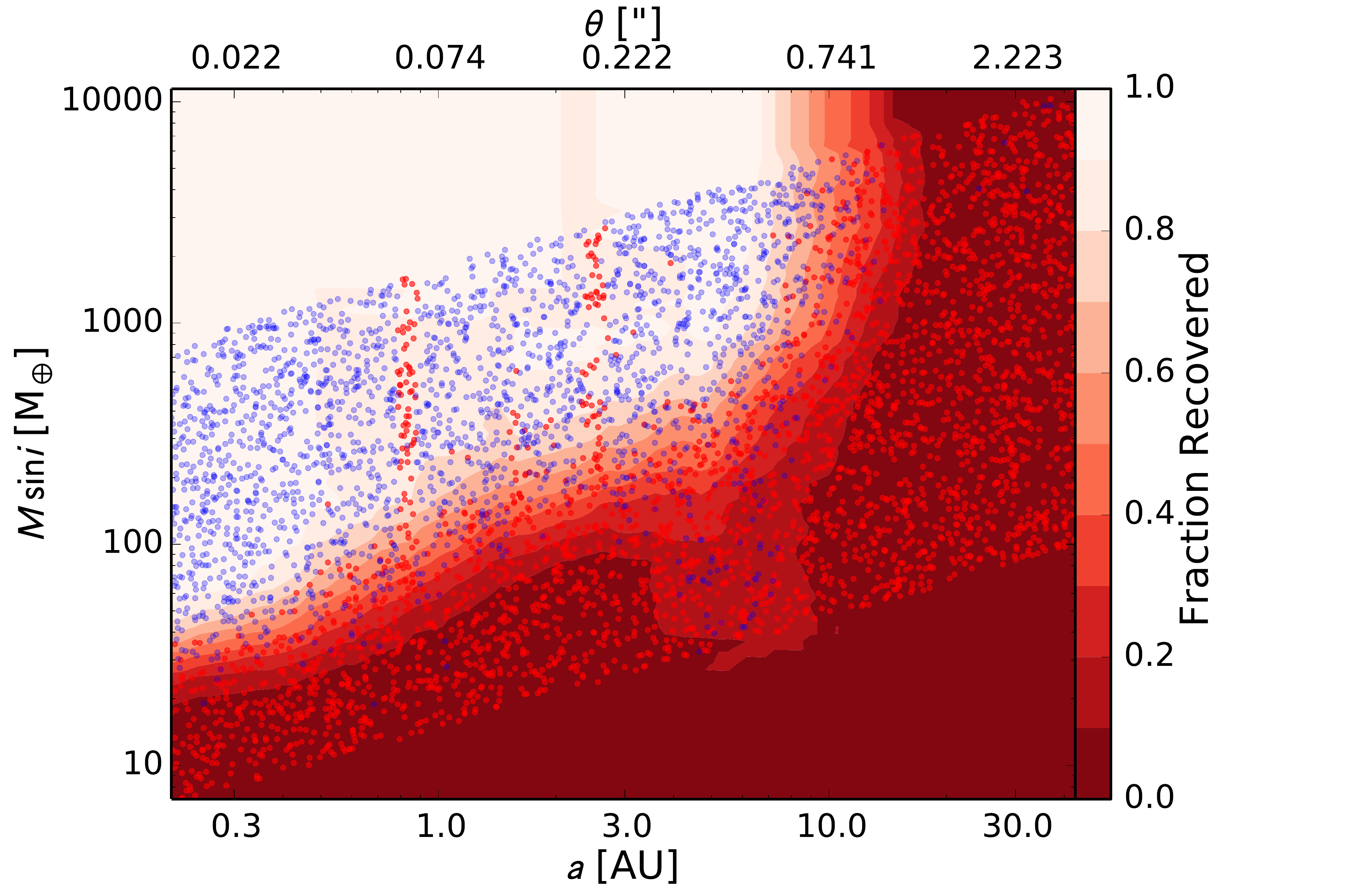}
\end{centering}
\caption{Results from an automated search for planets orbiting the star 
HD~9826 (HIP~7513; programs = C, A) 
based on RVs from Lick and/or Keck Observatory.
The set of plots on the left (analogous to Figures \ref{fig:search_example} and \ref{fig:search_example2}) 
show the planet search results 
and the plot on the right shows the completeness limits (analogous to Fig.\ \ref{fig:completeness_example}). 
See the captions of those figures for detailed descriptions.  
This star has three known planets.  We detect an additional periodicity at $\sim$4000 days and interpret this as the signature of a stellar magnetic activity cycle.
}
\label{fig:completeness_9826}
\end{figure}

\begin{figure}
\begin{centering}
\includegraphics[width=0.45\textwidth]{10476-finalfit.png}
\includegraphics[width=0.50\textwidth]{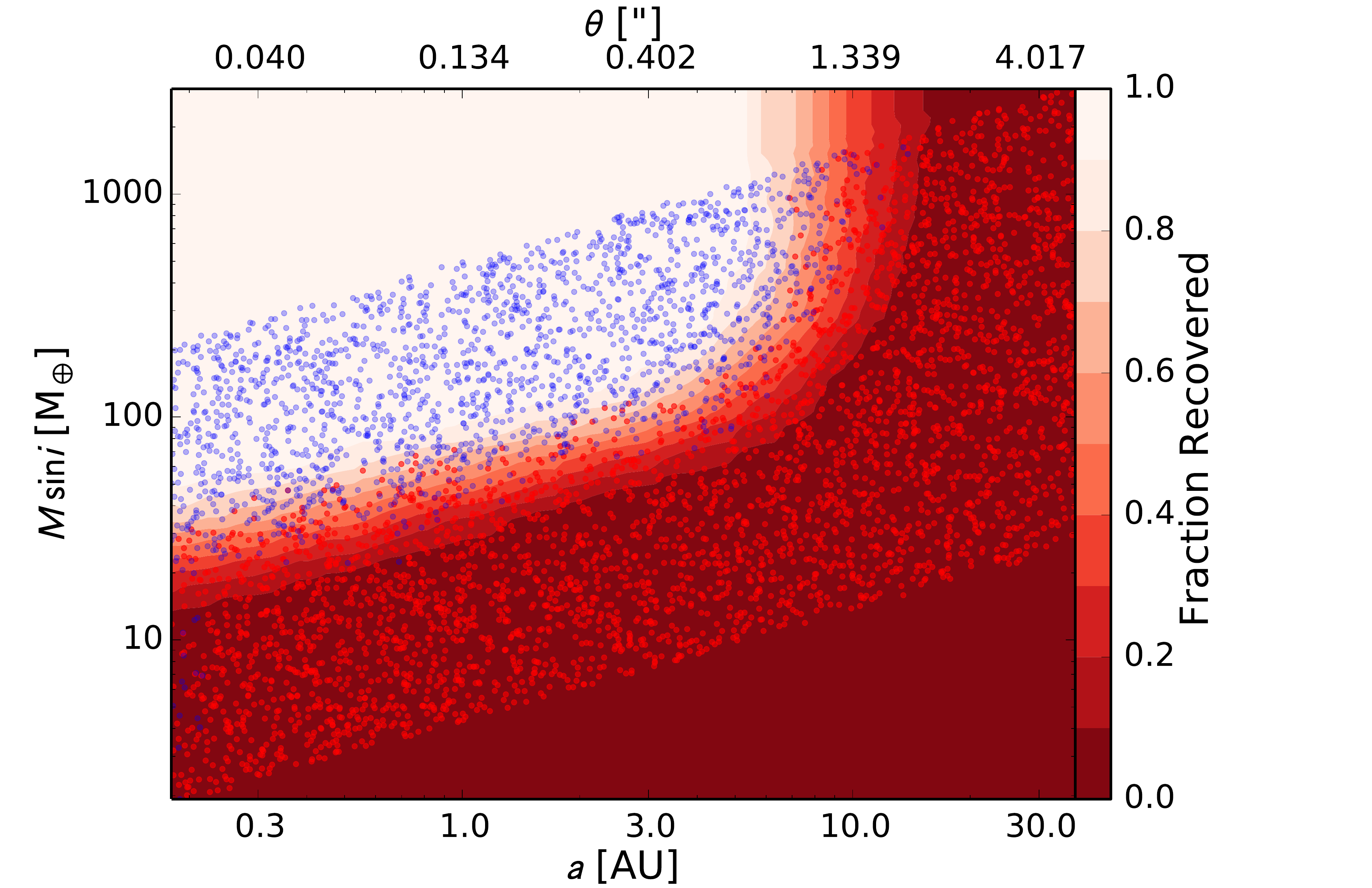}
\end{centering}
\caption{Results from an automated search for planets orbiting the star 
HD~10476 (HIP~7981; programs = S, C, A) 
based on RVs from Lick and/or Keck Observatory.
The set of plots on the left (analogous to Figures \ref{fig:search_example} and \ref{fig:search_example2}) 
show the planet search results 
and the plot on the right shows the completeness limits (analogous to Fig.\ \ref{fig:completeness_example}). 
See the captions of those figures for detailed descriptions.  
}
\label{fig:completeness_10476}
\end{figure}
\clearpage

\begin{figure}
\begin{centering}
\includegraphics[width=0.45\textwidth]{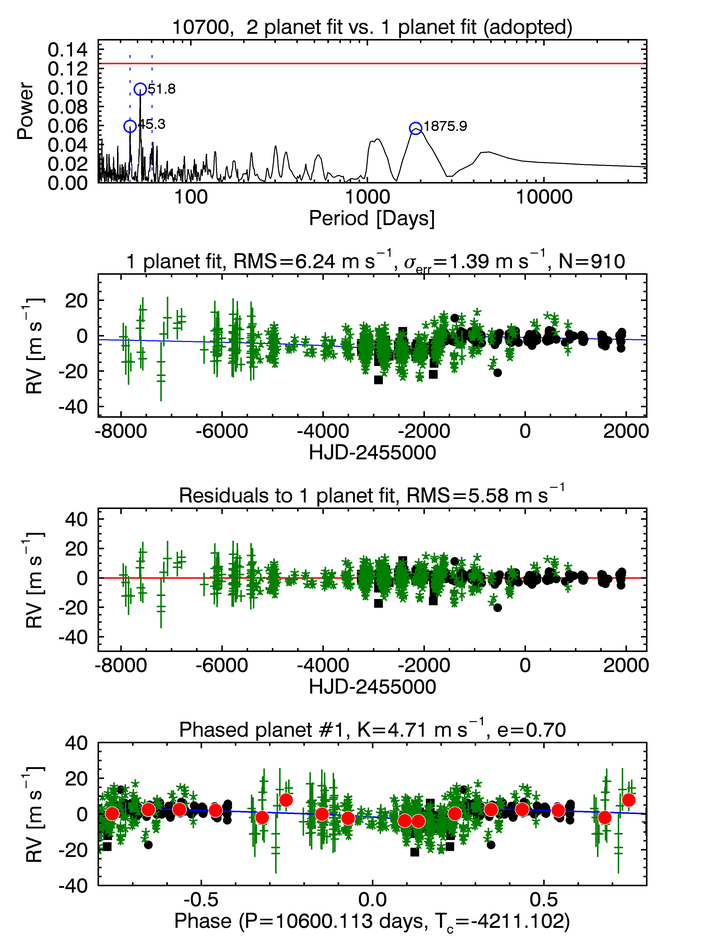}
\includegraphics[width=0.50\textwidth]{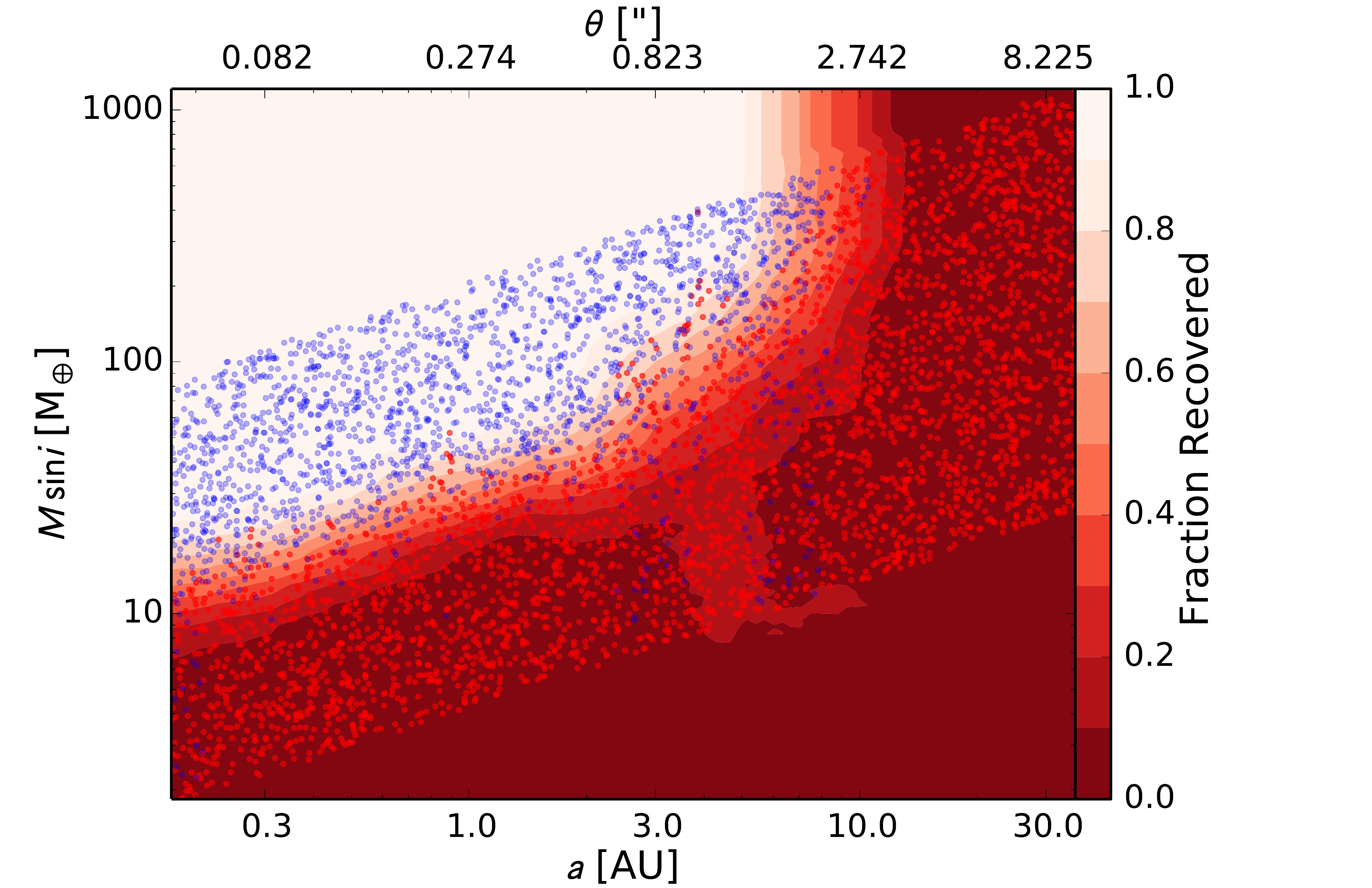}
\end{centering}
\caption{Results from an automated search for planets orbiting the star 
HD~10700 (HIP~8102; programs = S, C, A) 
based on RVs from Lick and/or Keck Observatory.
The set of plots on the left (analogous to Figures \ref{fig:search_example} and \ref{fig:search_example2}) 
show the planet search results 
and the plot on the right shows the completeness limits (analogous to Fig.\ \ref{fig:completeness_example}). 
See the captions of those figures for detailed descriptions.  
The automated pipeline  detects a long-period signal but it appears to be caused by a poorly constrained offset between datasets.  We do not detect to five controversial, low-mass planets claimed by \cite{Tuomi2013}.
}
\label{fig:completeness_10700}
\end{figure}

\begin{figure}
\begin{centering}
\includegraphics[width=0.45\textwidth]{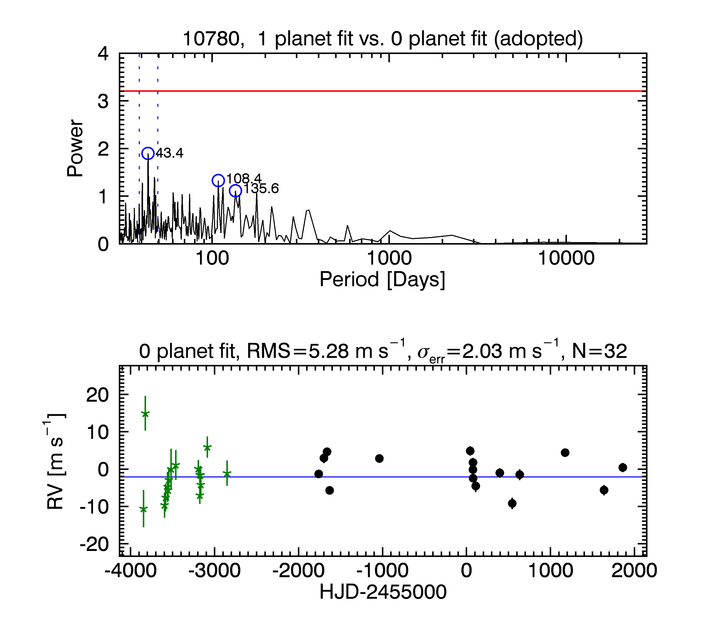}
\includegraphics[width=0.50\textwidth]{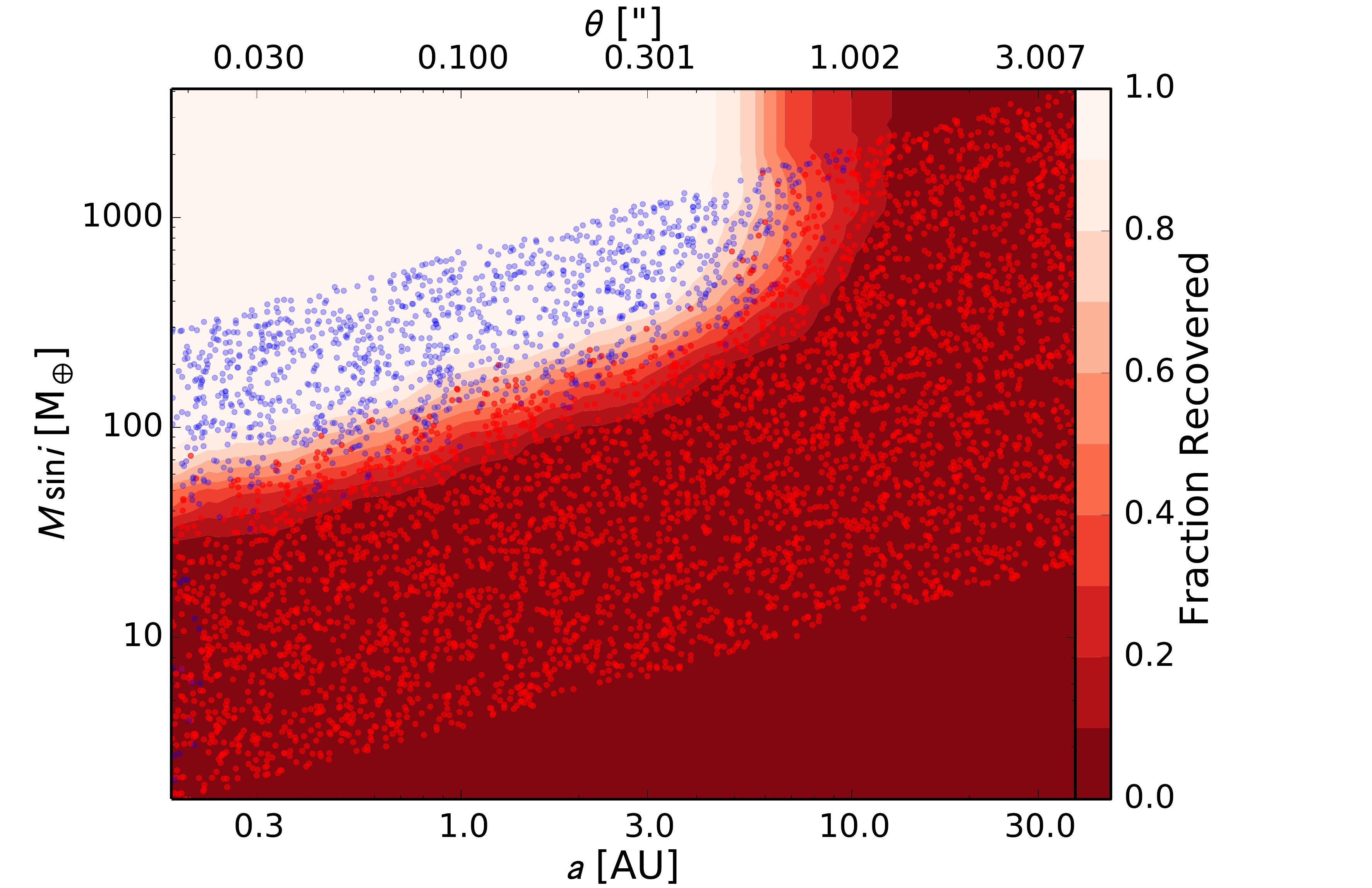}
\end{centering}
\caption{Results from an automated search for planets orbiting the star 
HD~10780 (HIP~8362; program = S) 
based on RVs from Lick and/or Keck Observatory.
The set of plots on the left (analogous to Figures \ref{fig:search_example} and \ref{fig:search_example2}) 
show the planet search results 
and the plot on the right shows the completeness limits (analogous to Fig.\ \ref{fig:completeness_example}). 
See the captions of those figures for detailed descriptions.  
}
\label{fig:completeness_10780}
\end{figure}
\clearpage

\begin{figure}
\begin{centering}
\includegraphics[width=0.45\textwidth]{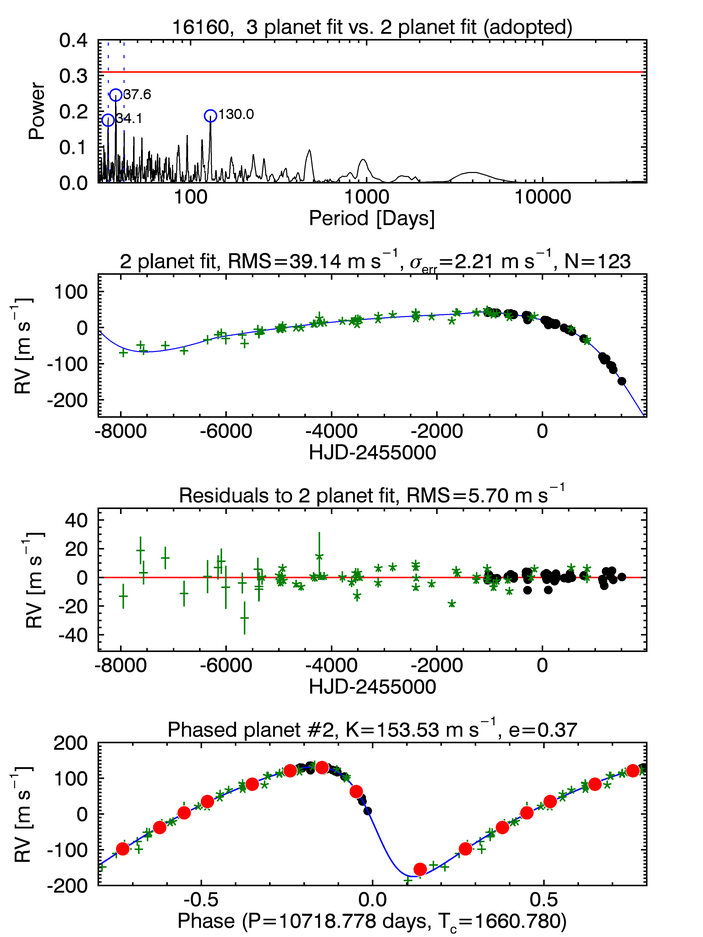}
\includegraphics[width=0.50\textwidth]{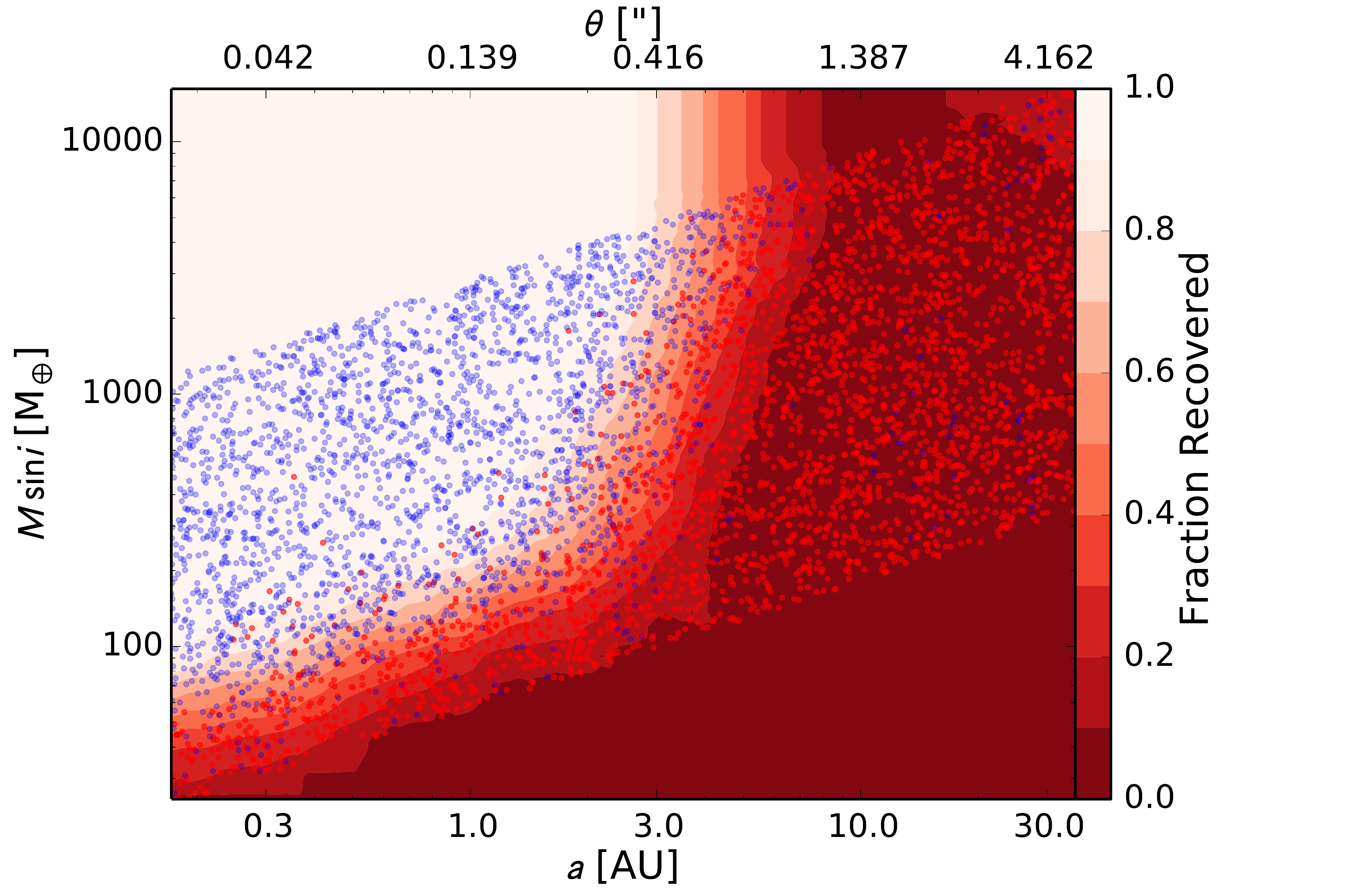}
\end{centering}
\caption{Results from an automated search for planets orbiting the star 
HD~16160 (HIP~12114; program = S) 
based on RVs from Lick and/or Keck Observatory.
The set of plots on the left (analogous to Figures \ref{fig:search_example} and \ref{fig:search_example2}) 
show the planet search results 
and the plot on the right shows the completeness limits (analogous to Fig.\ \ref{fig:completeness_example}). 
See the captions of those figures for detailed descriptions.  
The RV time-series for this star shows significant curvature from a late M-type companion. The RV curvature makes it difficult to detect long-period planets in this system.
}
\label{fig:completeness_16160}
\end{figure}

\begin{figure}
\begin{centering}
\includegraphics[width=0.45\textwidth]{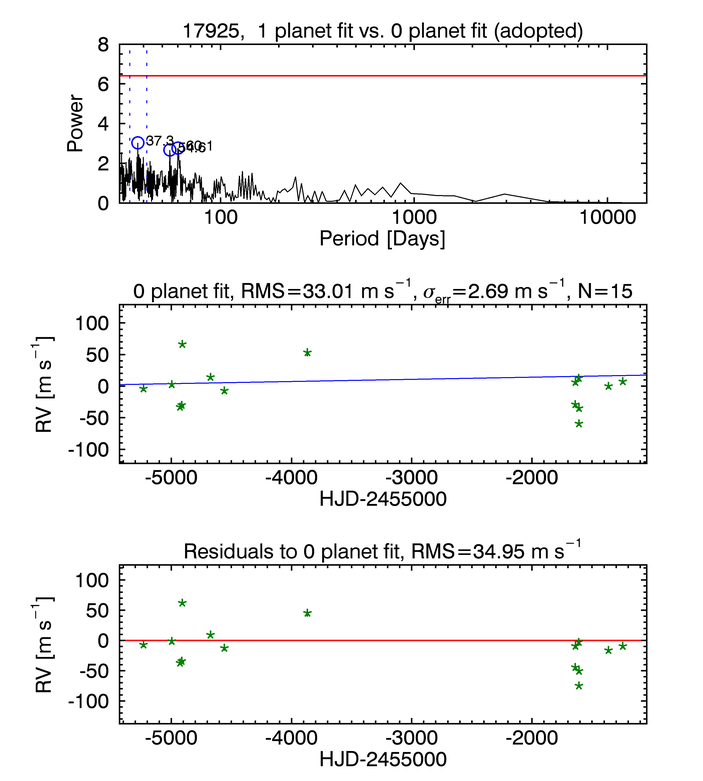}
\includegraphics[width=0.50\textwidth]{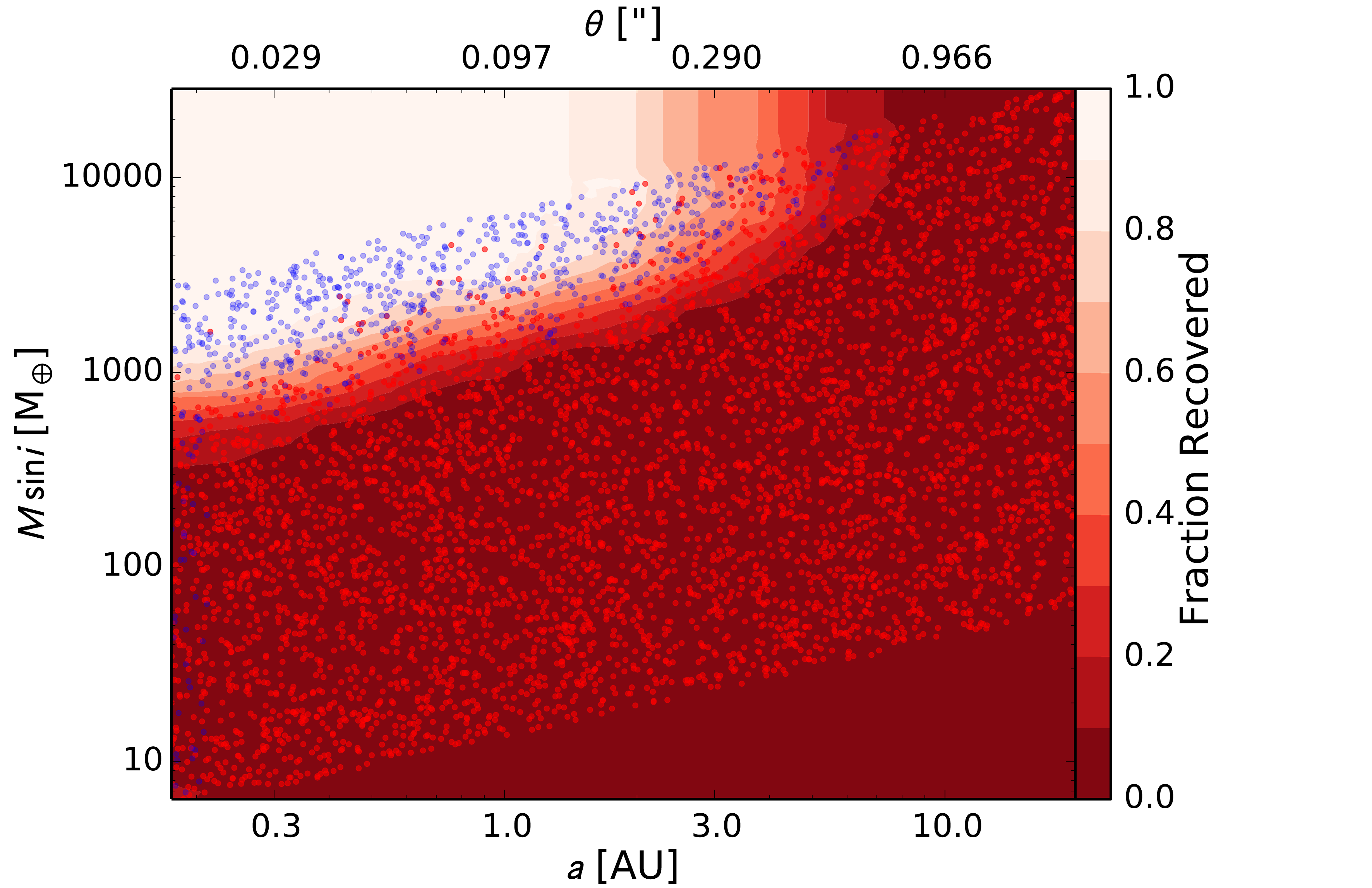}
\end{centering}
\caption{Results from an automated search for planets orbiting the star 
HD~17925 (HIP~13402; program = S) 
based on RVs from Lick and/or Keck Observatory.
The set of plots on the left (analogous to Figures \ref{fig:search_example} and \ref{fig:search_example2}) 
show the planet search results 
and the plot on the right shows the completeness limits (analogous to Fig.\ \ref{fig:completeness_example}). 
See the captions of those figures for detailed descriptions.  
The automated pipeline detects a marginally significant linear trend in the RV time series.
}
\label{fig:completeness_17925}
\end{figure}
\clearpage

\begin{figure}
\begin{centering}
\includegraphics[width=0.45\textwidth]{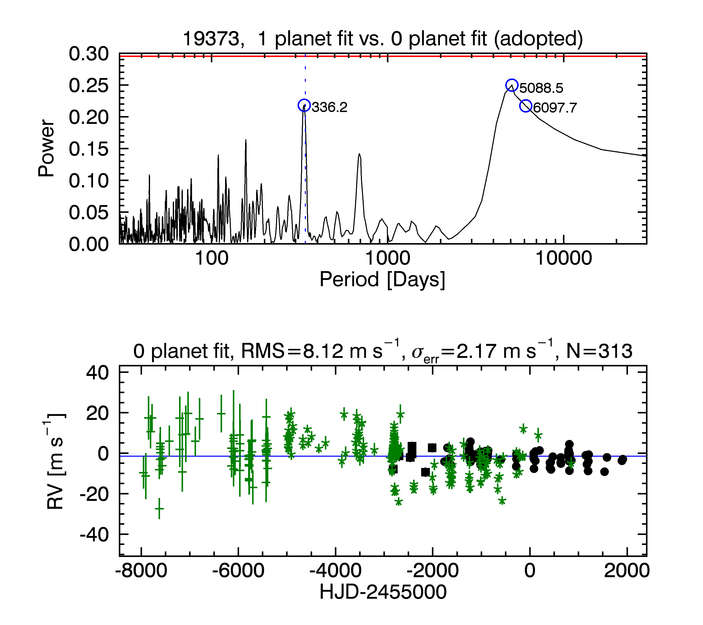}
\includegraphics[width=0.50\textwidth]{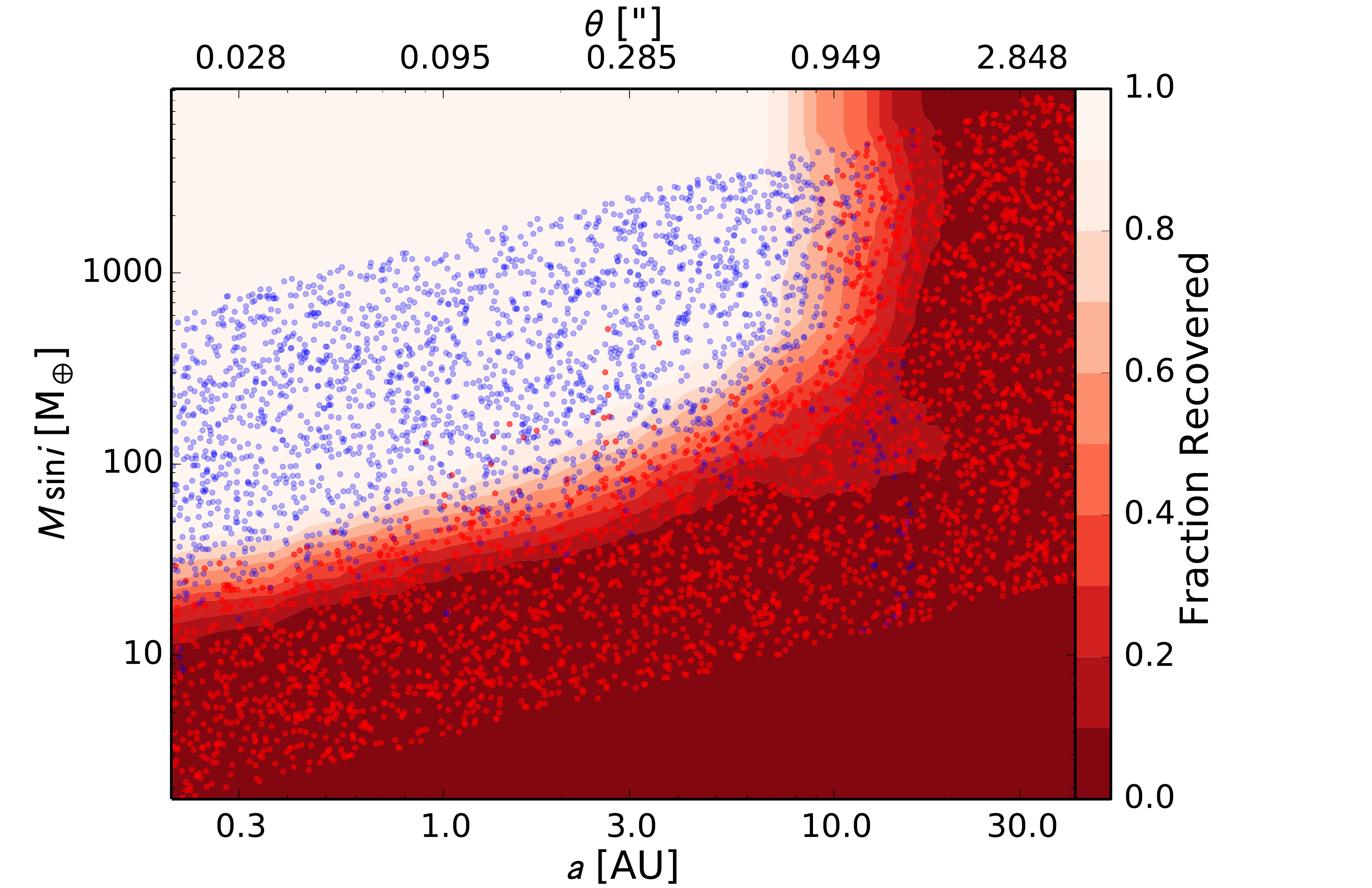}
\end{centering}
\caption{Results from an automated search for planets orbiting the star 
HD~19373 (HIP~14632; programs = S, C, A) 
based on RVs from Lick and/or Keck Observatory.
The set of plots on the left (analogous to Figures \ref{fig:search_example} and \ref{fig:search_example2}) 
show the planet search results 
and the plot on the right shows the completeness limits (analogous to Fig.\ \ref{fig:completeness_example}). 
See the captions of those figures for detailed descriptions.  
}
\label{fig:completeness_19373}
\end{figure}

\begin{figure}
\begin{centering}
\includegraphics[width=0.45\textwidth]{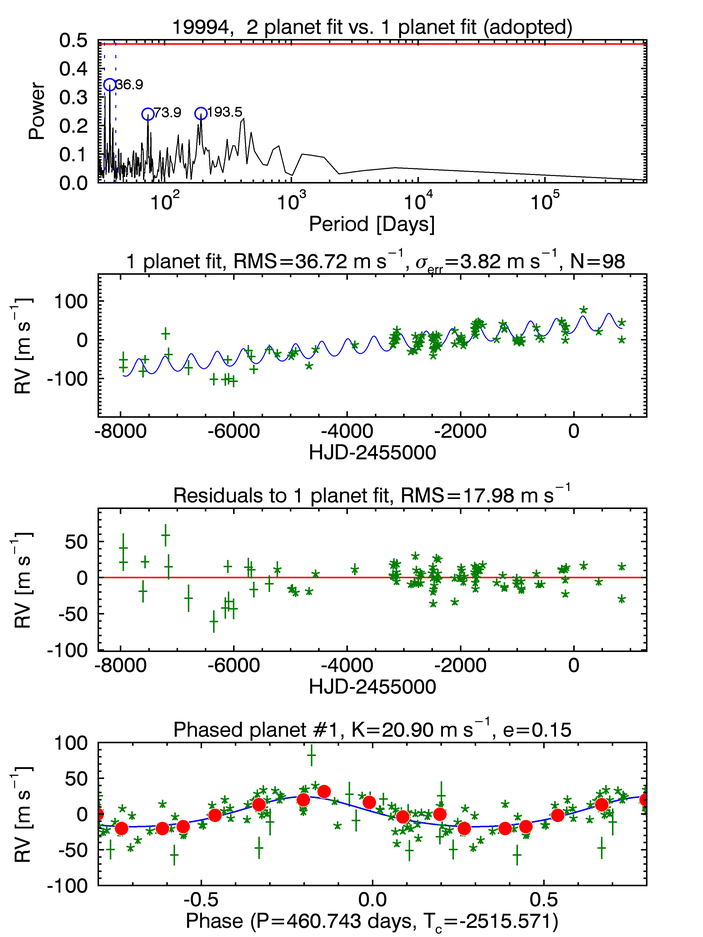}
\includegraphics[width=0.50\textwidth]{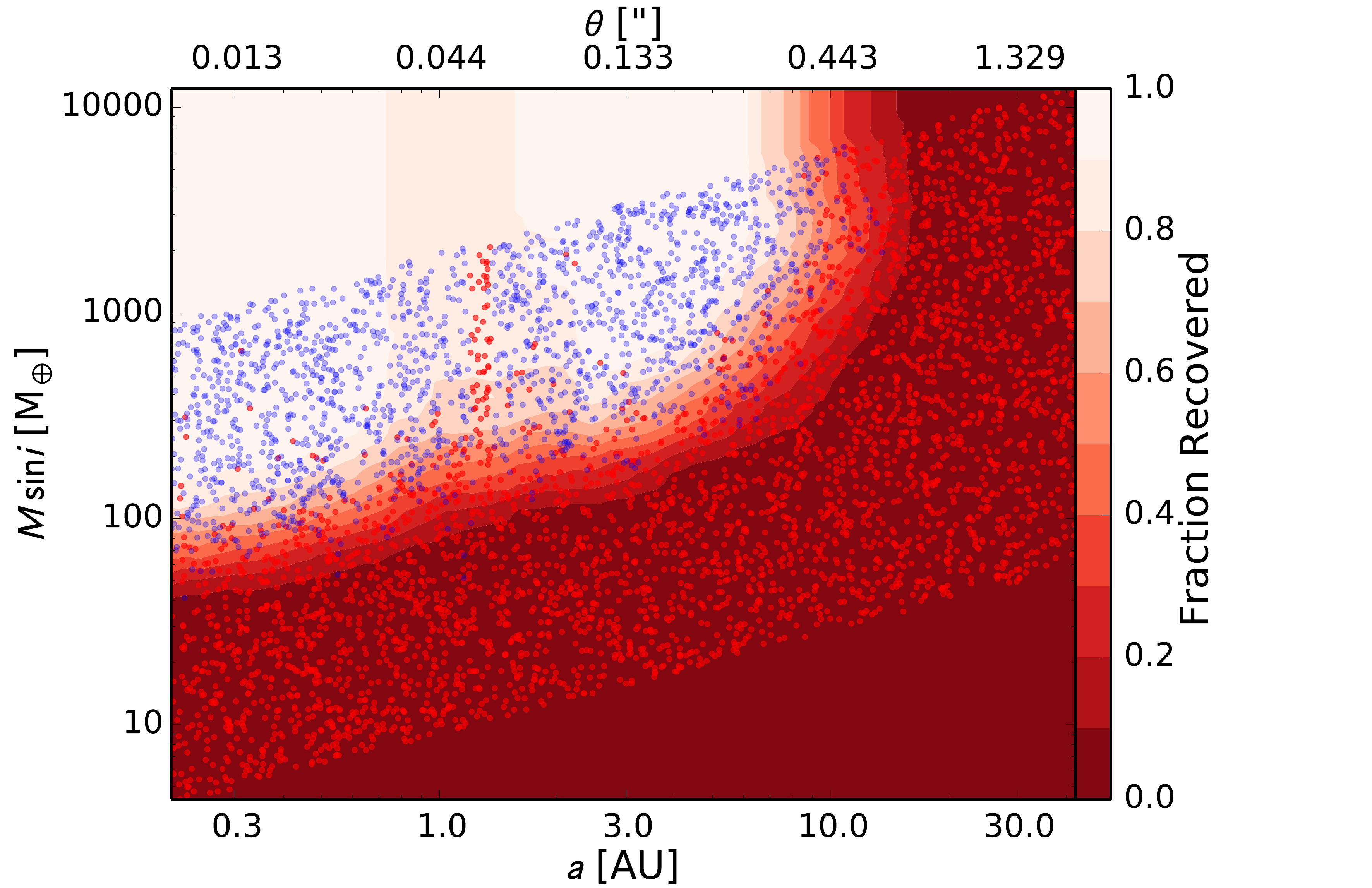}
\end{centering}
\caption{Results from an automated search for planets orbiting the star 
HD~19994 (HIP~14954; program = A) 
based on RVs from Lick and/or Keck Observatory.
The set of plots on the left (analogous to Figures \ref{fig:search_example} and \ref{fig:search_example2}) 
show the planet search results 
and the plot on the right shows the completeness limits (analogous to Fig.\ \ref{fig:completeness_example}). 
See the captions of those figures for detailed descriptions.  
This star has one known planet.  The pipeline also detects a linear trend in the RV time series.
}
\label{fig:completeness_19994}
\end{figure}
\clearpage

\begin{figure}
\begin{centering}
\includegraphics[width=0.45\textwidth]{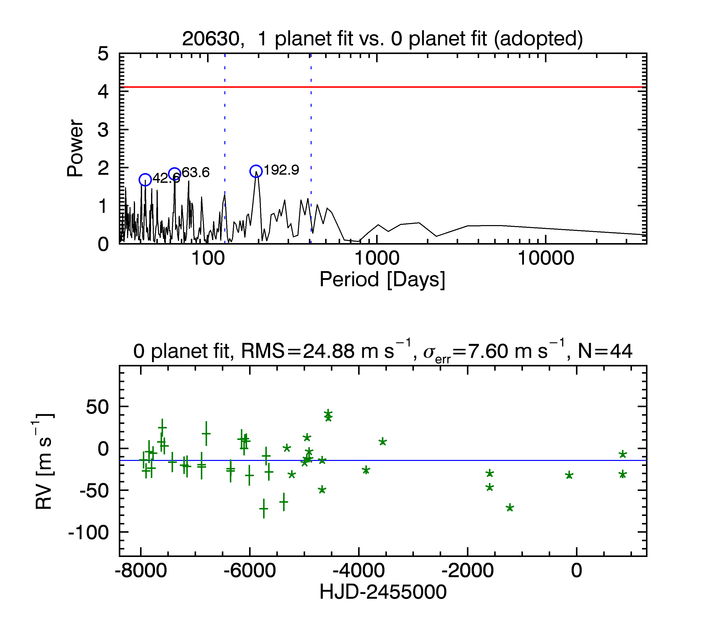}
\includegraphics[width=0.50\textwidth]{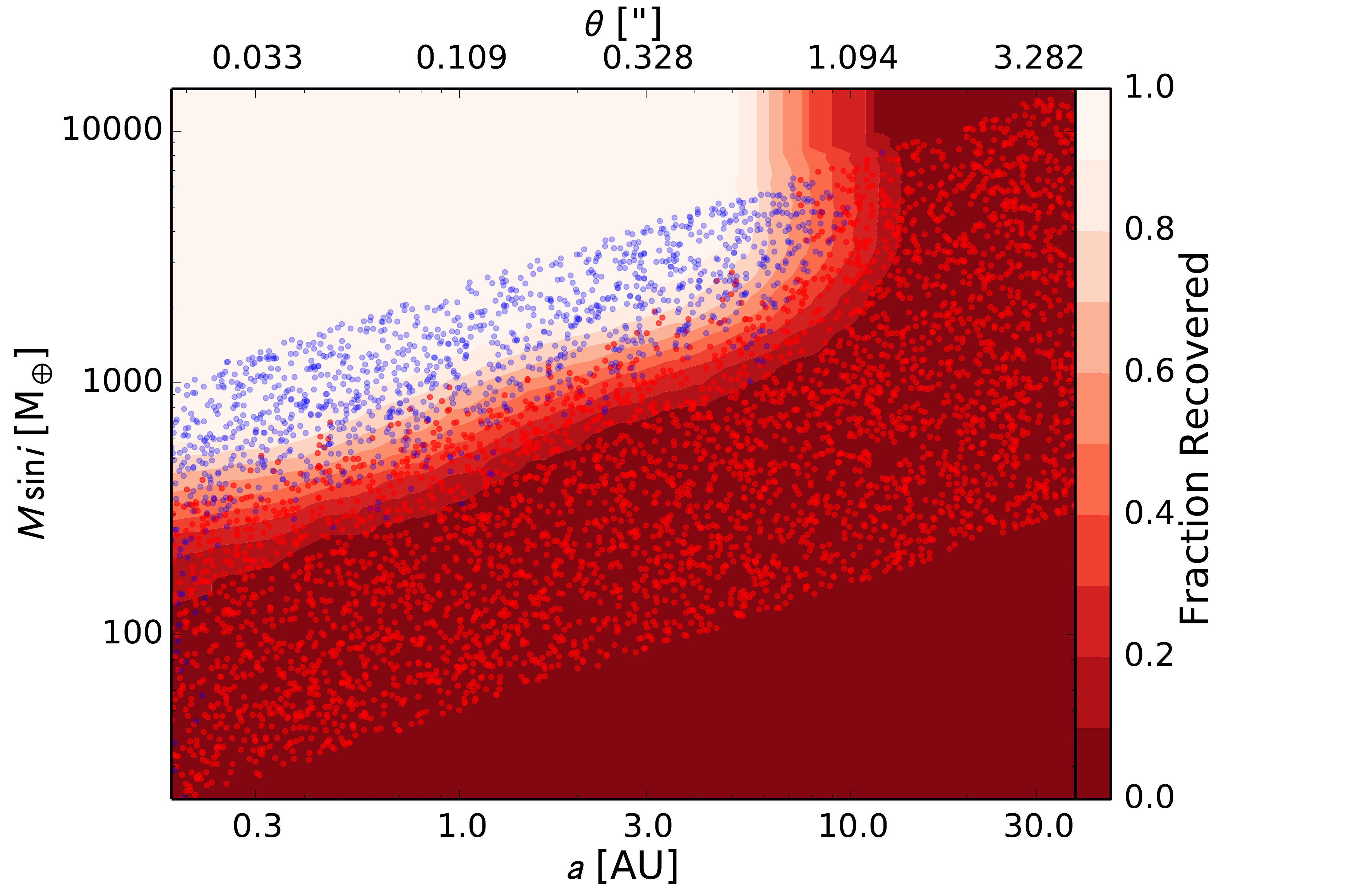}
\end{centering}
\caption{Results from an automated search for planets orbiting the star 
HD~20630 (HIP~15457; programs = S, C, A) 
based on RVs from Lick and/or Keck Observatory.
The set of plots on the left (analogous to Figures \ref{fig:search_example} and \ref{fig:search_example2}) 
show the planet search results 
and the plot on the right shows the completeness limits (analogous to Fig.\ \ref{fig:completeness_example}). 
See the captions of those figures for detailed descriptions.  
}
\label{fig:completeness_20630}
\end{figure}

\begin{figure}
\begin{centering}
\includegraphics[width=0.45\textwidth]{22049-finalfit.png}
\includegraphics[width=0.50\textwidth]{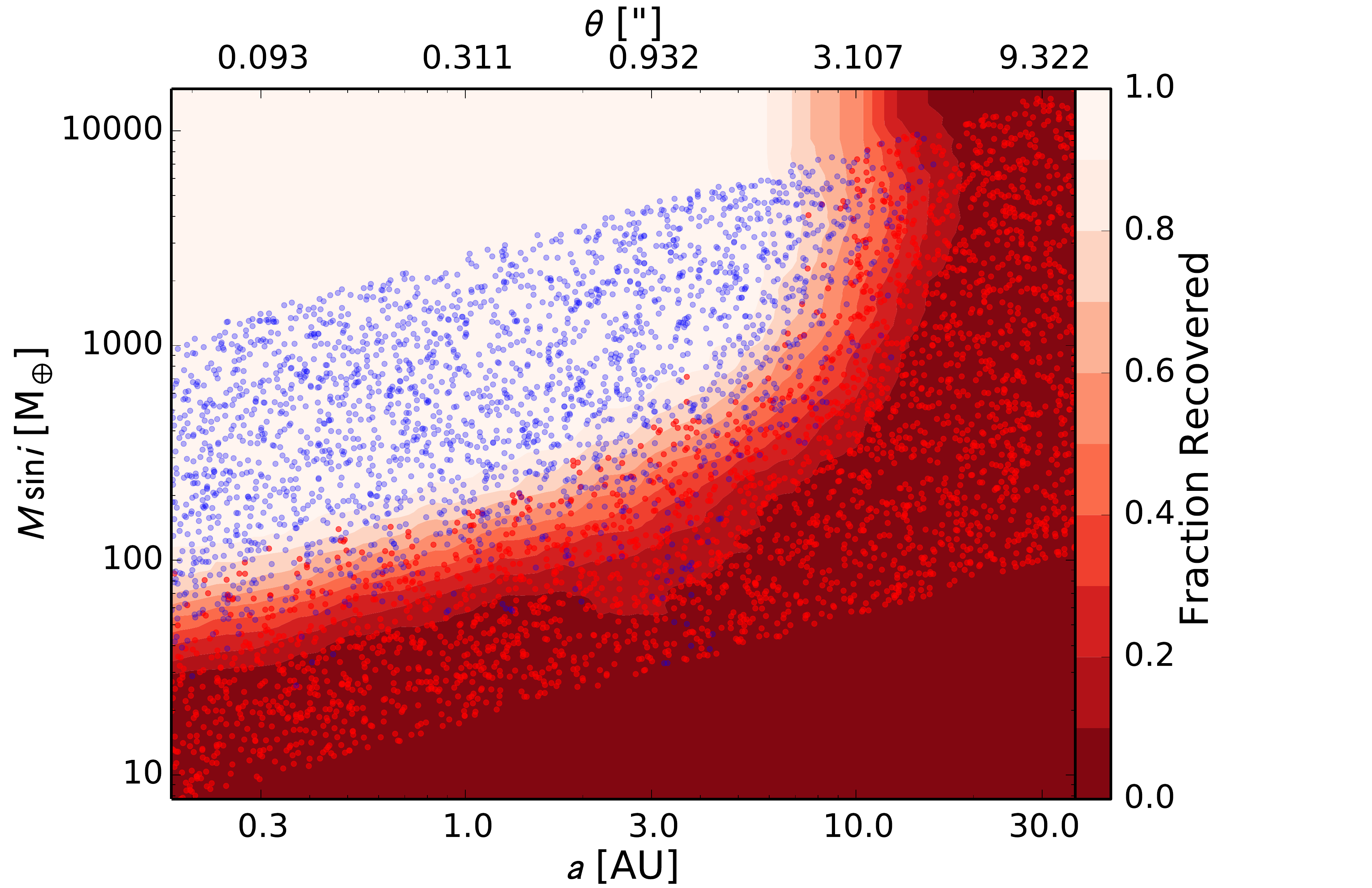}
\end{centering}
\caption{Results from an automated search for planets orbiting the star 
HD~22049 (HIP~16537; programs = C, A) 
based on RVs from Lick and/or Keck Observatory.
The set of plots on the left (analogous to Figures \ref{fig:search_example} and \ref{fig:search_example2}) 
show the planet search results 
and the plot on the right shows the completeness limits (analogous to Fig.\ \ref{fig:completeness_example}). 
See the captions of those figures for detailed descriptions.  
This star has one known, long-period planet, which we clearly detect in the Rv time series.  This claim has been controversial because the host star is young and active.  However, our measurements of the activity-sensitive \caii\ lines in Keck spectra do not correlate with the RVs.
}
\label{fig:completeness_22049}
\end{figure}
\clearpage

\begin{figure}
\begin{centering}
\includegraphics[width=0.45\textwidth]{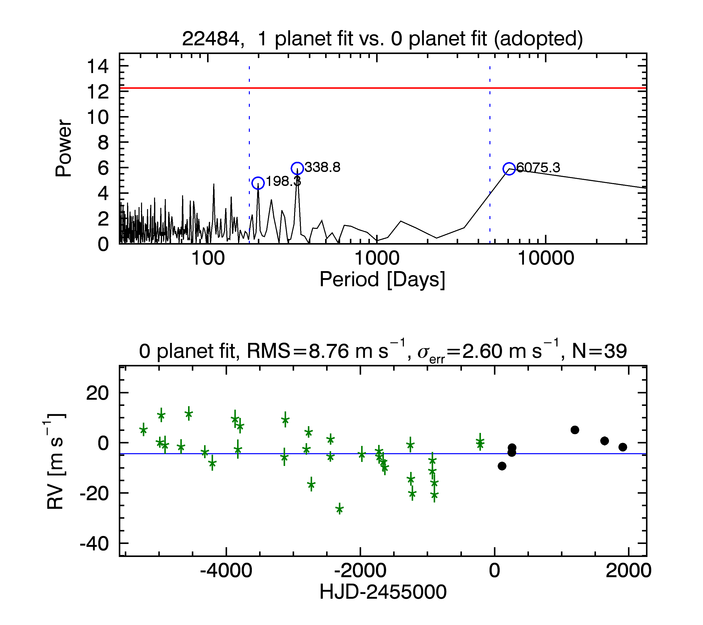}
\includegraphics[width=0.50\textwidth]{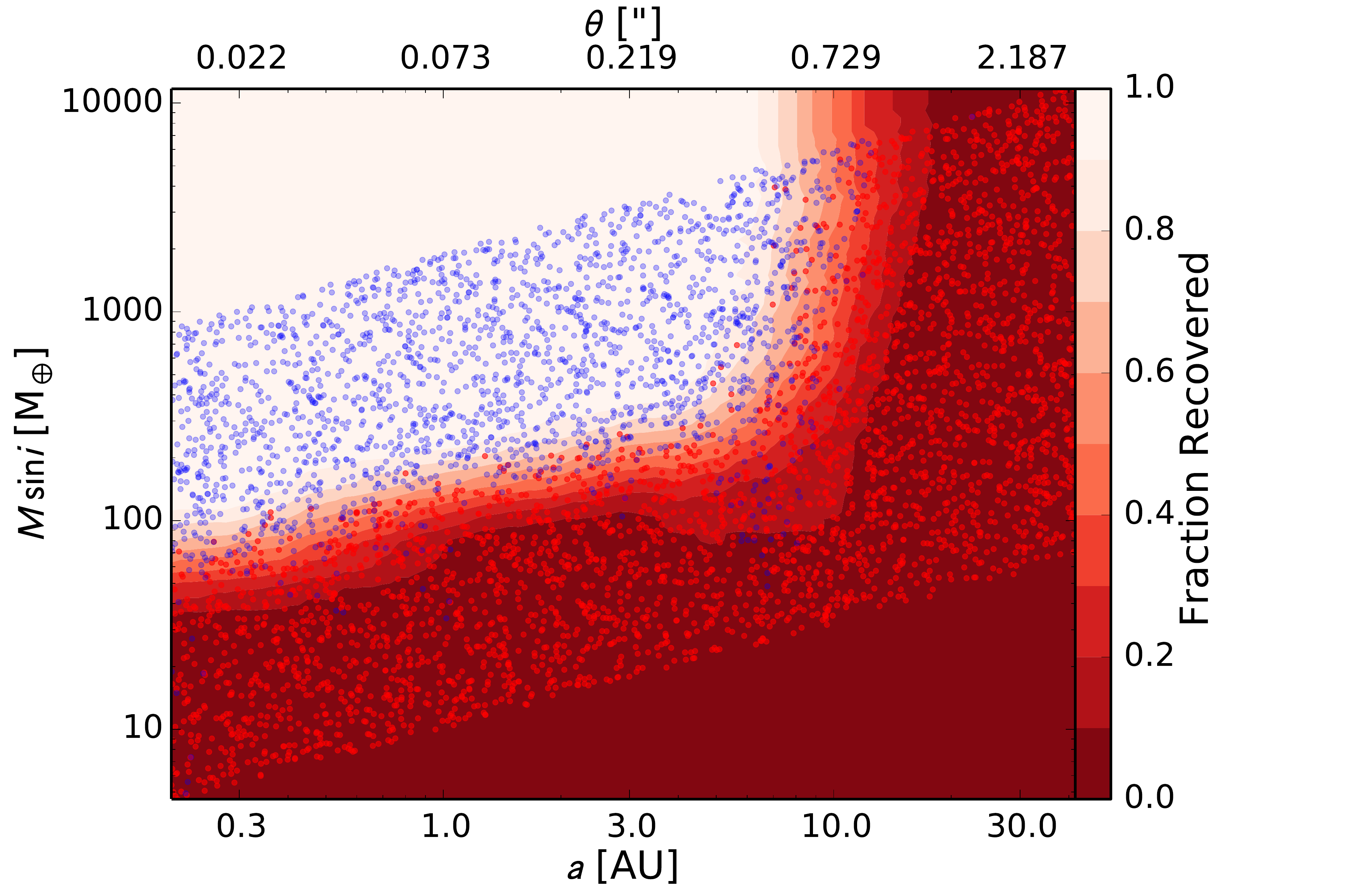}
\end{centering}
\caption{Results from an automated search for planets orbiting the star 
HD~22484 (HIP~16852; programs = S, C, A) 
based on RVs from Lick and/or Keck Observatory.
The set of plots on the left (analogous to Figures \ref{fig:search_example} and \ref{fig:search_example2}) 
show the planet search results 
and the plot on the right shows the completeness limits (analogous to Fig.\ \ref{fig:completeness_example}). 
See the captions of those figures for detailed descriptions.  
}
\label{fig:completeness_22484}
\end{figure}

\begin{figure}
\begin{centering}
\includegraphics[width=0.45\textwidth]{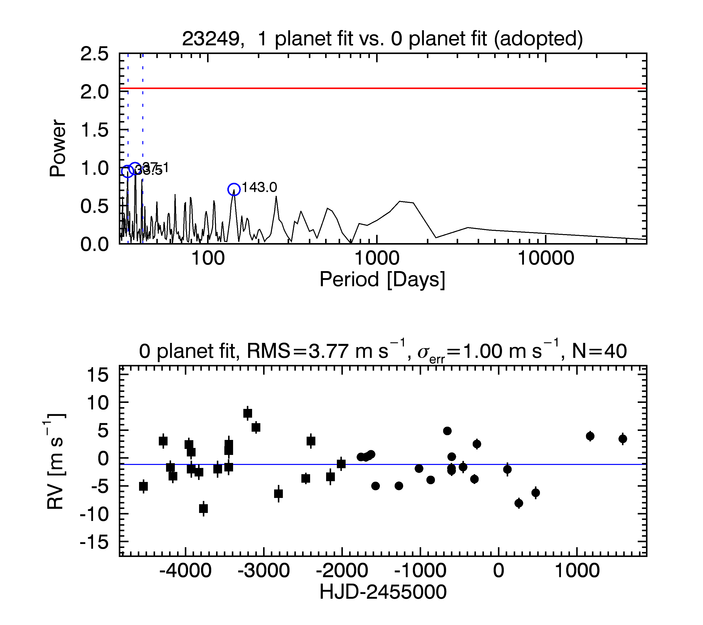}
\includegraphics[width=0.50\textwidth]{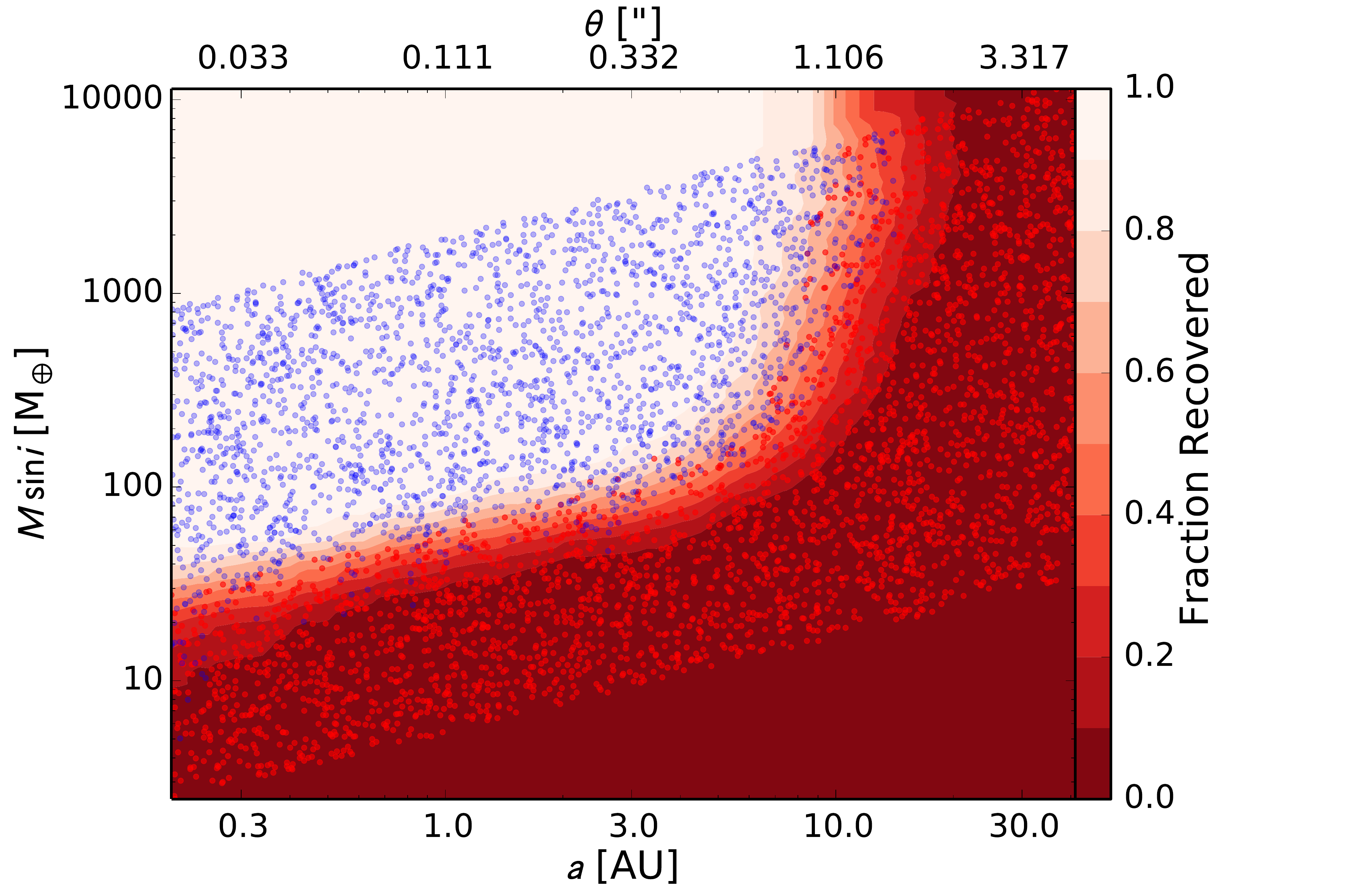}
\end{centering}
\caption{Results from an automated search for planets orbiting the star 
HD~23249 (HIP~17378; programs = S, C) 
based on RVs from Lick and/or Keck Observatory.
The set of plots on the left (analogous to Figures \ref{fig:search_example} and \ref{fig:search_example2}) 
show the planet search results 
and the plot on the right shows the completeness limits (analogous to Fig.\ \ref{fig:completeness_example}). 
See the captions of those figures for detailed descriptions.  
}
\label{fig:completeness_23249}
\end{figure}
\clearpage

\begin{figure}
\begin{centering}
\includegraphics[width=0.45\textwidth]{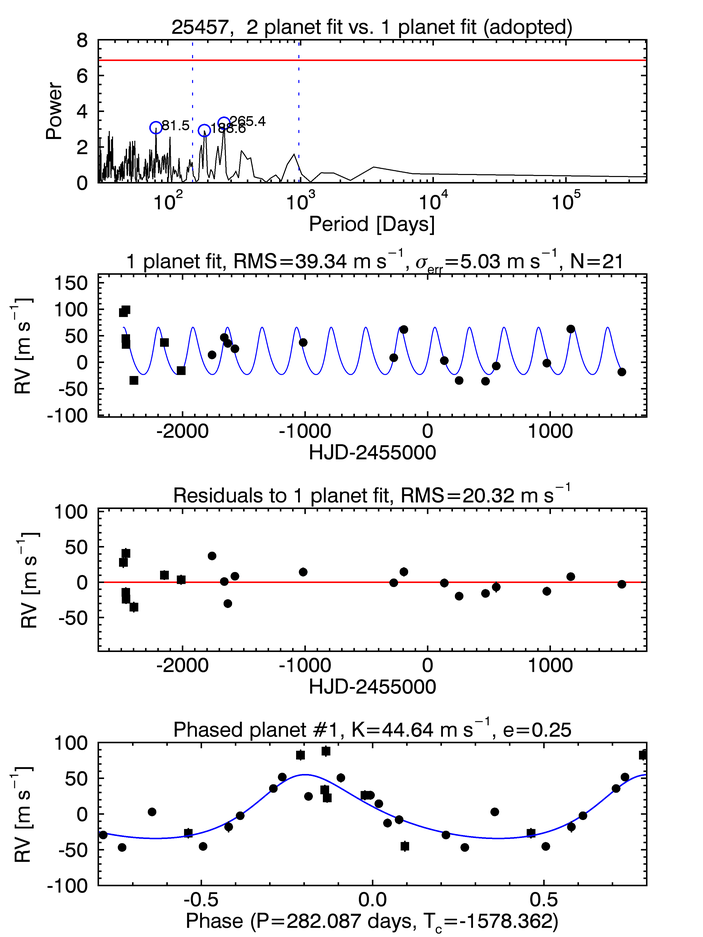}
\includegraphics[width=0.50\textwidth]{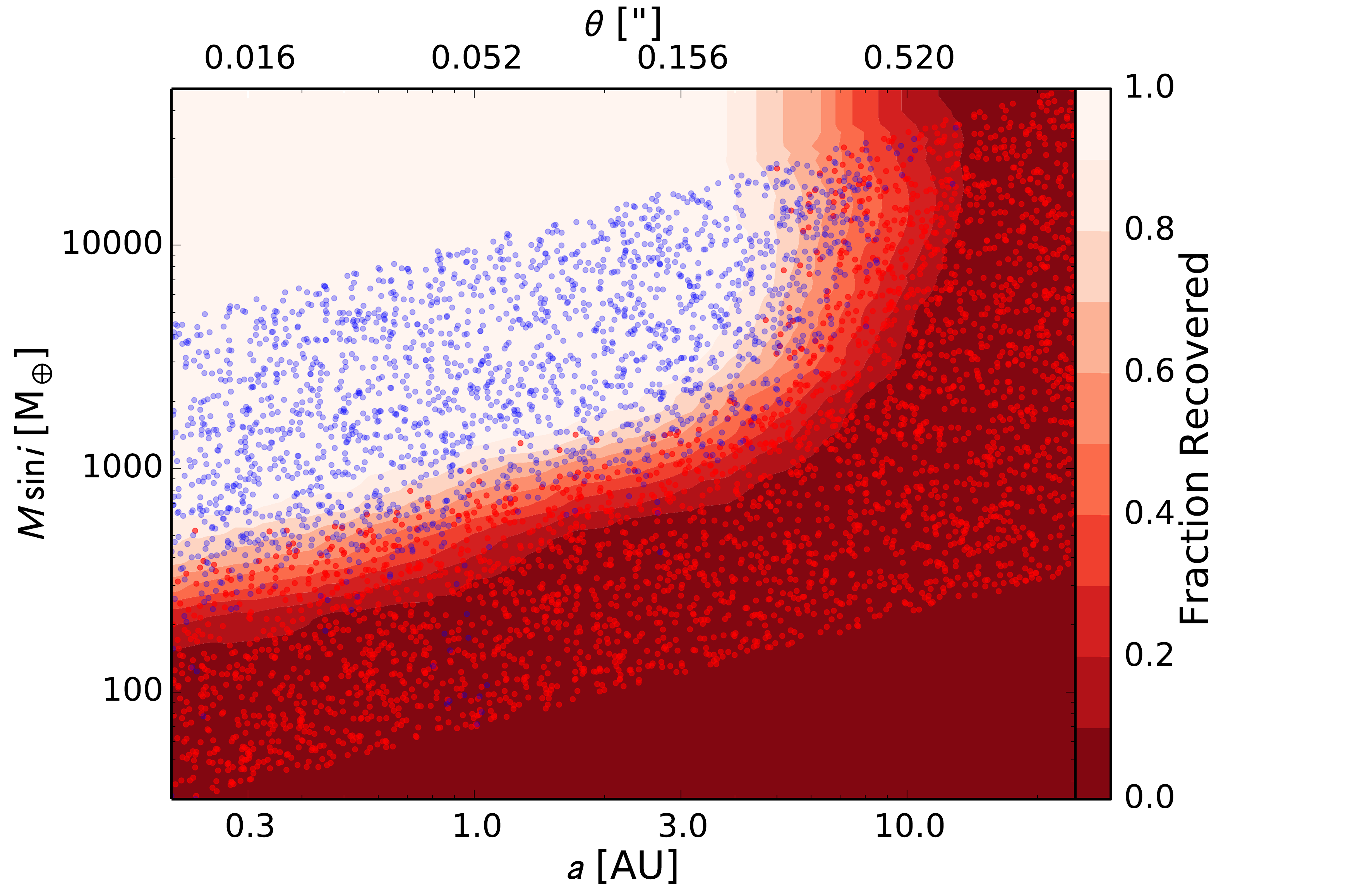}
\end{centering}
\caption{Results from an automated search for planets orbiting the star 
HD~25457 (HIP~18859; program = A) 
based on RVs from Lick and/or Keck Observatory.
The set of plots on the left (analogous to Figures \ref{fig:search_example} and \ref{fig:search_example2}) 
show the planet search results 
and the plot on the right shows the completeness limits (analogous to Fig.\ \ref{fig:completeness_example}). 
See the captions of those figures for detailed descriptions.  
The automated pipeline formally identifies a planet candidate based on 21 Keck RVs.  Given the high jitter and small number of observations, we deem this candidate not credible with the current data.
}
\label{fig:completeness_25457}
\end{figure}

\begin{figure}
\begin{centering}
\includegraphics[width=0.45\textwidth]{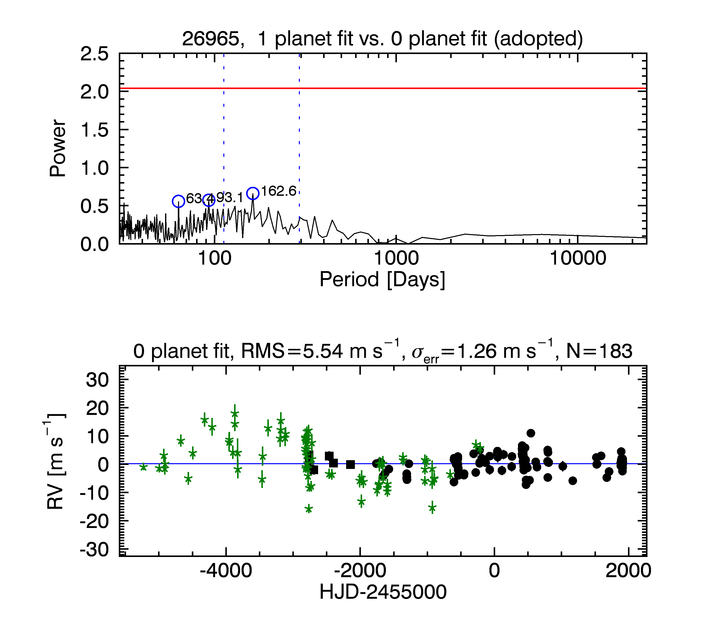}
\includegraphics[width=0.50\textwidth]{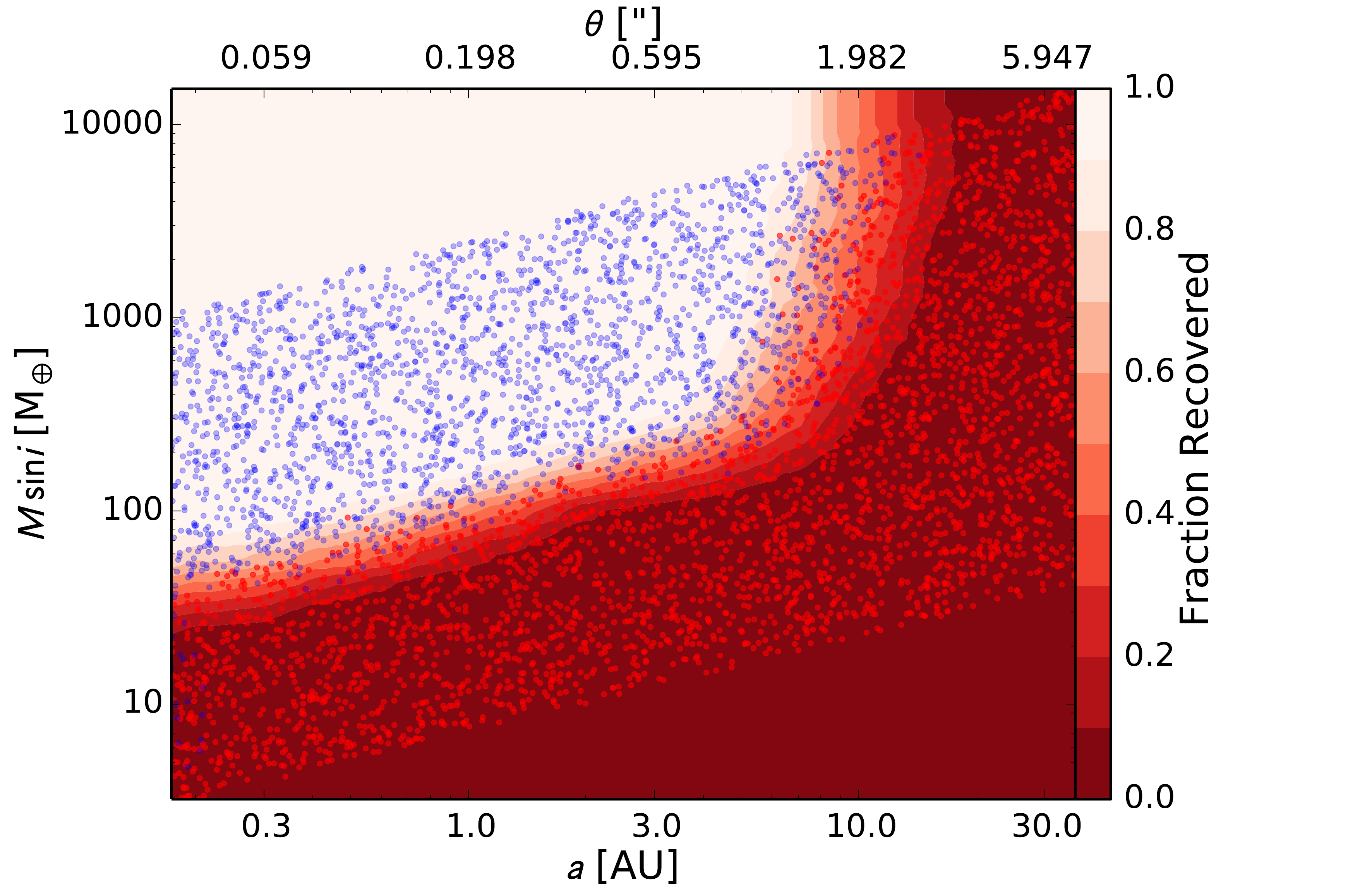}
\end{centering}
\caption{Results from an automated search for planets orbiting the star 
HD~26965 (HIP~19849; programs = S, C, A) 
based on RVs from Lick and/or Keck Observatory.
The set of plots on the left (analogous to Figures \ref{fig:search_example} and \ref{fig:search_example2}) 
show the planet search results 
and the plot on the right shows the completeness limits (analogous to Fig.\ \ref{fig:completeness_example}). 
See the captions of those figures for detailed descriptions.  
}
\label{fig:completeness_26965}
\end{figure}
\clearpage

\begin{figure}
\begin{centering}
\includegraphics[width=0.45\textwidth]{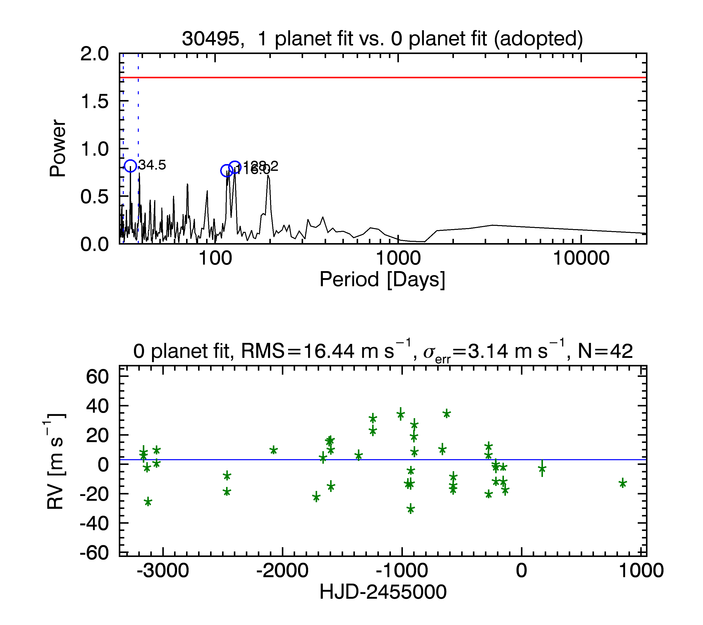}
\includegraphics[width=0.50\textwidth]{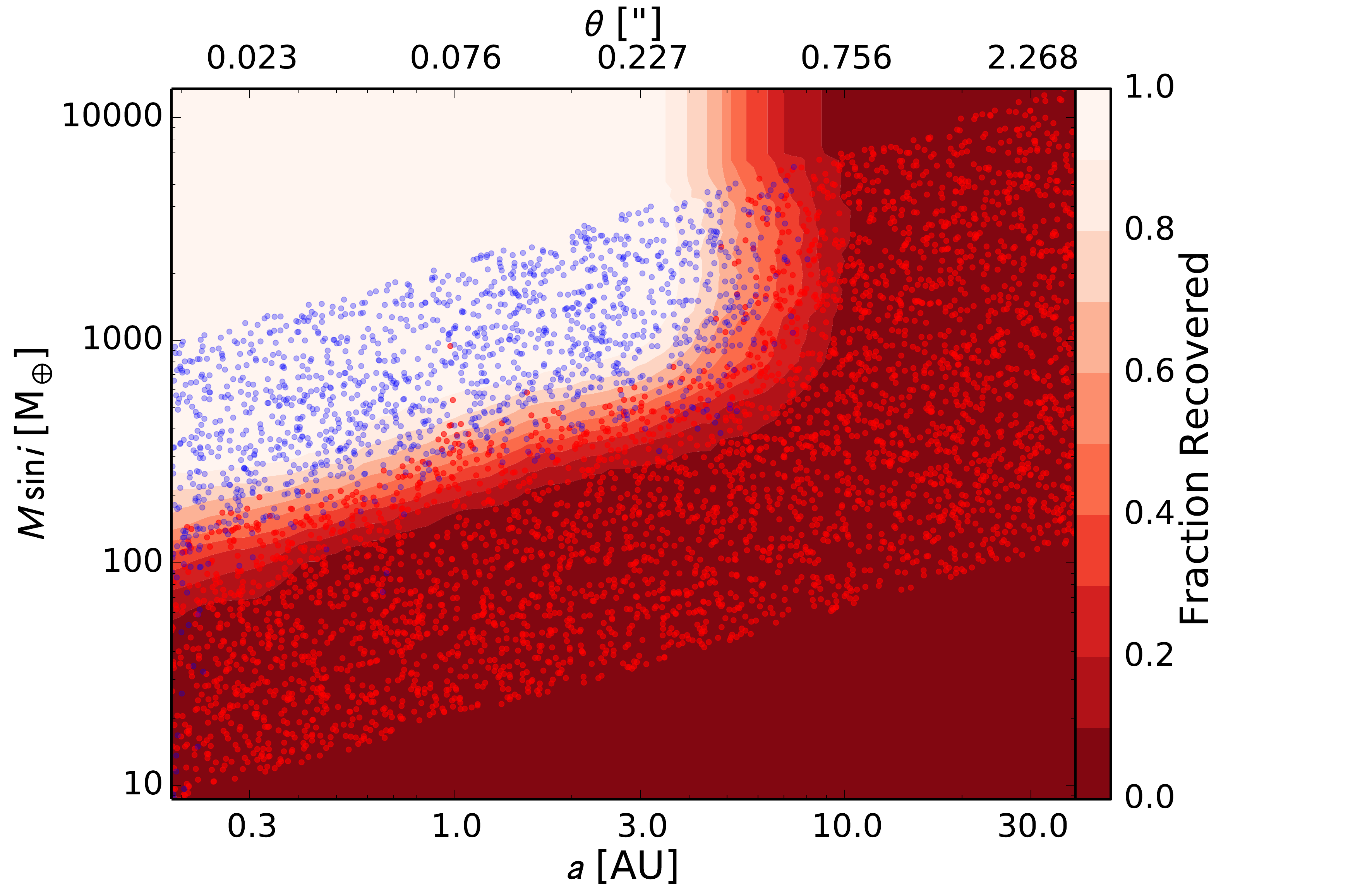}
\end{centering}
\caption{Results from an automated search for planets orbiting the star 
HD~30495 (HIP~22263; program = S) 
based on RVs from Lick and/or Keck Observatory.
The set of plots on the left (analogous to Figures \ref{fig:search_example} and \ref{fig:search_example2}) 
show the planet search results 
and the plot on the right shows the completeness limits (analogous to Fig.\ \ref{fig:completeness_example}). 
See the captions of those figures for detailed descriptions.  
}
\label{fig:completeness_30495}
\end{figure}

\begin{figure}
\begin{centering}
\includegraphics[width=0.45\textwidth]{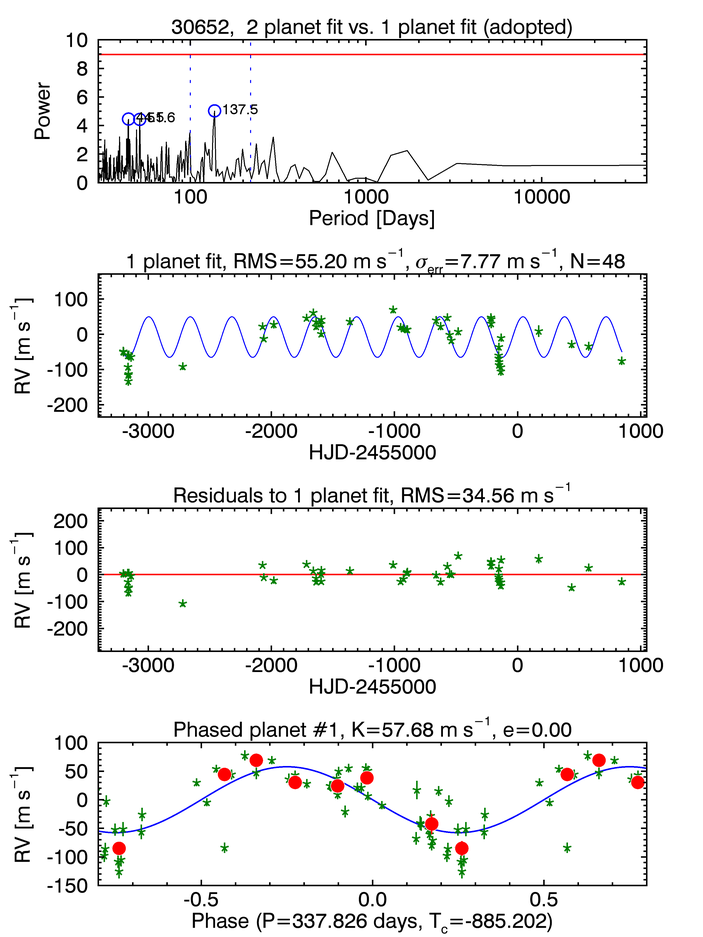}
\includegraphics[width=0.50\textwidth]{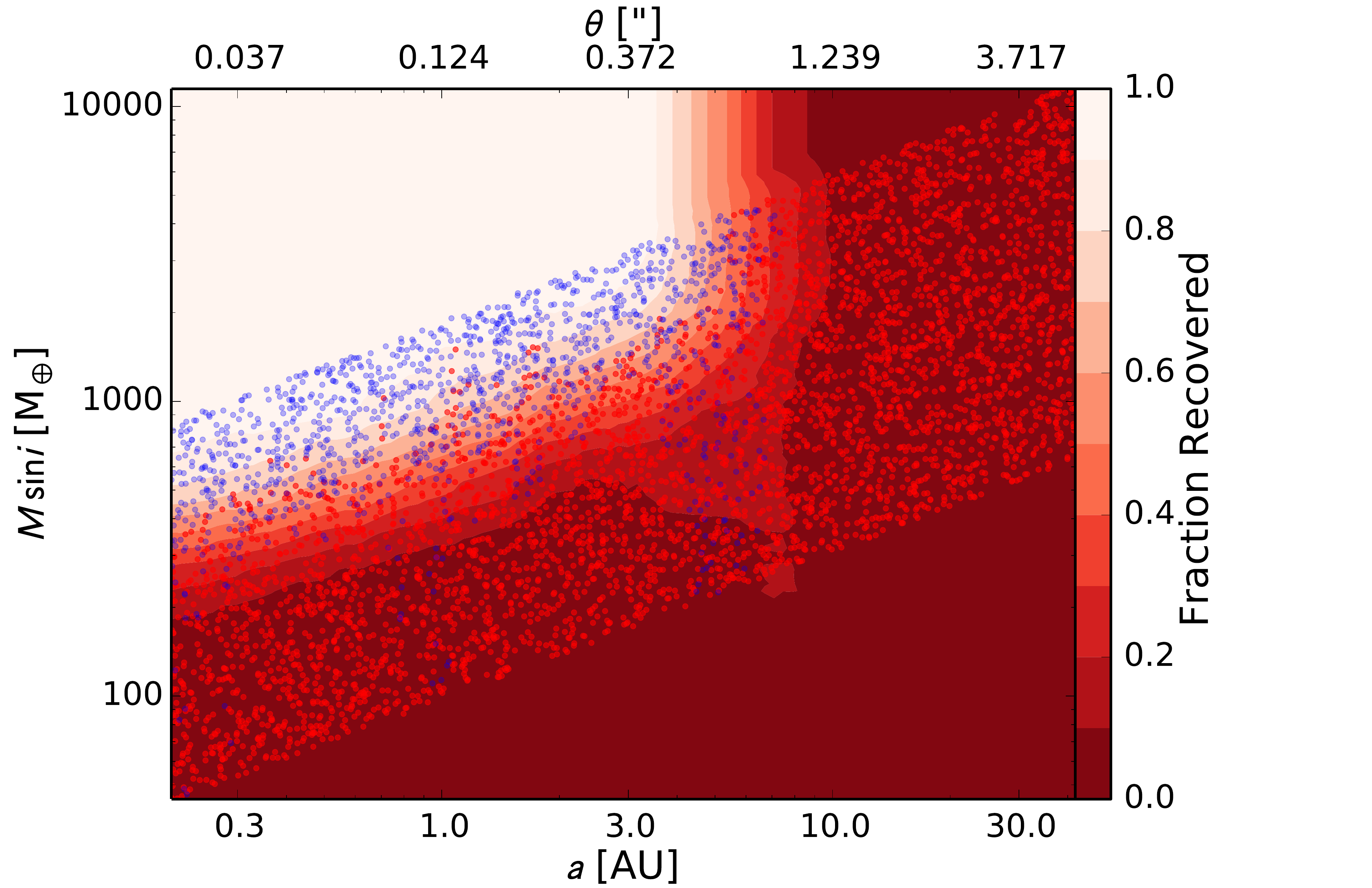}
\end{centering}
\caption{Results from an automated search for planets orbiting the star 
HD~30652 (HIP~22449; programs = S, C, A) 
based on RVs from Lick and/or Keck Observatory.
The set of plots on the left (analogous to Figures \ref{fig:search_example} and \ref{fig:search_example2}) 
show the planet search results 
and the plot on the right shows the completeness limits (analogous to Fig.\ \ref{fig:completeness_example}). 
See the captions of those figures for detailed descriptions.  
We detect a marginally significant periodic signal with a period of 338 days.  However, the poor observing history and the proximity of the period to one year, we conclude that this signal is not caused by a real planetary companion.  The nature of the signal is most likely non-astrophysical in nature.
}
\label{fig:completeness_30652}
\end{figure}
\clearpage

\begin{figure}
\begin{centering}
\includegraphics[width=0.45\textwidth]{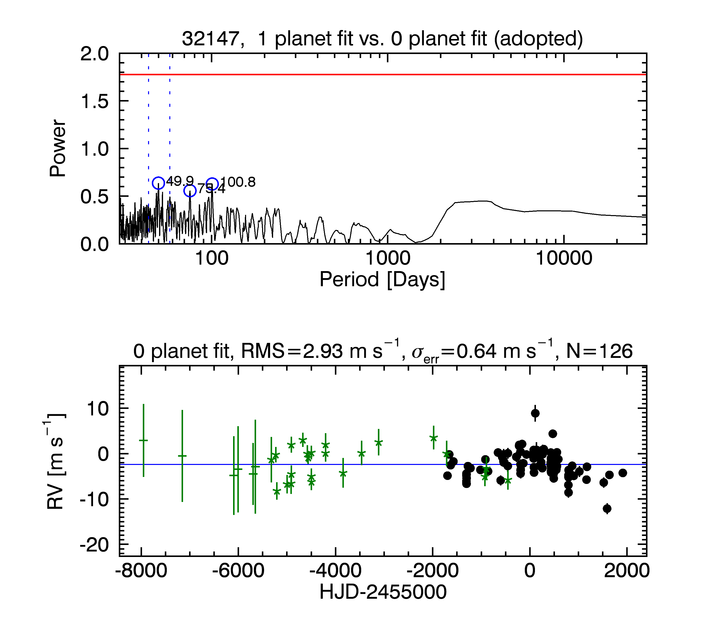}
\includegraphics[width=0.50\textwidth]{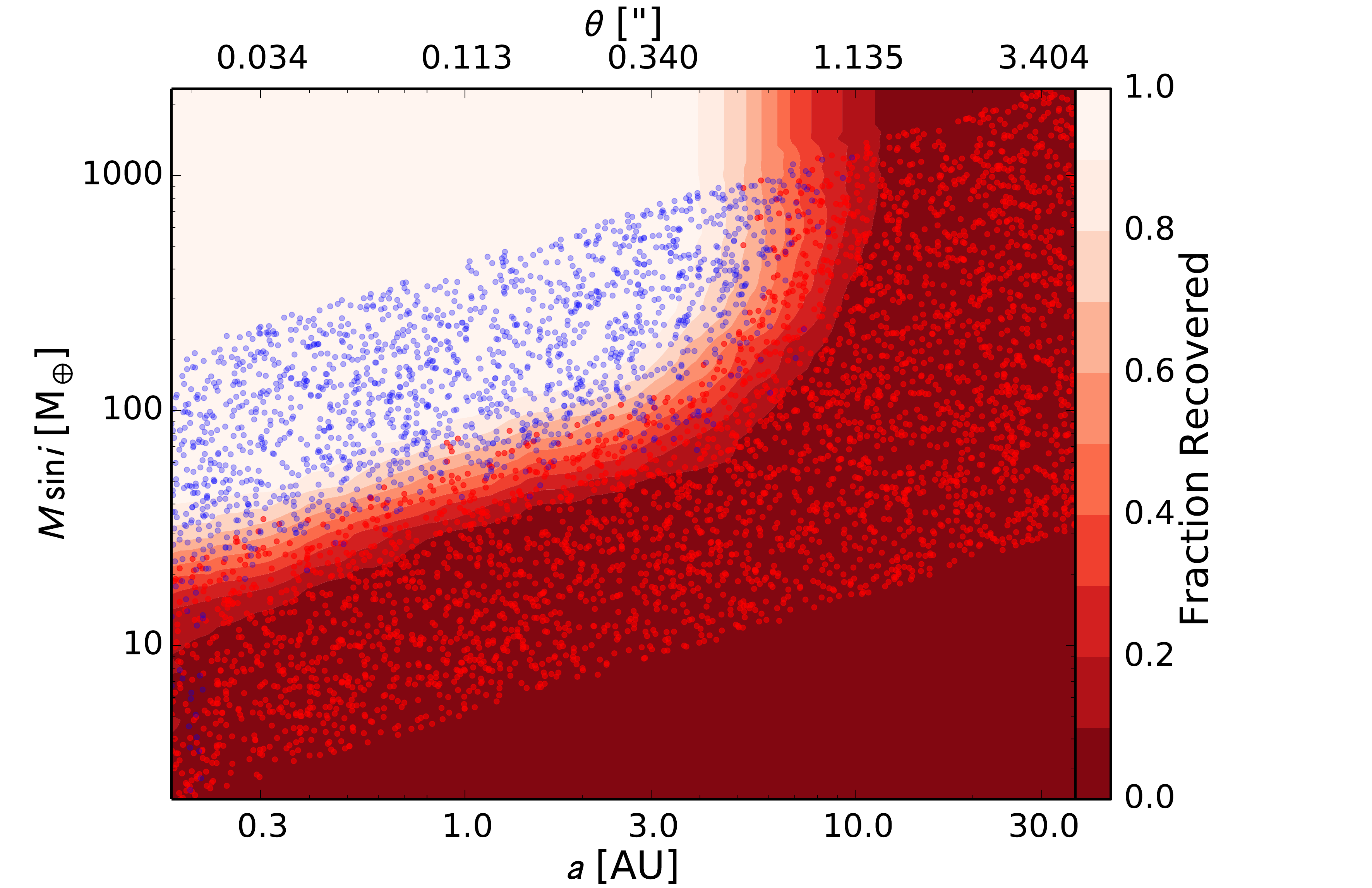}
\end{centering}
\caption{Results from an automated search for planets orbiting the star 
HD~32147 (HIP~23311; program = S) 
based on RVs from Lick and/or Keck Observatory.
The set of plots on the left (analogous to Figures \ref{fig:search_example} and \ref{fig:search_example2}) 
show the planet search results 
and the plot on the right shows the completeness limits (analogous to Fig.\ \ref{fig:completeness_example}). 
See the captions of those figures for detailed descriptions.  
}
\label{fig:completeness_32147}
\end{figure}

\begin{figure}
\begin{centering}
\includegraphics[width=0.45\textwidth]{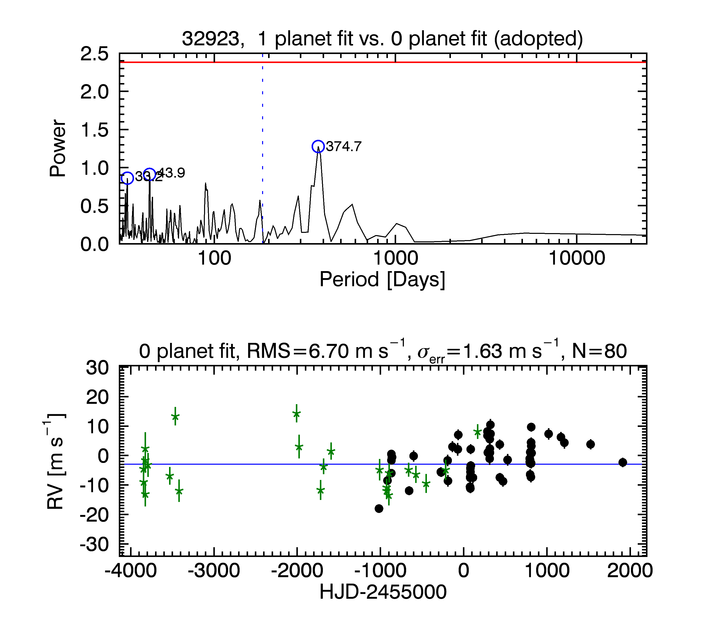}
\includegraphics[width=0.50\textwidth]{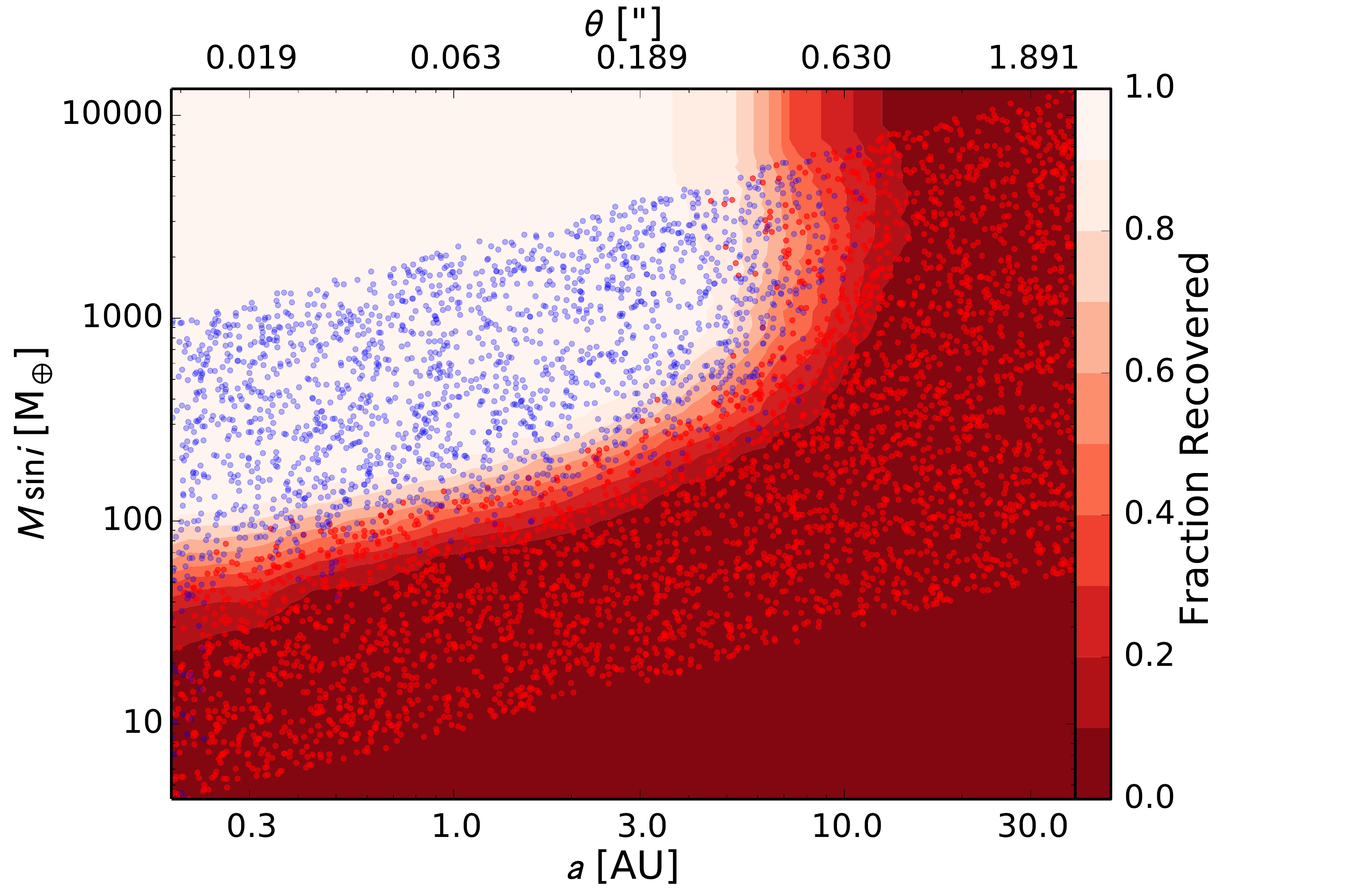}
\end{centering}
\caption{Results from an automated search for planets orbiting the star 
HD~32923 (HIP~23835; programs = S, C, A) 
based on RVs from Lick and/or Keck Observatory.
The set of plots on the left (analogous to Figures \ref{fig:search_example} and \ref{fig:search_example2}) 
show the planet search results 
and the plot on the right shows the completeness limits (analogous to Fig.\ \ref{fig:completeness_example}). 
See the captions of those figures for detailed descriptions.  
}
\label{fig:completeness_32923}
\end{figure}
\clearpage

\begin{figure}
\begin{centering}
\includegraphics[width=0.45\textwidth]{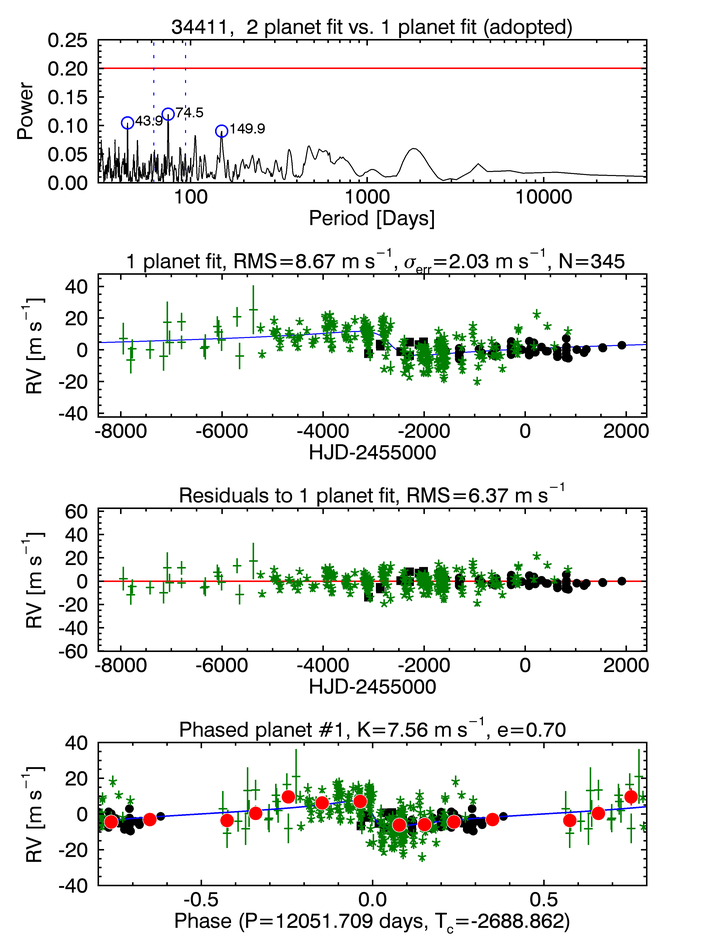}
\includegraphics[width=0.50\textwidth]{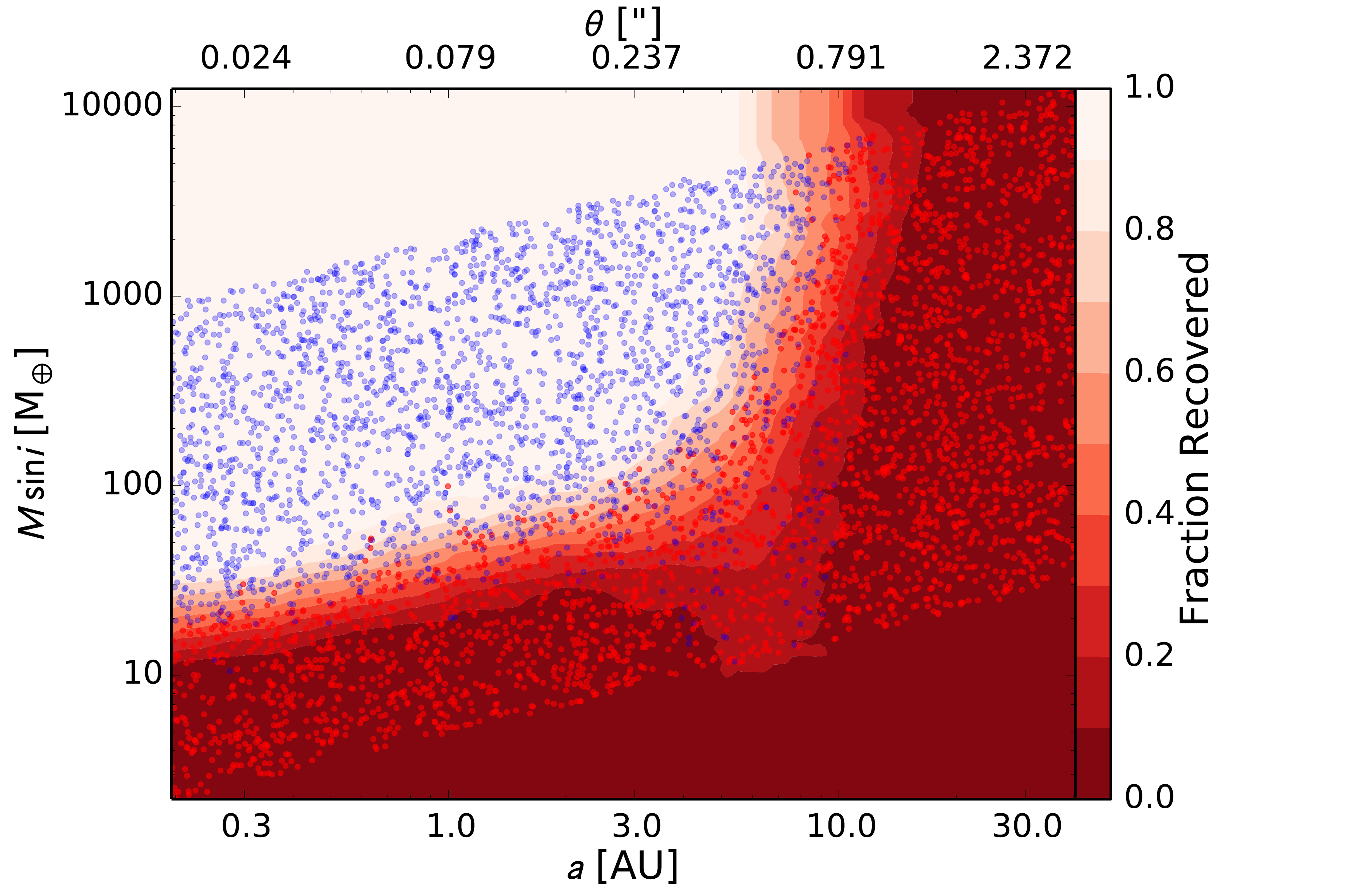}
\end{centering}
\caption{Results from an automated search for planets orbiting the star 
HD~34411 (HIP~24813; programs = S, C, A) 
based on RVs from Lick and/or Keck Observatory.
The set of plots on the left (analogous to Figures \ref{fig:search_example} and \ref{fig:search_example2}) 
show the planet search results 
and the plot on the right shows the completeness limits (analogous to Fig.\ \ref{fig:completeness_example}). 
See the captions of those figures for detailed descriptions.  
The automated pipeline picks up a long-period, high-eccentricity signal that is likely due to poorly constrained offsets between instruments.
}
\label{fig:completeness_34411}
\end{figure}

\begin{figure}
\begin{centering}
\includegraphics[width=0.45\textwidth]{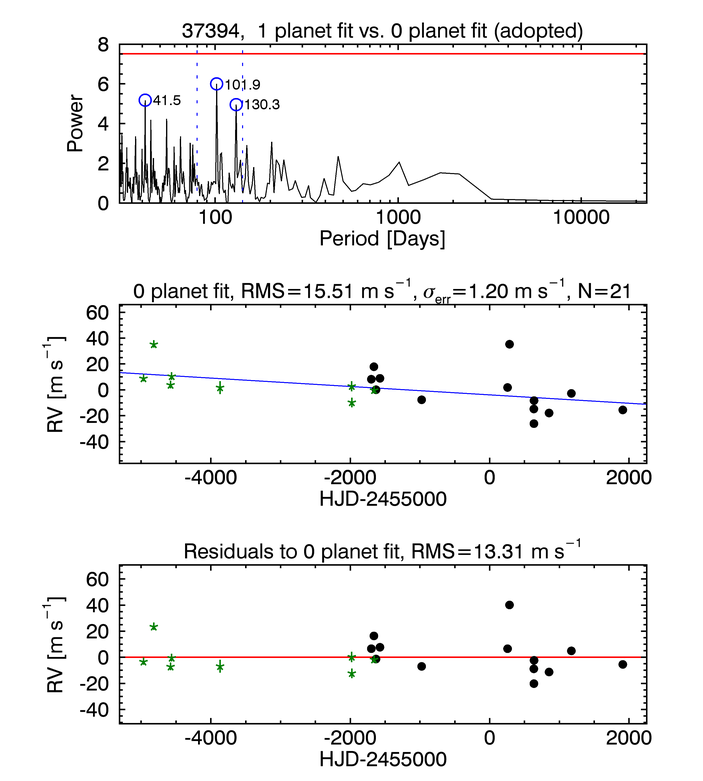}
\includegraphics[width=0.50\textwidth]{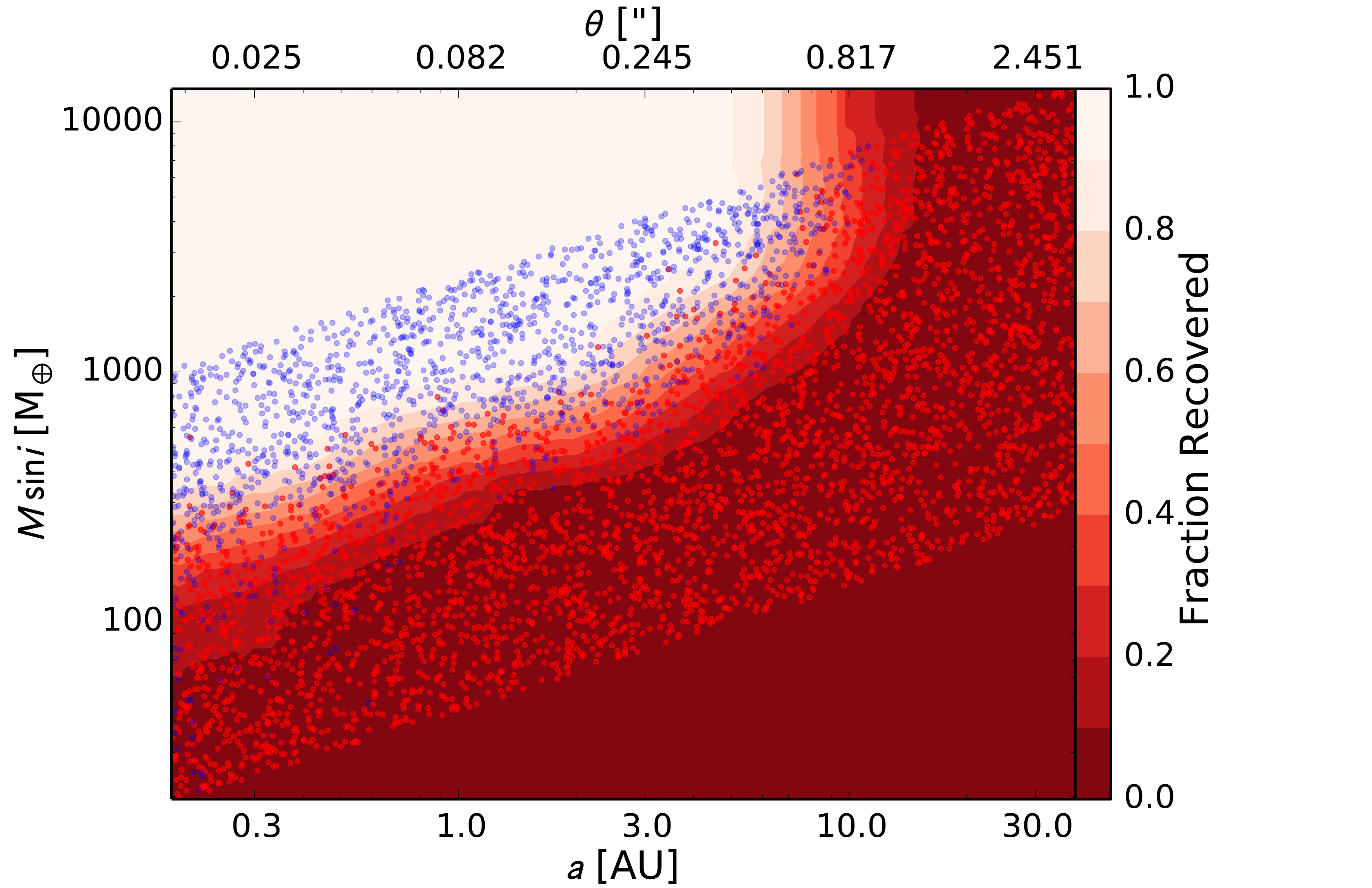}
\end{centering}
\caption{Results from an automated search for planets orbiting the star 
HD~37394 (HIP~26779; program = S) 
based on RVs from Lick and/or Keck Observatory.
The set of plots on the left (analogous to Figures \ref{fig:search_example} and \ref{fig:search_example2}) 
show the planet search results 
and the plot on the right shows the completeness limits (analogous to Fig.\ \ref{fig:completeness_example}). 
See the captions of those figures for detailed descriptions.  
This young, active star shows a small, marginally significant linear trend.
}
\label{fig:completeness_37394}
\end{figure}
\clearpage

\begin{figure}
\begin{centering}
\includegraphics[width=0.45\textwidth]{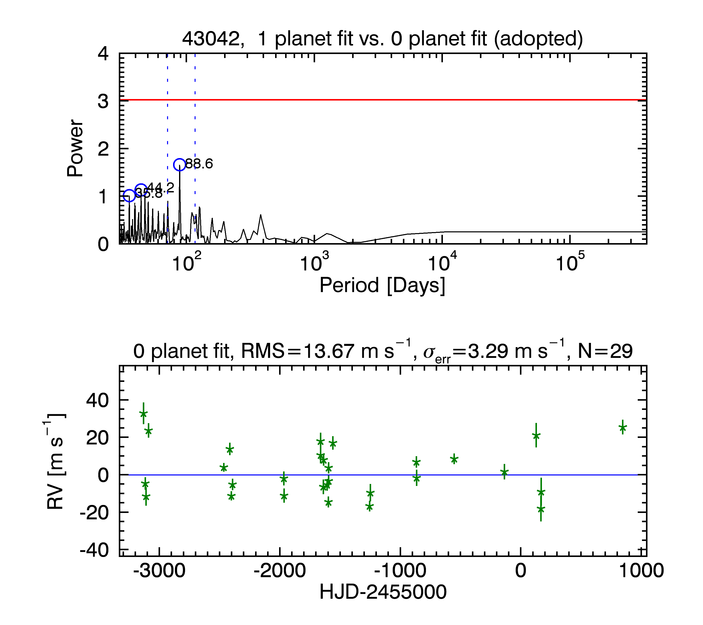}
\includegraphics[width=0.50\textwidth]{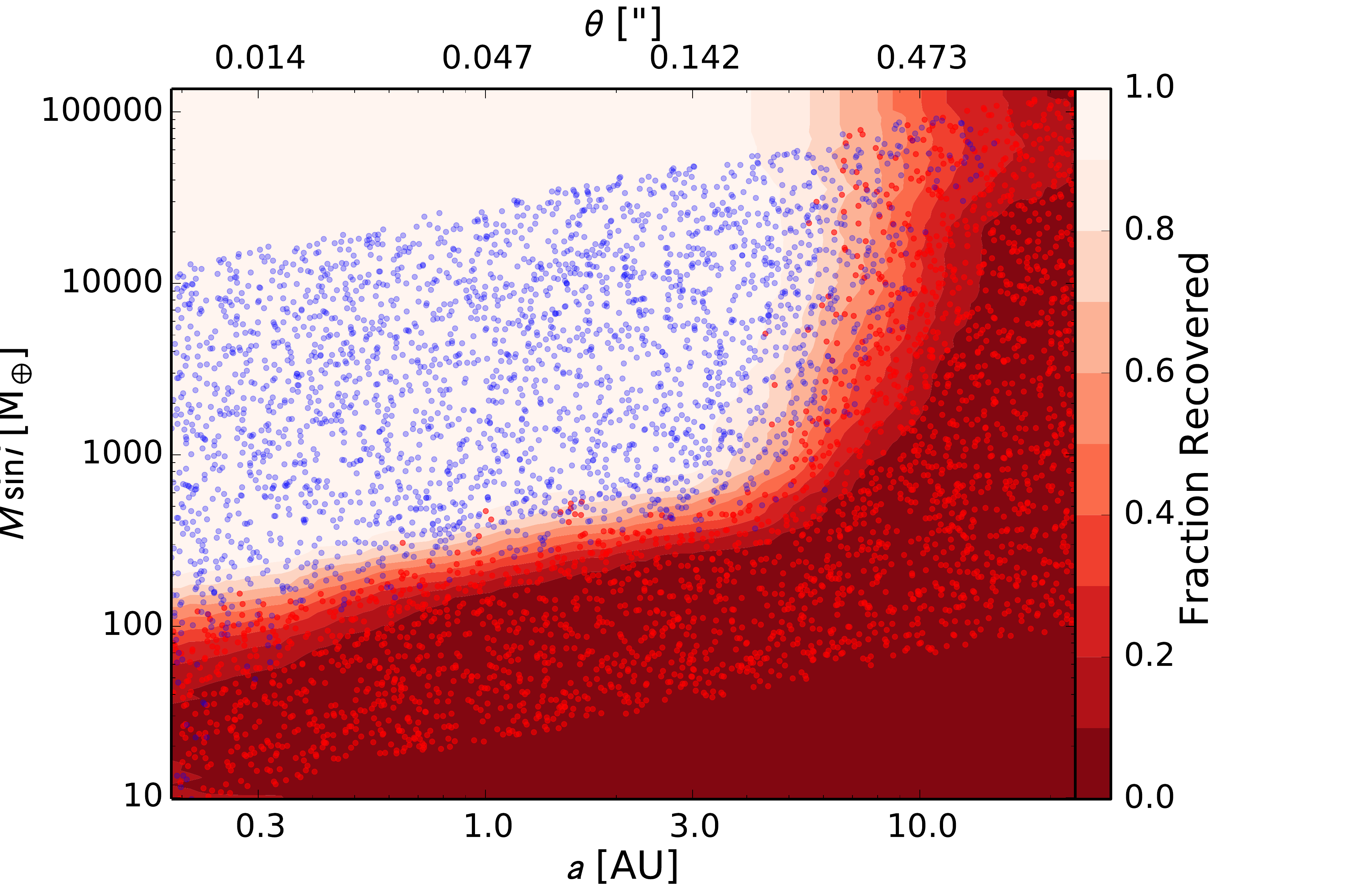}
\end{centering}
\caption{Results from an automated search for planets orbiting the star 
HD~43042 (HIP~29650; program = A) 
based on RVs from Lick and/or Keck Observatory.
The set of plots on the left (analogous to Figures \ref{fig:search_example} and \ref{fig:search_example2}) 
show the planet search results 
and the plot on the right shows the completeness limits (analogous to Fig.\ \ref{fig:completeness_example}). 
See the captions of those figures for detailed descriptions.  
}
\label{fig:completeness_43042}
\end{figure}

\begin{figure}
\begin{centering}
\includegraphics[width=0.45\textwidth]{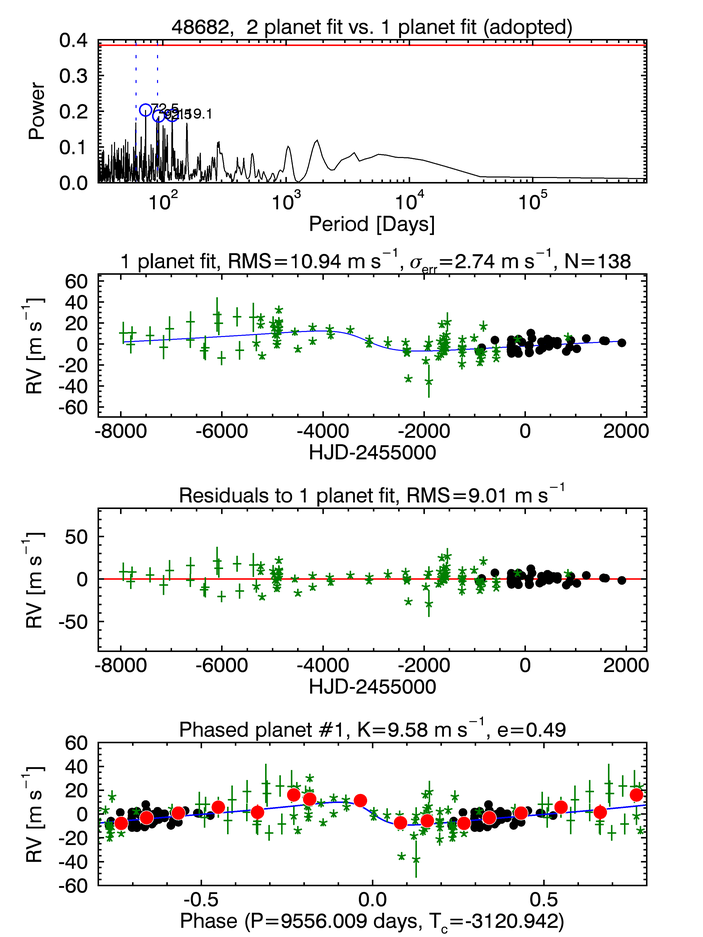}
\includegraphics[width=0.50\textwidth]{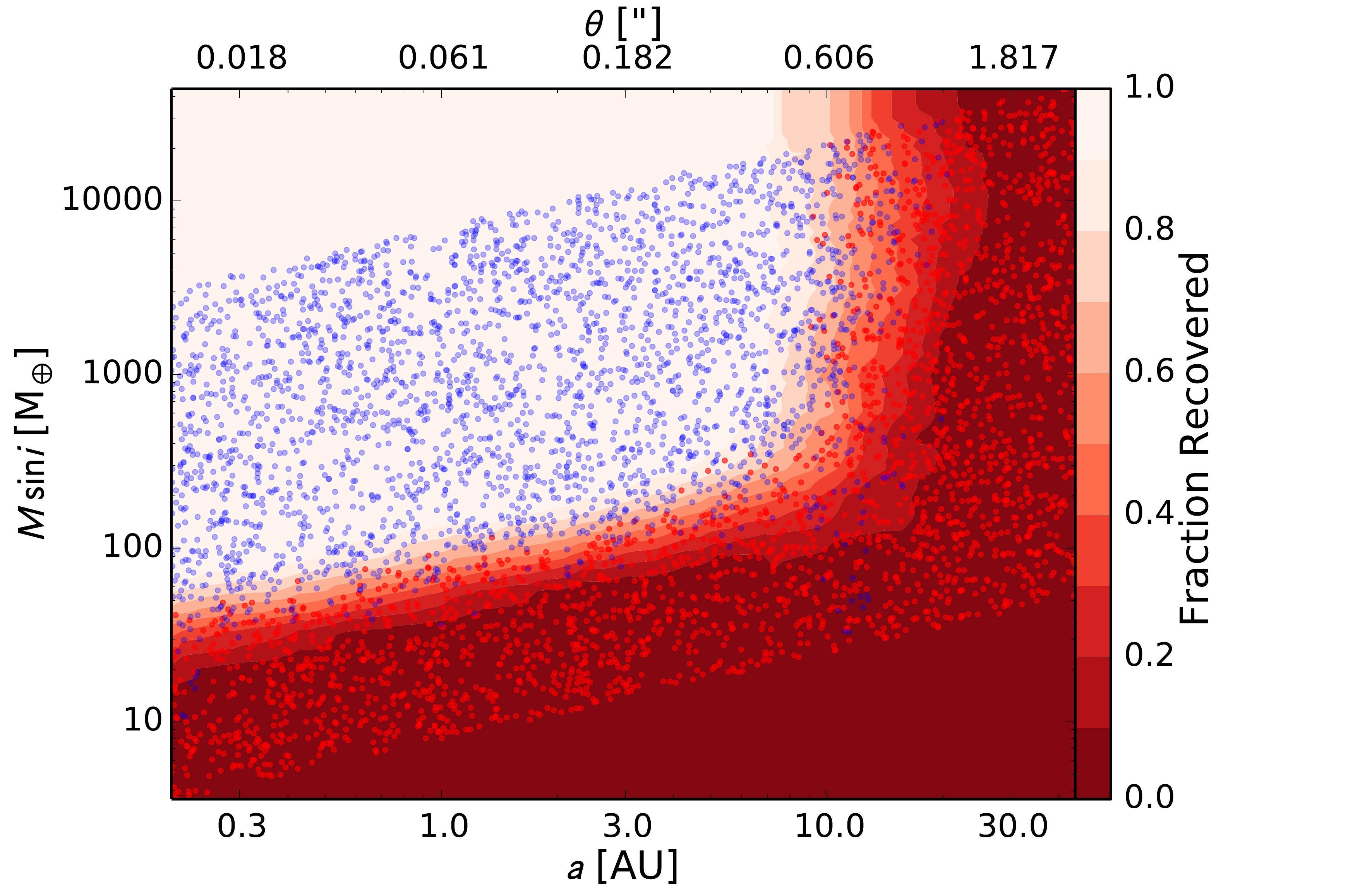}
\end{centering}
\caption{Results from an automated search for planets orbiting the star 
HD~48682 (HIP~32480; program = A) 
based on RVs from Lick and/or Keck Observatory.
The set of plots on the left (analogous to Figures \ref{fig:search_example} and \ref{fig:search_example2}) 
show the planet search results 
and the plot on the right shows the completeness limits (analogous to Fig.\ \ref{fig:completeness_example}). 
See the captions of those figures for detailed descriptions.  
This star has a formally adopted signal that appears to be due to uncorrected zero-point offsets in the Lick RVs and not due a planet.
}
\label{fig:completeness_48682}
\end{figure}
\clearpage

\begin{figure}
\begin{centering}
\includegraphics[width=0.45\textwidth]{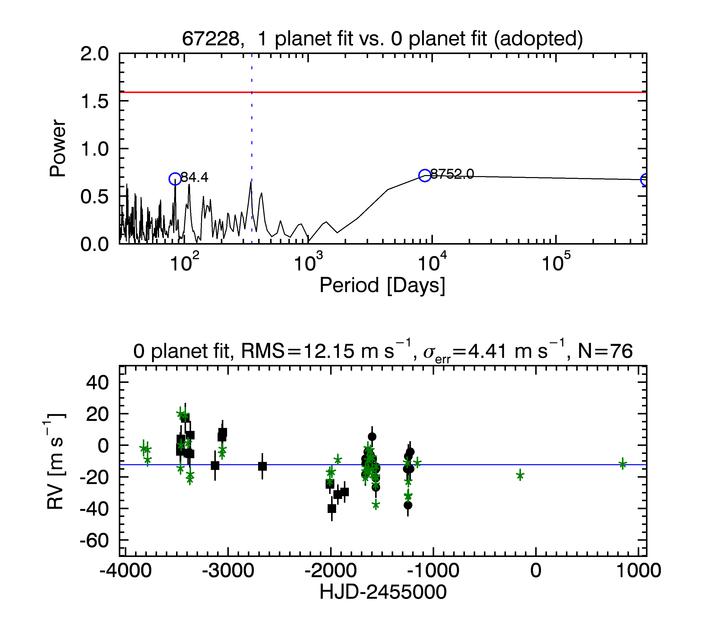}
\includegraphics[width=0.50\textwidth]{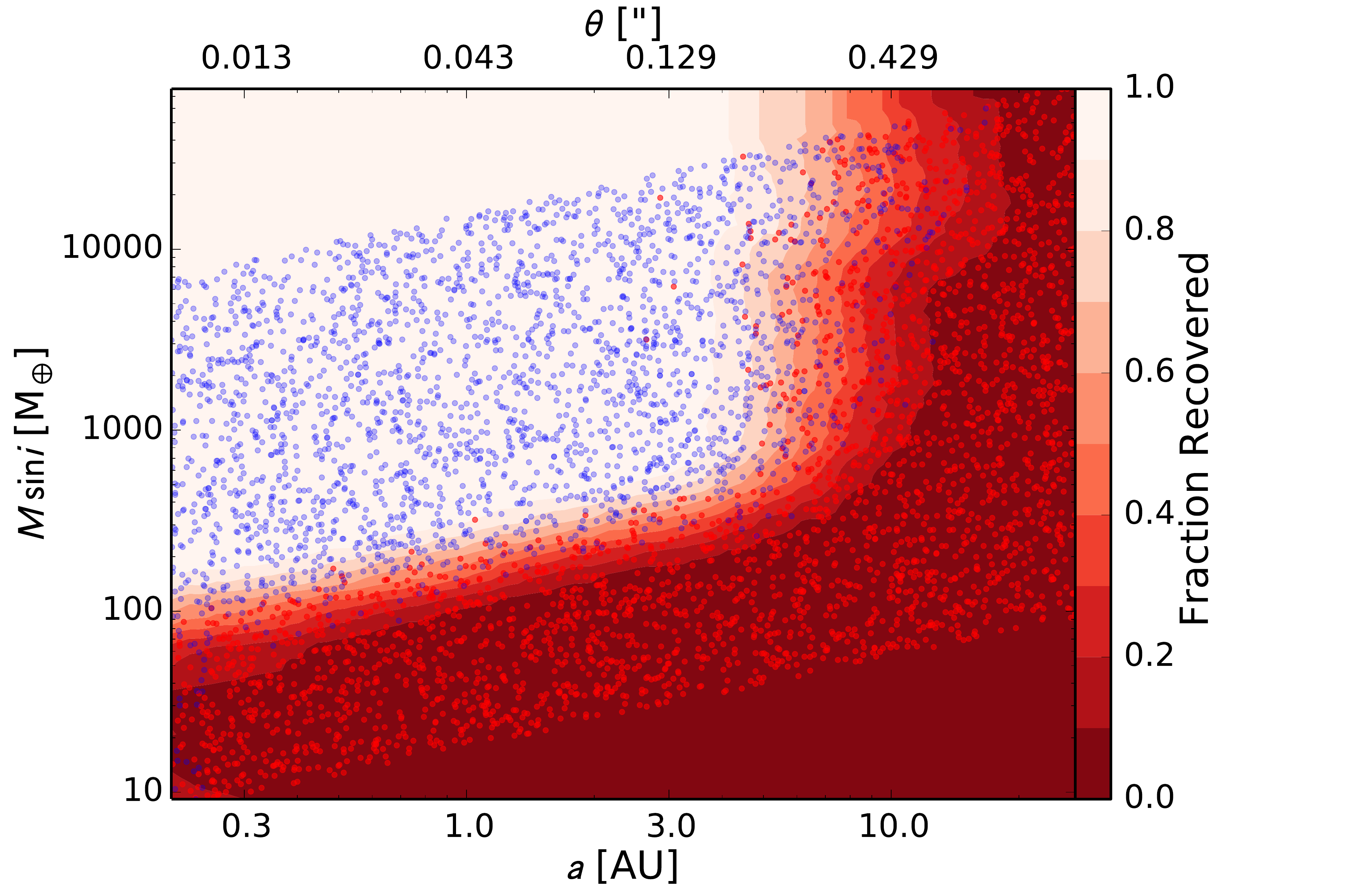}
\end{centering}
\caption{Results from an automated search for planets orbiting the star 
HD~67228 (HIP~39780; program = A) 
based on RVs from Lick and/or Keck Observatory.
The set of plots on the left (analogous to Figures \ref{fig:search_example} and \ref{fig:search_example2}) 
show the planet search results 
and the plot on the right shows the completeness limits (analogous to Fig.\ \ref{fig:completeness_example}). 
See the captions of those figures for detailed descriptions.  
}
\label{fig:completeness_67228}
\end{figure}

\begin{figure}
\begin{centering}
\includegraphics[width=0.45\textwidth]{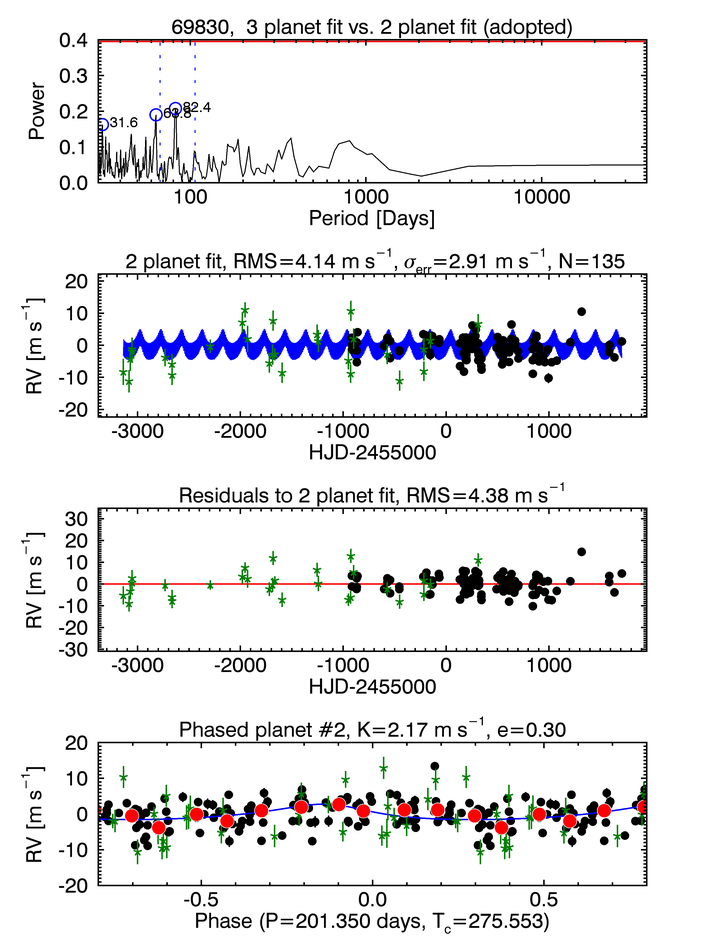}
\includegraphics[width=0.50\textwidth]{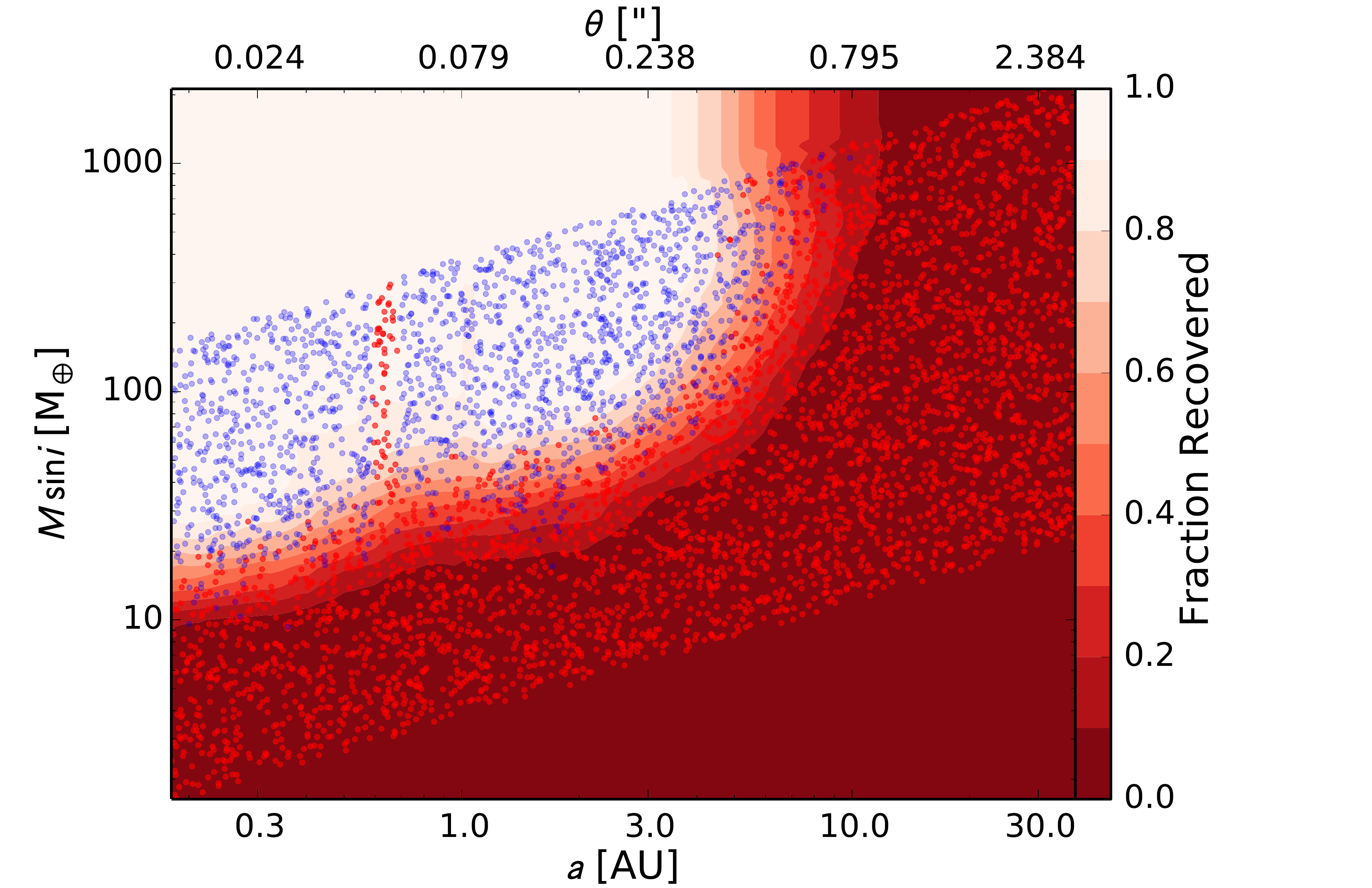}
\end{centering}
\caption{Results from an automated search for planets orbiting the star 
HD~69830 (HIP~40693; programs = S) 
based on RVs from Lick and/or Keck Observatory.
The set of plots on the left (analogous to Figures \ref{fig:search_example} and \ref{fig:search_example2}) 
show the planet search results 
and the plot on the right shows the completeness limits (analogous to Fig.\ \ref{fig:completeness_example}). 
See the captions of those figures for detailed descriptions.  
This star hosts three Neptune-mass planets, two of which we detect with our Keck and Lick RVs.  While we do not detect the 31.6 day  signal, we have not performed an analysis to show that our non-detection is dispositive.
}
\label{fig:completeness_69830}
\end{figure}
\clearpage

\begin{figure}
\begin{centering}
\includegraphics[width=0.45\textwidth]{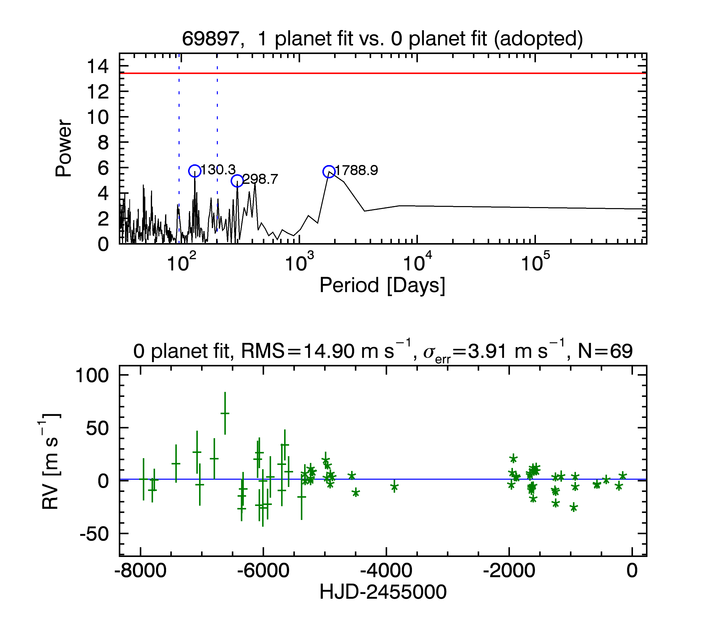}
\includegraphics[width=0.50\textwidth]{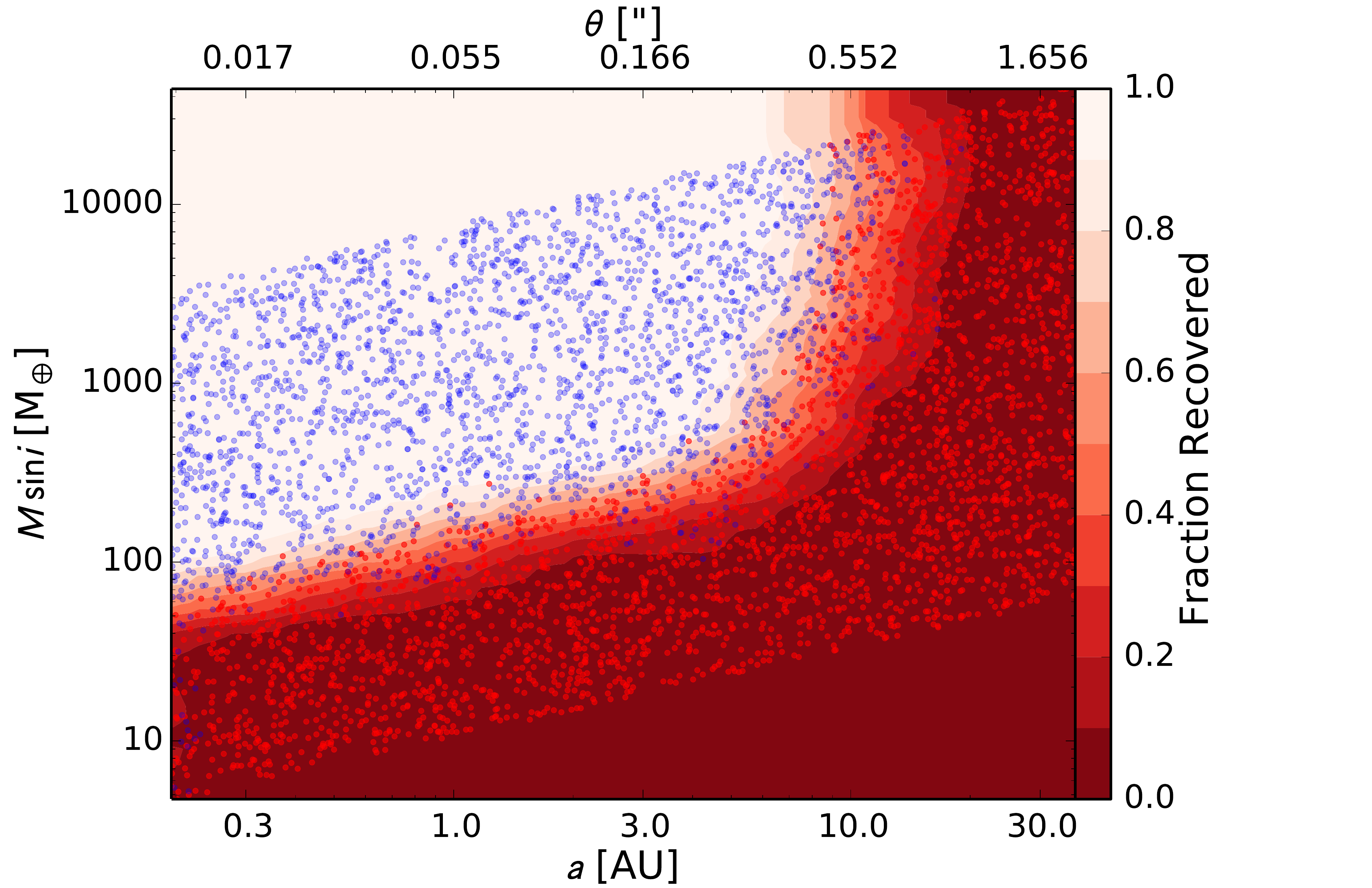}
\end{centering}
\caption{Results from an automated search for planets orbiting the star 
HD~69897 (HIP~40843; program = A) 
based on RVs from Lick and/or Keck Observatory.
The set of plots on the left (analogous to Figures \ref{fig:search_example} and \ref{fig:search_example2}) 
show the planet search results 
and the plot on the right shows the completeness limits (analogous to Fig.\ \ref{fig:completeness_example}). 
See the captions of those figures for detailed descriptions.  
}
\label{fig:completeness_69897}
\end{figure}

\begin{figure}
\begin{centering}
\includegraphics[width=0.45\textwidth]{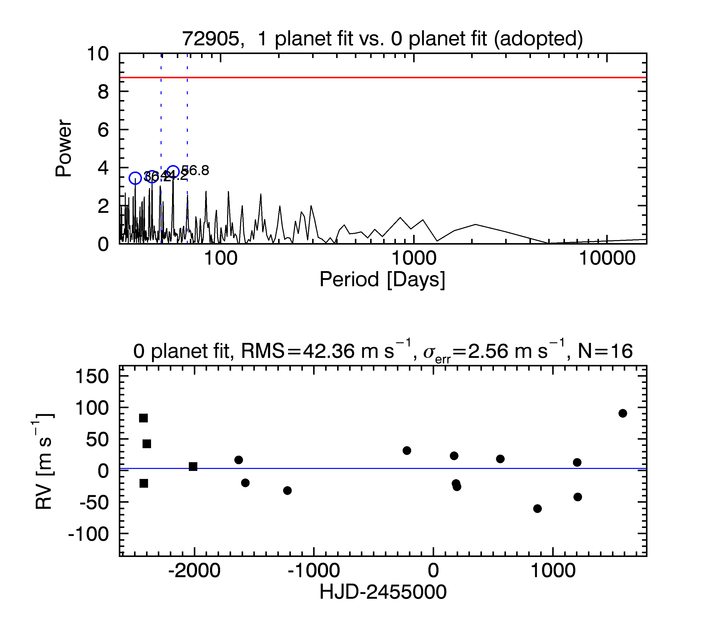}
\includegraphics[width=0.50\textwidth]{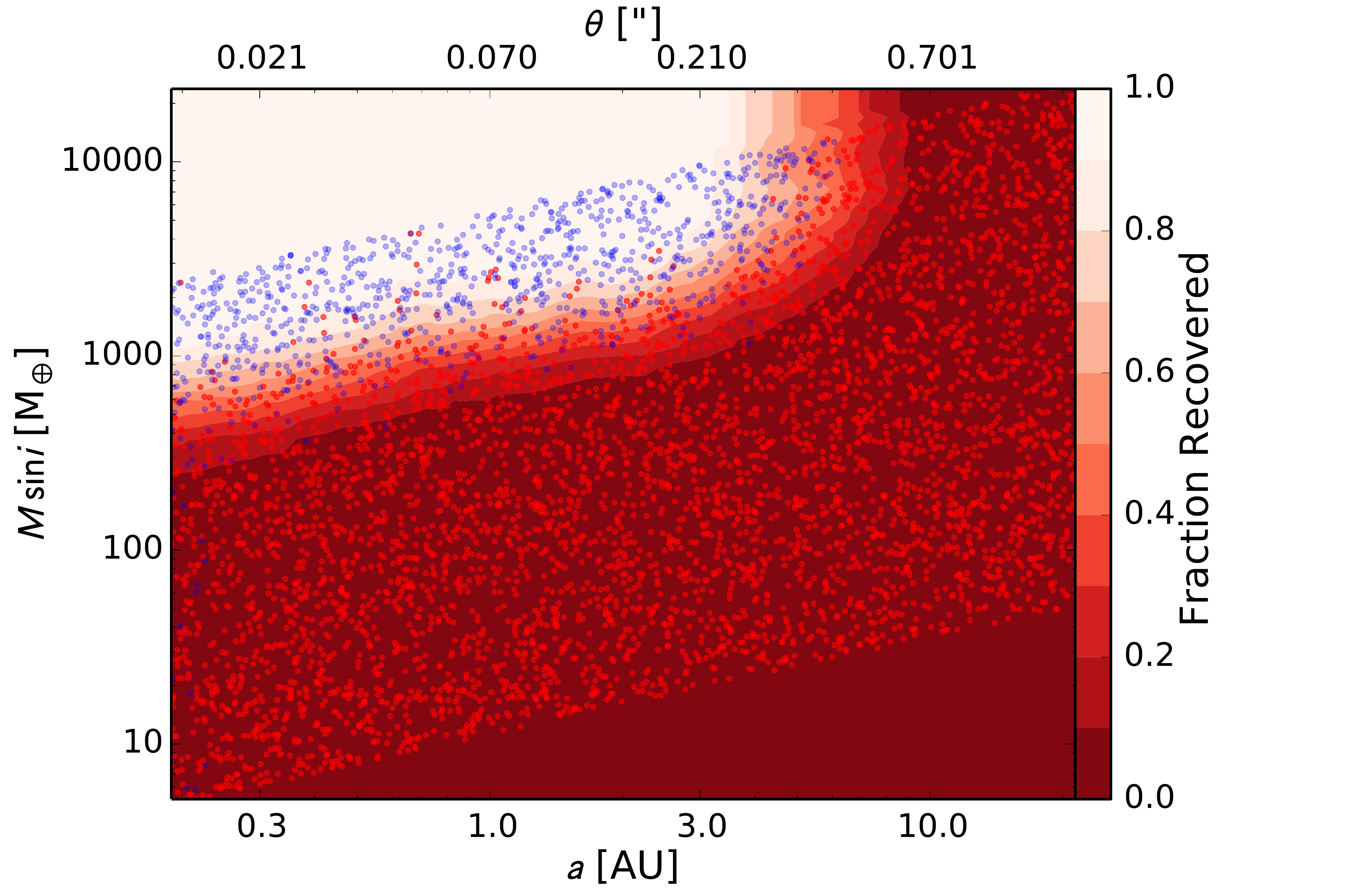}
\end{centering}
\caption{Results from an automated search for planets orbiting the star 
HD~72905 (HIP~42438; program = S) 
based on RVs from Lick and/or Keck Observatory.
The set of plots on the left (analogous to Figures \ref{fig:search_example} and \ref{fig:search_example2}) 
show the planet search results 
and the plot on the right shows the completeness limits (analogous to Fig.\ \ref{fig:completeness_example}). 
See the captions of those figures for detailed descriptions.  
}
\label{fig:completeness_72905}
\end{figure}
\clearpage

\begin{figure}
\begin{centering}
\includegraphics[width=0.45\textwidth]{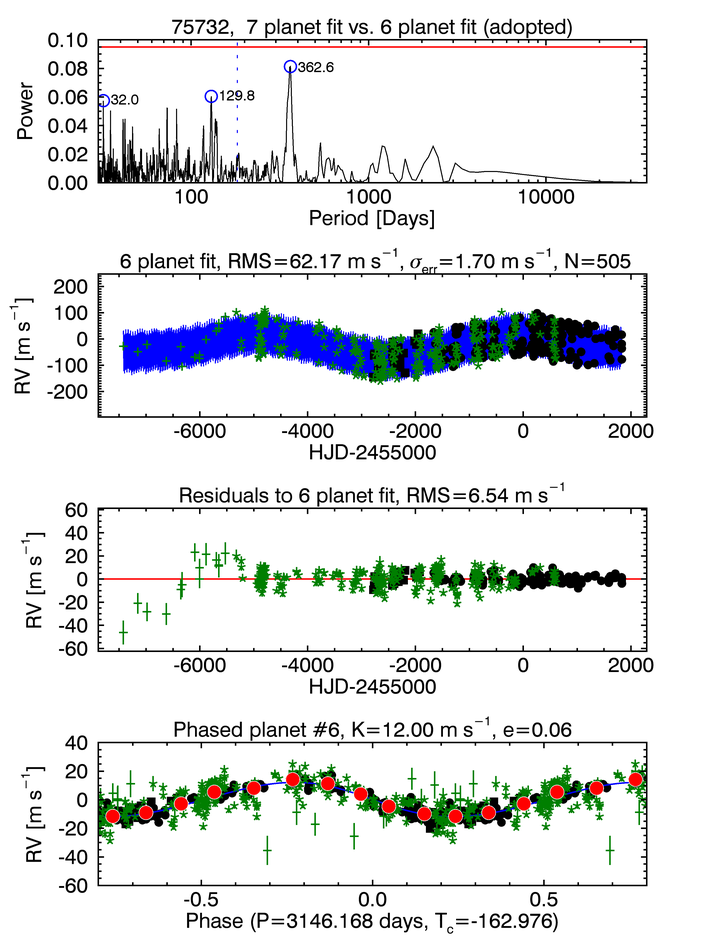}
\includegraphics[width=0.50\textwidth]{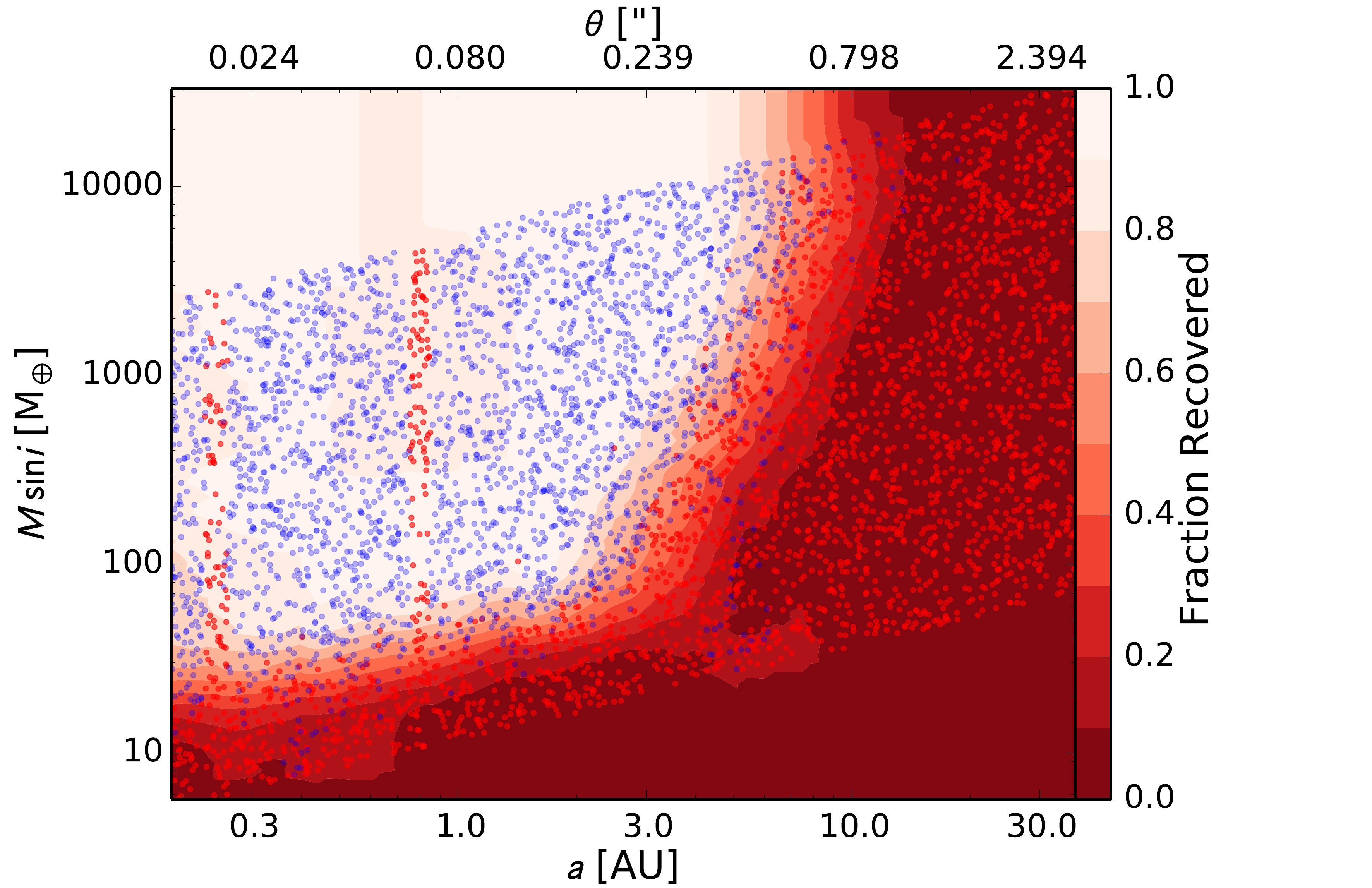}
\end{centering}
\caption{Results from an automated search for planets orbiting the star 
HD~75732 (HIP~43587; program = S) 
based on RVs from Lick and/or Keck Observatory.
The set of plots on the left (analogous to Figures \ref{fig:search_example} and \ref{fig:search_example2}) 
show the planet search results 
and the plot on the right shows the completeness limits (analogous to Fig.\ \ref{fig:completeness_example}). 
See the captions of those figures for detailed descriptions.  
This star hosts five known planets.
}
\label{fig:completeness_75732}
\end{figure}

\begin{figure}
\begin{centering}
\includegraphics[width=0.45\textwidth]{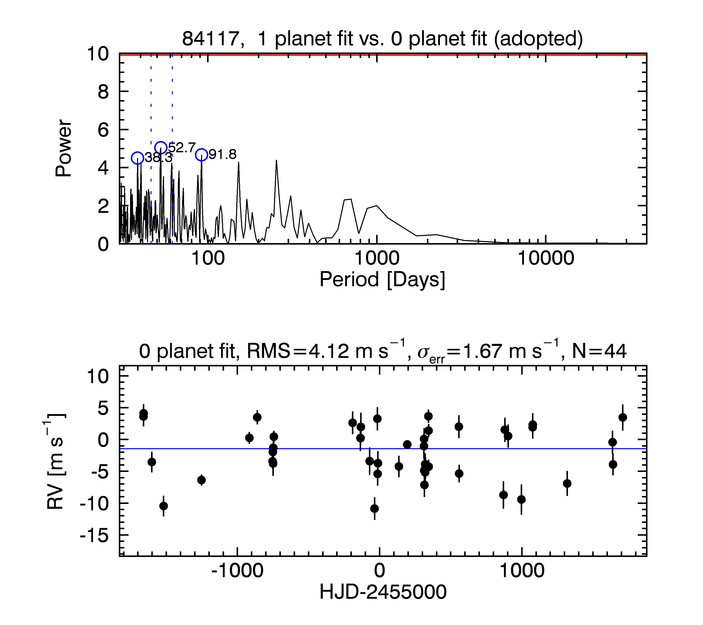}
\includegraphics[width=0.50\textwidth]{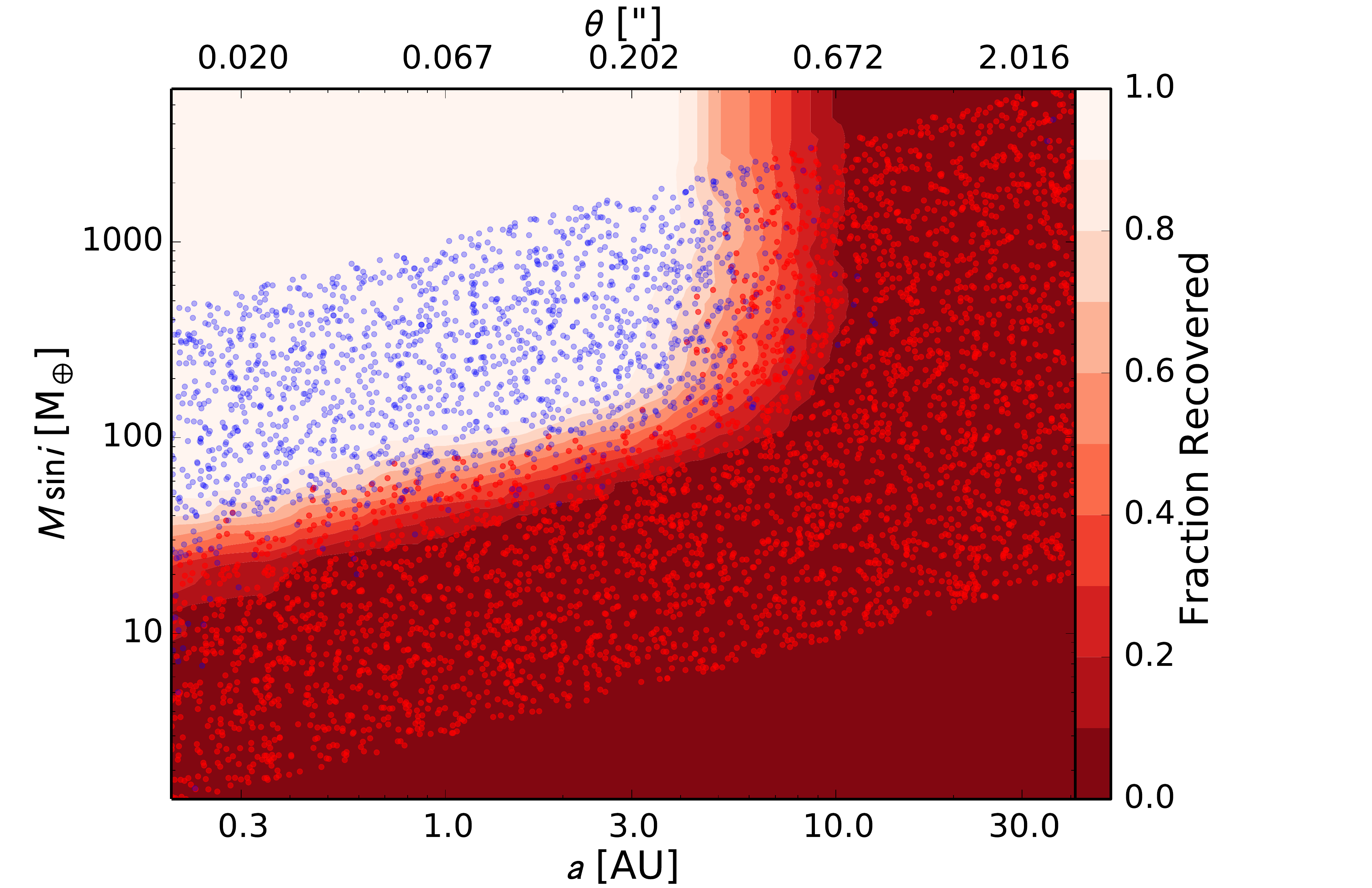}
\end{centering}
\caption{Results from an automated search for planets orbiting the star 
HD~84117 (HIP~47592; programs = S, C, A) 
based on RVs from Lick and/or Keck Observatory.
The set of plots on the left (analogous to Figures \ref{fig:search_example} and \ref{fig:search_example2}) 
show the planet search results 
and the plot on the right shows the completeness limits (analogous to Fig.\ \ref{fig:completeness_example}). 
See the captions of those figures for detailed descriptions.  
}
\label{fig:completeness_84117}
\end{figure}
\clearpage

\begin{figure}
\begin{centering}
\includegraphics[width=0.45\textwidth]{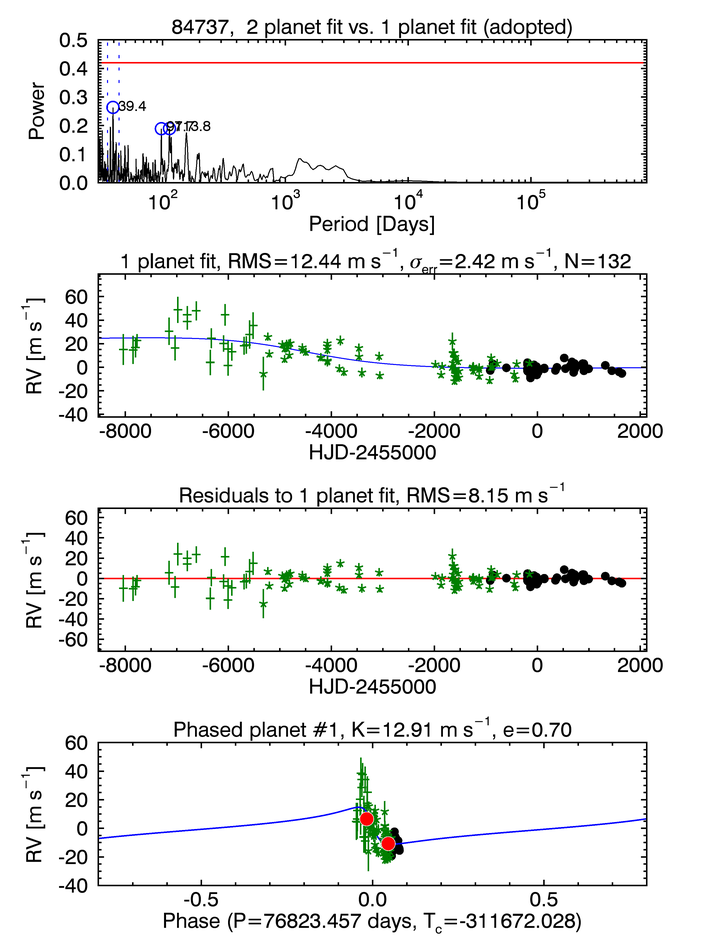}
\includegraphics[width=0.50\textwidth]{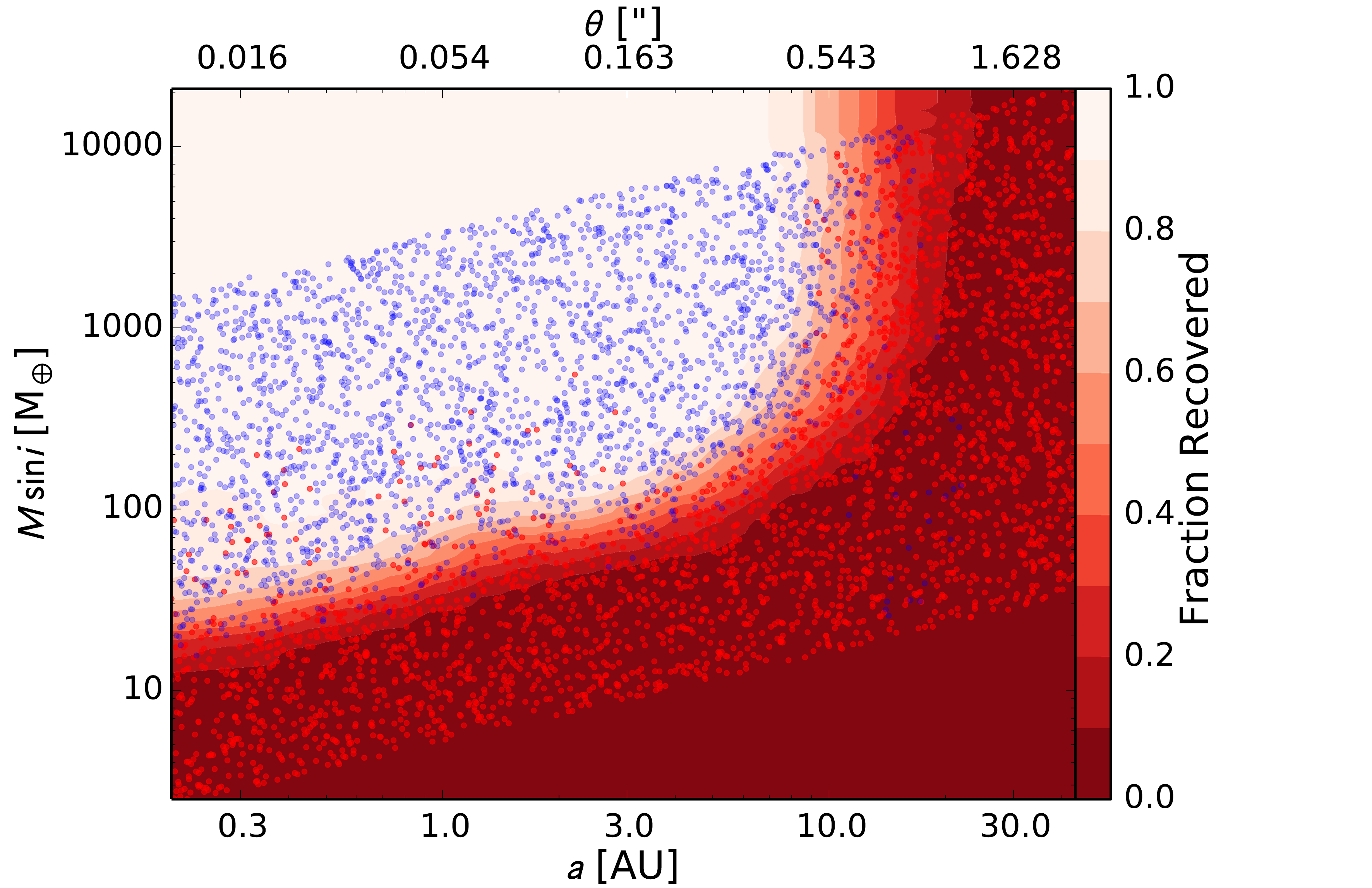}
\end{centering}
\caption{Results from an automated search for planets orbiting the star 
HD~84737 (HIP~48113; program = A) 
based on RVs from Lick and/or Keck Observatory.
The set of plots on the left (analogous to Figures \ref{fig:search_example} and \ref{fig:search_example2}) 
show the planet search results 
and the plot on the right shows the completeness limits (analogous to Fig.\ \ref{fig:completeness_example}). 
See the captions of those figures for detailed descriptions.  
This star has a formally adopted signal that appears to be due to uncorrected zero-point offsets in the Lick RVs and not due a planet.
}
\label{fig:completeness_84737}
\end{figure}

\begin{figure}
\begin{centering}
\includegraphics[width=0.45\textwidth]{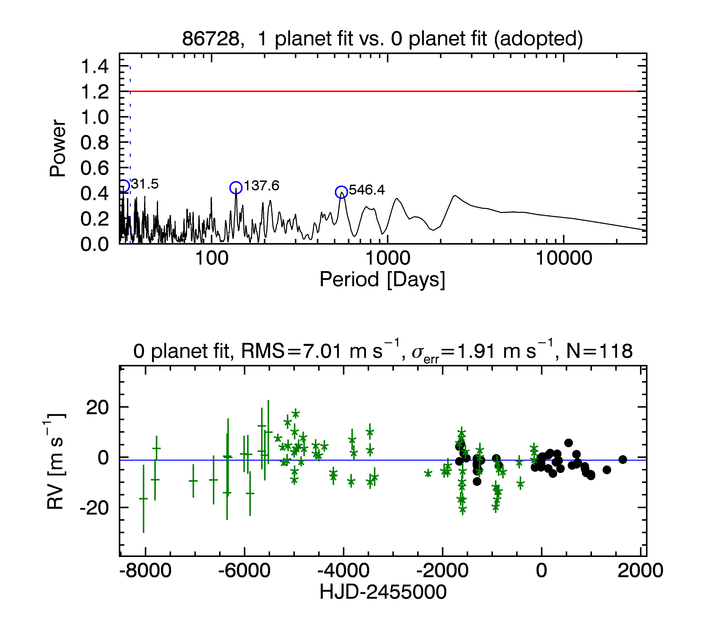}
\includegraphics[width=0.50\textwidth]{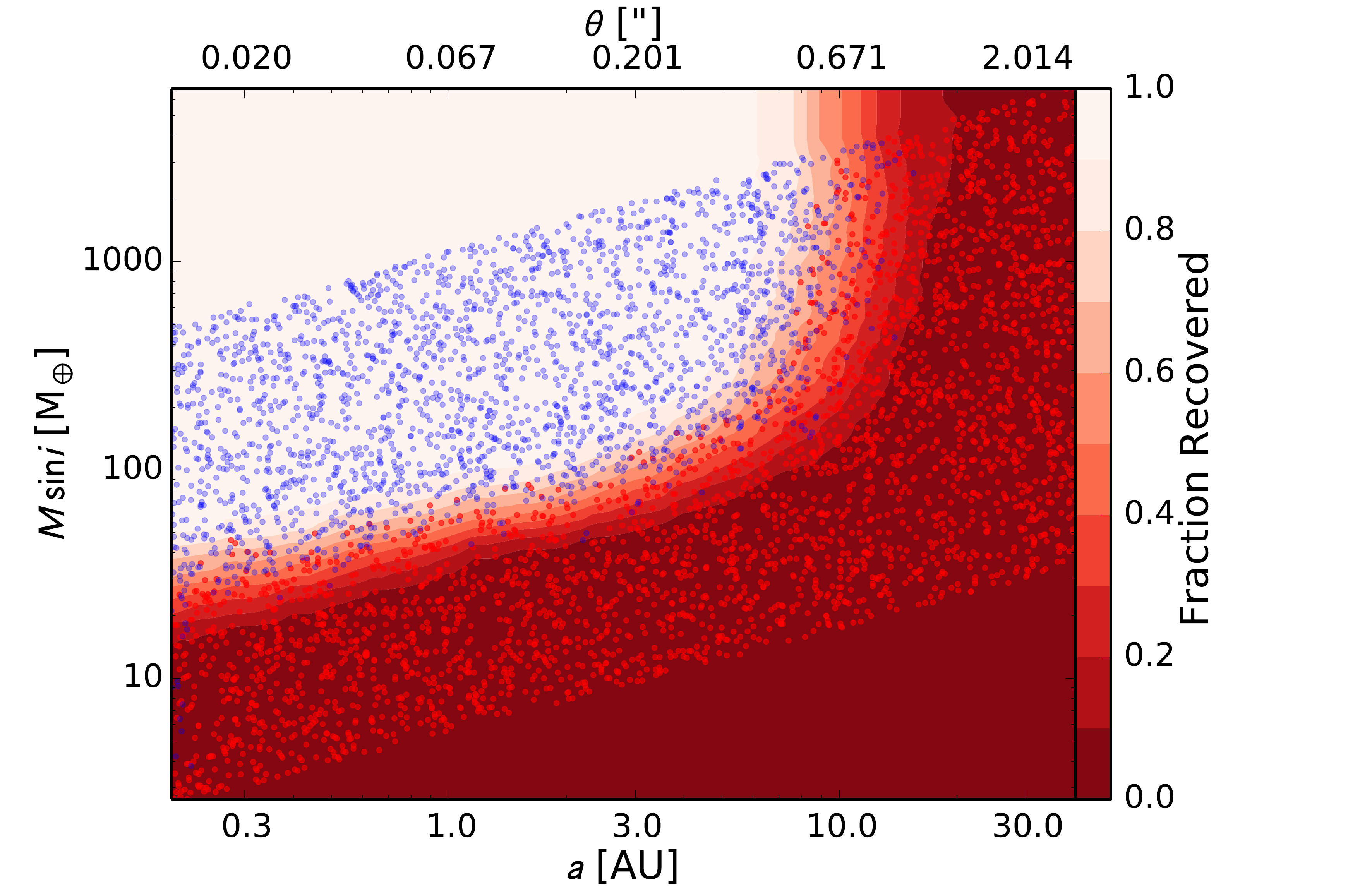}
\end{centering}
\caption{Results from an automated search for planets orbiting the star 
HD~86728 (HIP~49081; programs = S, A) 
based on RVs from Lick and/or Keck Observatory.
The set of plots on the left (analogous to Figures \ref{fig:search_example} and \ref{fig:search_example2}) 
show the planet search results 
and the plot on the right shows the completeness limits (analogous to Fig.\ \ref{fig:completeness_example}). 
See the captions of those figures for detailed descriptions.  
}
\label{fig:completeness_86728}
\end{figure}
\clearpage

\begin{figure}
\begin{centering}
\includegraphics[width=0.45\textwidth]{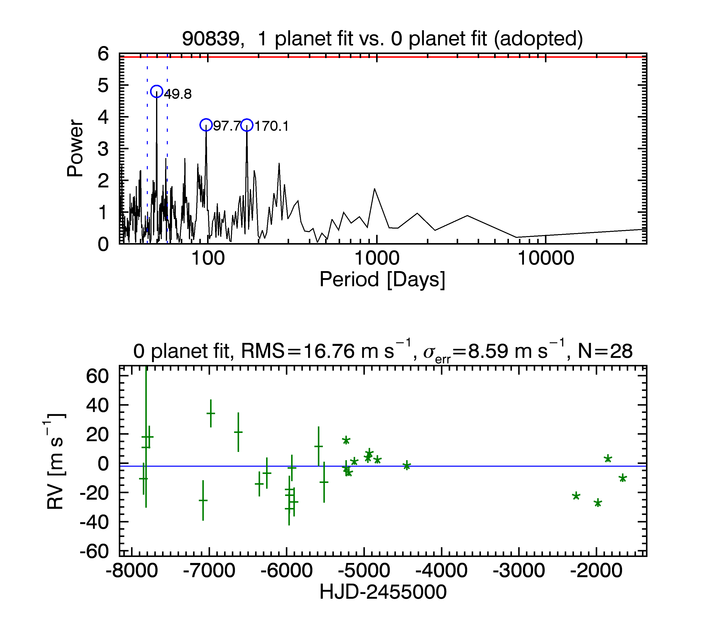}
\includegraphics[width=0.50\textwidth]{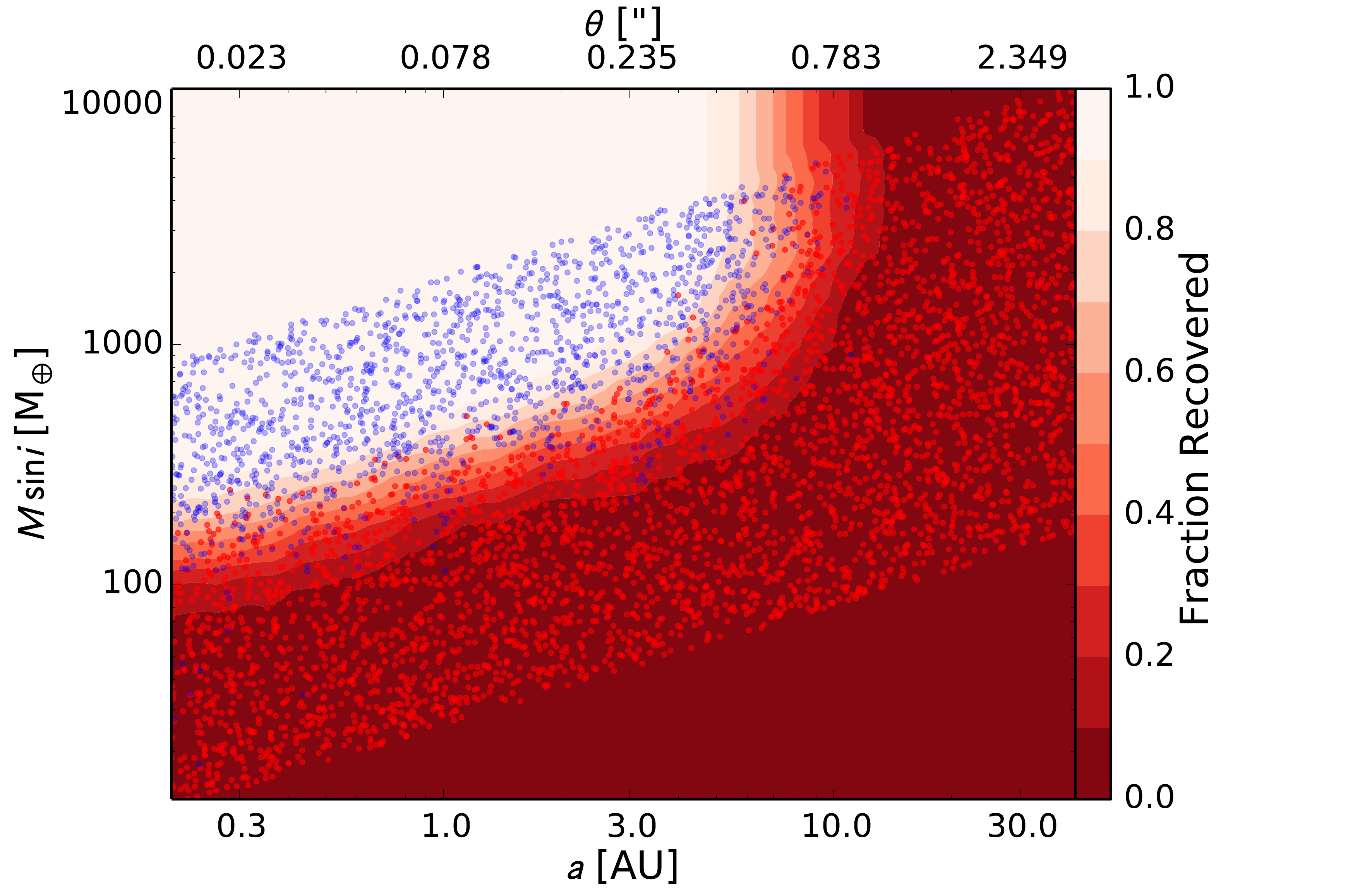}
\end{centering}
\caption{Results from an automated search for planets orbiting the star 
HD~90839 (HIP~51459; programs = S, C, A) 
based on RVs from Lick and/or Keck Observatory.
The set of plots on the left (analogous to Figures \ref{fig:search_example} and \ref{fig:search_example2}) 
show the planet search results 
and the plot on the right shows the completeness limits (analogous to Fig.\ \ref{fig:completeness_example}). 
See the captions of those figures for detailed descriptions.  
}
\label{fig:completeness_90839}
\end{figure}

\begin{figure}
\begin{centering}
\includegraphics[width=0.45\textwidth]{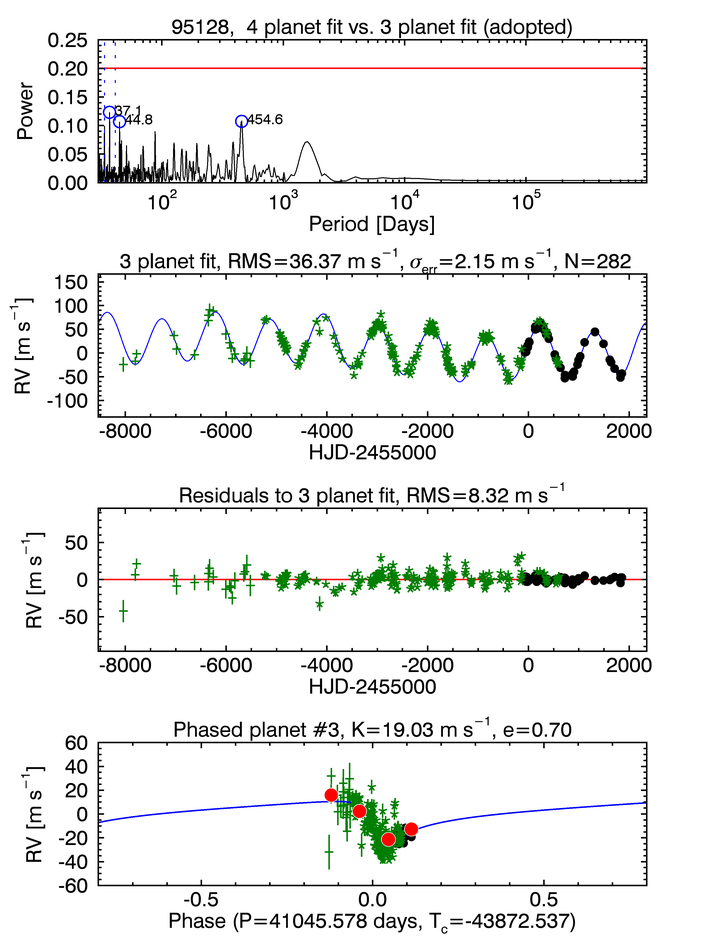}
\includegraphics[width=0.50\textwidth]{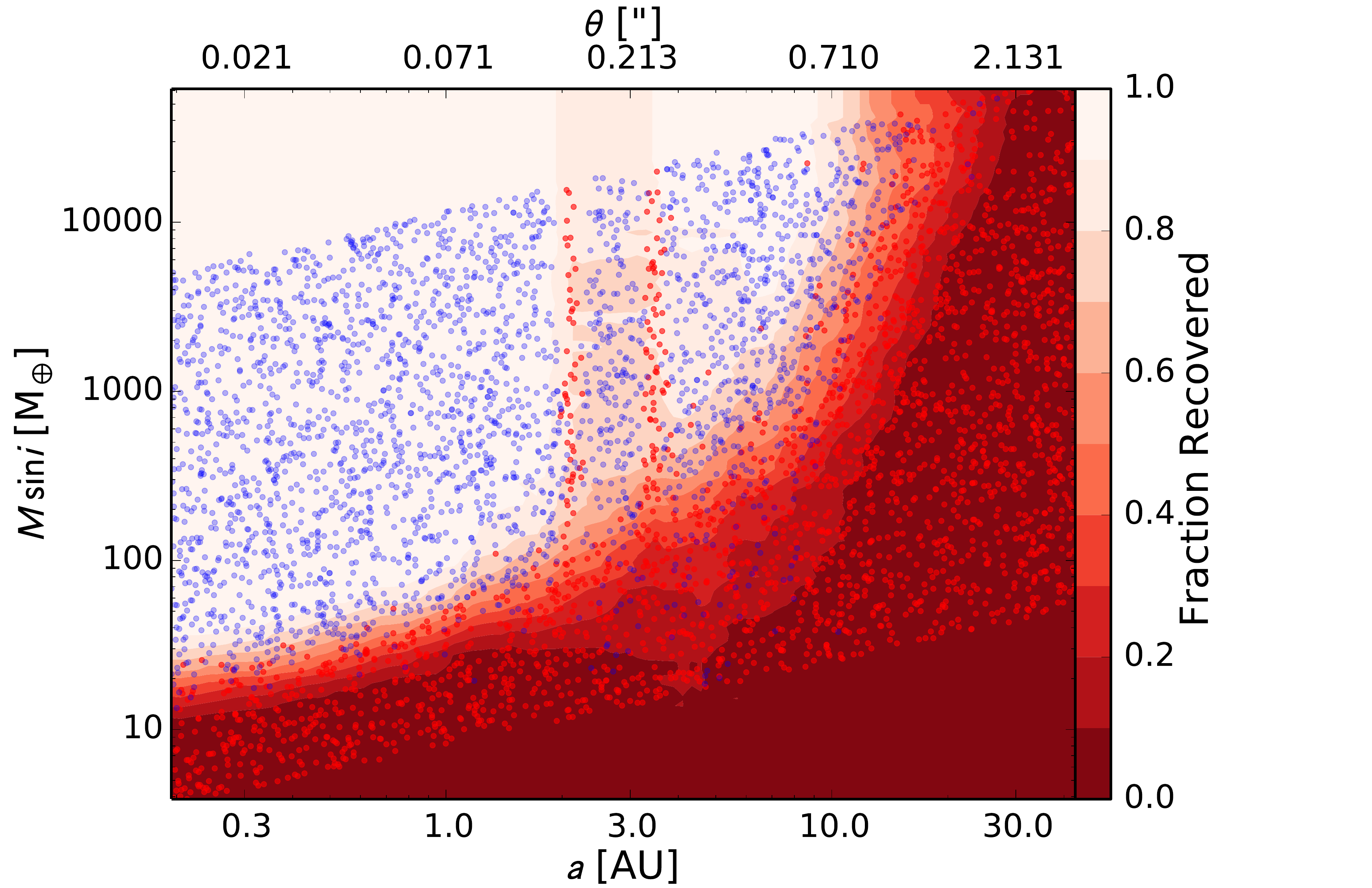}
\end{centering}
\caption{Results from an automated search for planets orbiting the star 
HD~95128 (HIP~53721; programs = A) 
based on RVs from Lick and/or Keck Observatory.
The set of plots on the left (analogous to Figures \ref{fig:search_example} and \ref{fig:search_example2}) 
show the planet search results 
and the plot on the right shows the completeness limits (analogous to Fig.\ \ref{fig:completeness_example}). 
See the captions of those figures for detailed descriptions.  
This star hosts two well-known giant planets with semi-major axes of 2 and 3.6 AU, and possibly a third planet at $\sim$11 AU.  Our automated search prefers a model with three planets, although the outer most planet has a poorly constrained orbit.
}
\label{fig:completeness_95128}
\end{figure}
\clearpage

\begin{figure}
\begin{centering}
\includegraphics[width=0.45\textwidth]{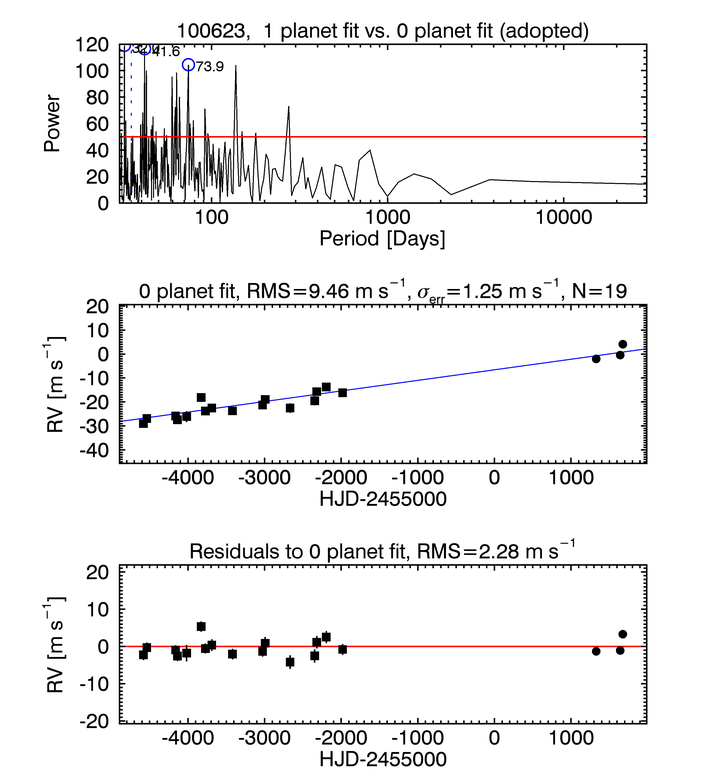}
\includegraphics[width=0.50\textwidth]{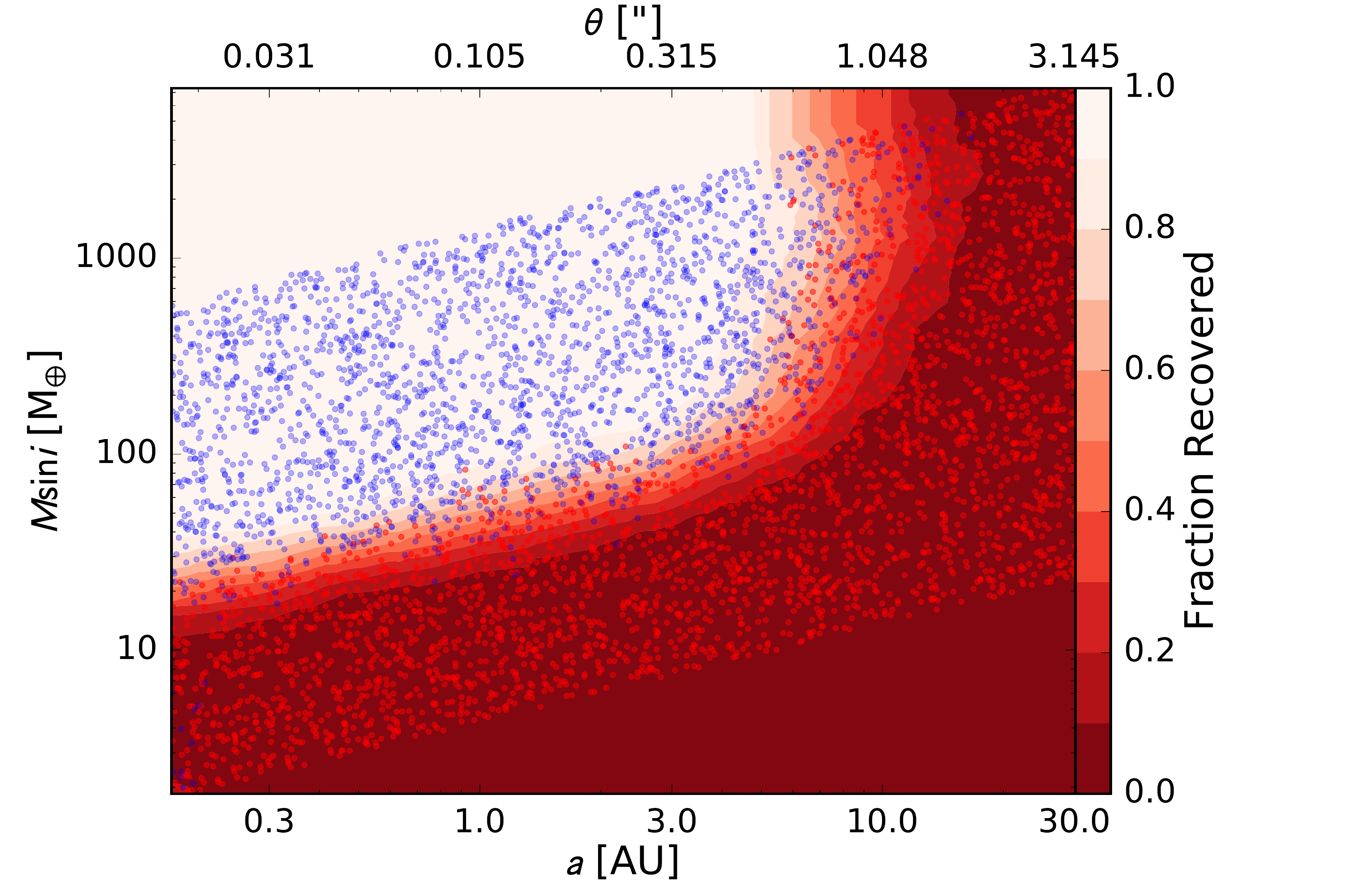}
\end{centering}
\caption{Results from an automated search for planets orbiting the star 
HD~100623 (HIP~56452; programs = S) 
based on RVs from Lick and/or Keck Observatory.
The set of plots on the left (analogous to Figures \ref{fig:search_example} and \ref{fig:search_example2}) 
show the planet search results 
and the plot on the right shows the completeness limits (analogous to Fig.\ \ref{fig:completeness_example}). 
See the captions of those figures for detailed descriptions.  
This star shows a significant linear trend with no detectable curvature.
}
\label{fig:completeness_100623}
\end{figure}

\begin{figure}
\begin{centering}
\includegraphics[width=0.45\textwidth]{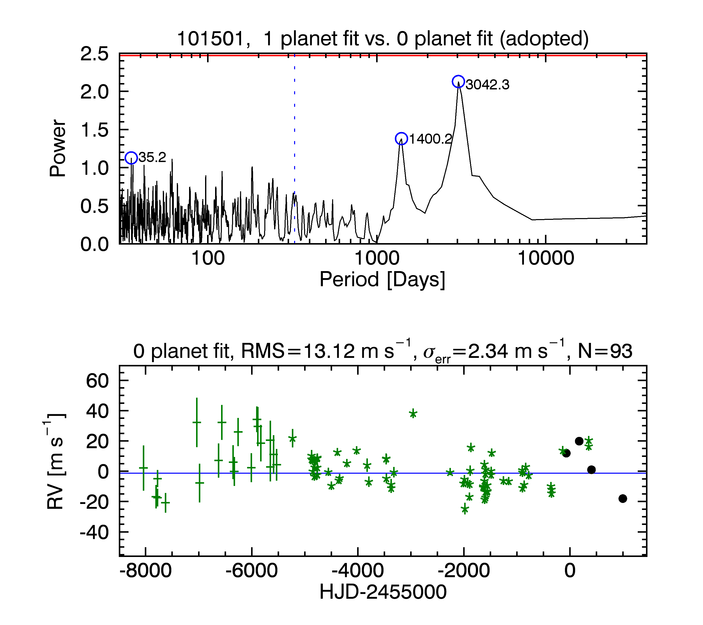}
\includegraphics[width=0.50\textwidth]{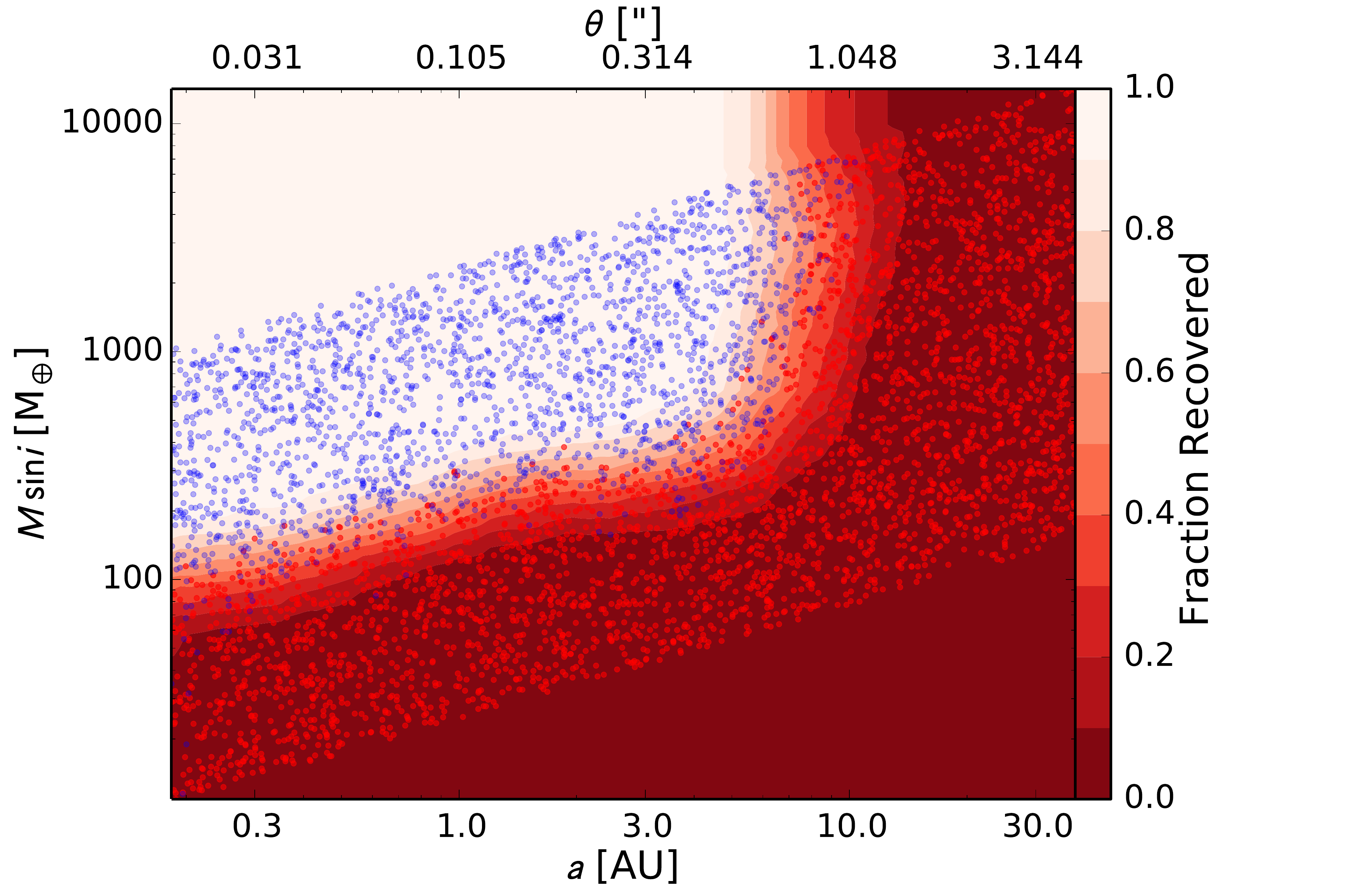}
\end{centering}
\caption{Results from an automated search for planets orbiting the star 
HD~101501 (HIP~56997; programs = S, C, A) 
based on RVs from Lick and/or Keck Observatory.
The set of plots on the left (analogous to Figures \ref{fig:search_example} and \ref{fig:search_example2}) 
show the planet search results 
and the plot on the right shows the completeness limits (analogous to Fig.\ \ref{fig:completeness_example}). 
See the captions of those figures for detailed descriptions.  
}
\label{fig:completeness_101501}
\end{figure}
\clearpage

\begin{figure}
\begin{centering}
\includegraphics[width=0.45\textwidth]{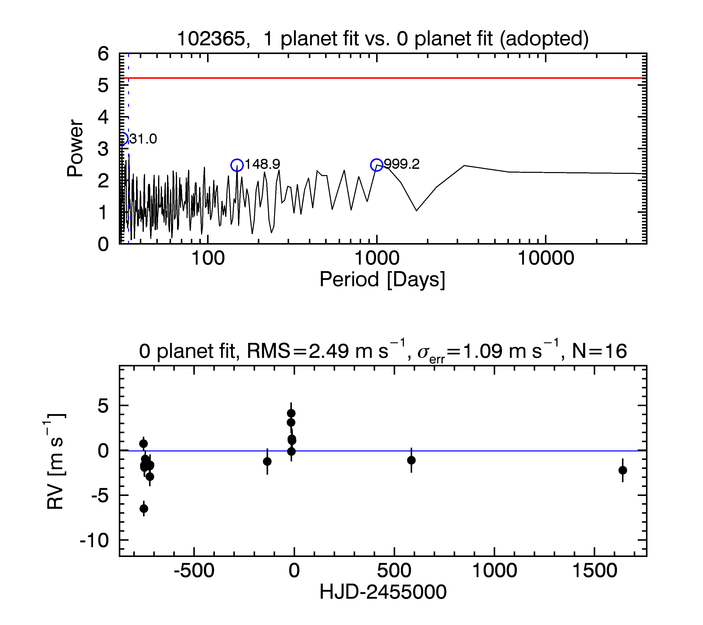}
\includegraphics[width=0.50\textwidth]{102365-recovery.pdf}
\end{centering}
\caption{Results from an automated search for planets orbiting the star 
HD~102365 (HIP~57443; programs = S, C, A) 
based on RVs from Lick and/or Keck Observatory.
The set of plots on the left (analogous to Figures \ref{fig:search_example} and \ref{fig:search_example2}) 
show the planet search results 
and the plot on the right shows the completeness limits (analogous to Fig.\ \ref{fig:completeness_example}). 
See the captions of those figures for detailed descriptions.  
}
\label{fig:completeness_102365}
\end{figure}

\begin{figure}
\begin{centering}
\includegraphics[width=0.45\textwidth]{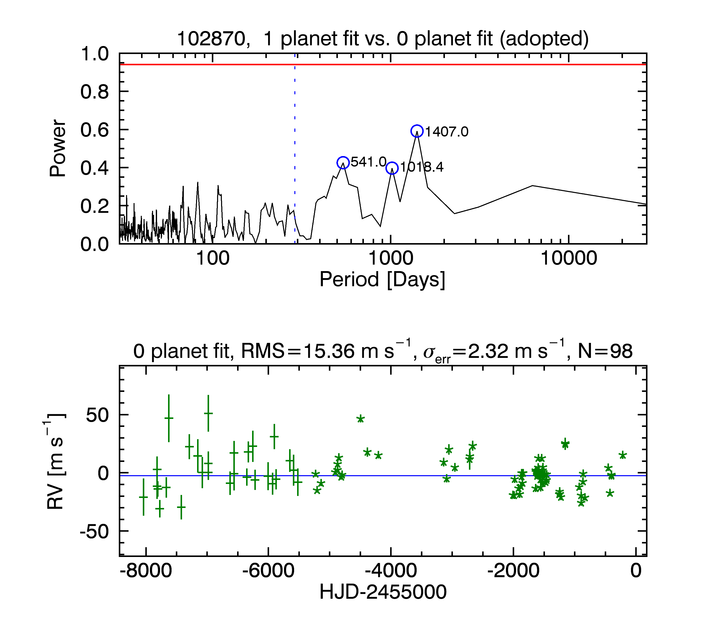}
\includegraphics[width=0.50\textwidth]{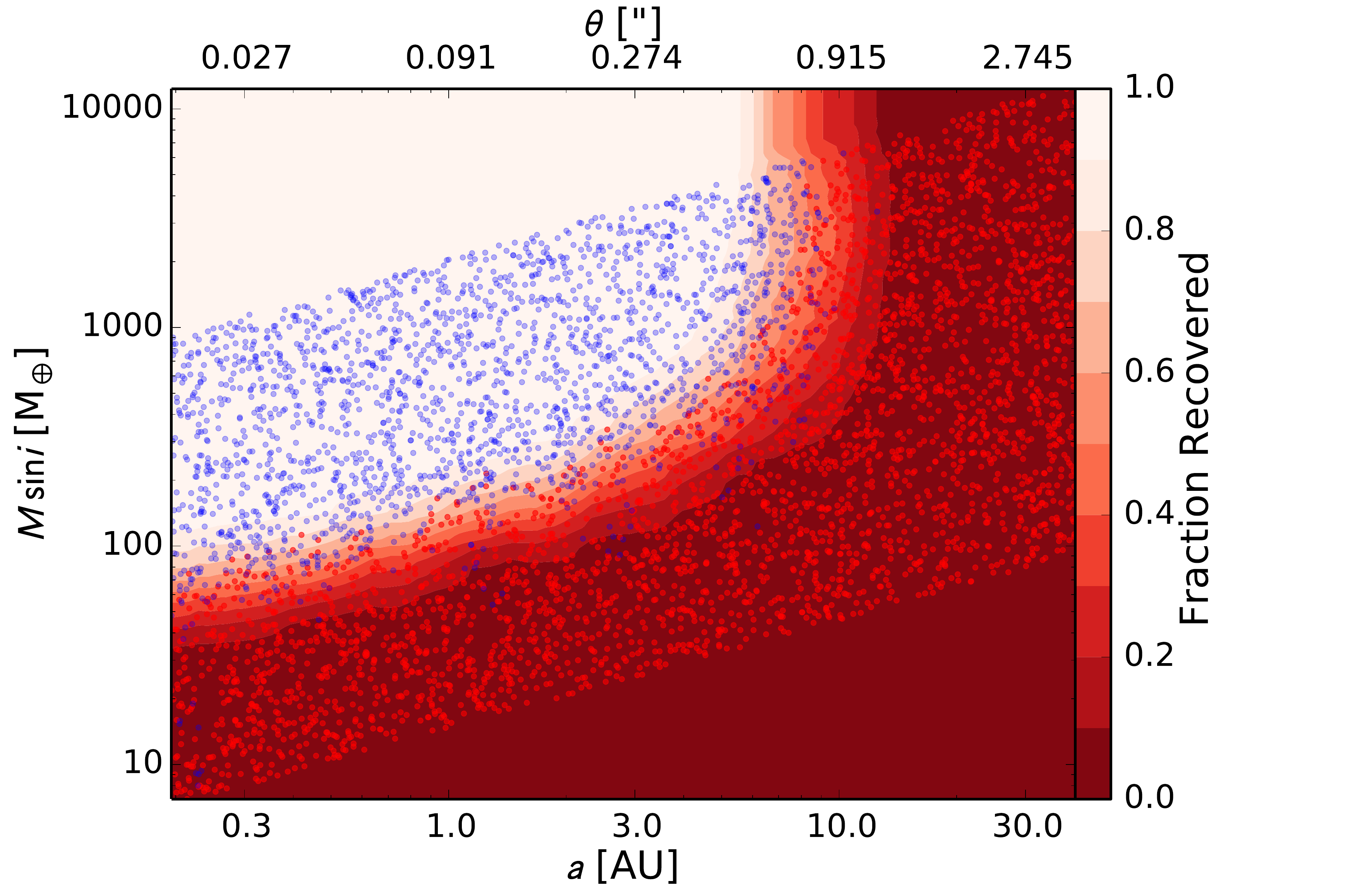}
\end{centering}
\caption{Results from an automated search for planets orbiting the star 
HD~102870 (HIP~57757; programs = S, C, A) 
based on RVs from Lick and/or Keck Observatory.
The set of plots on the left (analogous to Figures \ref{fig:search_example} and \ref{fig:search_example2}) 
show the planet search results 
and the plot on the right shows the completeness limits (analogous to Fig.\ \ref{fig:completeness_example}). 
See the captions of those figures for detailed descriptions.  
}
\label{fig:completeness_102870}
\end{figure}
\clearpage

\begin{figure}
\begin{centering}
\includegraphics[width=0.45\textwidth]{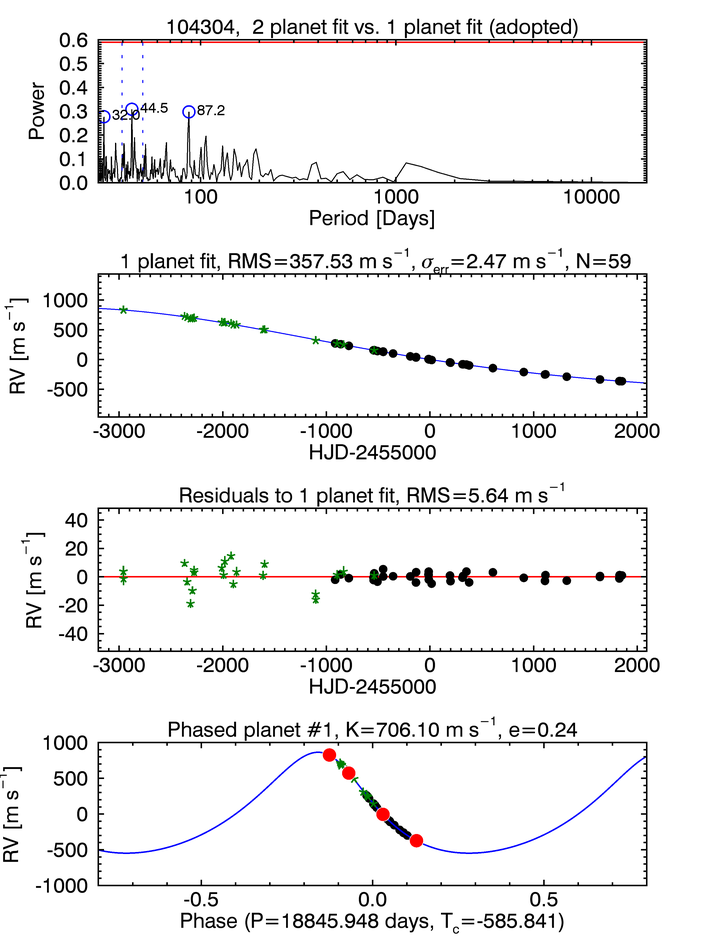}
\includegraphics[width=0.50\textwidth]{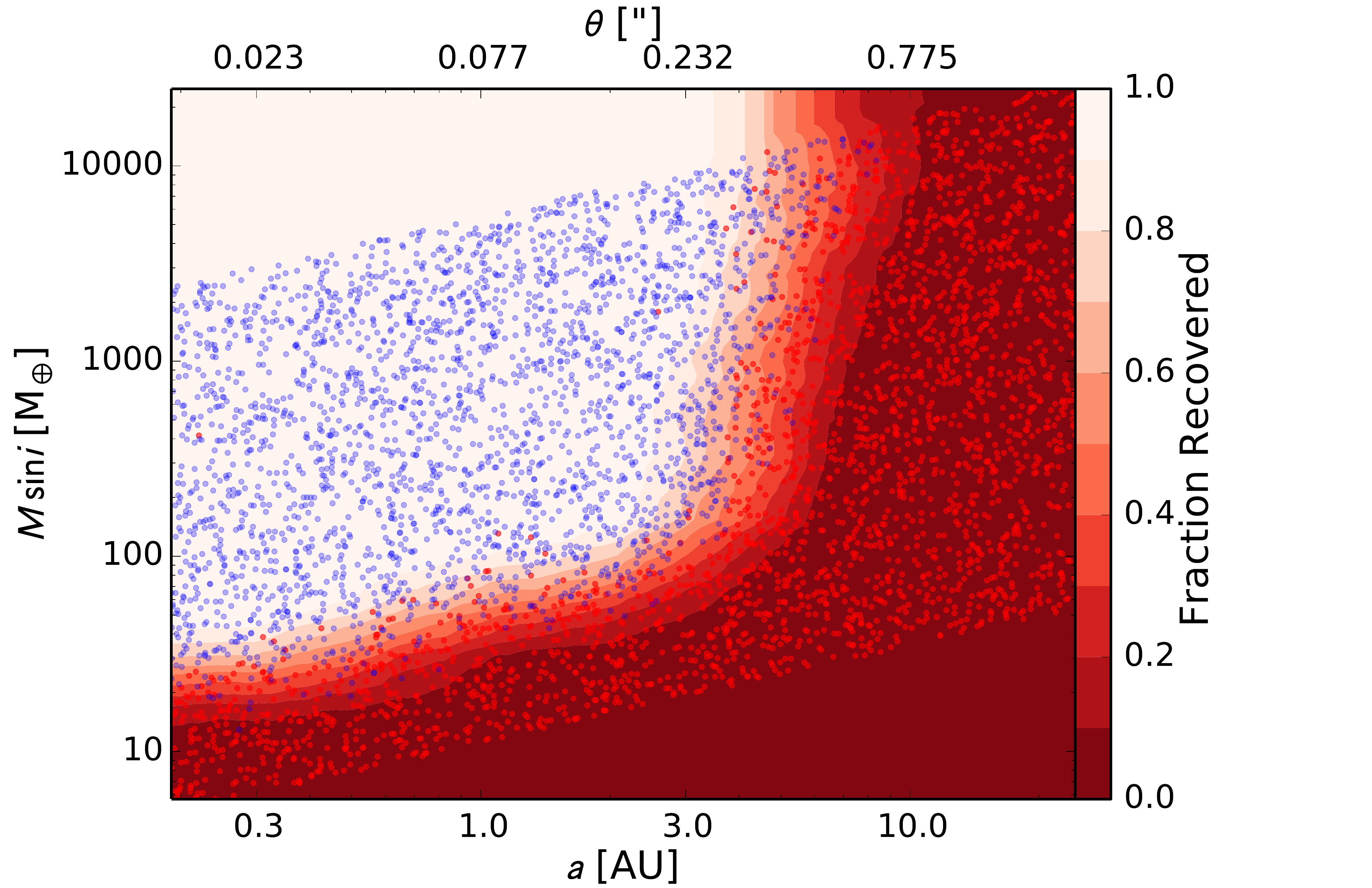}
\end{centering}
\caption{Results from an automated search for planets orbiting the star 
HD~104304 (HIP~58576; program = S) 
based on RVs from Lick and/or Keck Observatory.
The set of plots on the left (analogous to Figures \ref{fig:search_example} and \ref{fig:search_example2}) 
show the planet search results 
and the plot on the right shows the completeness limits (analogous to Fig.\ \ref{fig:completeness_example}). 
See the captions of those figures for detailed descriptions.  
This system shows a strong  long-term linear trend with curvature, likely due to a detected low-mass, stellar companion.  Our automated search prefers a model with a linear velocity trend (constant acceleration) in addition to the orbit segment from the companion (three bodies total), with considerable model degeneracy.
}
\label{fig:completeness_104304}
\end{figure}

\begin{figure}
\begin{centering}
\includegraphics[width=0.45\textwidth]{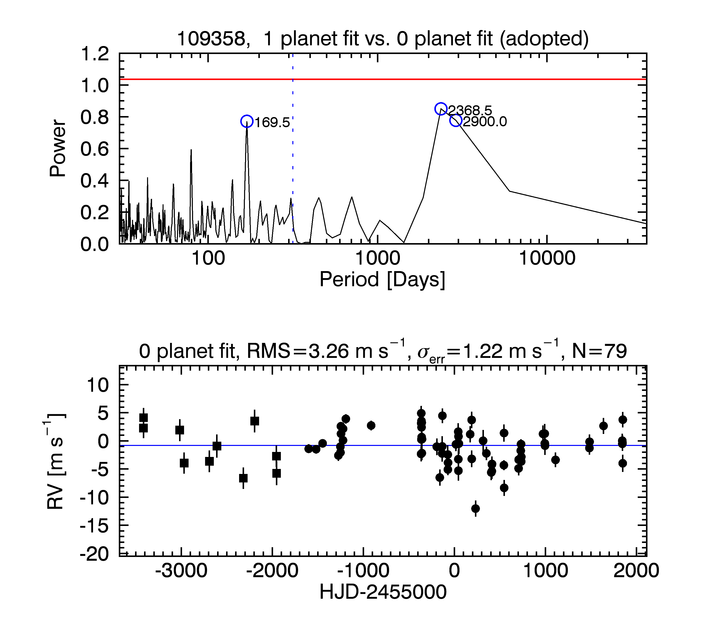}
\includegraphics[width=0.50\textwidth]{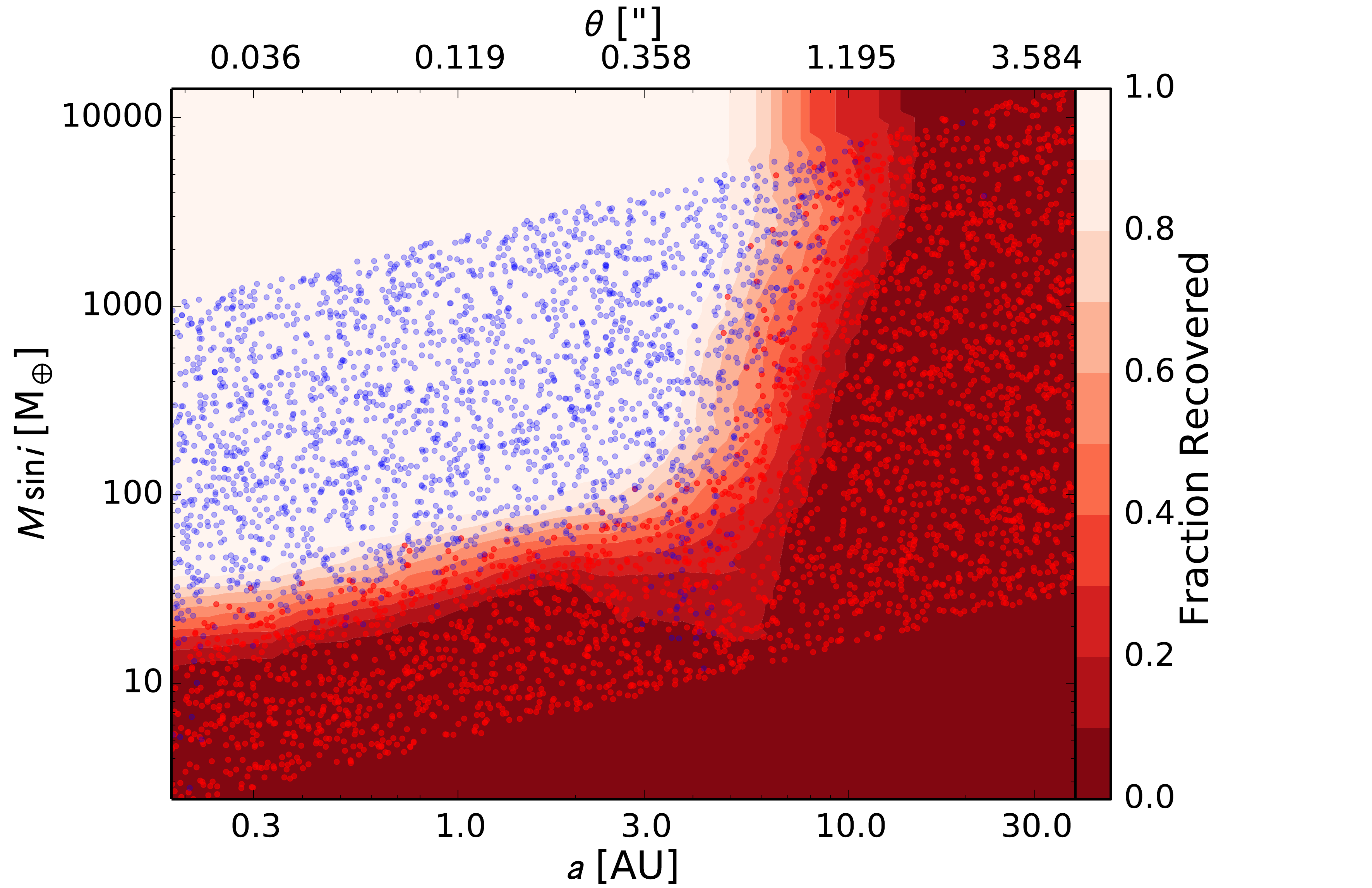}
\end{centering}
\caption{Results from an automated search for planets orbiting the star 
HD~109358 (HIP~61317; programs = S, C, A) 
based on RVs from Lick and/or Keck Observatory.
The set of plots on the left (analogous to Figures \ref{fig:search_example} and \ref{fig:search_example2}) 
show the planet search results 
and the plot on the right shows the completeness limits (analogous to Fig.\ \ref{fig:completeness_example}). 
See the captions of those figures for detailed descriptions.  
}
\label{fig:completeness_109358}
\end{figure}
\clearpage

\begin{figure}
\begin{centering}
\includegraphics[width=0.45\textwidth]{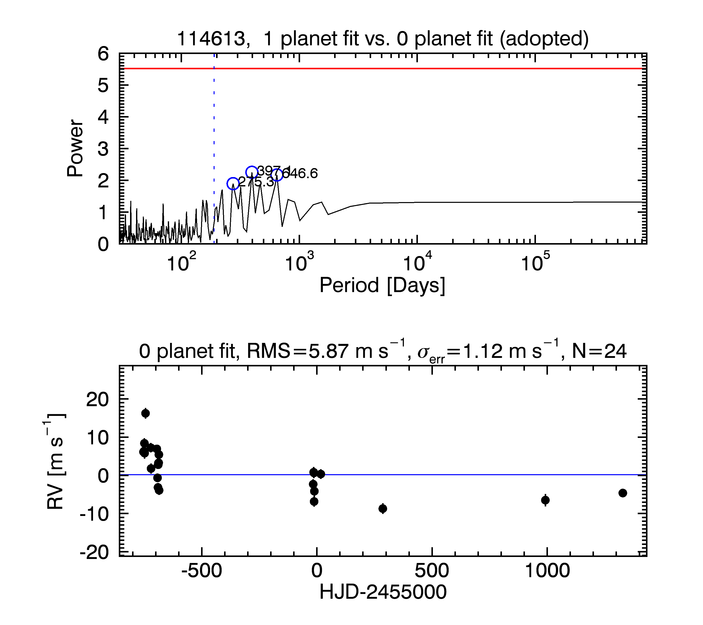}
\includegraphics[width=0.50\textwidth]{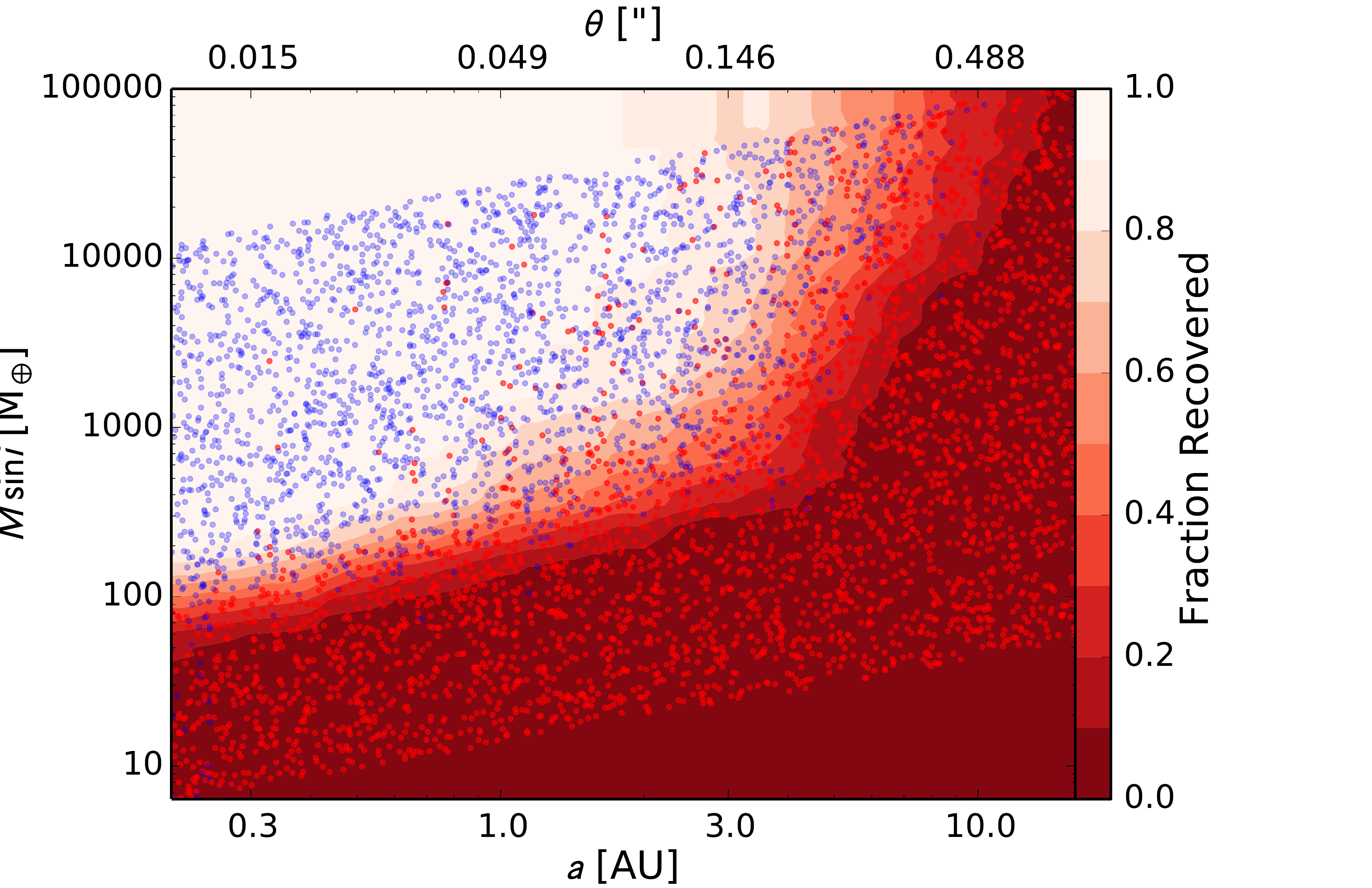}
\end{centering}
\caption{Results from an automated search for planets orbiting the star 
HD~114613 (HIP~64408; program = A) 
based on RVs from Lick and/or Keck Observatory.
The set of plots on the left (analogous to Figures \ref{fig:search_example} and \ref{fig:search_example2}) 
show the planet search results 
and the plot on the right shows the completeness limits (analogous to Fig.\ \ref{fig:completeness_example}). 
See the captions of those figures for detailed descriptions.  
}
\label{fig:completeness_114613}
\end{figure}

\begin{figure}
\begin{centering}
\includegraphics[width=0.45\textwidth]{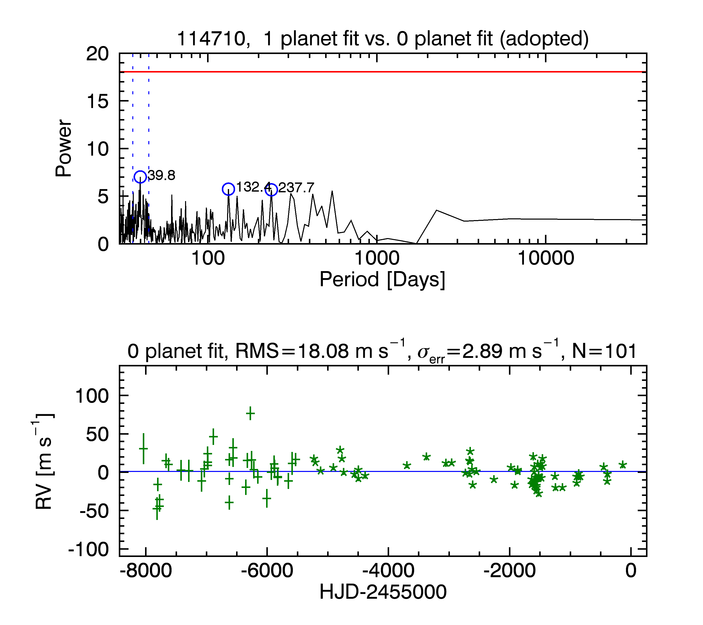}
\includegraphics[width=0.50\textwidth]{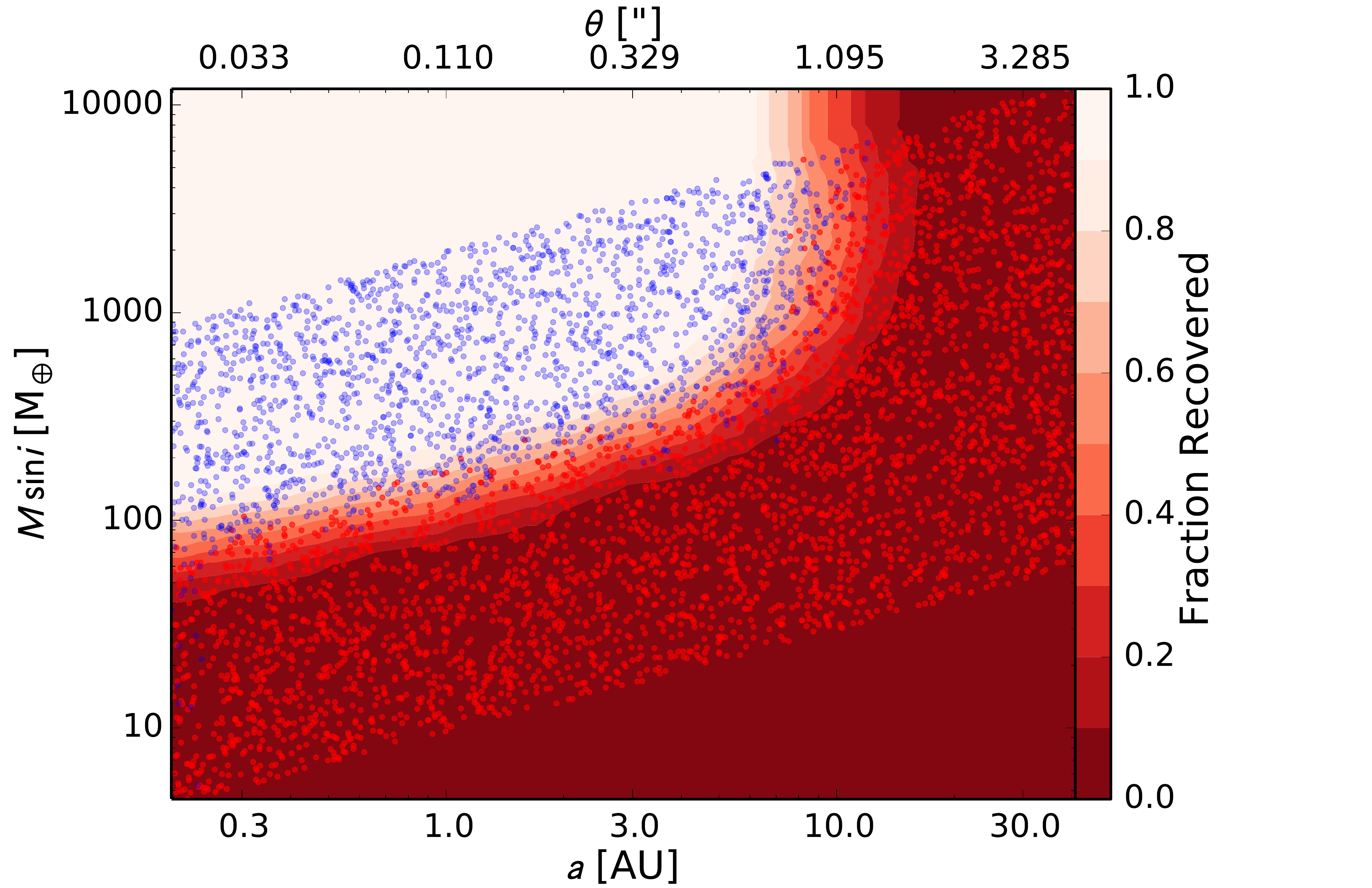}
\end{centering}
\caption{Results from an automated search for planets orbiting the star 
HD~114710 (HIP~64394; programs = S, C, A) 
based on RVs from Lick and/or Keck Observatory.
The set of plots on the left (analogous to Figures \ref{fig:search_example} and \ref{fig:search_example2}) 
show the planet search results 
and the plot on the right shows the completeness limits (analogous to Fig.\ \ref{fig:completeness_example}). 
See the captions of those figures for detailed descriptions.  
}
\label{fig:completeness_114710}
\end{figure}
\clearpage

\begin{figure}
\begin{centering}
\includegraphics[width=0.45\textwidth]{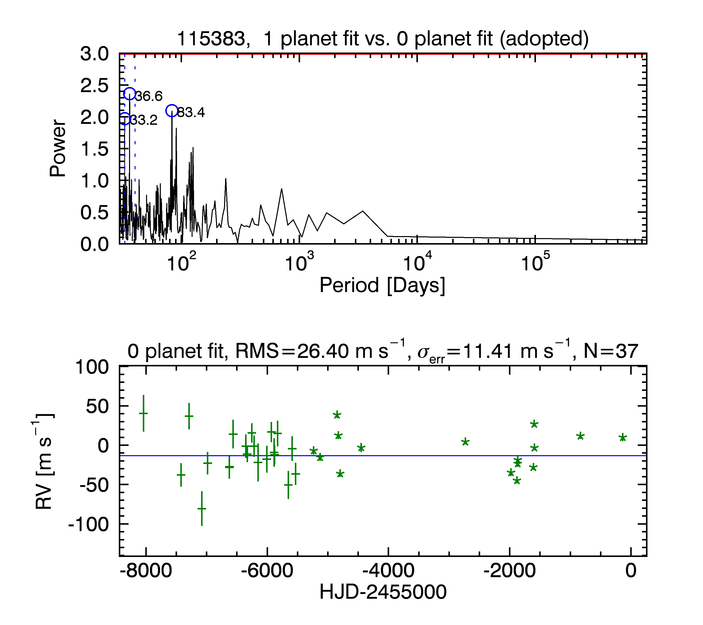}
\includegraphics[width=0.50\textwidth]{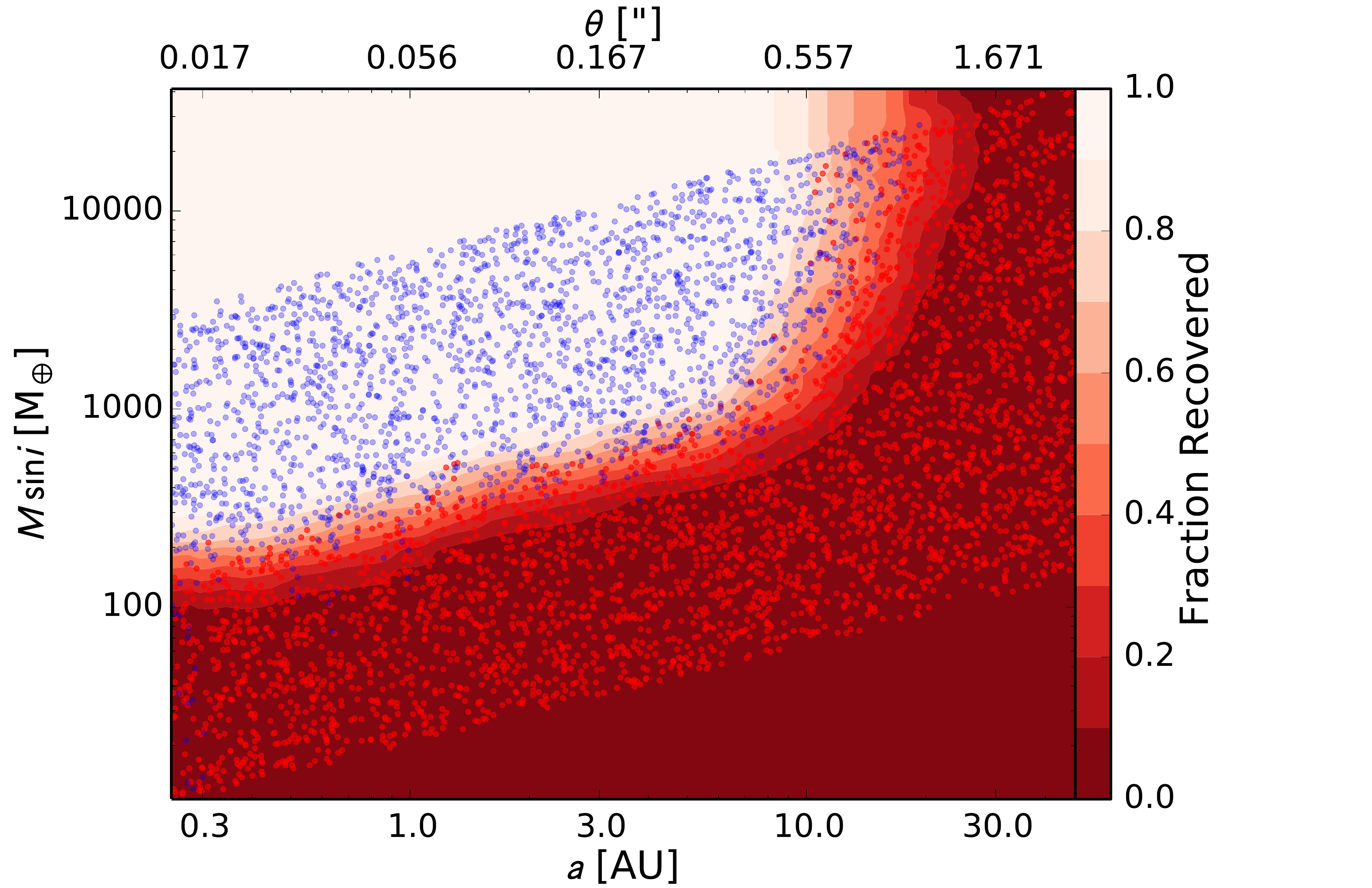}
\end{centering}
\caption{Results from an automated search for planets orbiting the star 
HD~115383 (HIP~64792; program = A) 
based on RVs from Lick and/or Keck Observatory.
The set of plots on the left (analogous to Figures \ref{fig:search_example} and \ref{fig:search_example2}) 
show the planet search results 
and the plot on the right shows the completeness limits (analogous to Fig.\ \ref{fig:completeness_example}). 
See the captions of those figures for detailed descriptions.  
}
\label{fig:completeness_115383}
\end{figure}

\begin{figure}
\begin{centering}
\includegraphics[width=0.45\textwidth]{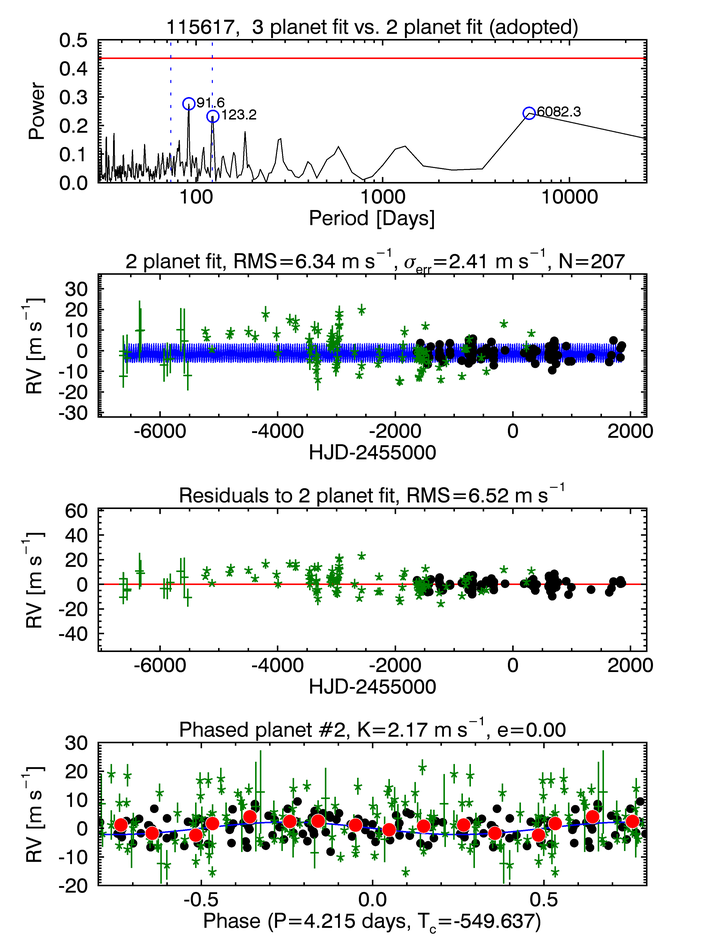}
\includegraphics[width=0.50\textwidth]{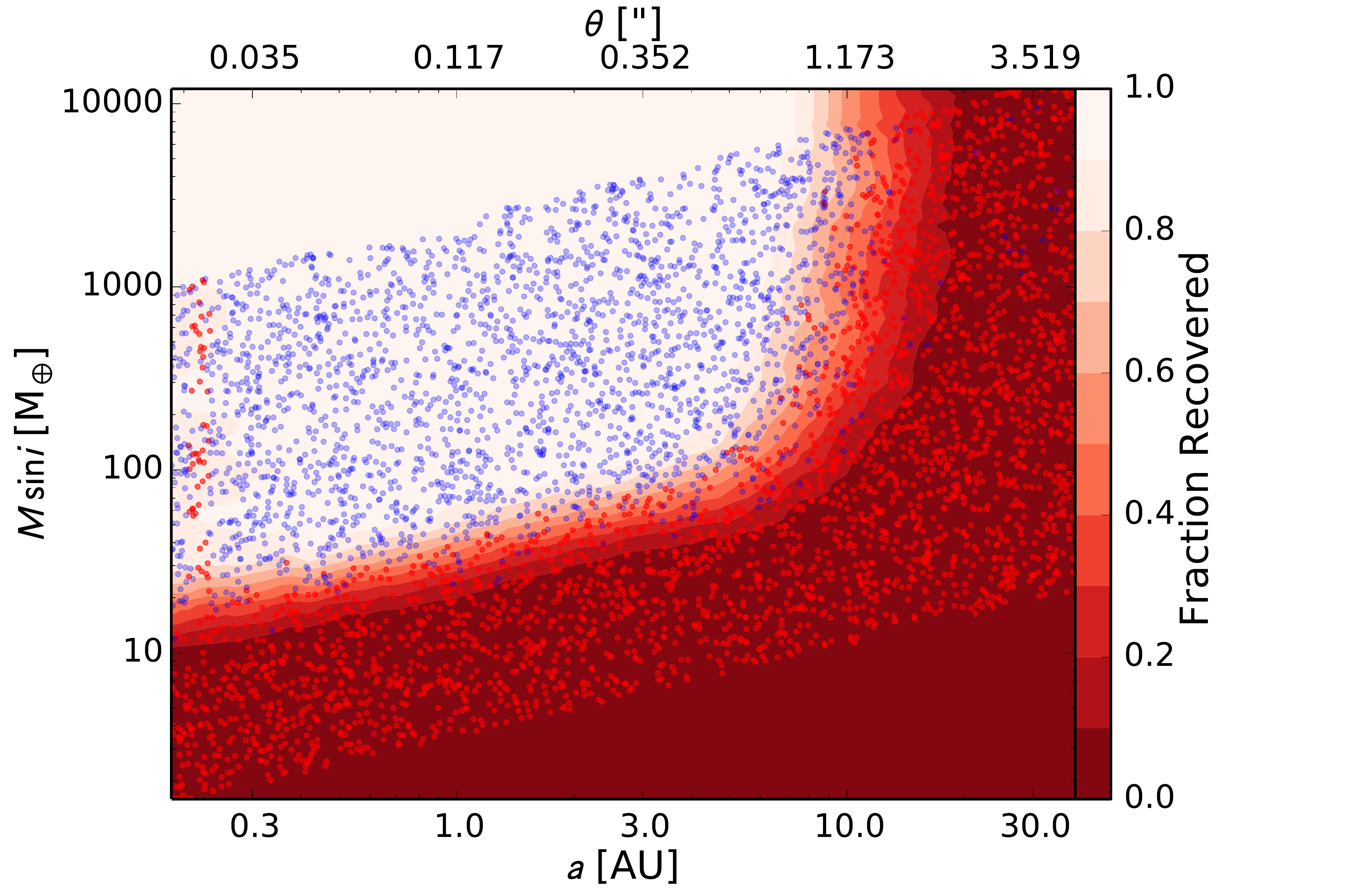}
\end{centering}
\caption{Results from an automated search for planets orbiting the star 
HD~115617 (HIP~64924; program = S, C, A) 
based on RVs from Lick and/or Keck Observatory.
The set of plots on the left (analogous to Figures \ref{fig:search_example} and \ref{fig:search_example2}) 
show the planet search results 
and the plot on the right shows the completeness limits (analogous to Fig.\ \ref{fig:completeness_example}). 
See the captions of those figures for detailed descriptions.  
This star has three reported small planets, two of which we detect in our Keck RVs.
}
\label{fig:completeness_115617}
\end{figure}
\clearpage

\begin{figure}
\begin{centering}
\includegraphics[width=0.45\textwidth]{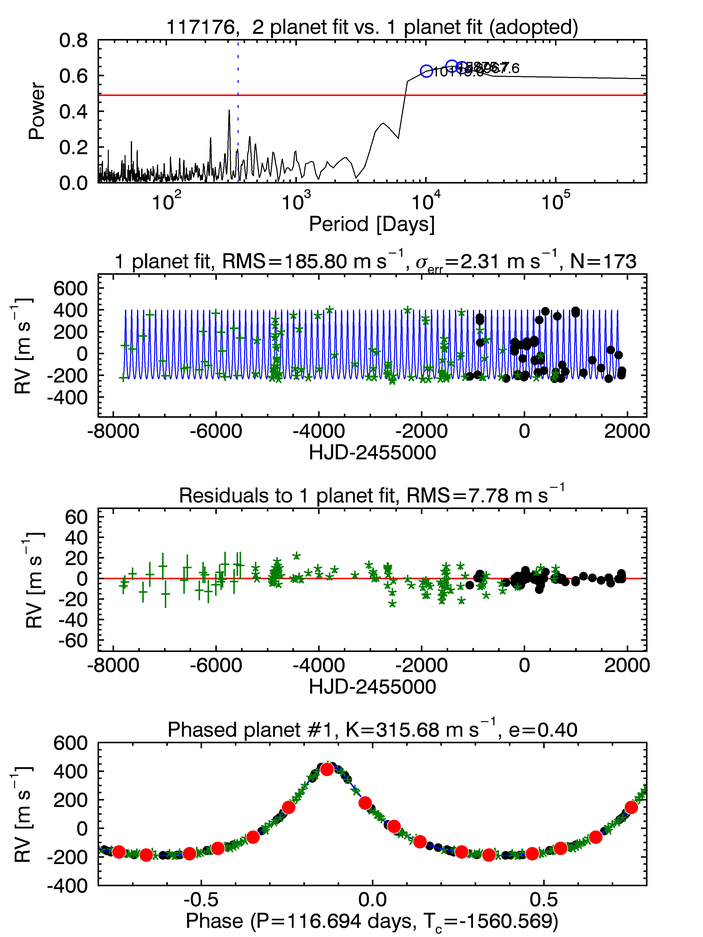}
\includegraphics[width=0.50\textwidth]{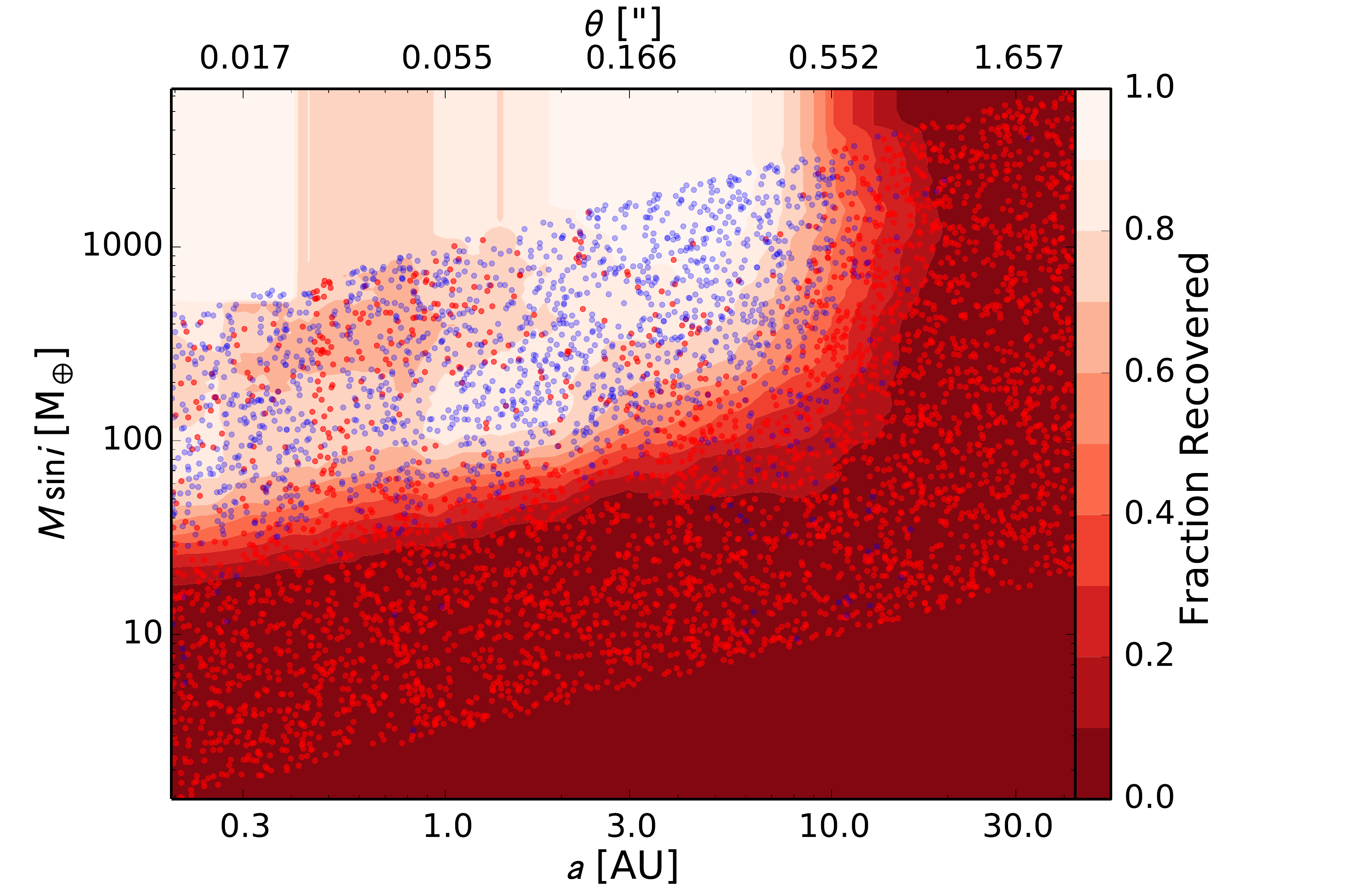}
\end{centering}
\caption{Results from an automated search for planets orbiting the star 
HD~117176 (HIP~65721; program = A) 
based on RVs from Lick and/or Keck Observatory.
The set of plots on the left (analogous to Figures \ref{fig:search_example} and \ref{fig:search_example2}) 
show the planet search results 
and the plot on the right shows the completeness limits (analogous to Fig.\ \ref{fig:completeness_example}). 
See the captions of those figures for detailed descriptions.  
This star hosts a giant planet.
}
\label{fig:completeness_117176}
\end{figure}

\begin{figure}
\begin{centering}
\includegraphics[width=0.45\textwidth]{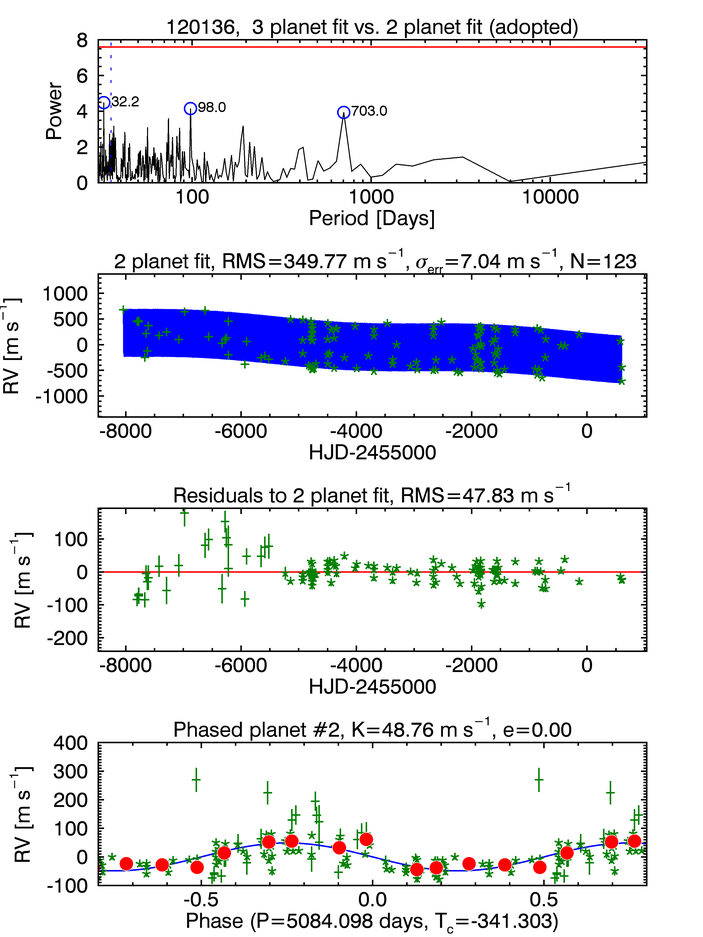}
\includegraphics[width=0.50\textwidth]{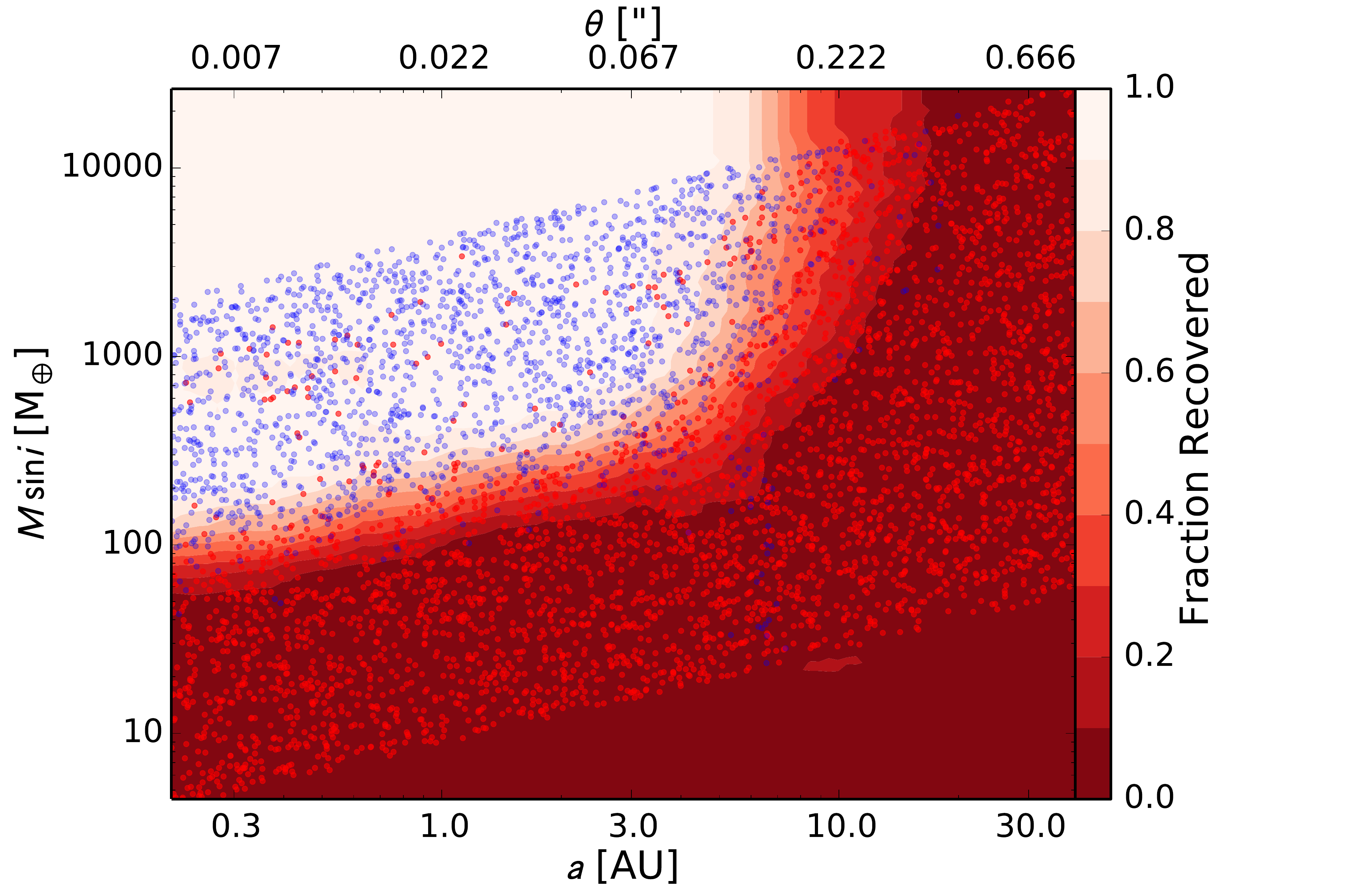}
\end{centering}
\caption{Results from an automated search for planets orbiting the star 
HD~120136 (HIP~67275; programs = S, C, A) 
based on RVs from Lick and/or Keck Observatory.
The set of plots on the left (analogous to Figures \ref{fig:search_example} and \ref{fig:search_example2}) 
show the planet search results 
and the plot on the right shows the completeness limits (analogous to Fig.\ \ref{fig:completeness_example}). 
See the captions of those figures for detailed descriptions.  
This star hosts a massive, hot Jupiter planet.  We also detect a significant linear trend in the RV time series and a periodicity at $\sim$5000 days. The trend is likely real, but the 5000 day periodicity could be due to instrumental offsets in the Lick-only data or a stellar magnetic activity cycle.
}
\label{fig:completeness_120136}
\end{figure}
\clearpage

\begin{figure}
\begin{centering}
\includegraphics[width=0.45\textwidth]{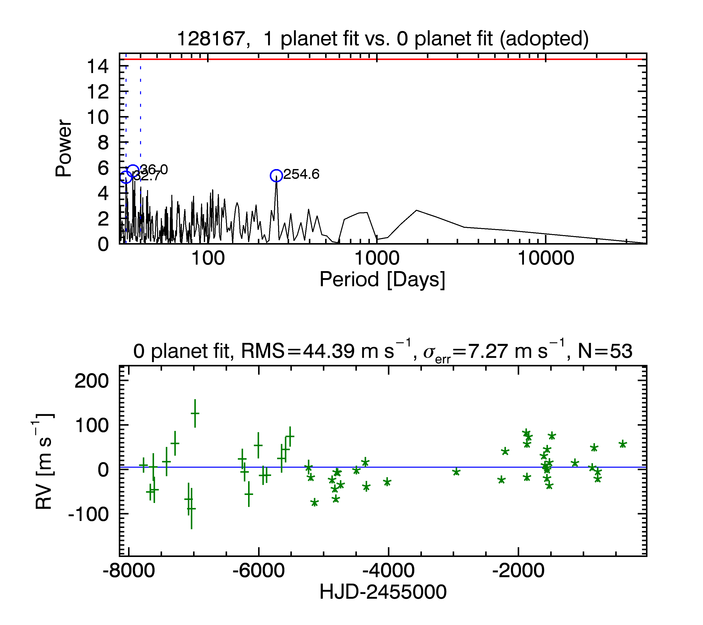}
\includegraphics[width=0.50\textwidth]{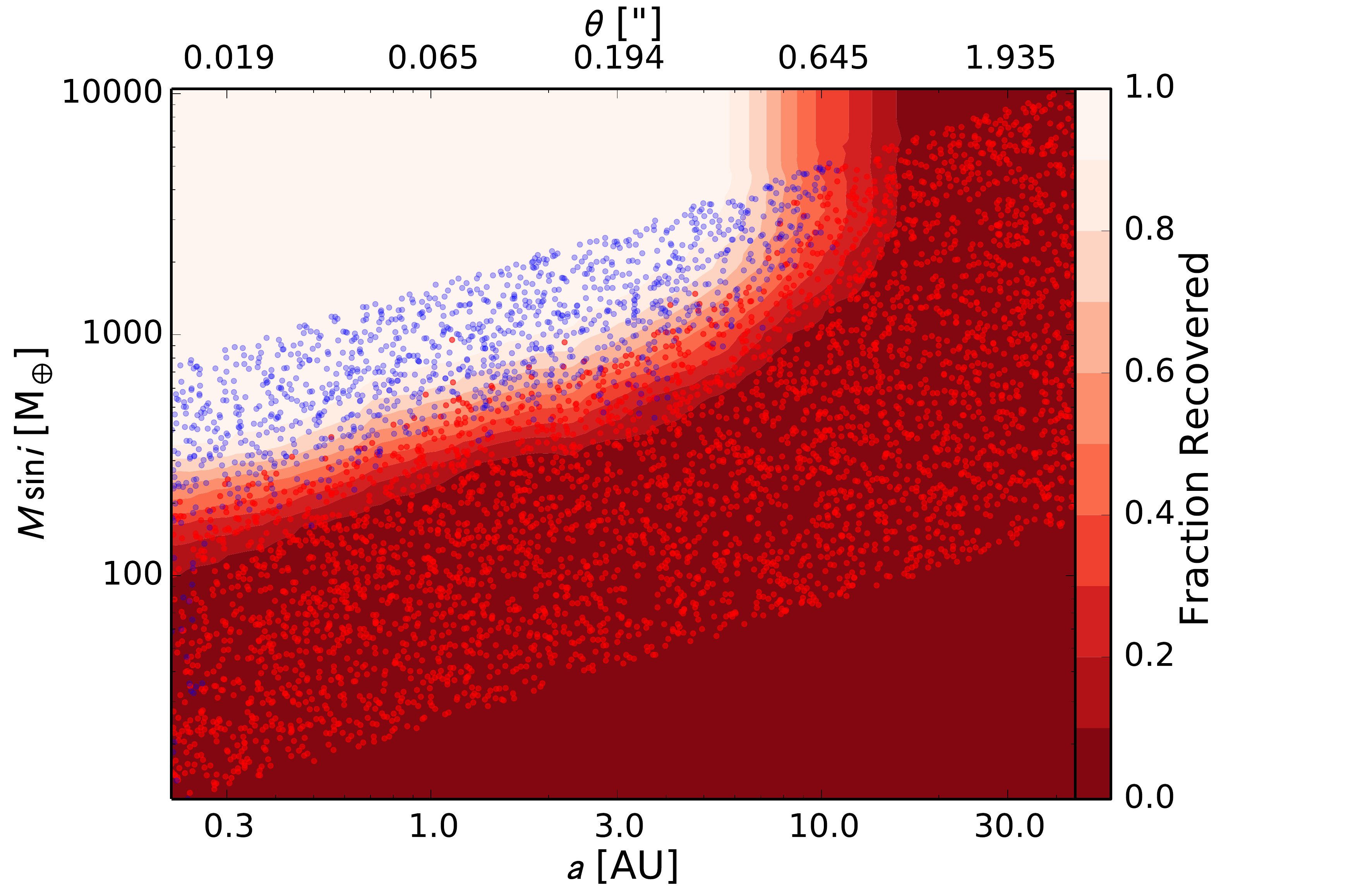}
\end{centering}
\caption{Results from an automated search for planets orbiting the star 
HD~128167 (HIP~71284; programs = S, C, A) 
based on RVs from Lick and/or Keck Observatory.
The set of plots on the left (analogous to Figures \ref{fig:search_example} and \ref{fig:search_example2}) 
show the planet search results 
and the plot on the right shows the completeness limits (analogous to Fig.\ \ref{fig:completeness_example}). 
See the captions of those figures for detailed descriptions.  
}
\label{fig:completeness_128167}
\end{figure}

\begin{figure}
\begin{centering}
\includegraphics[width=0.45\textwidth]{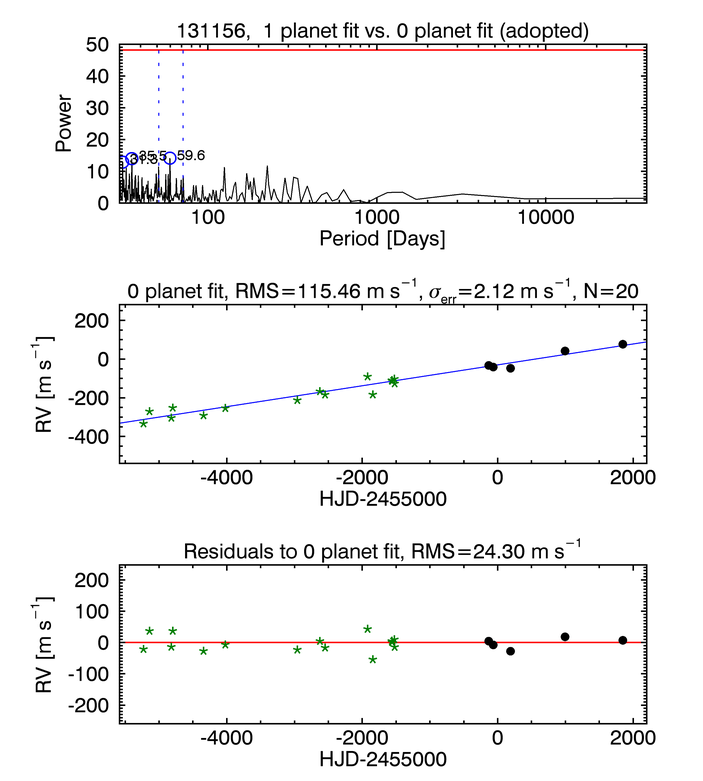}
\includegraphics[width=0.50\textwidth]{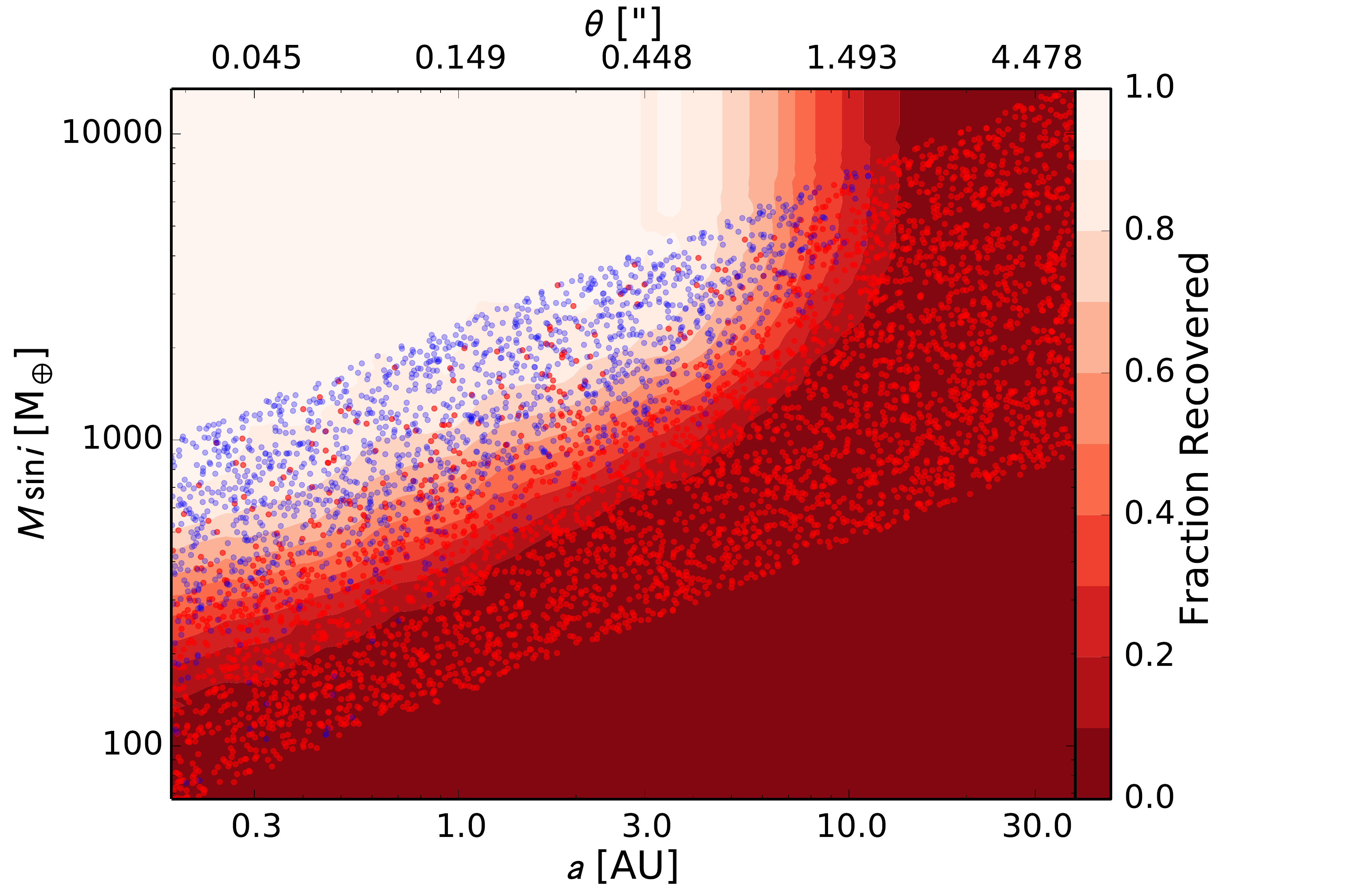}
\end{centering}
\caption{Results from an automated search for planets orbiting the star 
HD~131156 (HIP~72659; programs = S, C, A) 
based on RVs from Lick and/or Keck Observatory.
The set of plots on the left (analogous to Figures \ref{fig:search_example} and \ref{fig:search_example2}) 
show the planet search results 
and the plot on the right shows the completeness limits (analogous to Fig.\ \ref{fig:completeness_example}). 
See the captions of those figures for detailed descriptions.  
This star has a significant linear trend in the RV time series, likely due to a binary companion.
}
\label{fig:completeness_131156}
\end{figure}
\clearpage

\begin{figure}
\begin{centering}
\includegraphics[width=0.45\textwidth]{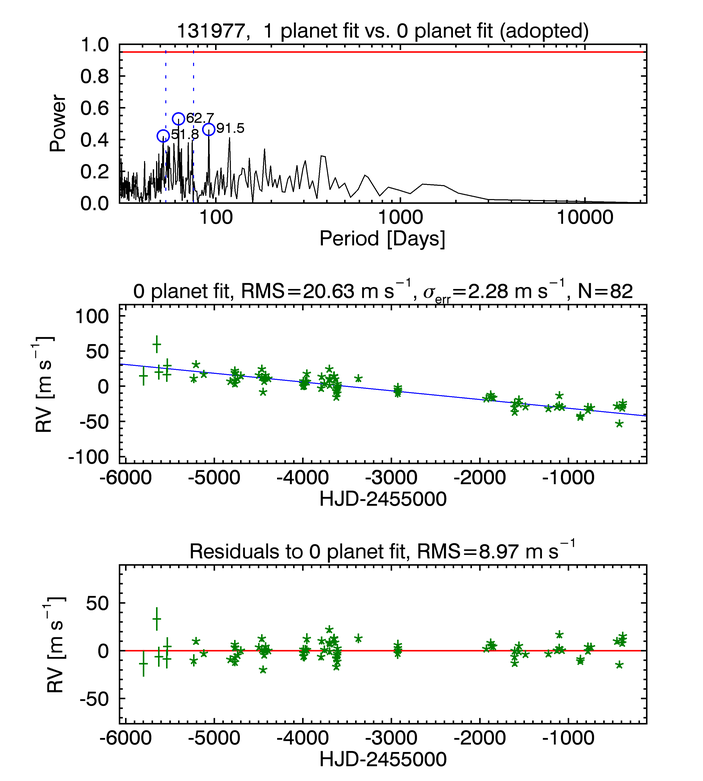}
\includegraphics[width=0.50\textwidth]{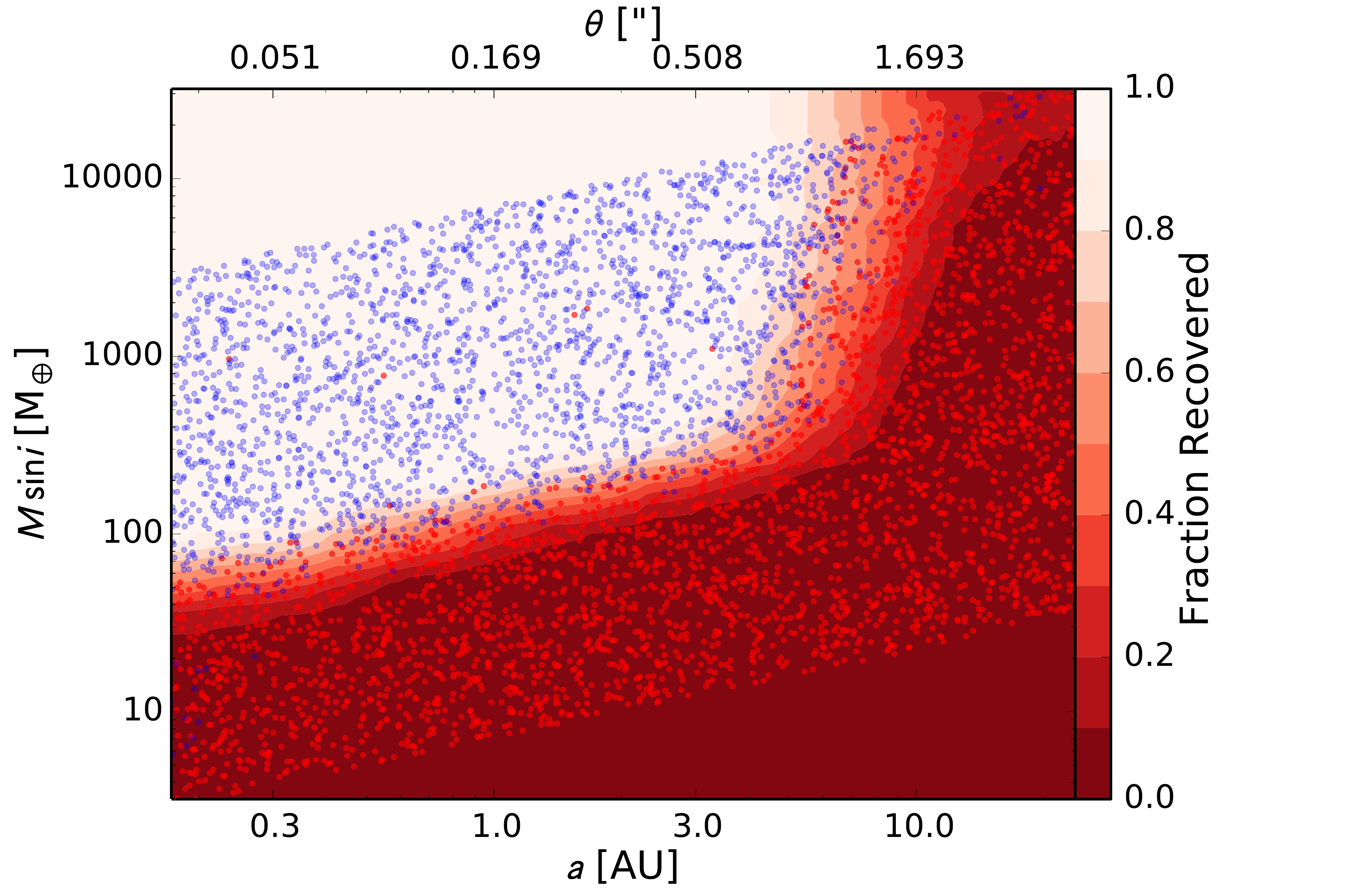}
\end{centering}
\caption{Results from an automated search for planets orbiting the star 
HD~131977 (HIP~73184; programs = S, C) 
based on RVs from Lick and/or Keck Observatory.
The set of plots on the left (analogous to Figures \ref{fig:search_example} and \ref{fig:search_example2}) 
show the planet search results 
and the plot on the right shows the completeness limits (analogous to Fig.\ \ref{fig:completeness_example}). 
See the captions of those figures for detailed descriptions.  
This star has a significant linear trend in the RV time series, likely due to a binary companion.
}
\label{fig:completeness_131977}
\end{figure}

\begin{figure}
\begin{centering}
\includegraphics[width=0.45\textwidth]{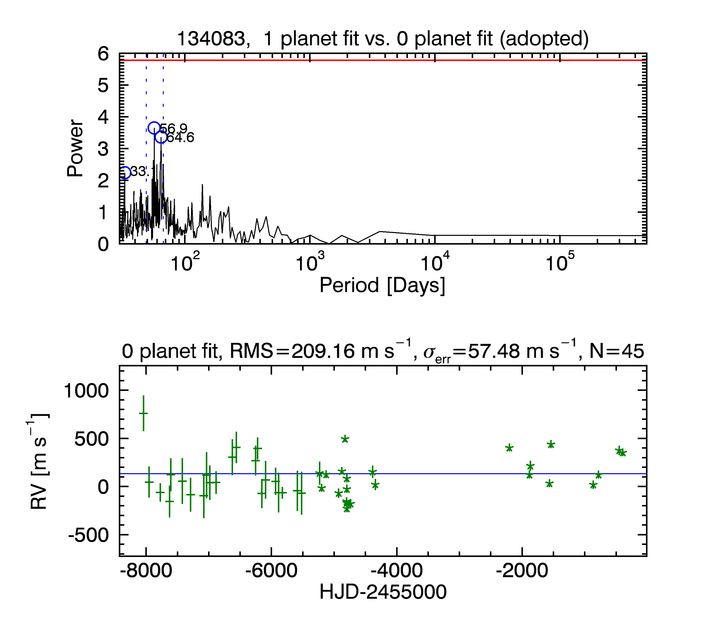}
\includegraphics[width=0.50\textwidth]{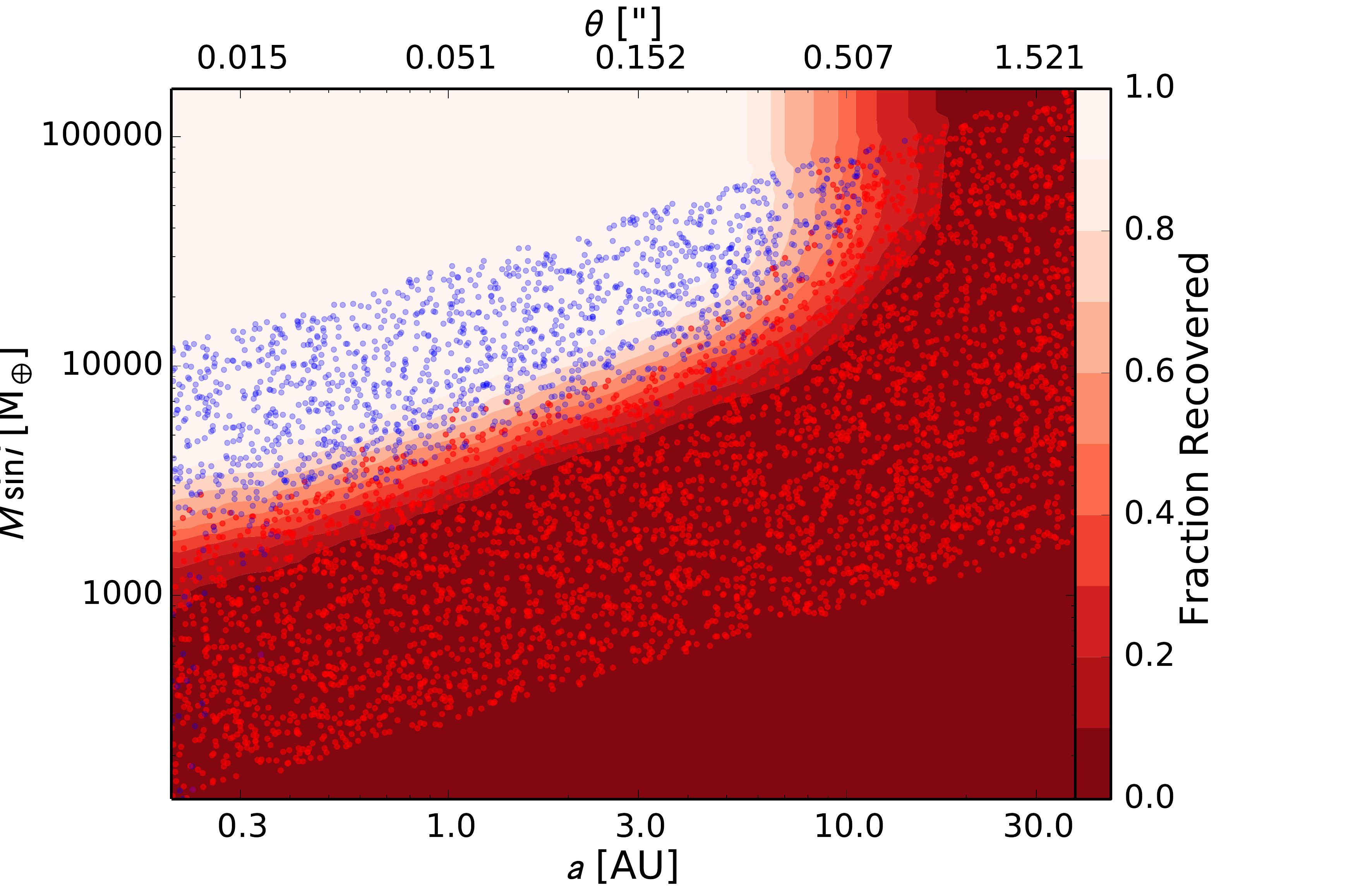}
\end{centering}
\caption{Results from an automated search for planets orbiting the star 
HD~134083 (HIP~73996; program = A) 
based on RVs from Lick and/or Keck Observatory.
The set of plots on the left (analogous to Figures \ref{fig:search_example} and \ref{fig:search_example2}) 
show the planet search results 
and the plot on the right shows the completeness limits (analogous to Fig.\ \ref{fig:completeness_example}). 
See the captions of those figures for detailed descriptions.  
}
\end{figure}
\label{fig:completeness_134083}
\clearpage

\begin{figure}
\begin{centering}
\includegraphics[width=0.45\textwidth]{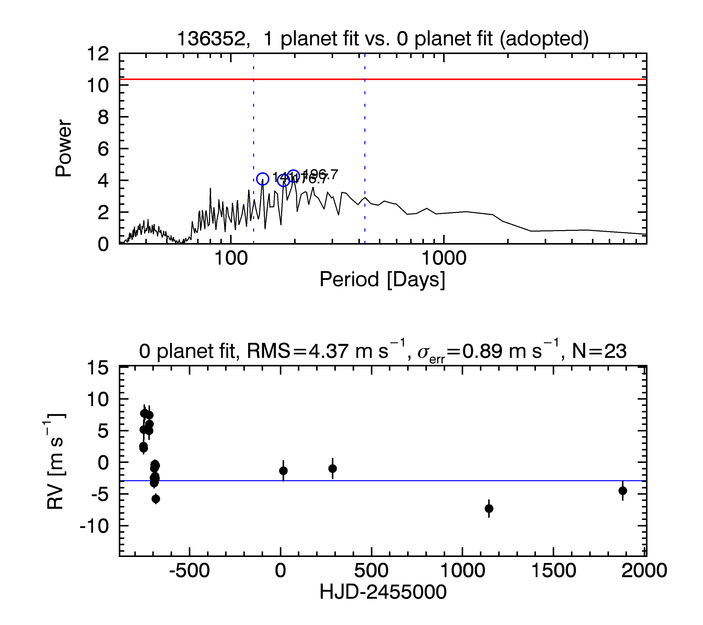}
\includegraphics[width=0.50\textwidth]{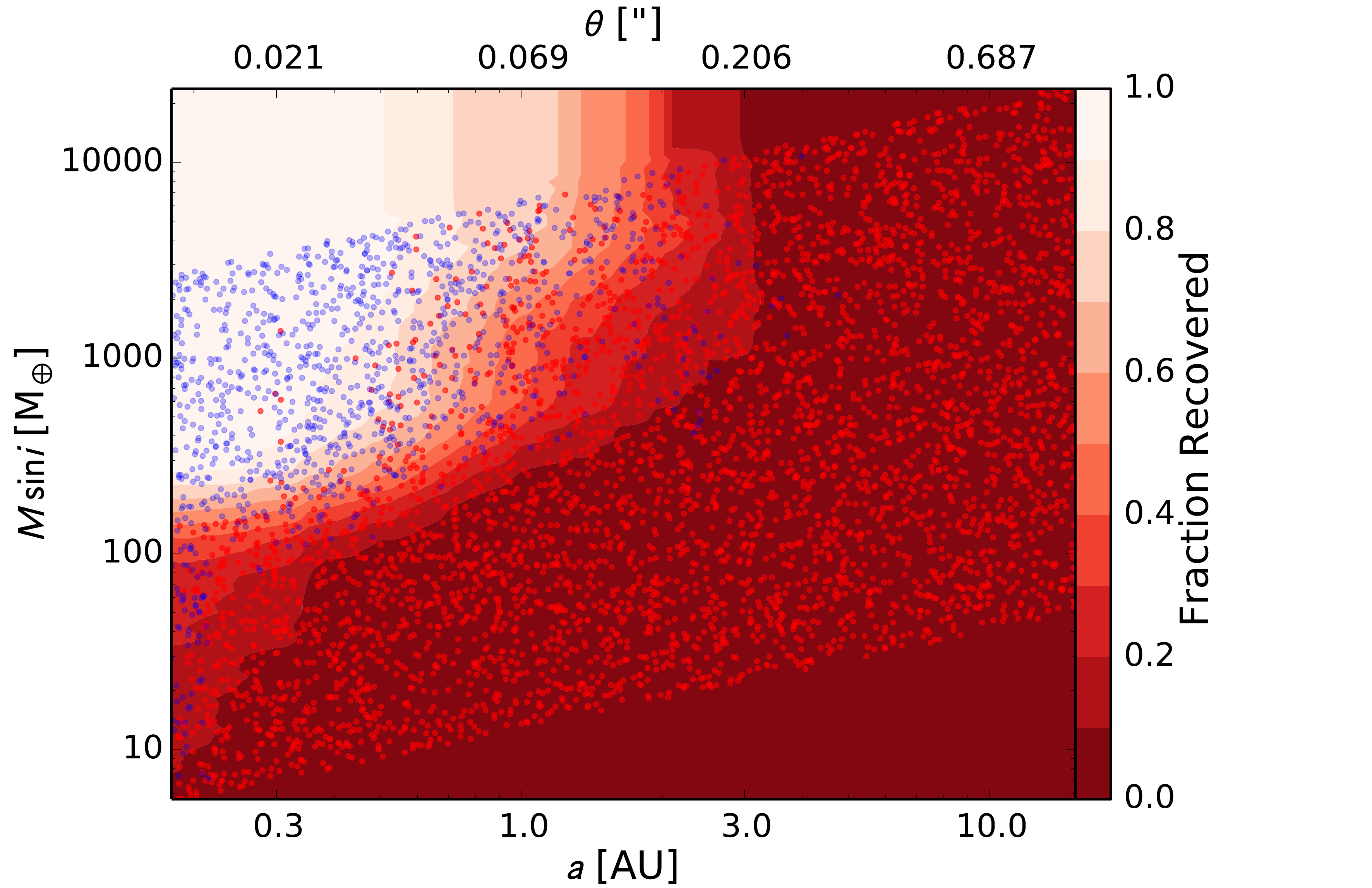}
\end{centering}
\caption{Results from an automated search for planets orbiting the star 
HD~136352 (HIP~75181; program = S) 
based on RVs from Lick and/or Keck Observatory.
The set of plots on the left (analogous to Figures \ref{fig:search_example} and \ref{fig:search_example2}) 
show the planet search results 
and the plot on the right shows the completeness limits (analogous to Fig.\ \ref{fig:completeness_example}). 
See the captions of those figures for detailed descriptions.  
}
\label{fig:completeness_136352}
\end{figure}

\begin{figure}
\begin{centering}
\includegraphics[width=0.45\textwidth]{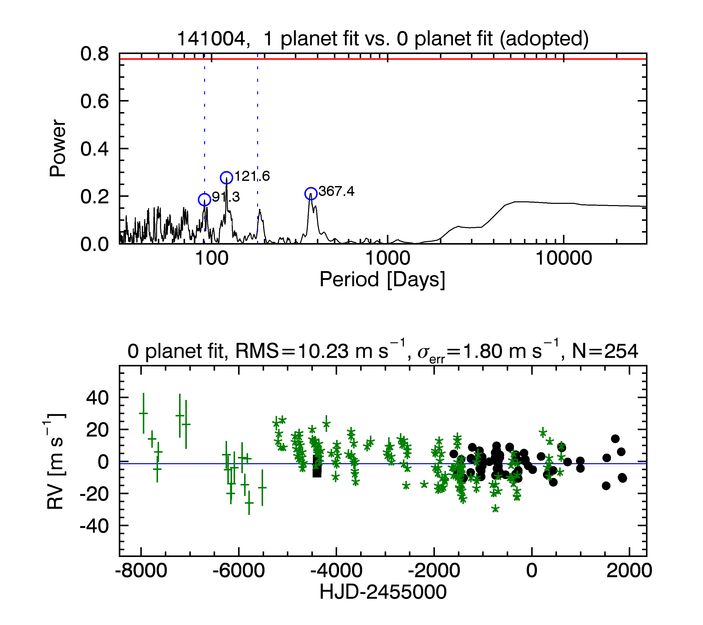}
\includegraphics[width=0.50\textwidth]{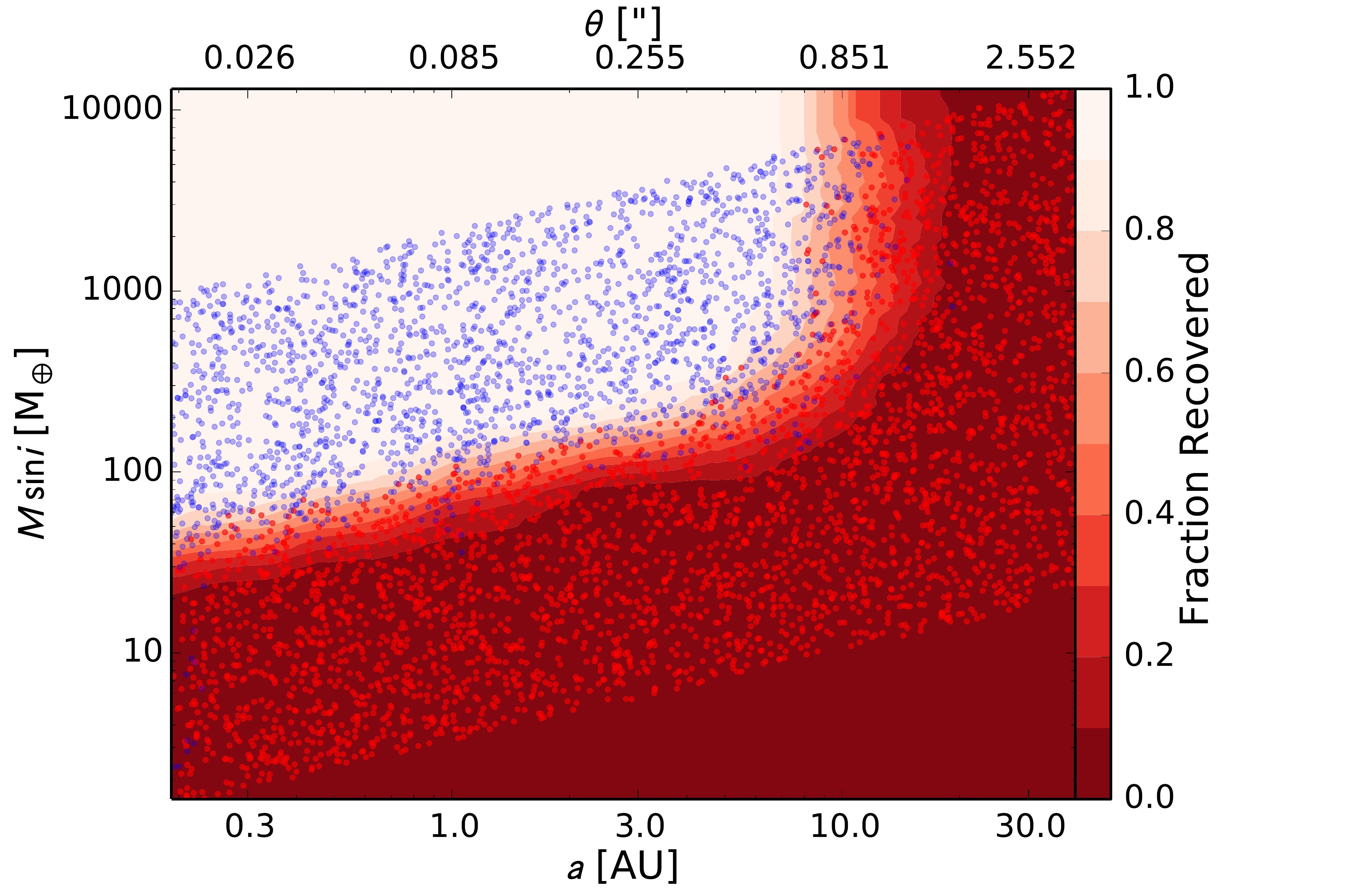}
\end{centering}
\caption{Results from an automated search for planets orbiting the star 
HD~141004 (HIP~77257; programs = S, C, A) 
based on RVs from Lick and/or Keck Observatory.
The set of plots on the left (analogous to Figures \ref{fig:search_example} and \ref{fig:search_example2}) 
show the planet search results 
and the plot on the right shows the completeness limits (analogous to Fig.\ \ref{fig:completeness_example}). 
See the captions of those figures for detailed descriptions.  
}
\end{figure}
\label{fig:completeness_141004}
\clearpage

\begin{figure}
\begin{centering}
\includegraphics[width=0.45\textwidth]{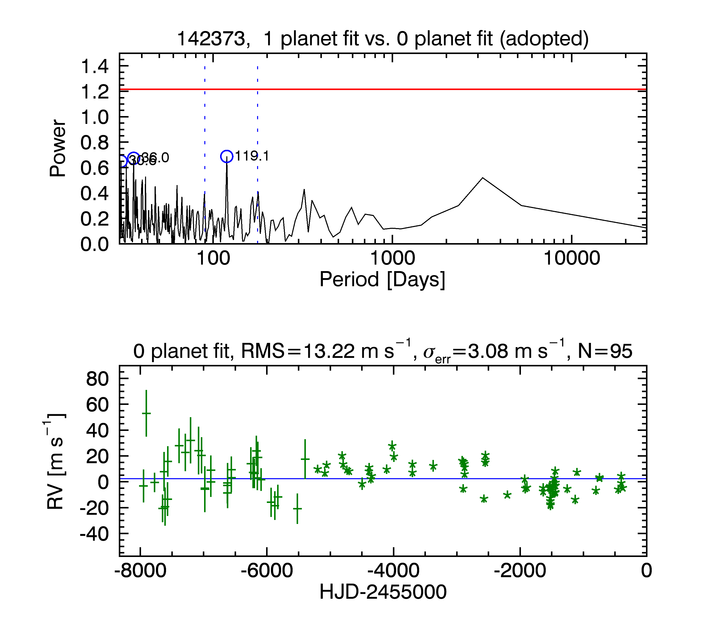}
\includegraphics[width=0.50\textwidth]{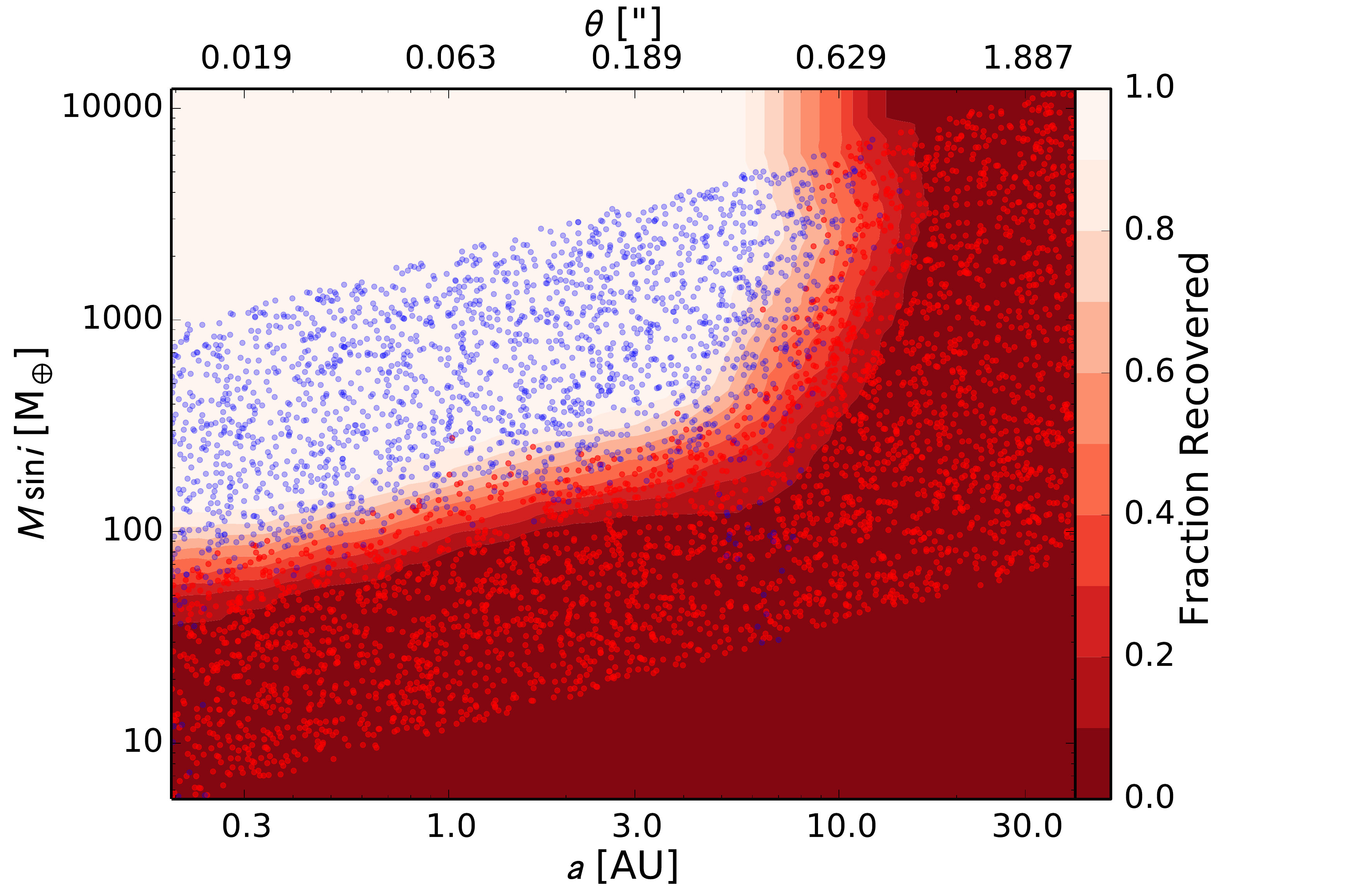}
\end{centering}
\caption{Results from an automated search for planets orbiting the star 
HD~142373 (HIP~77760; programs = S, C, A) 
based on RVs from Lick and/or Keck Observatory.
The set of plots on the left (analogous to Figures \ref{fig:search_example} and \ref{fig:search_example2}) 
show the planet search results 
and the plot on the right shows the completeness limits (analogous to Fig.\ \ref{fig:completeness_example}). 
See the captions of those figures for detailed descriptions.  
}
\label{fig:completeness_142373}
\end{figure}

\begin{figure}
\begin{centering}
\includegraphics[width=0.45\textwidth]{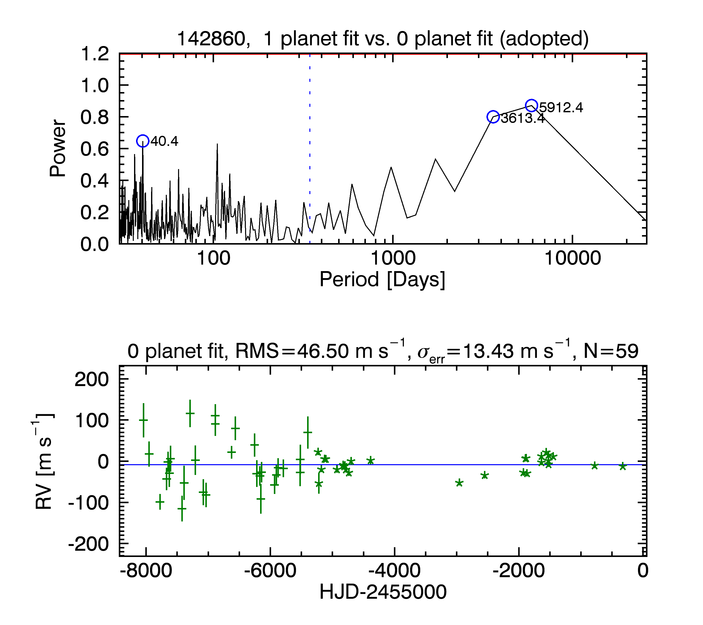}
\includegraphics[width=0.50\textwidth]{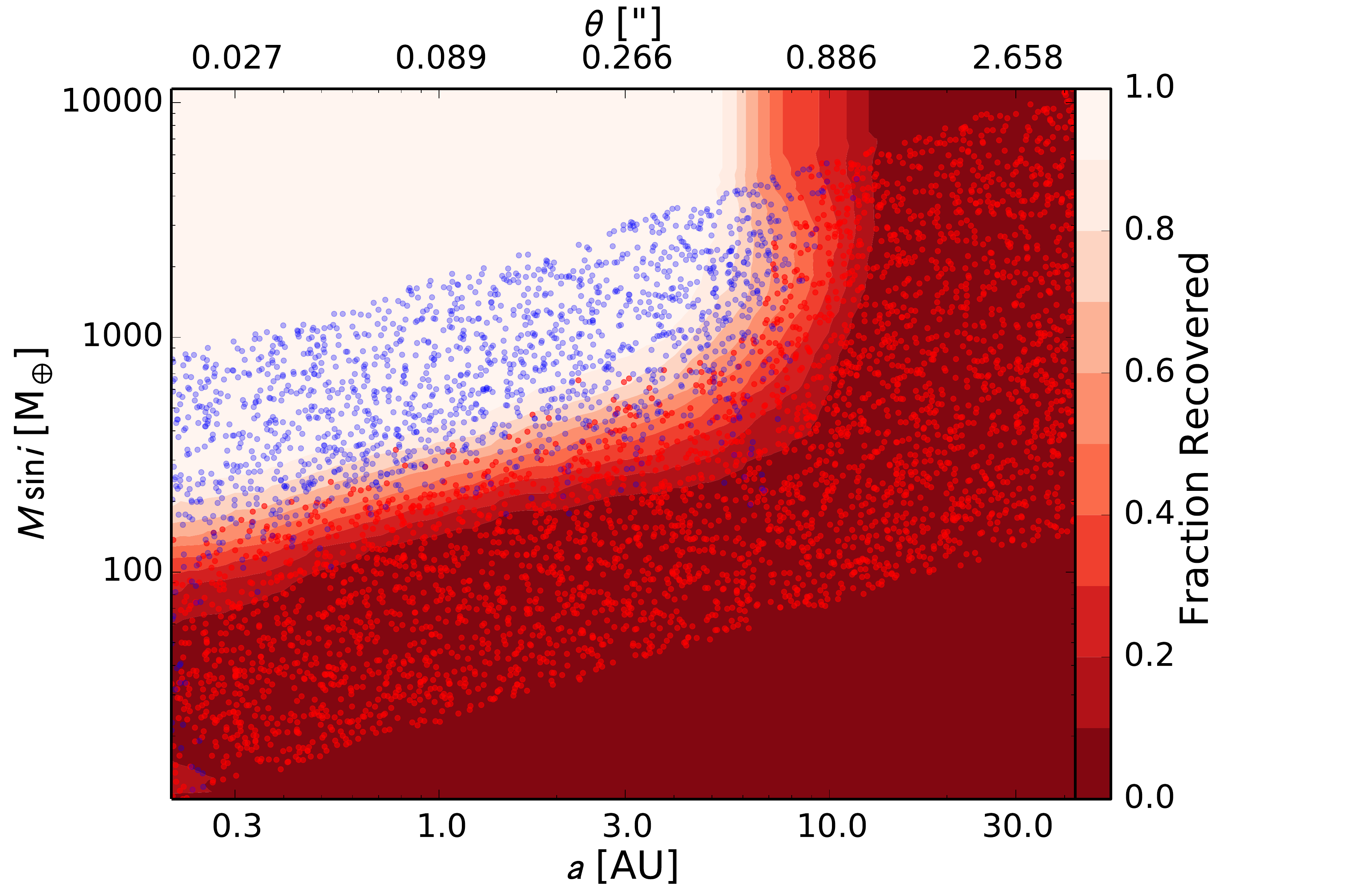}
\end{centering}
\caption{Results from an automated search for planets orbiting the star 
HD~142860 (HIP~78072; programs = S, C, A) 
based on RVs from Lick and/or Keck Observatory.
The set of plots on the left (analogous to Figures \ref{fig:search_example} and \ref{fig:search_example2}) 
show the planet search results 
and the plot on the right shows the completeness limits (analogous to Fig.\ \ref{fig:completeness_example}). 
See the captions of those figures for detailed descriptions.  
}
\end{figure}
\label{fig:completeness_142860}
\clearpage

\begin{figure}
\begin{centering}
\includegraphics[width=0.45\textwidth]{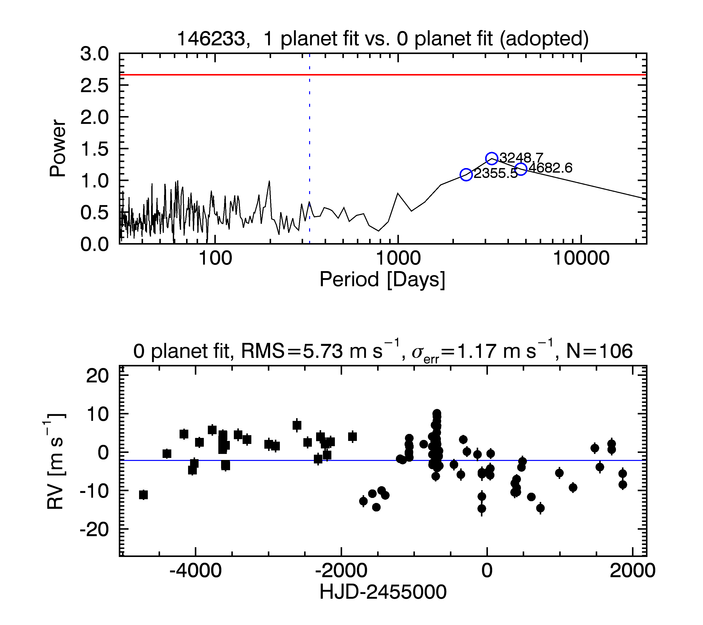}
\includegraphics[width=0.50\textwidth]{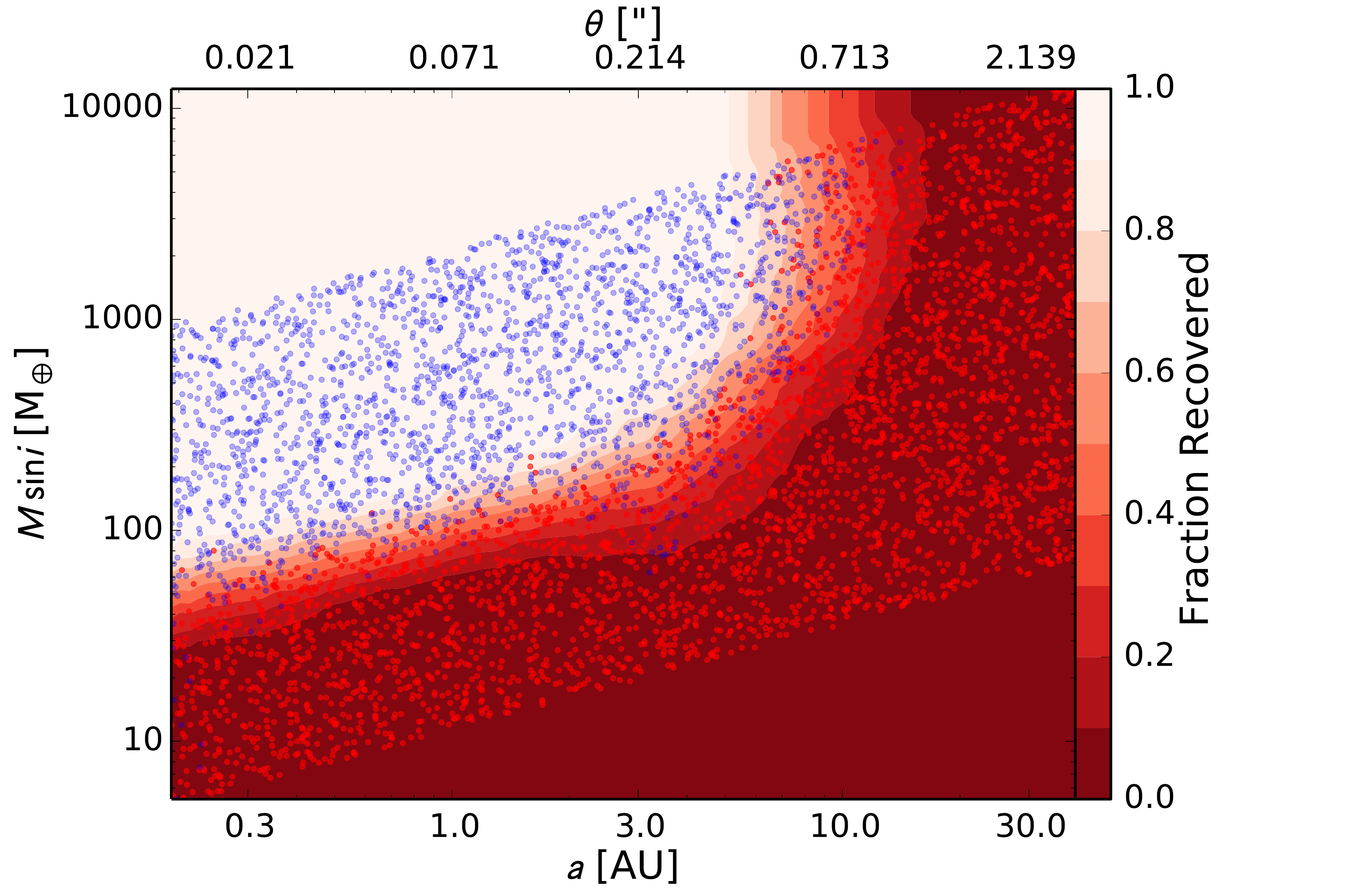}
\end{centering}
\caption{Results from an automated search for planets orbiting the star 
HD~146233 (HIP~79672; program = S) 
based on RVs from Lick and/or Keck Observatory.
The set of plots on the left (analogous to Figures \ref{fig:search_example} and \ref{fig:search_example2}) 
show the planet search results 
and the plot on the right shows the completeness limits (analogous to Fig.\ \ref{fig:completeness_example}). 
See the captions of those figures for detailed descriptions.  
}
\label{fig:completeness_146233}
\end{figure}

\begin{figure}
\begin{centering}
\includegraphics[width=0.45\textwidth]{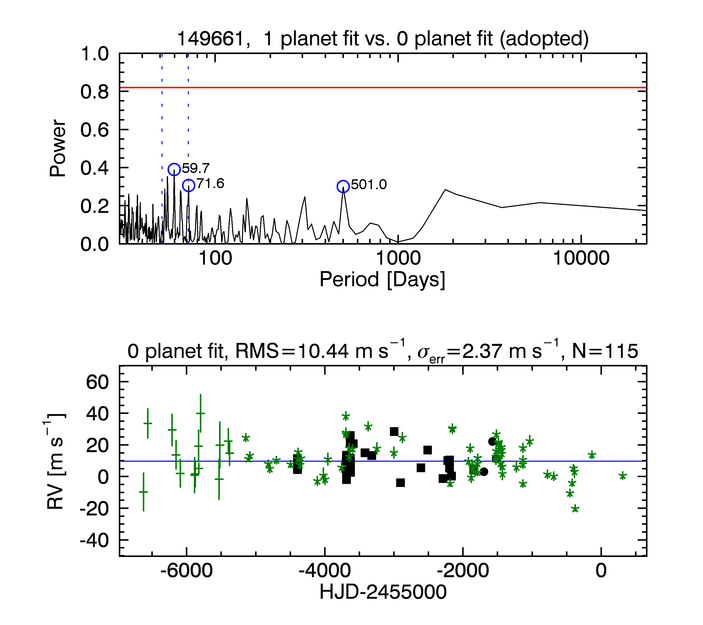}
\includegraphics[width=0.50\textwidth]{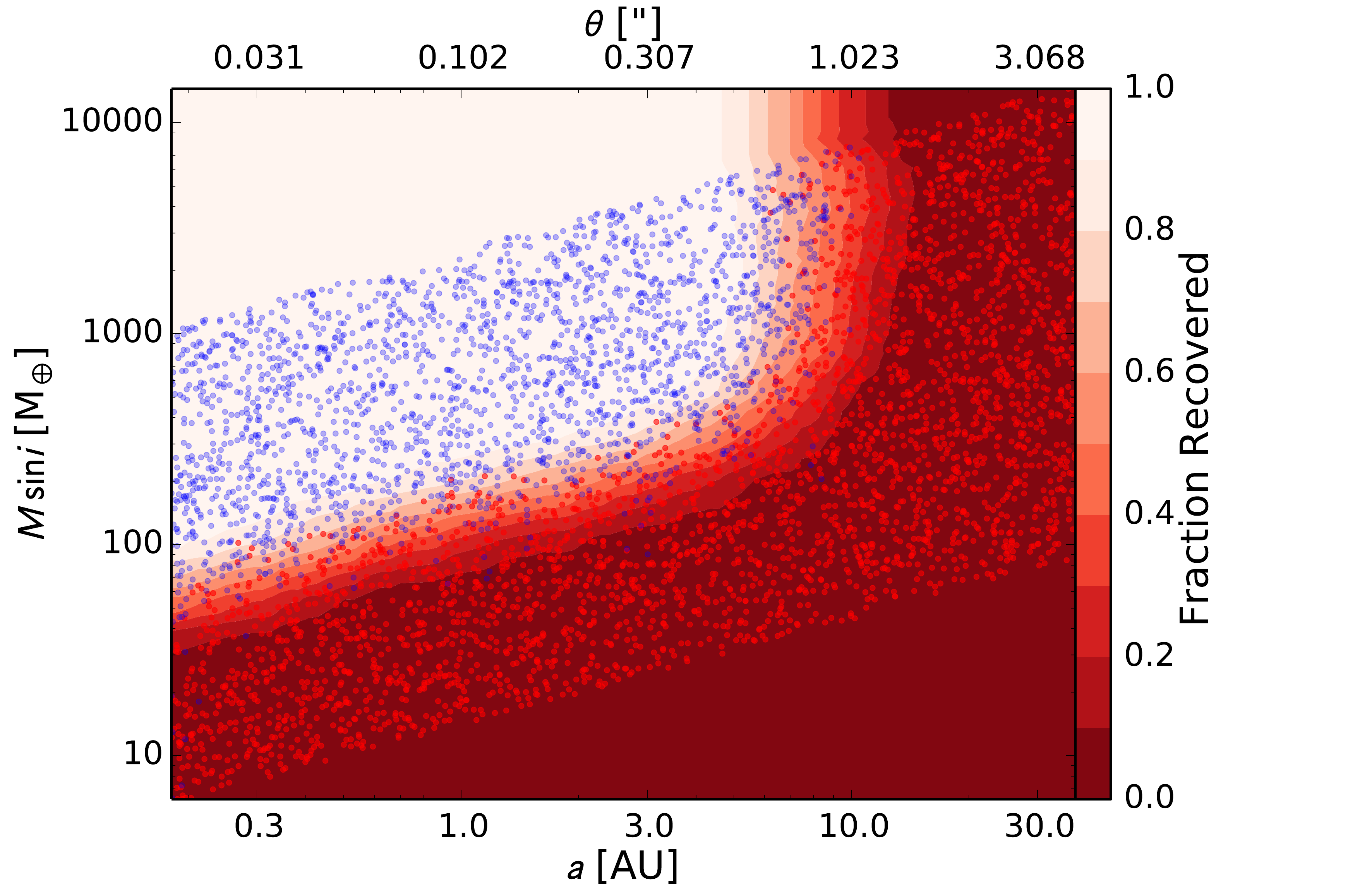}
\end{centering}
\caption{Results from an automated search for planets orbiting the star 
HD~149661 (HIP~81300; program = S) 
based on RVs from Lick and/or Keck Observatory.
The set of plots on the left (analogous to Figures \ref{fig:search_example} and \ref{fig:search_example2}) 
show the planet search results 
and the plot on the right shows the completeness limits (analogous to Fig.\ \ref{fig:completeness_example}). 
See the captions of those figures for detailed descriptions.  
}
\end{figure}
\label{fig:completeness_149661}
\clearpage

\begin{figure}
\begin{centering}
\includegraphics[width=0.45\textwidth]{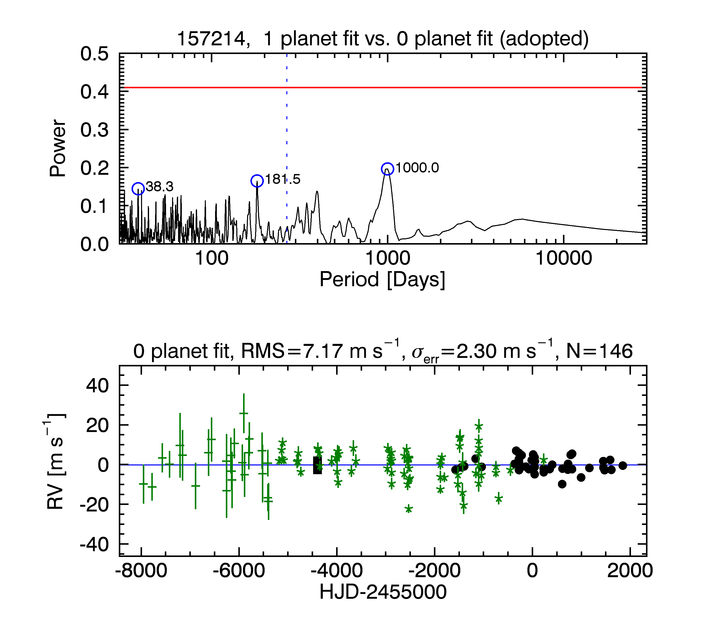}
\includegraphics[width=0.50\textwidth]{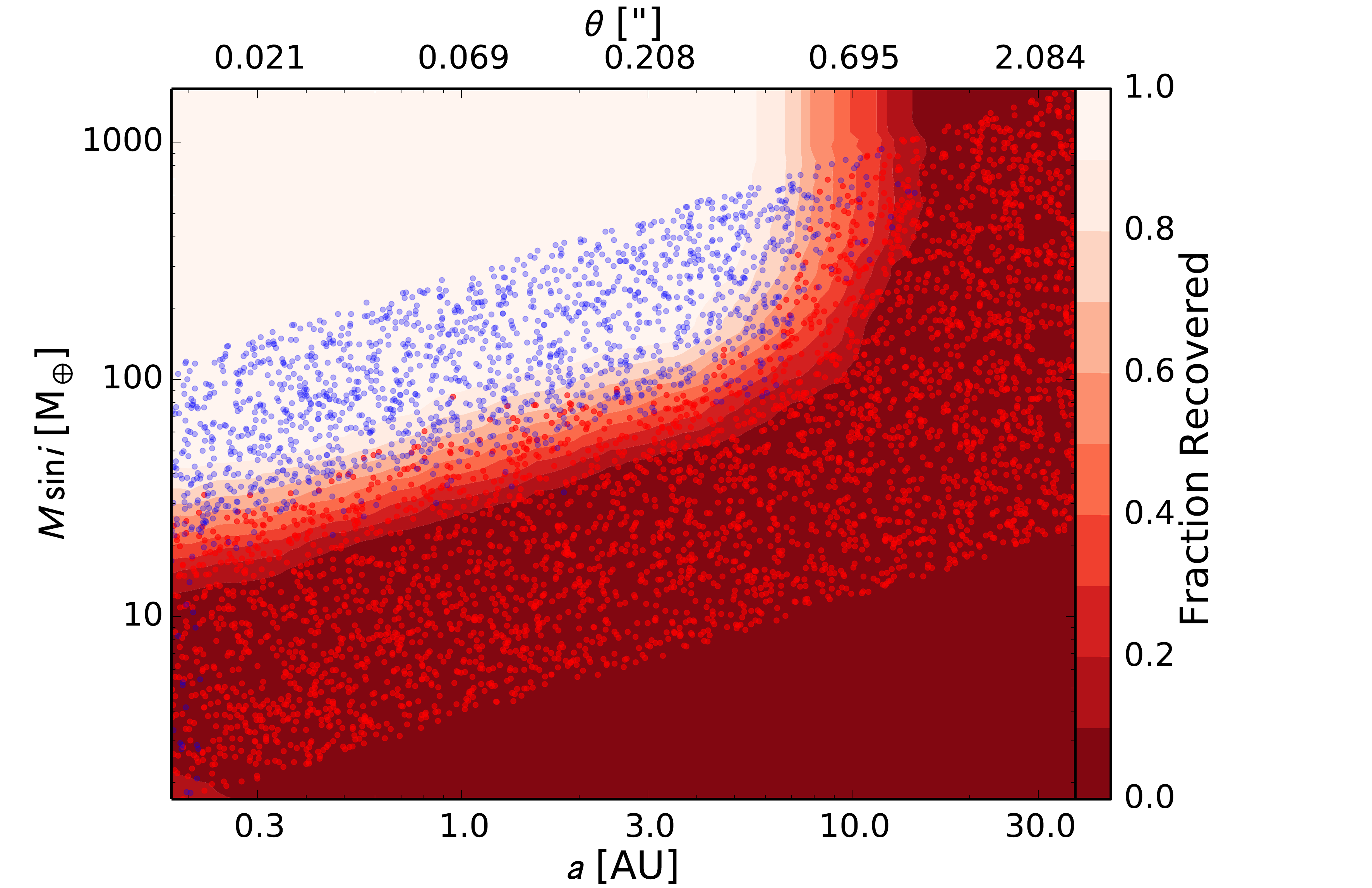}
\end{centering}
\caption{Results from an automated search for planets orbiting the star 
HD~157214 (HIP~84862; programs = S, A) 
based on RVs from Lick and/or Keck Observatory.
The set of plots on the left (analogous to Figures \ref{fig:search_example} and \ref{fig:search_example2}) 
show the planet search results 
and the plot on the right shows the completeness limits (analogous to Fig.\ \ref{fig:completeness_example}). 
See the captions of those figures for detailed descriptions.  
}
\label{fig:completeness_157214}
\end{figure}

\begin{figure}
\begin{centering}
\includegraphics[width=0.45\textwidth]{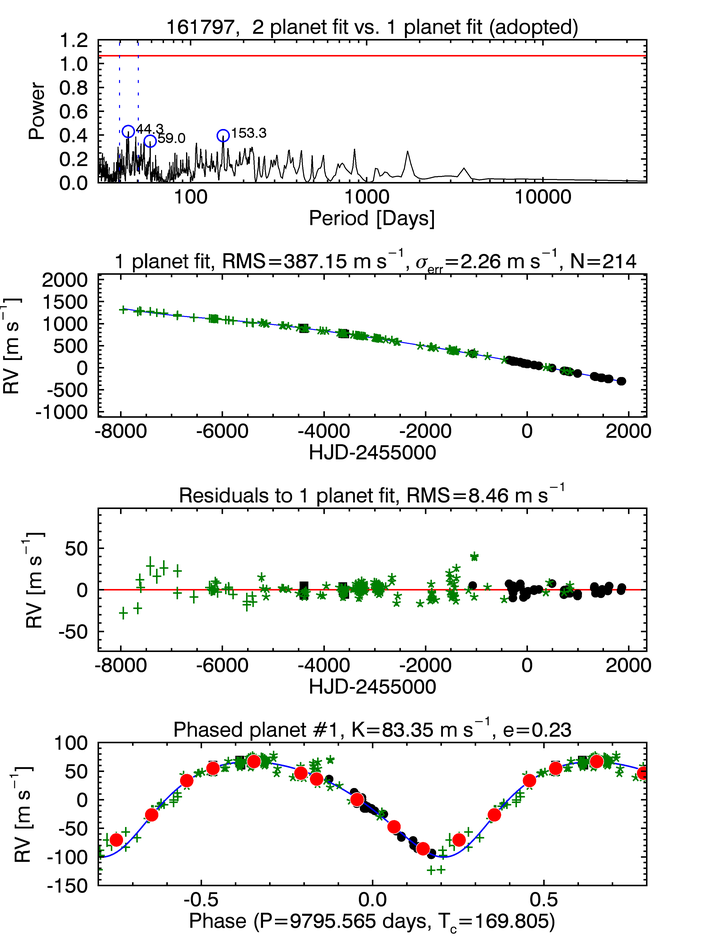}
\includegraphics[width=0.50\textwidth]{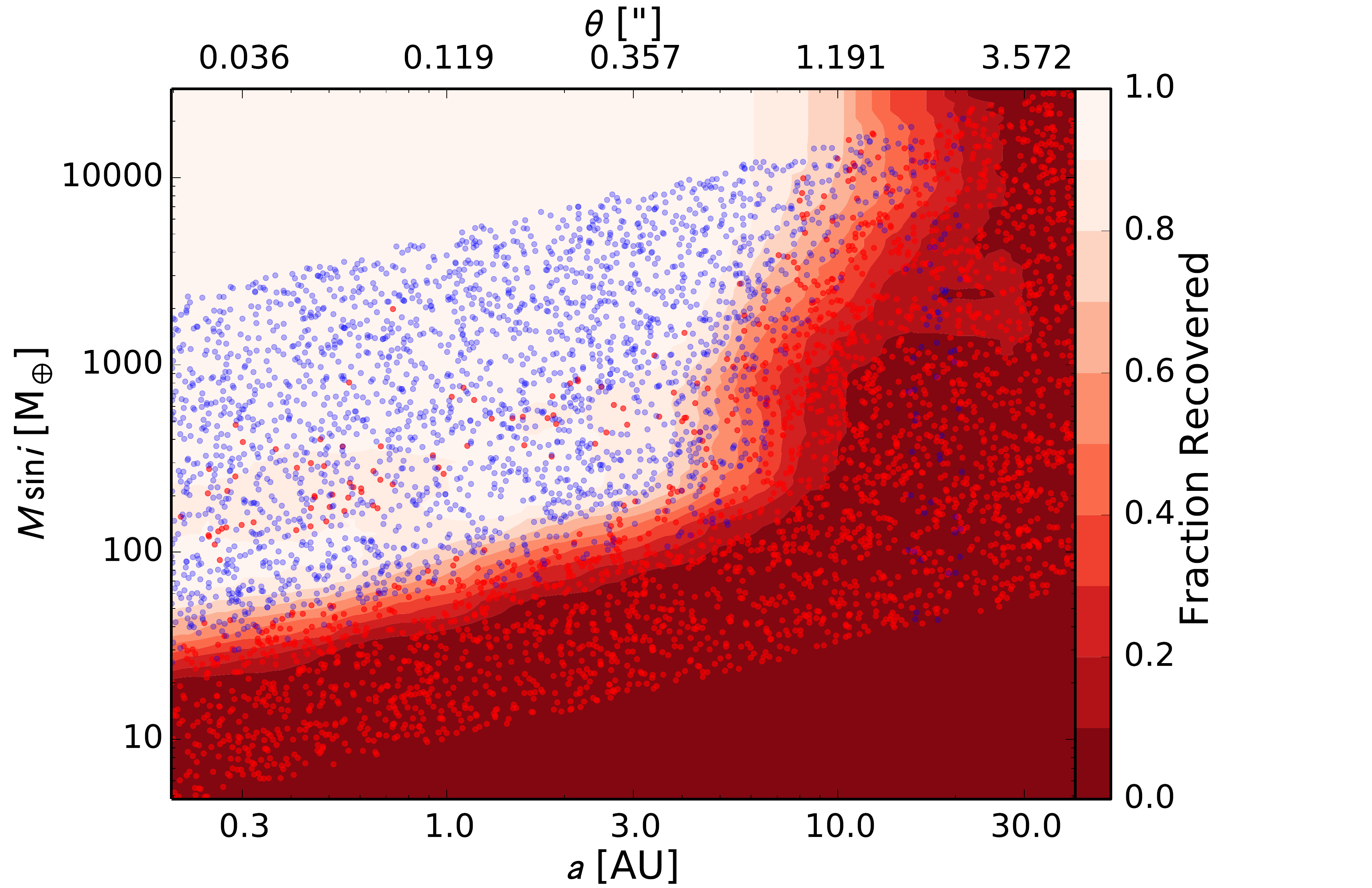}
\end{centering}
\caption{Results from an automated search for planets orbiting the star 
HD~161797 (HIP~86974; programs = S, C, A) 
based on RVs from Lick and/or Keck Observatory.
The set of plots on the left (analogous to Figures \ref{fig:search_example} and \ref{fig:search_example2}) 
show the planet search results 
and the plot on the right shows the completeness limits (analogous to Fig.\ \ref{fig:completeness_example}). 
See the captions of those figures for detailed descriptions.  
This star is in a hierarchical triple system and shows a strong linear trend and significant curvature or a possibly closed orbit in the RV time series.
}
\label{fig:completeness_161797}
\end{figure}
\clearpage

\begin{figure}
\begin{centering}
\includegraphics[width=0.45\textwidth]{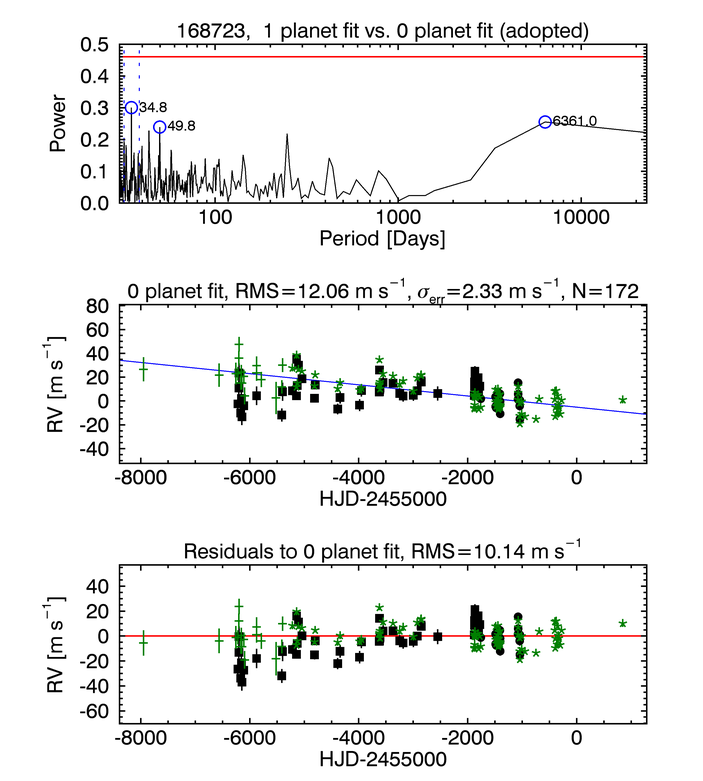}
\includegraphics[width=0.50\textwidth]{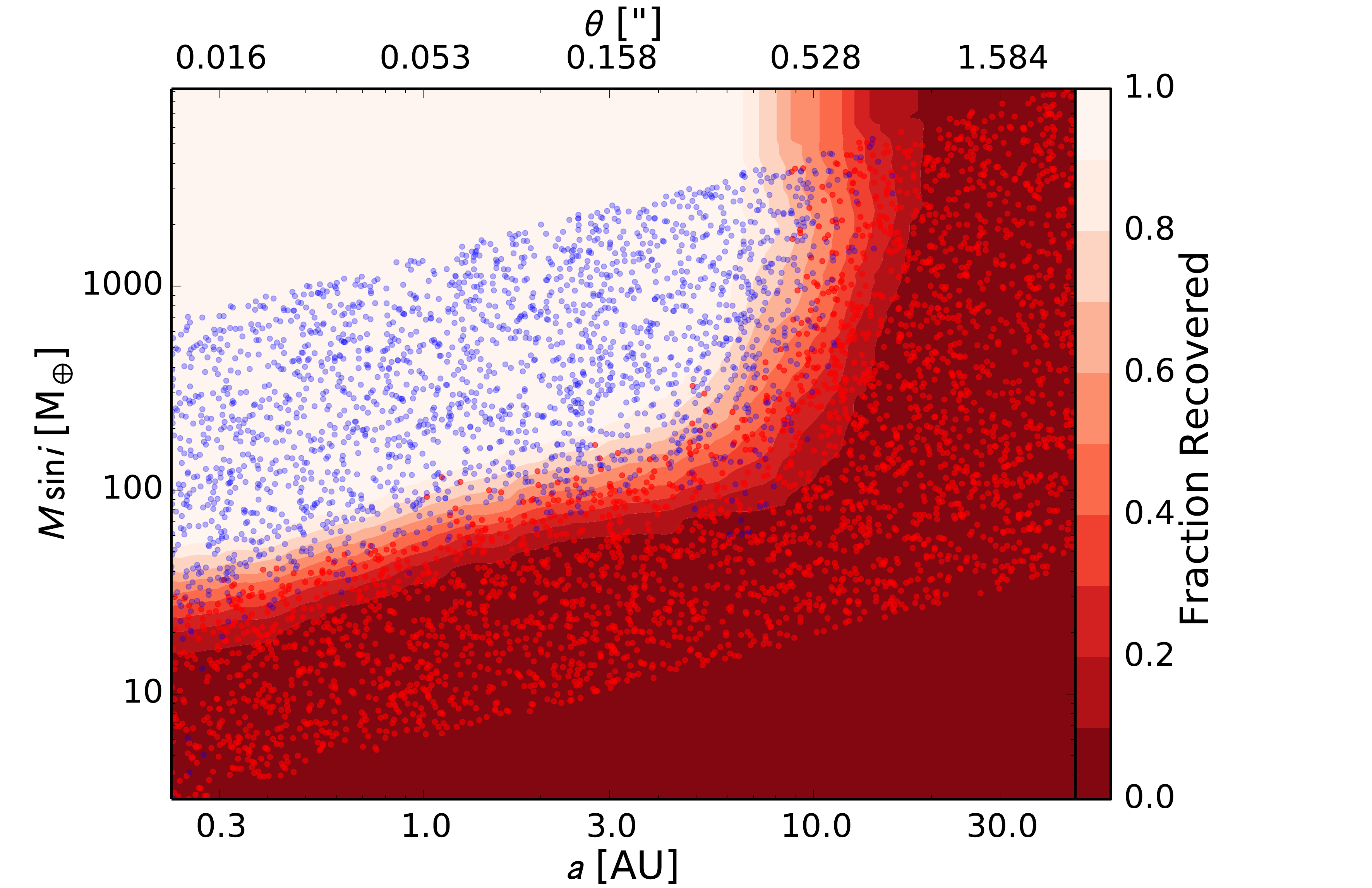}
\end{centering}
\caption{Results from an automated search for planets orbiting the star 
HD~168723 (HIP~89962; program = C) 
based on RVs from Lick and/or Keck Observatory.
The set of plots on the left (analogous to Figures \ref{fig:search_example} and \ref{fig:search_example2}) 
show the planet search results 
and the plot on the right shows the completeness limits (analogous to Fig.\ \ref{fig:completeness_example}). 
See the captions of those figures for detailed descriptions.  
The Lick data for this star show a marginal linear trend in the RV time series while the long-baseline pre-upgrade Keck data (code = k) does not show the trend.
}
\label{fig:completeness_168723}
\end{figure}

\begin{figure}
\begin{centering}
\includegraphics[width=0.45\textwidth]{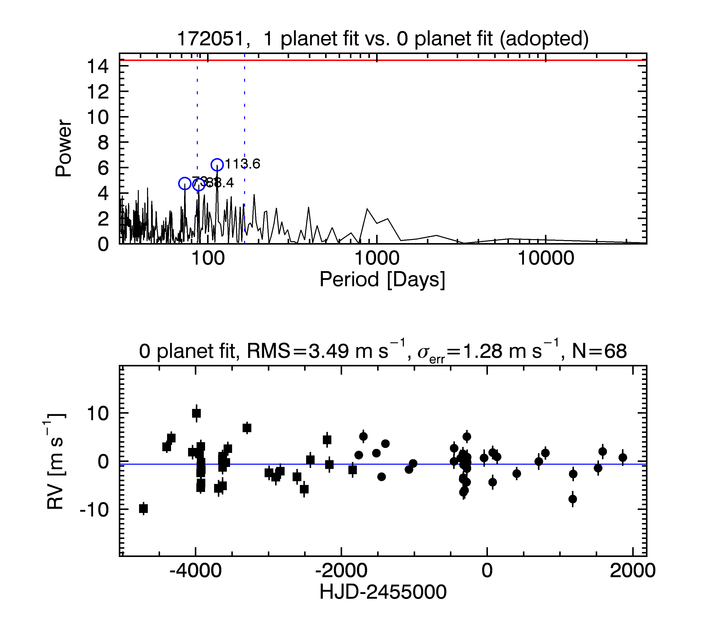}
\includegraphics[width=0.50\textwidth]{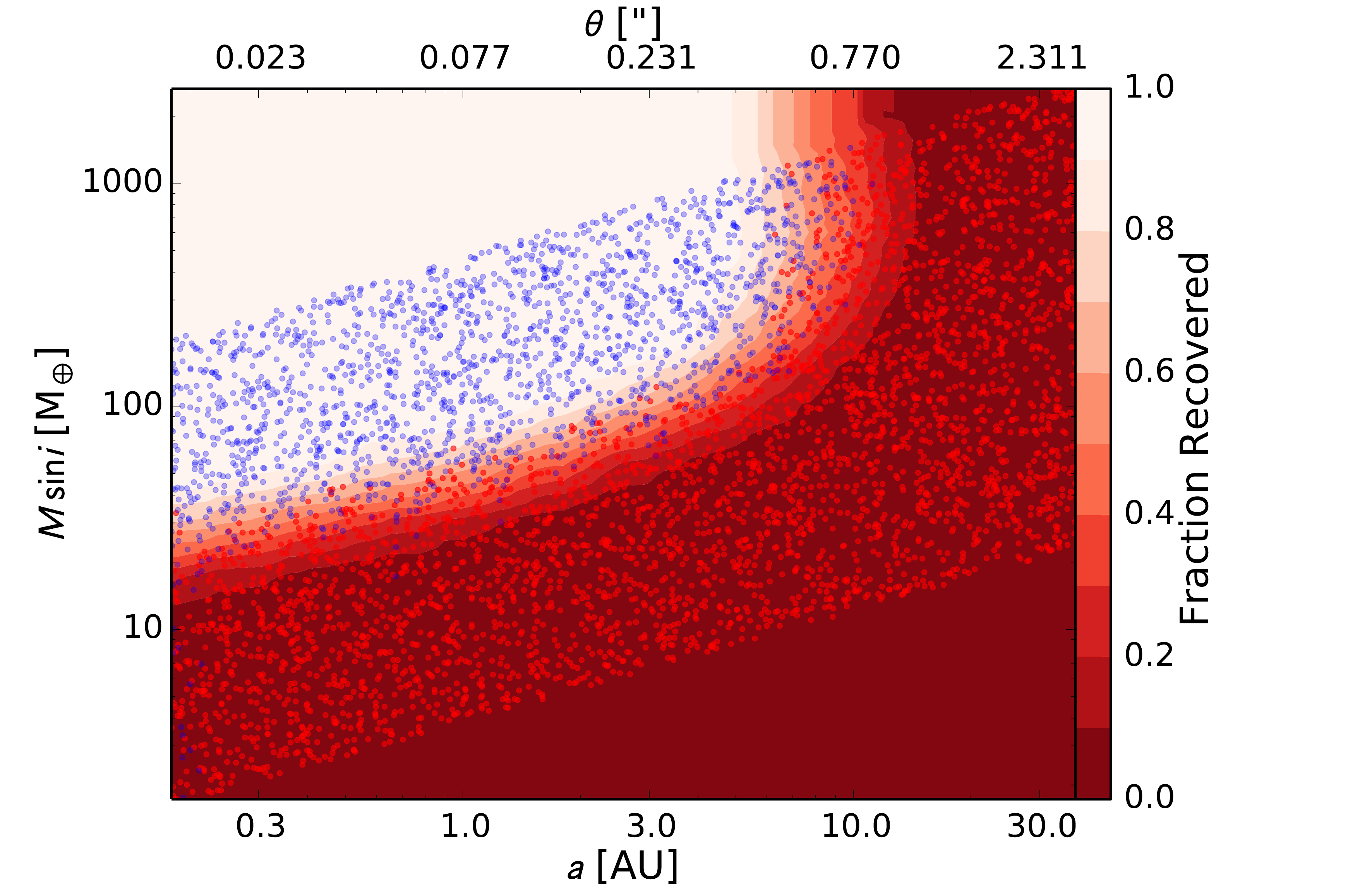}
\end{centering}
\caption{Results from an automated search for planets orbiting the star 
HD~172051 (HIP~91438; program = S) 
based on RVs from Lick and/or Keck Observatory.
The set of plots on the left (analogous to Figures \ref{fig:search_example} and \ref{fig:search_example2}) 
show the planet search results 
and the plot on the right shows the completeness limits (analogous to Fig.\ \ref{fig:completeness_example}). 
See the captions of those figures for detailed descriptions.  
}
\label{fig:completeness_172051}
\end{figure}
\clearpage

\begin{figure}
\begin{centering}
\includegraphics[width=0.45\textwidth]{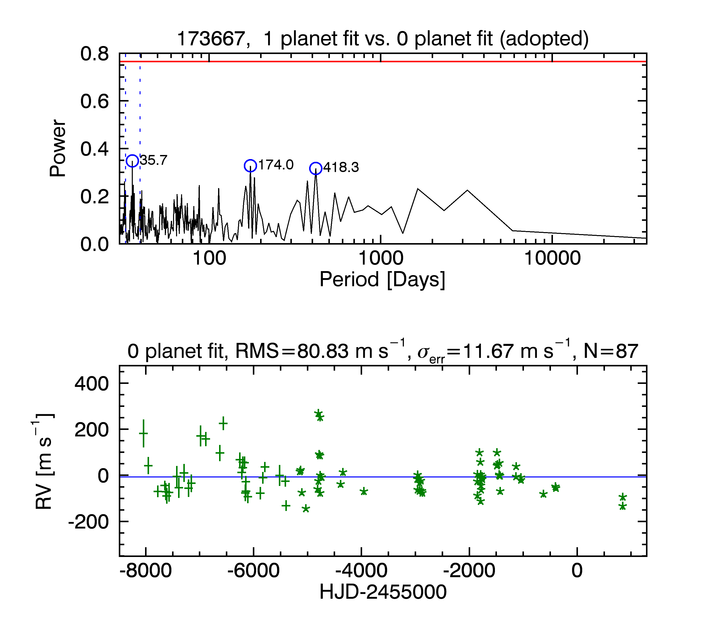}
\includegraphics[width=0.50\textwidth]{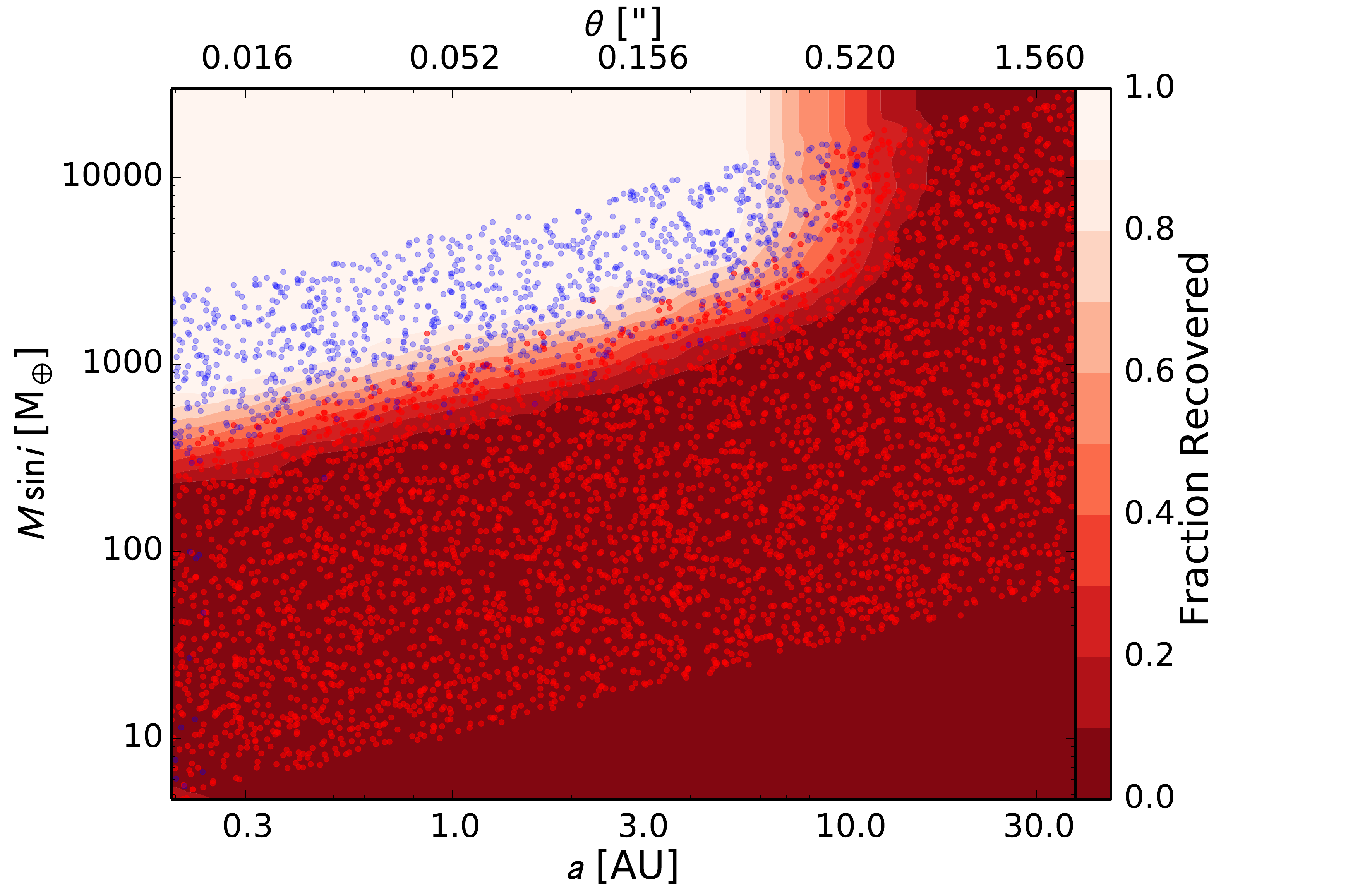}
\end{centering}
\caption{Results from an automated search for planets orbiting the star 
HD~173667 (HIP~92043; programs = C, A) 
based on RVs from Lick and/or Keck Observatory.
The set of plots on the left (analogous to Figures \ref{fig:search_example} and \ref{fig:search_example2}) 
show the planet search results 
and the plot on the right shows the completeness limits (analogous to Fig.\ \ref{fig:completeness_example}). 
See the captions of those figures for detailed descriptions.  
}
\label{fig:completeness_173667}
\end{figure}

\begin{figure}
\begin{centering}
\includegraphics[width=0.45\textwidth]{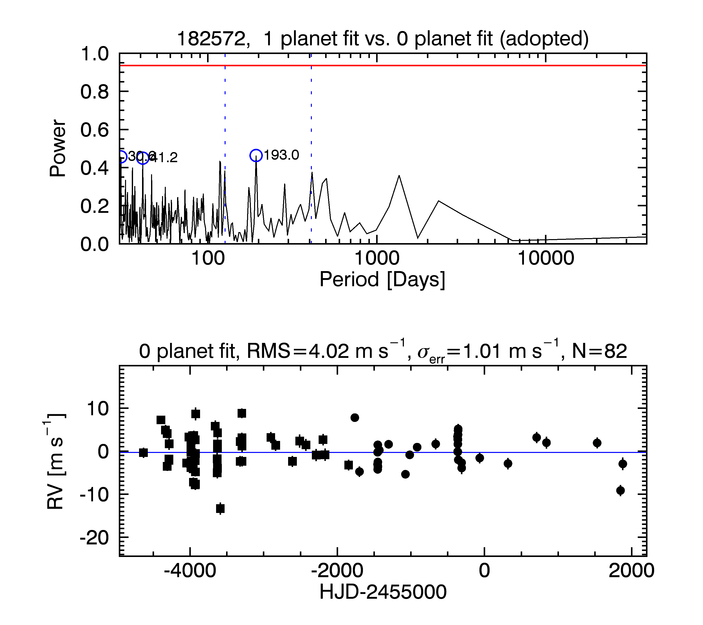}
\includegraphics[width=0.50\textwidth]{182572-recovery.pdf}
\end{centering}
\caption{Results from an automated search for planets orbiting the star 
HD~182572 (HIP~95447; programs = S, A) 
based on RVs from Lick and/or Keck Observatory.
The set of plots on the left (analogous to Figures \ref{fig:search_example} and \ref{fig:search_example2}) 
show the planet search results 
and the plot on the right shows the completeness limits (analogous to Fig.\ \ref{fig:completeness_example}). 
See the captions of those figures for detailed descriptions.  
}
\label{fig:completeness_182572}
\end{figure}
\clearpage

\begin{figure}
\begin{centering}
\includegraphics[width=0.45\textwidth]{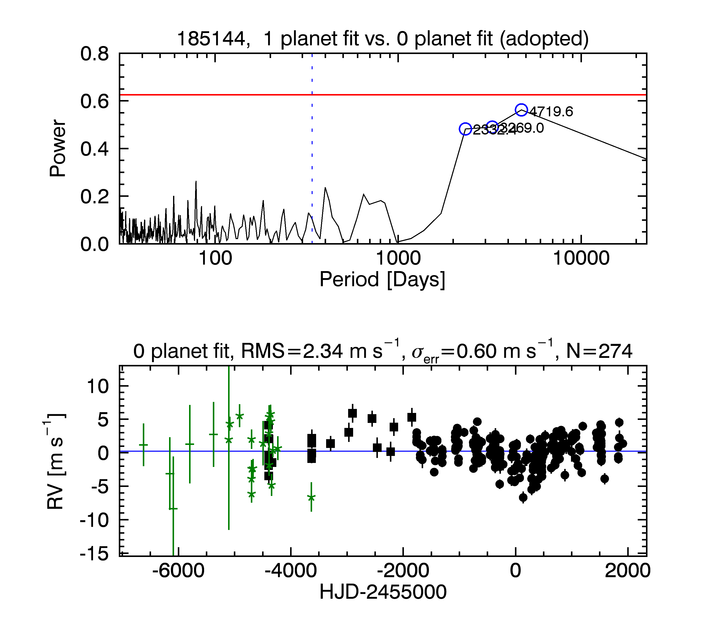}
\includegraphics[width=0.50\textwidth]{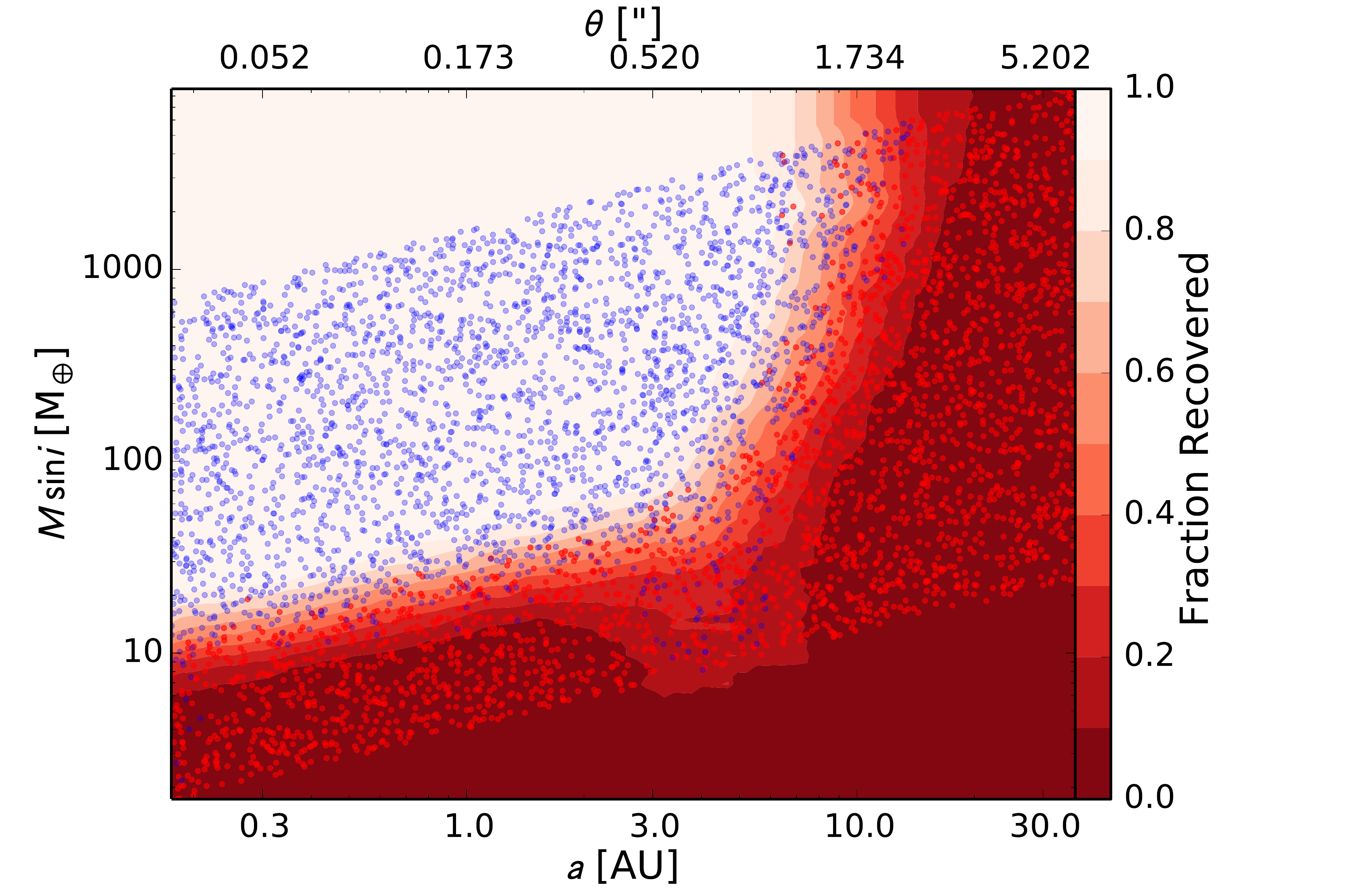}
\end{centering}
\caption{Results from an automated search for planets orbiting the star 
HD~185144 (HIP~96100; programs = S, C, A) 
based on RVs from Lick and/or Keck Observatory.
The set of plots on the left (analogous to Figures \ref{fig:search_example} and \ref{fig:search_example2}) 
show the planet search results 
and the plot on the right shows the completeness limits (analogous to Fig.\ \ref{fig:completeness_example}). 
See the captions of those figures for detailed descriptions.  
}
\label{fig:completeness_185144}
\end{figure}

\begin{figure}
\begin{centering}
\includegraphics[width=0.45\textwidth]{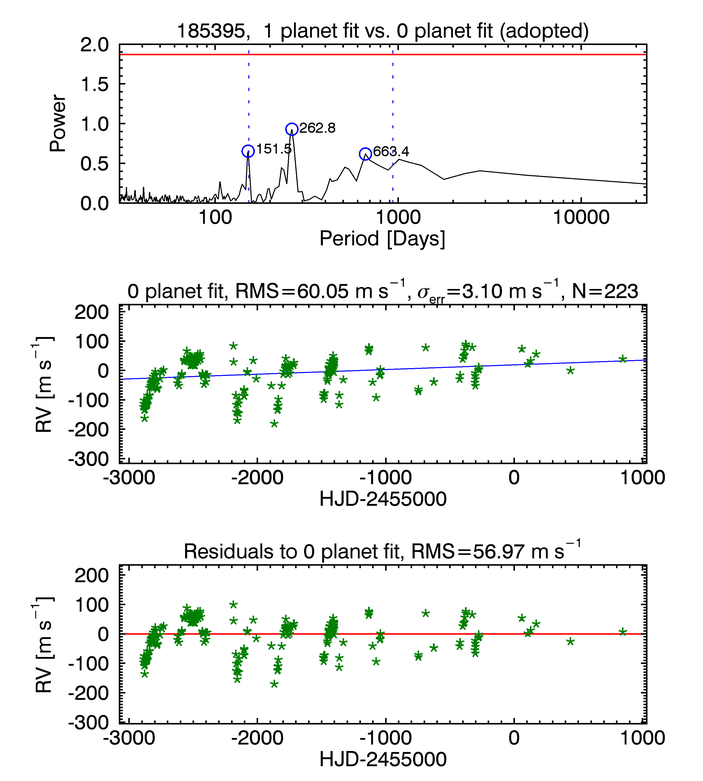}
\includegraphics[width=0.50\textwidth]{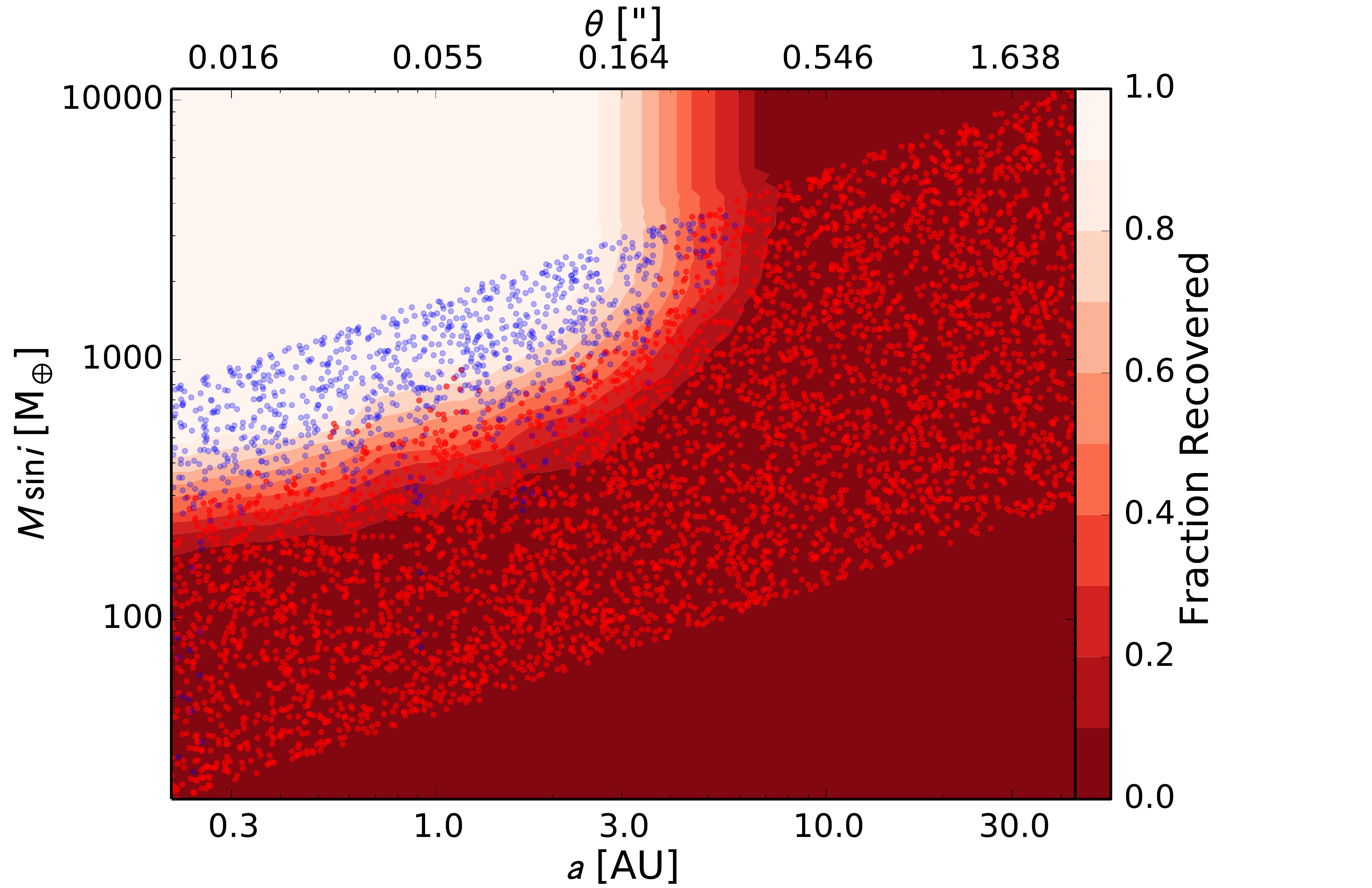}
\end{centering}
\caption{Results from an automated search for planets orbiting the star 
HD~185395 (HIP~96441; programs = C, A) 
based on RVs from Lick and/or Keck Observatory.
The set of plots on the left (analogous to Figures \ref{fig:search_example} and \ref{fig:search_example2}) 
show the planet search results 
and the plot on the right shows the completeness limits (analogous to Fig.\ \ref{fig:completeness_example}). 
See the captions of those figures for detailed descriptions.  
This early-type star (F4 V) has high jitter and a claimed, controversial planet with $P \approx 150$~days.  We see evidence in our Lick data for RV variation at this period and other periods related by the yearly alias.  Our pipeline formally adopts a linear trend model for the RVs.
}
\label{fig:completeness_185395}
\end{figure}
\clearpage

\begin{figure}
\begin{centering}
\includegraphics[width=0.45\textwidth]{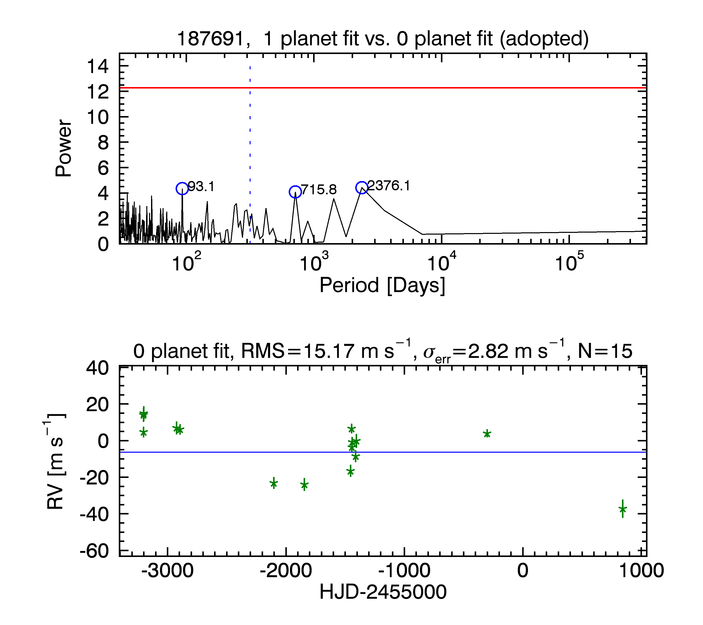}
\includegraphics[width=0.50\textwidth]{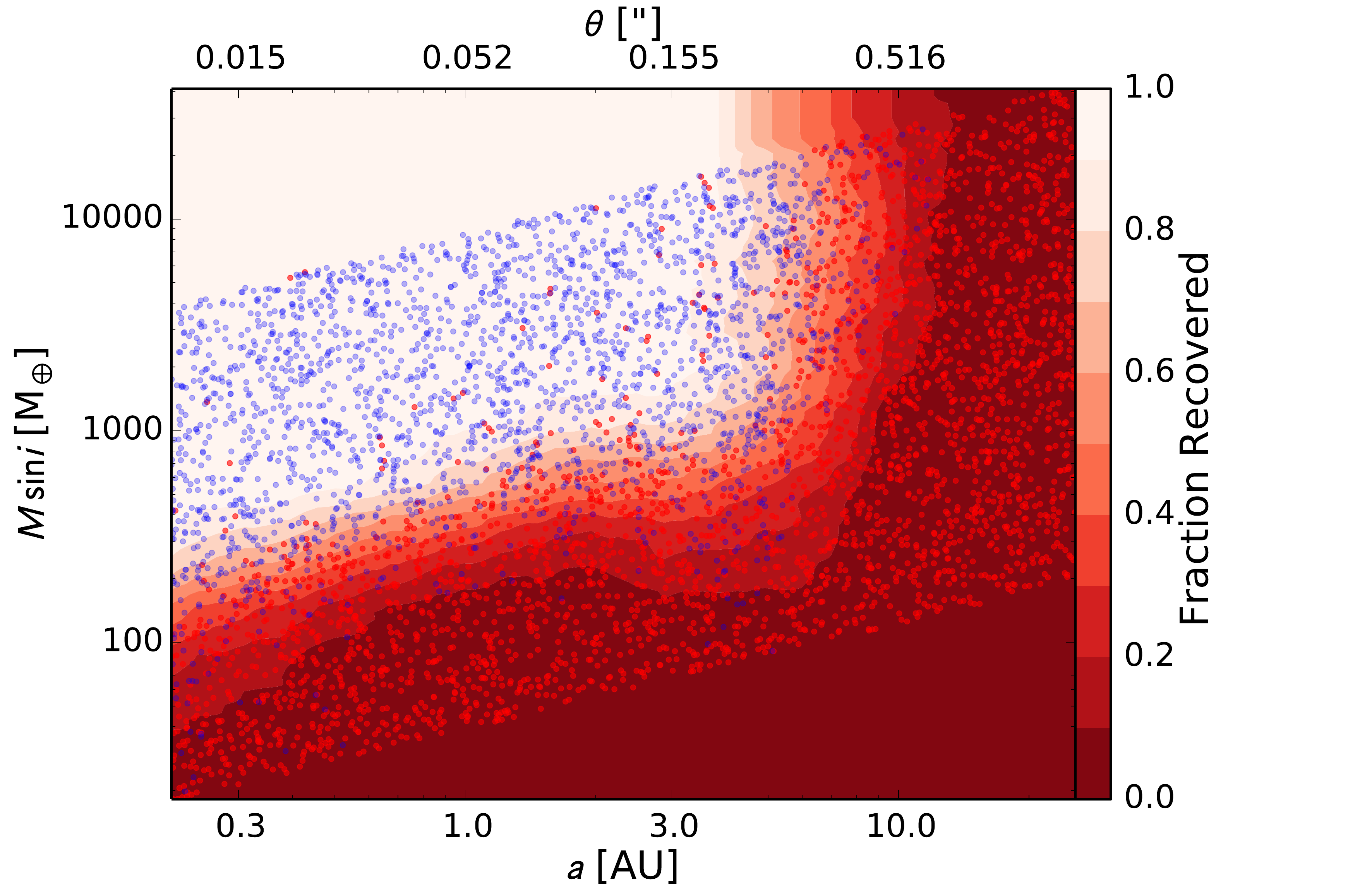}
\end{centering}
\caption{Results from an automated search for planets orbiting the star 
HD~187691 (HIP~97675; program = A) 
based on RVs from Lick and/or Keck Observatory.
The set of plots on the left (analogous to Figures \ref{fig:search_example} and \ref{fig:search_example2}) 
show the planet search results 
and the plot on the right shows the completeness limits (analogous to Fig.\ \ref{fig:completeness_example}). 
See the captions of those figures for detailed descriptions.  
}
\label{fig:completeness_187691}
\end{figure}

\begin{figure}
\begin{centering}
\includegraphics[width=0.45\textwidth]{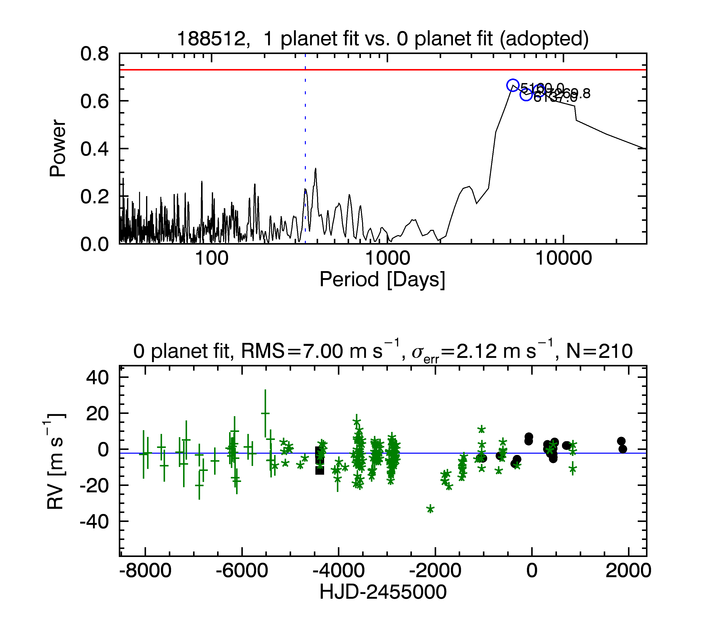}
\includegraphics[width=0.50\textwidth]{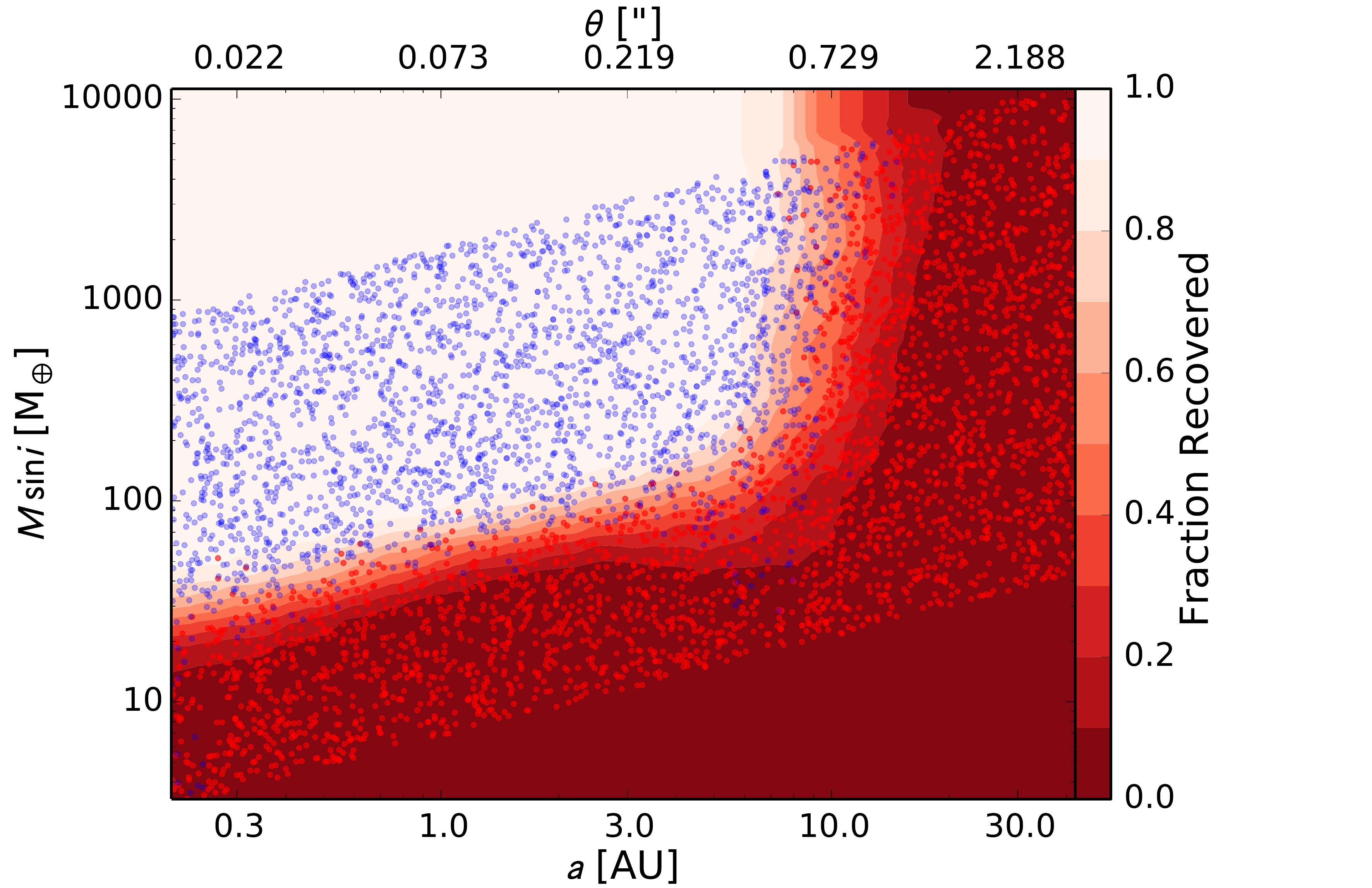}
\end{centering}
\caption{Results from an automated search for planets orbiting the star 
HD~188512 (HIP~98036; programs = S, C) 
based on RVs from Lick and/or Keck Observatory.
The set of plots on the left (analogous to Figures \ref{fig:search_example} and \ref{fig:search_example2}) 
show the planet search results 
and the plot on the right shows the completeness limits (analogous to Fig.\ \ref{fig:completeness_example}). 
See the captions of those figures for detailed descriptions.  
}
\label{fig:completeness_188512}
\end{figure}
\clearpage

\begin{figure}
\begin{centering}
\includegraphics[width=0.45\textwidth]{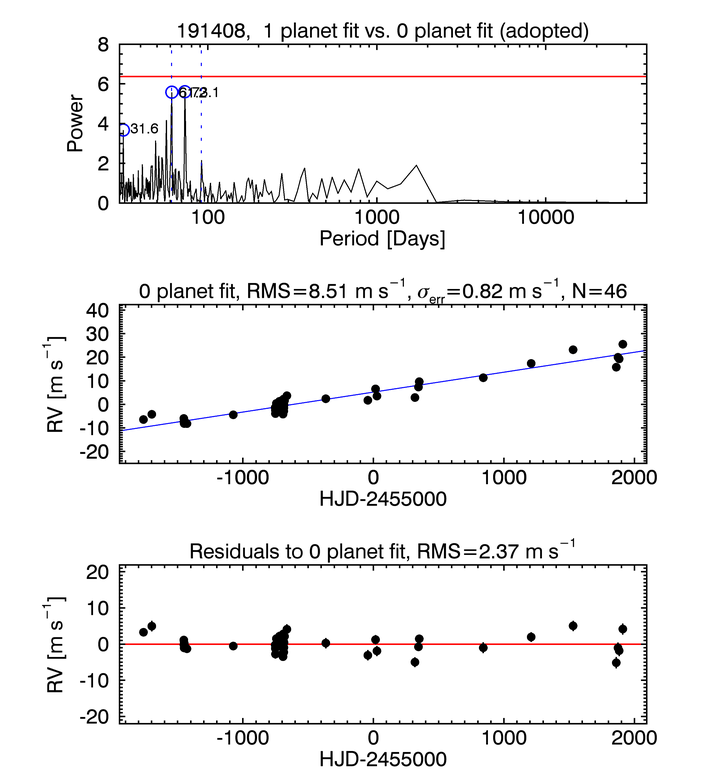}
\includegraphics[width=0.50\textwidth]{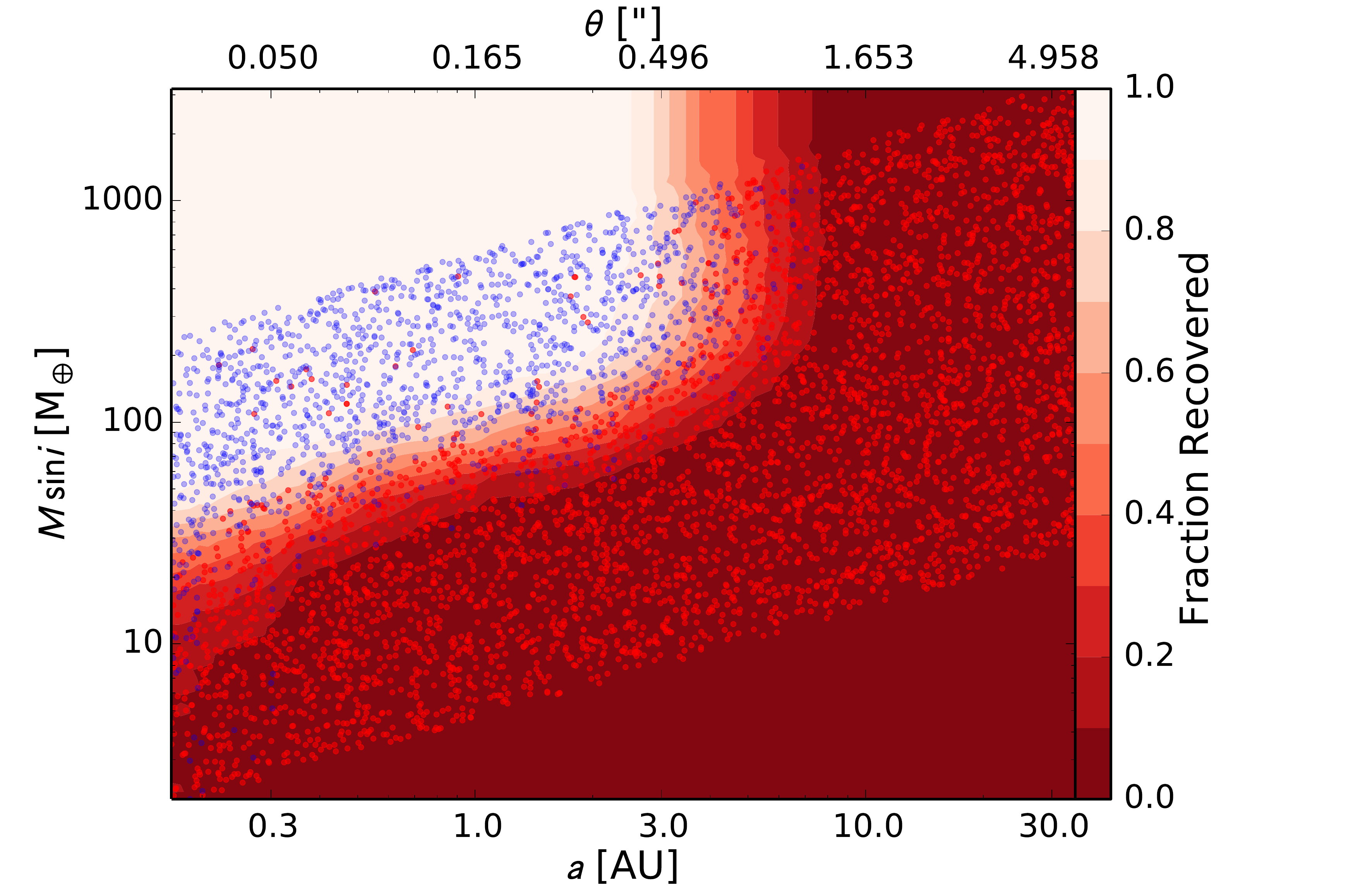}
\end{centering}
\caption{Results from an automated search for planets orbiting the star 
HD~191408 (HIP~99461; programs = S, C, A) 
based on RVs from Lick and/or Keck Observatory.
The set of plots on the left (analogous to Figures \ref{fig:search_example} and \ref{fig:search_example2}) 
show the planet search results 
and the plot on the right shows the completeness limits (analogous to Fig.\ \ref{fig:completeness_example}). 
See the captions of those figures for detailed descriptions.  
This star shows a slight linear trend with no detectable curvature, presumably due to its common proper motion companion.
}
\label{fig:completeness_191408}
\end{figure}

\begin{figure}
\begin{centering}
\includegraphics[width=0.45\textwidth]{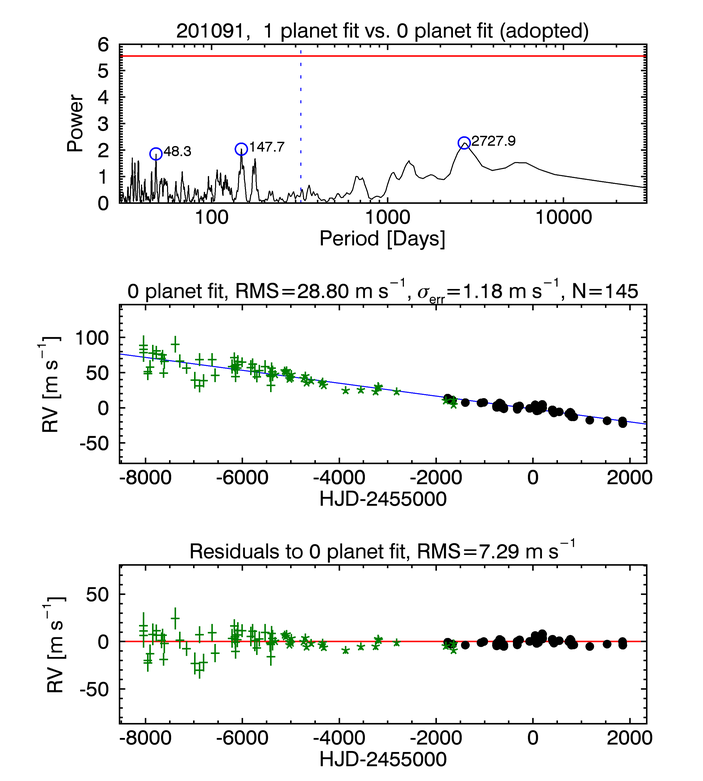}
\includegraphics[width=0.50\textwidth]{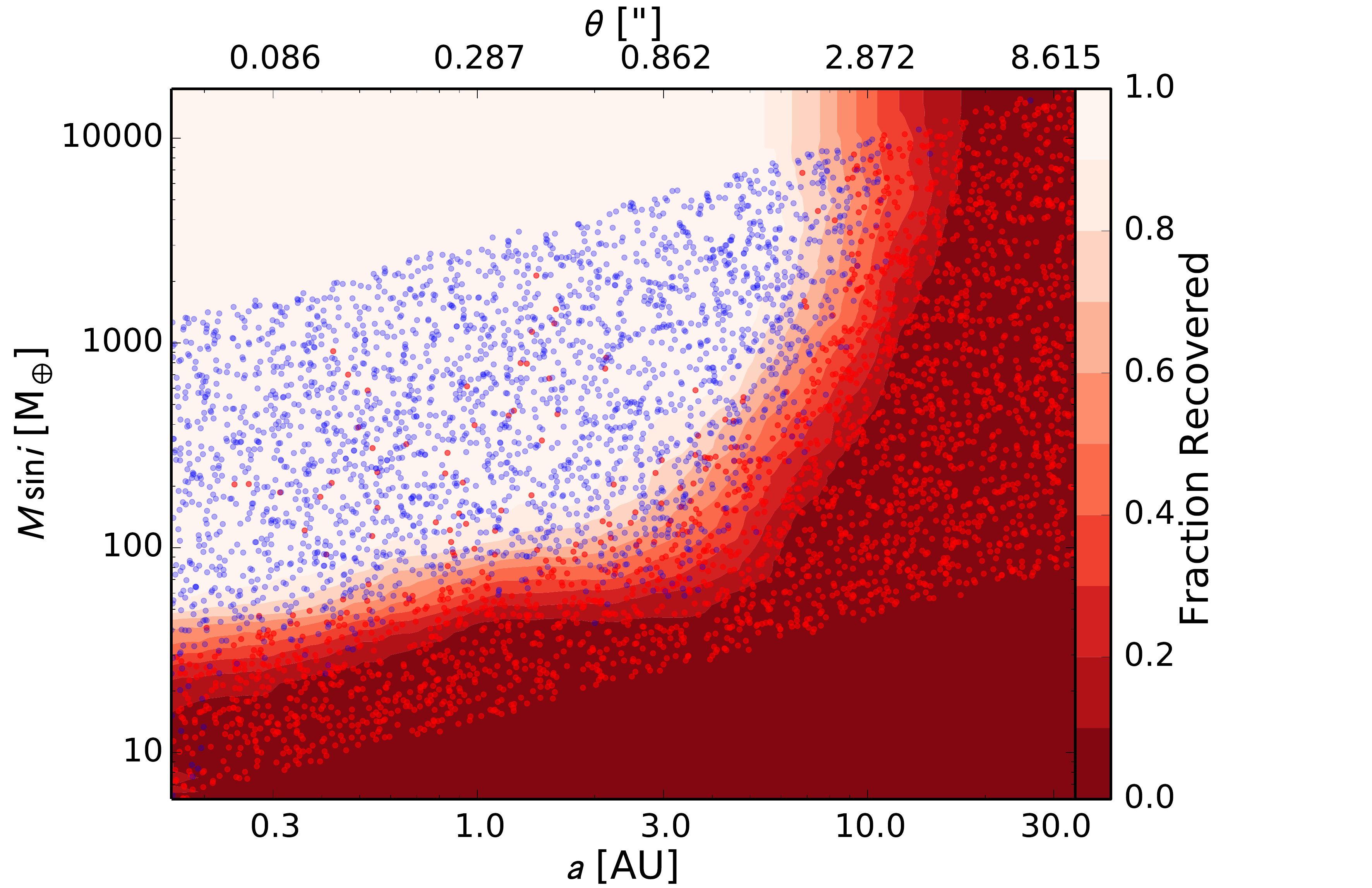}
\end{centering}
\caption{Results from an automated search for planets orbiting the star 
HD~201091 (HIP~104214; program = S, C, A) 
based on RVs from Lick and/or Keck Observatory.
The set of plots on the left (analogous to Figures \ref{fig:search_example} and \ref{fig:search_example2}) 
show the planet search results 
and the plot on the right shows the completeness limits (analogous to Fig.\ \ref{fig:completeness_example}). 
See the captions of those figures for detailed descriptions.  
This star shows a significant linear trend with no detectable curvature, presumably due to its known stellar companion.
}
\label{fig:completeness_201091}
\end{figure}
\clearpage

\begin{figure}
\begin{centering}
\includegraphics[width=0.45\textwidth]{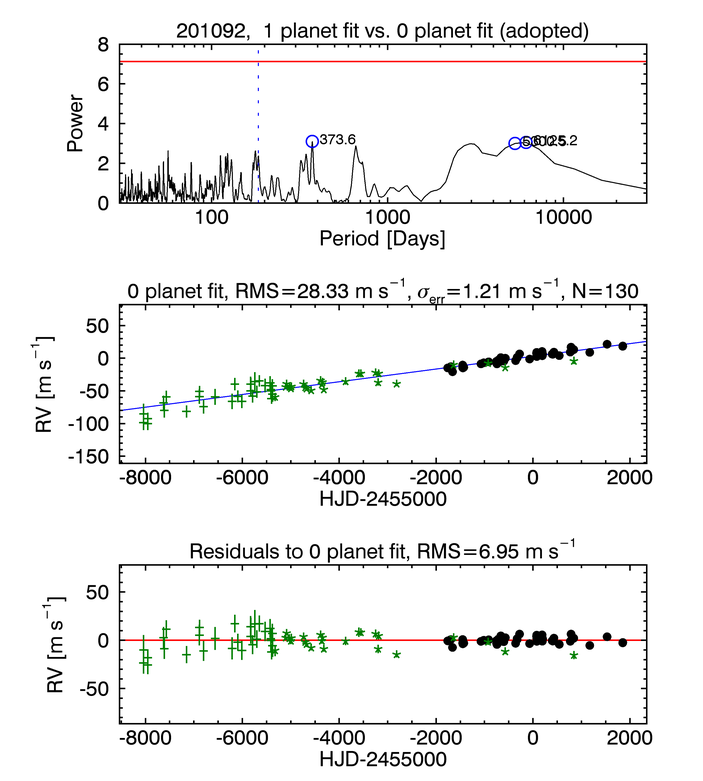}
\includegraphics[width=0.50\textwidth]{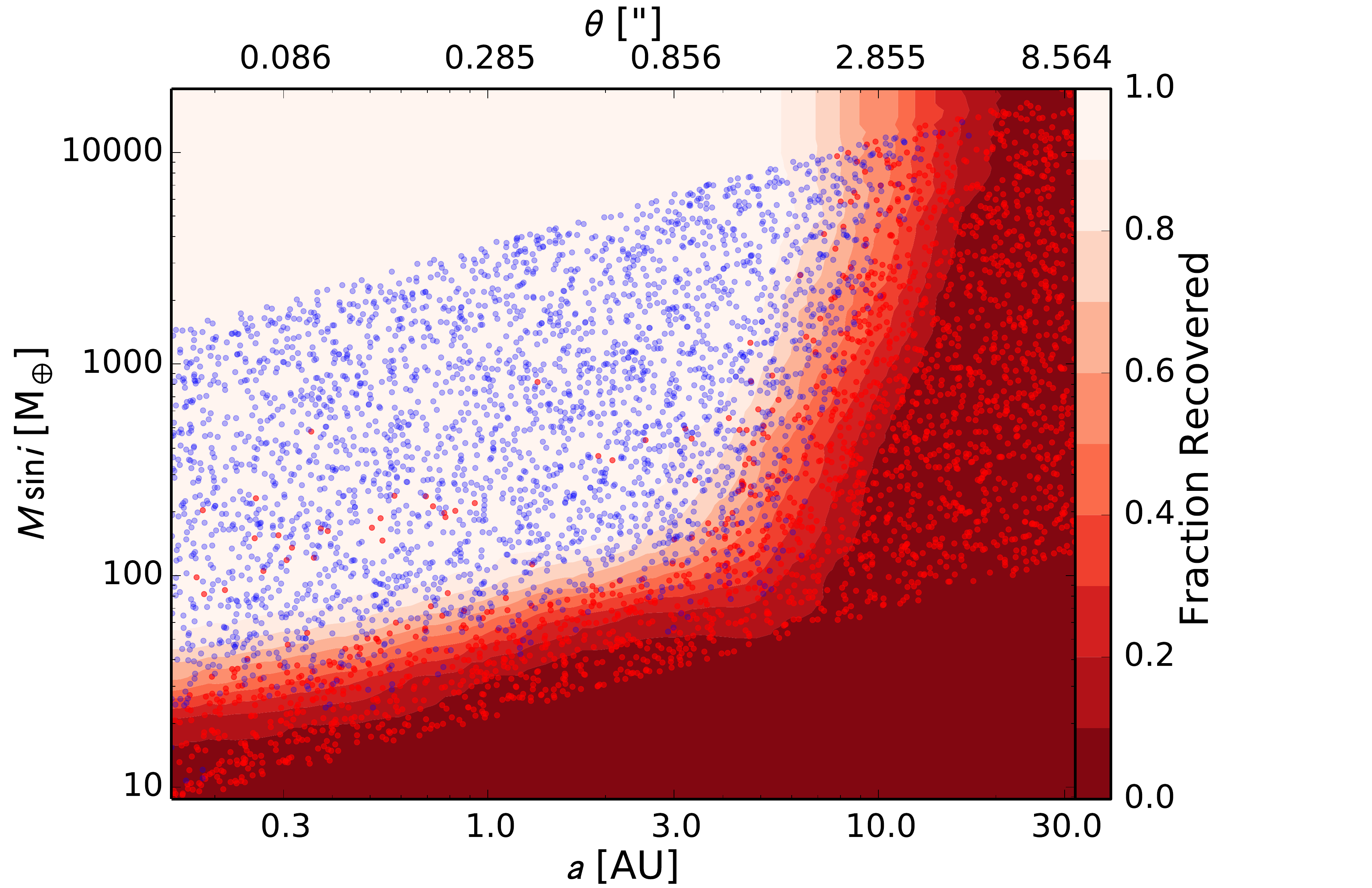}
\end{centering}
\caption{Results from an automated search for planets orbiting the star 
HD~201092 (HIP~104217; program = S) 
based on RVs from Lick and/or Keck Observatory.
The set of plots on the left (analogous to Figures \ref{fig:search_example} and \ref{fig:search_example2}) 
show the planet search results 
and the plot on the right shows the completeness limits (analogous to Fig.\ \ref{fig:completeness_example}). 
See the captions of those figures for detailed descriptions.  
This star shows a significant linear trend with no detectable curvature, presumably due to its known stellar companion.
}
\label{fig:completeness_201092}
\end{figure}

\begin{figure}
\begin{centering}
\includegraphics[width=0.45\textwidth]{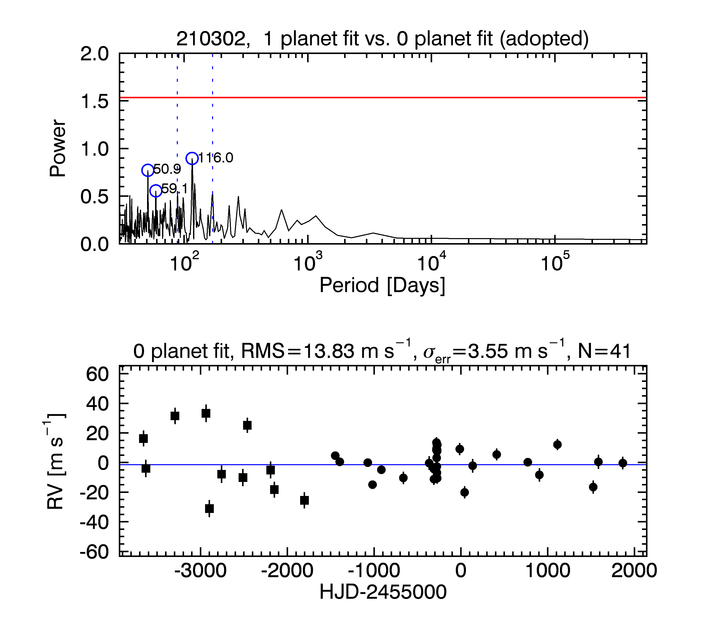}
\includegraphics[width=0.50\textwidth]{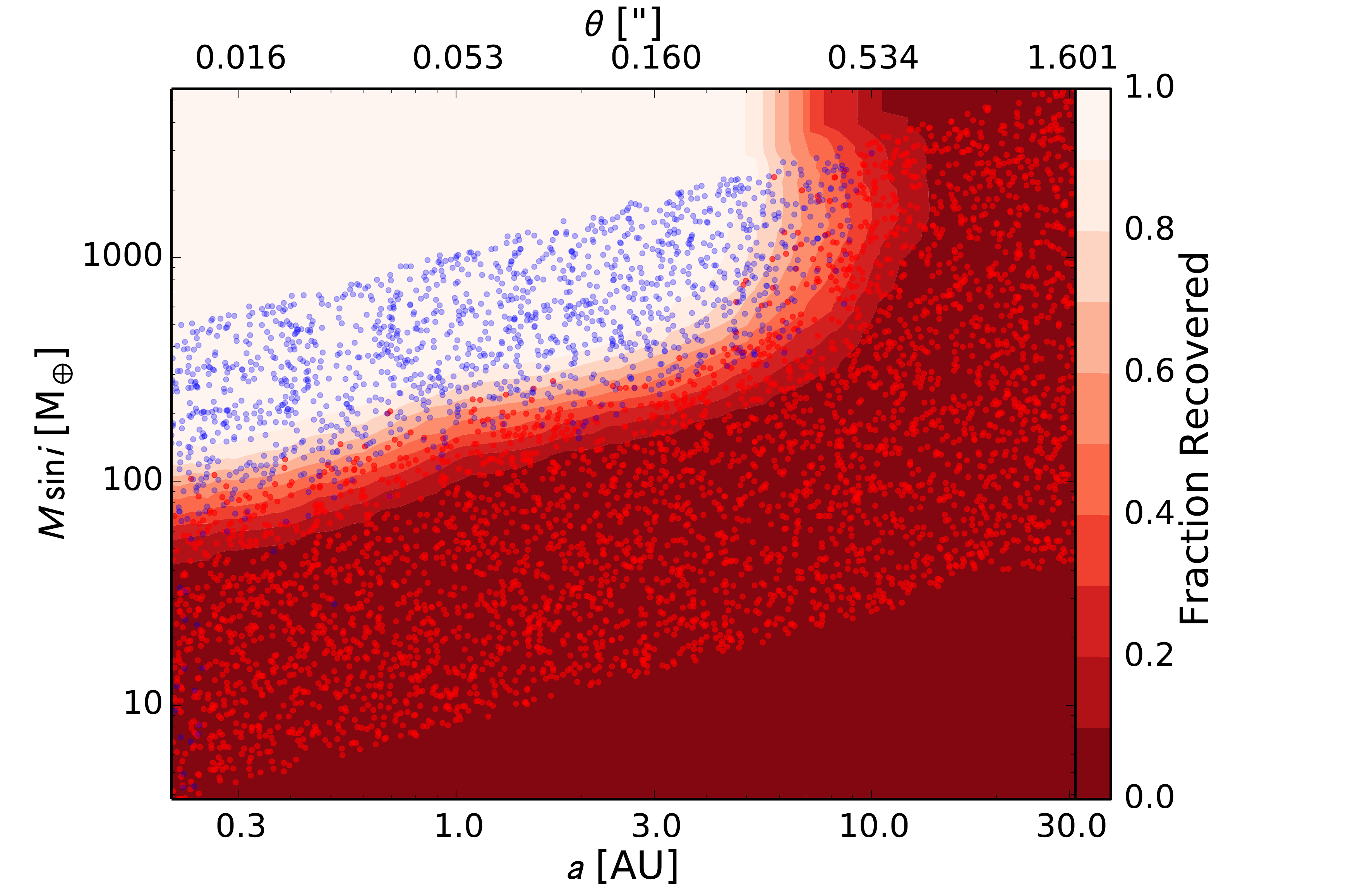}
\end{centering}
\caption{Results from an automated search for planets orbiting the star 
HD~210302 (HIP~109422; program = A) 
based on RVs from Lick and/or Keck Observatory.
The set of plots on the left (analogous to Figures \ref{fig:search_example} and \ref{fig:search_example2}) 
show the planet search results 
and the plot on the right shows the completeness limits (analogous to Fig.\ \ref{fig:completeness_example}). 
See the captions of those figures for detailed descriptions.  
}
\label{fig:completeness_210302}
\end{figure}
\clearpage

\begin{figure}
\begin{centering}
\includegraphics[width=0.45\textwidth]{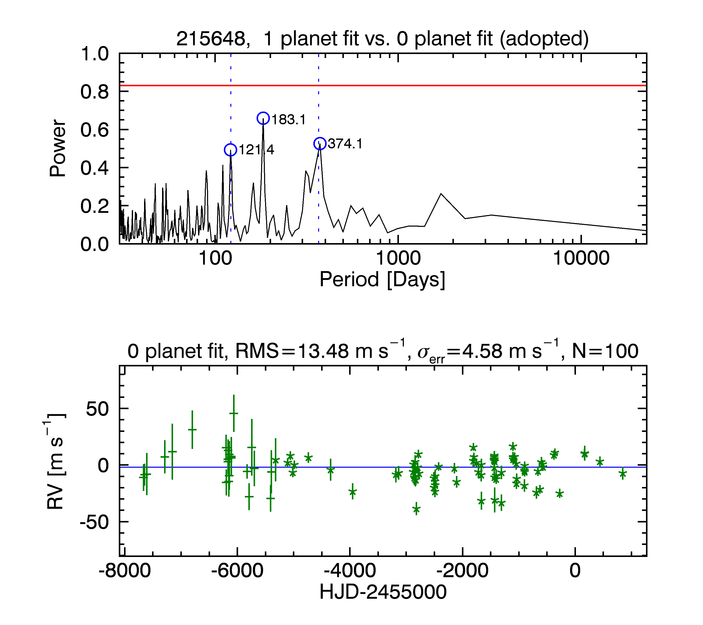}
\includegraphics[width=0.50\textwidth]{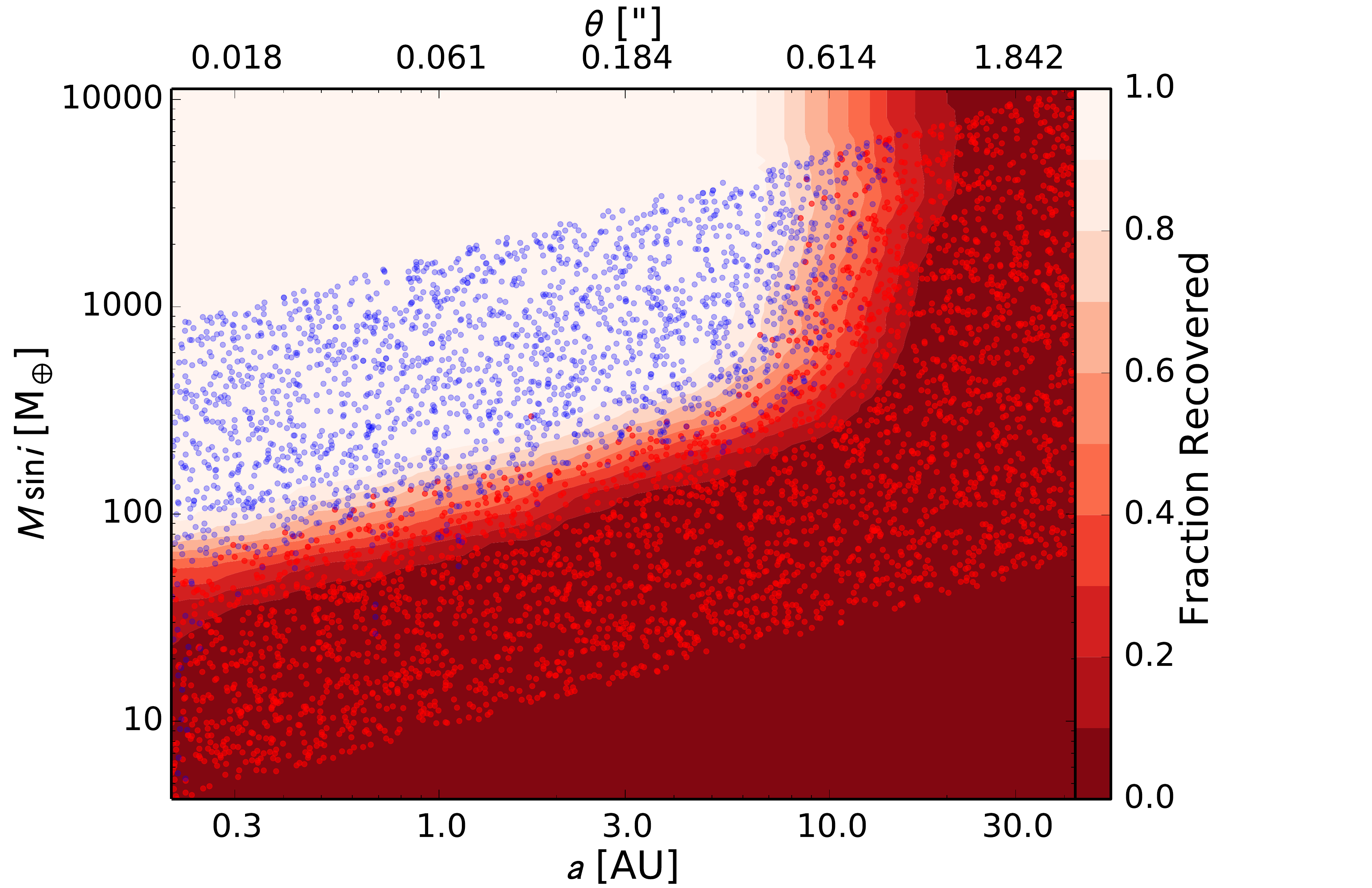}
\end{centering}
\caption{Results from an automated search for planets orbiting the star 
HD~215648 (HIP~112447; programs = S, C, A) 
based on RVs from Lick and/or Keck Observatory.
The set of plots on the left (analogous to Figures \ref{fig:search_example} and \ref{fig:search_example2}) 
show the planet search results 
and the plot on the right shows the completeness limits (analogous to Fig.\ \ref{fig:completeness_example}). 
See the captions of those figures for detailed descriptions.  
}
\label{fig:completeness_215648}
\end{figure}

\begin{figure}
\begin{centering}
\includegraphics[width=0.45\textwidth]{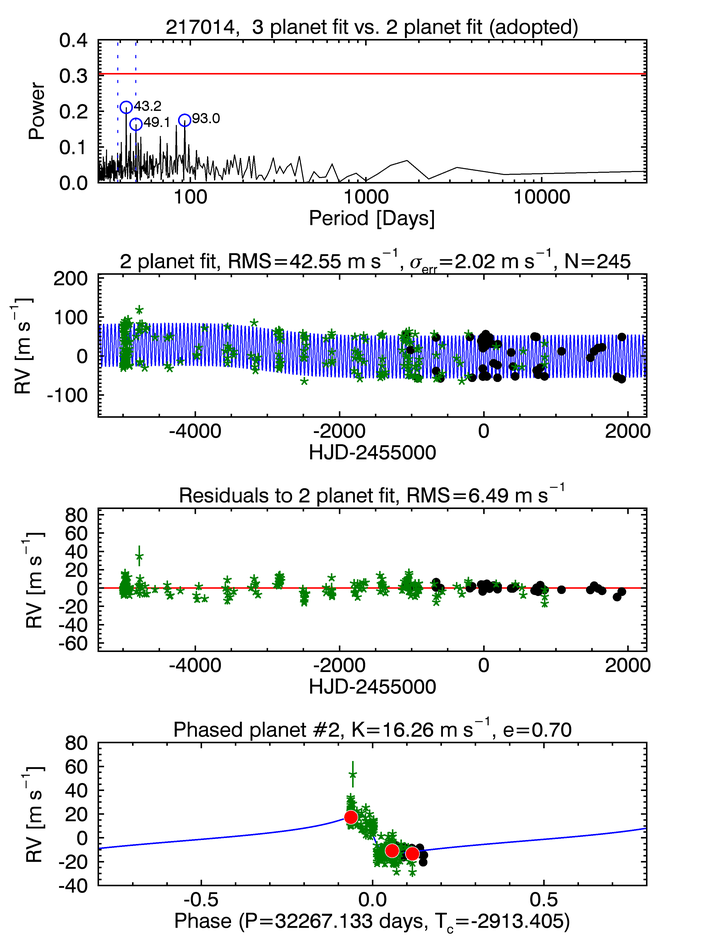}
\includegraphics[width=0.50\textwidth]{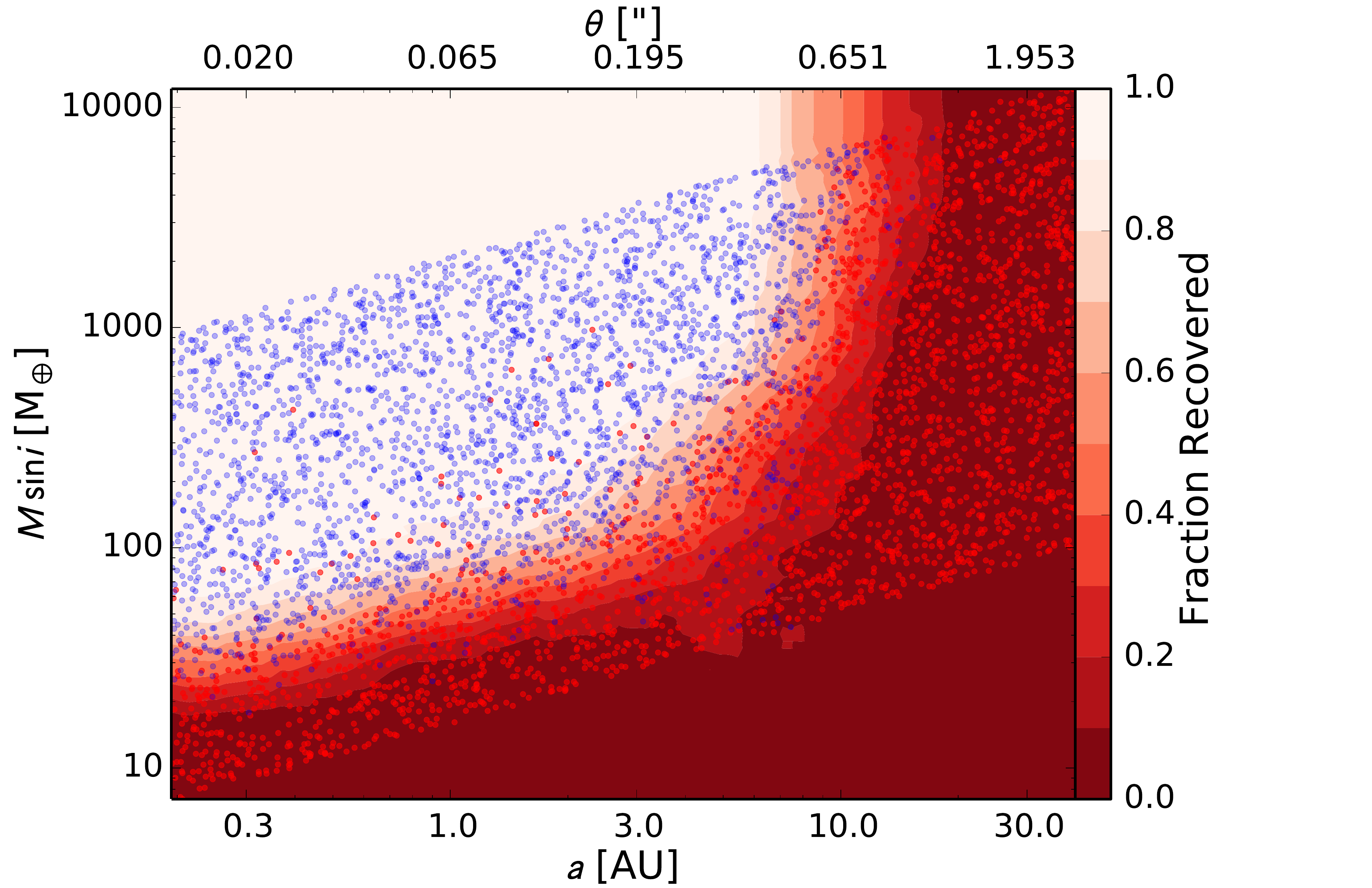}
\end{centering}
\caption{Results from an automated search for planets orbiting the star 
HD~217014 (HIP~113357; program = S) 
based on RVs from Lick and/or Keck Observatory.
The set of plots on the left (analogous to Figures \ref{fig:search_example} and \ref{fig:search_example2}) 
show the planet search results 
and the plot on the right shows the completeness limits (analogous to Fig.\ \ref{fig:completeness_example}). 
See the captions of those figures for detailed descriptions.  
This star hosts a hot Jupiter.  We also detect a long-period signal but this is likely caused by instrumental offsets within the Lick dataset.
}
\label{fig:completeness_217014}
\end{figure}
\clearpage

\begin{figure}
\begin{centering}
\includegraphics[width=0.45\textwidth]{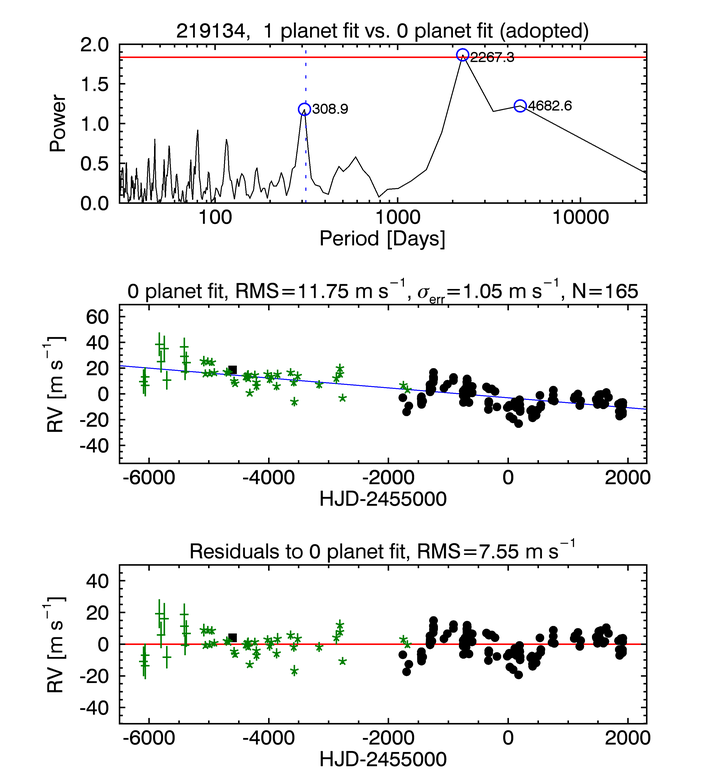}
\includegraphics[width=0.50\textwidth]{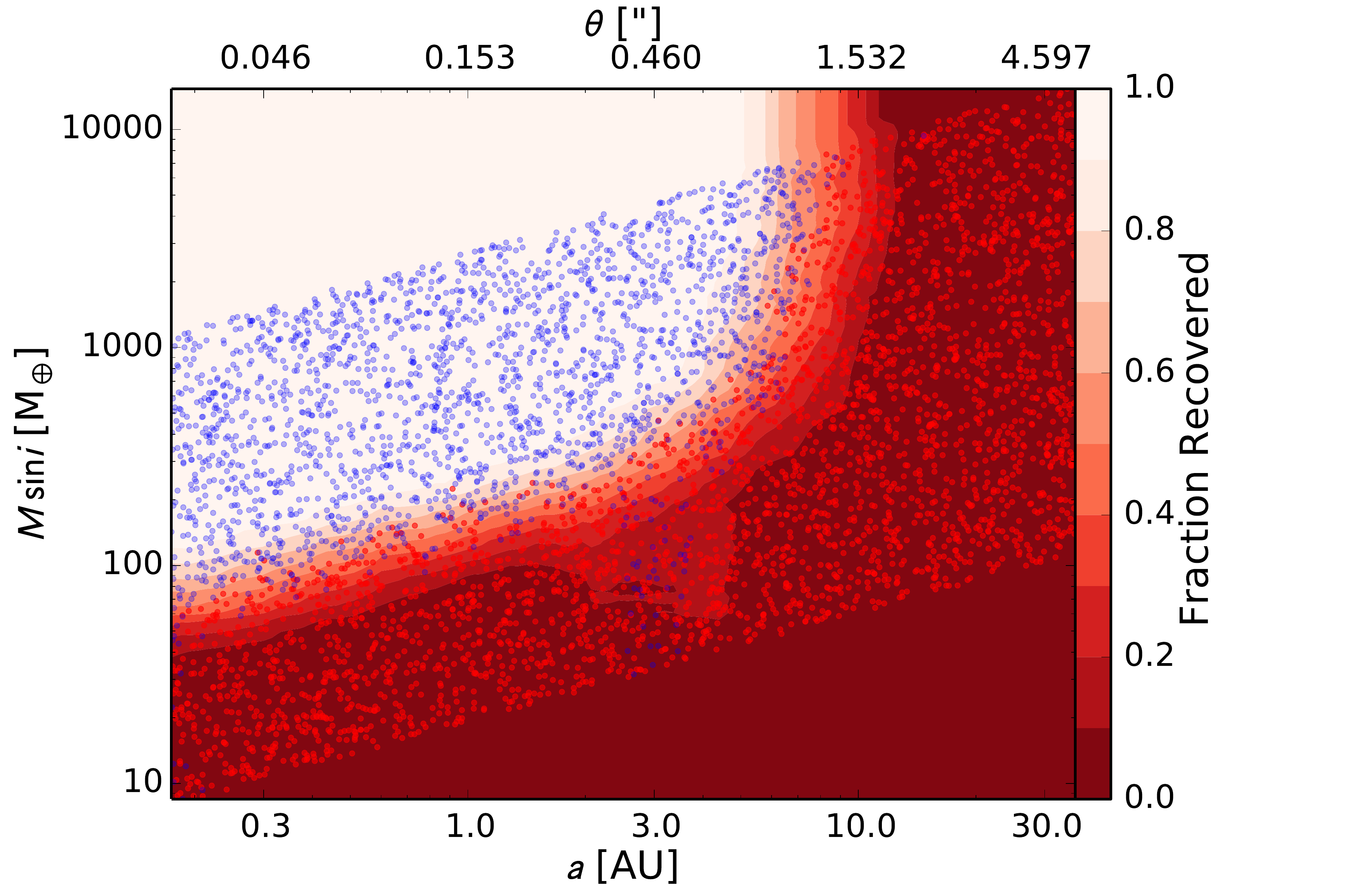}
\end{centering}
\caption{Results from an automated search for planets orbiting the star 
HD~219134 (HIP~114622; programs = S, C) 
based on RVs from Lick and/or Keck Observatory.
The set of plots on the left (analogous to Figures \ref{fig:search_example} and \ref{fig:search_example2}) 
show the planet search results 
and the plot on the right shows the completeness limits (analogous to Fig.\ \ref{fig:completeness_example}). 
See the captions of those figures for detailed descriptions.  
This star has candidate planets including a giant planet in a 3 AU orbit that we will continue to examine as more RVs are gathered.  Our automated pipeline formally prefers a model with a linear trend in the RV time series.
}
\label{fig:completeness_219134}
\end{figure}

\begin{figure}
\begin{centering}
\includegraphics[width=0.45\textwidth]{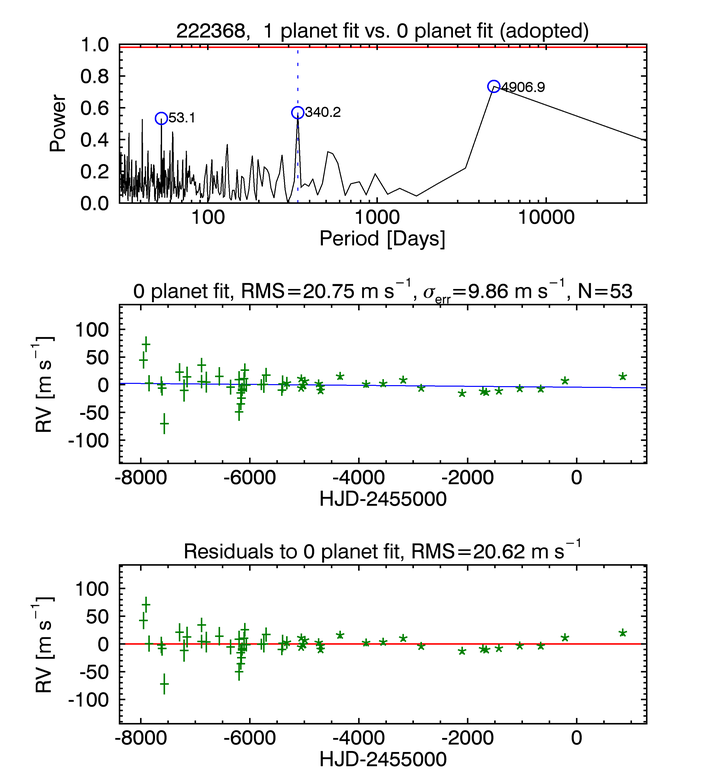}
\includegraphics[width=0.50\textwidth]{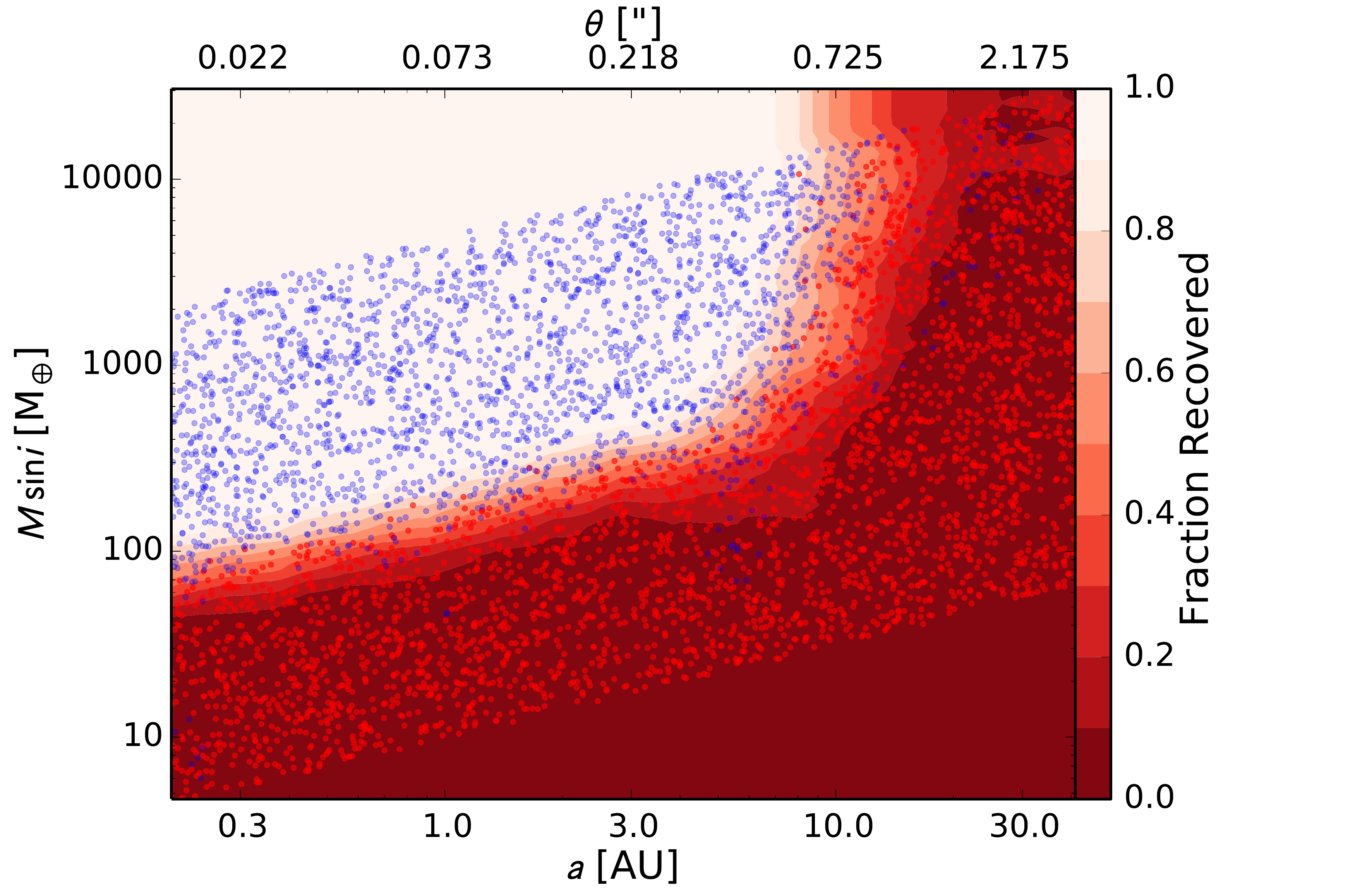}
\end{centering}
\caption{Results from an automated search for planets orbiting the star 
HD~222368 (HIP~116771; programs = S, C, A) 
based on RVs from Lick and/or Keck Observatory.
The set of plots on the left (analogous to Figures \ref{fig:search_example} and \ref{fig:search_example2}) 
show the planet search results 
and the plot on the right shows the completeness limits (analogous to Fig.\ \ref{fig:completeness_example}). 
See the captions of those figures for detailed descriptions.  
}
\label{fig:completeness_222368}
\end{figure}

\end{document}